\documentclass[a4paper,12pt]{amsart}
\usepackage[hmarginratio={1:1},vmarginratio={1:1},heightrounded,textwidth=455pt]{geometry}
\usepackage{amsfonts}
\usepackage{amsmath}
\usepackage[utf8]{inputenc}
\usepackage{amssymb, colonequals}
\usepackage{algorithmicx}
\usepackage{algpseudocode}
\usepackage{mathtools}
\usepackage{algorithm}
\usepackage{tikz}
\usetikzlibrary{arrows}
\usetikzlibrary{patterns}
\usetikzlibrary{shapes,snakes}
\usetikzlibrary{arrows}
\usetikzlibrary{patterns}
\usetikzlibrary{shapes,snakes}
\usetikzlibrary{calc}

\usepackage[nocompress]{cite}

\usepackage[textsize=small]{todonotes}
\usepackage{url}
\usepackage{enumitem}
\usepackage{multirow}
\usepackage{chngcntr}
\counterwithin{figure}{section}
\setlist[itemize]{leftmargin=0.3in}
\setlist[enumerate]{leftmargin=0.3in}
\usepackage{subcaption}
\usetikzlibrary{shapes.geometric}
\usetikzlibrary{decorations.pathreplacing,calligraphy}
\usepackage{standalone}

\definecolor{fuchsia}{rgb}{1.0, 0.0, 1.0}
\definecolor{jazzberryjam}{rgb}{0.65, 0.04, 0.37}
\definecolor{brightturquoise}{rgb}{0.03, 0.91, 0.87}
\definecolor{brilliantlavender}{rgb}{0.96, 0.73, 1.0}
\definecolor{chartreuse(traditional)}{rgb}{0.87, 1.0, 0.0}
\definecolor{slateblue}{rgb}{0.42, 0.35, 0.8}
\definecolor{darkorchid}{rgb}{0.6, 0.2, 0.8}

\def\centerarc[#1](#2)(#3:#4:#5)% Syntax: [draw options] (center) (initial angle:final angle:radius)
{ \draw[#1] ($(#2)+({#5*cos(#3)},{#5*sin(#3)})$) arc (#3:#4:#5); }

\let\leq\leqslant
\let\geq\geqslant
\let\setminus\smallsetminus

\let\rho\varrho

\newcommand{\classfont}{\mathsf}
\newcommand{\NP}{\ensuremath{\classfont{NP}}}
\newcommand{\PP}{\ensuremath{\classfont{P}}}
\newcommand{\FPT}{\ensuremath{\classfont{FPT}}}
\newcommand{\XP}{\ensuremath{\classfont{XP}}}

\renewcommand{\circle}{O}
\newcommand{\stick}{S}
\newcommand{\Cl}{\operatorname{Cl}}
\newcommand{\LLL}{\mathbb{L}}
\newcommand{\TTT}{\mathbb{T}}
\newcommand{\LL}{\mathcal{L}}
\newcommand{\NNN}{\mathbb{N}}

\makeatletter
\newcommand\ie{i.e\@ifnextchar.{}{.\@}}
\newcommand\etc{etc\@ifnextchar.{}{.\@}}
\newcommand\etal{et~al\@ifnextchar.{}{.\@}}
\def\namedlabel#1#2{\begingroup
    #2%
    \def\@currentlabel{#2}%
    \phantomsection\label{#1}\endgroup
}
\makeatother

\newcounter{dummy} 
\numberwithin{dummy}{section}
\newtheorem{theorem}[dummy]{Theorem}
\newtheorem{claim}[dummy]{Claim}

\newtheorem{lemma}[dummy]{Lemma}

\newtheorem{observation}[dummy]{Observation}
\newtheorem{definition}[dummy]{Definition}

\newcounter{hackcount}

\title[Lollipop graphs and medusa graphs]{Beyond circular-arc graphs -- recognizing lollipop graphs and medusa graphs.}

\author[A{\u{g}}ao{\u{g}}lu {\c{C}}a{\u{g}}{\i}r{\i}c{\i} et al.]{Deniz A{\u{g}}ao{\u{g}}lu {\c{C}}a{\u{g}}{\i}r{\i}c{\i}$^1$$\ddagger$, Onur {\c{C}}a{\u{g}}{\i}r{\i}c{\i}$^2$, Jan Derbisz$^3$$^\dagger$, Tim A.\ Hartmann$^4$, Petr Hlin{\v{e}}n{\'y}$^1$$\ddagger$, 
Jan Kratochv{\'{\i}}l$^5$, Tomasz Krawczyk$^3$, Peter Zeman$^6$$^{*}$}

\address{$^1$Masaryk University, Brno, Czech Republic}
\email{agaoglu@mail.muni.cz, hlineny@fi.muni.cz}

\address{$^2$Toronto Metropolitan University, Toronto, Canada}
\email{cagirici@ryerson.ca}

\address{$^3$Theoretical Computer Science Department, 
Faculty of Mathematics and Computer Science, Jagiellonian University in Krak\'ow, Poland}
\email{jan.derbisz@doctoral.uj.edu.pl, tomasz.krawczyk@uj.edu.pl}

\address{$^4$RWTH Aachen University, Aachen, Germany}
\email{hartmann@algo.rwth-aachen.de}

\address{$^5$Department of Applied Mathematics,
Faculty of Mathematics and Physics, Charles University, Prague, Czech Republic}
\email{honza@kam.mff.cuni.cz}

\address{$^6$Institut de math\'{e}matiques, University of Neuch\^{a}tel, Switzerland}
\email{zeman.peter.sk@gmail.com}

\thanks{$^\dagger$Research of this author was partially funded by Polish National Science Center (NCN) grant 2021/41/N/ST6/03671 and by the Priority Research Area SciMat under the program Excellence Initiative – Research University at the Jagiellonian University in Krak\'{o}w}
\thanks{$^{*}$Research of this author is supported by Swiss National Science Foundation project PP00P2-202667}
\thanks{$\ddagger$Research of these authors is supported by the Czech Science Foundation project no. 20-04567S}
\begin{document}

\thispagestyle{empty}

\begin{abstract}
In 1992 Bir\'{o}, Hujter and Tuza introduced, for every fixed connected graph~$H$, the class of $H$-graphs, 
defined as the intersection graphs of connected subgraphs of some subdivision of~$H$.
Such classes of graphs are related to many known graph classes: for example, $K_2$-graphs correspond to interval graphs, $K_3$-graphs correspond to circular-arc graphs, 
the union of $T$-graphs, where $T$ ranges over all trees, 
correspond to chordal graphs.  
Recently, quite a lot of research has been devoted to understanding the tractability border for various computational problems, such as recognition 
or isomorphism testing, in classes of $H$-graphs for different graphs $H$.

In this work we undertake this research topic, focusing on the recognition problem.
Chaplick, T{\"{o}}pfer, Voborn{\'{\i}}k, and Zeman showed an $\XP$-algorithm testing whether a given graph is a $T$-graph, where the parameter is the size of the tree $T$.
In particular, for every fixed tree $T$ the recognition of $T$-graphs can be solved in polynomial time. 
Tucker showed a polynomial time algorithm recognizing $K_3$-graphs (circular-arc graphs).
On the other hand, Chaplick at al.\ showed also that recognition of $H$-graphs 
is \NP-hard if $H$ contains two different cycles sharing an edge. 

%\medskip
The main two results of this work narrow the gap between the $\NP$-hard and $\PP$ cases of $H$-graphs recognition.
First, we show that recognition of $H$-graphs is $\NP$-hard when $H$ contains two different cycles.
On the other hand, we show a polynomial-time algorithm recognizing $L$-graphs, where 
$L$ is a graph containing a cycle and an edge attached to it ($L$~is called a \emph{lollipop} and $L$-graphs are called \emph{lollipop graphs}).
Clearly, the class of lollipop graphs extends the class of circular-arc graphs.
Our work leaves open the recognition problems of $M$-graphs for every unicyclic graph $M$ different from
a cycle and a~lollipop.

Other results of this work, which shed some light on the cases that remain open, are as follows:
\begin{itemize}
\item recognition of $M$-graphs, where $M$ is a fixed unicyclic graph, admits a polynomial time algorithm 
if we restrict the input to graphs containing particular holes (hence the recognition of $M$-graphs is probably most difficult for chordal graphs),
\item recognition of \emph{medusa graphs}, which are defined as the union of $M$-graphs, where $M$ runs over all unicyclic graphs, is $\NP$-complete.
 Note that medusa graphs extend both circular-arc graphs and chordal graphs.
\end{itemize}
%Both of these graph classes extend the class of circular-arc graphs and
%the latter extends also the class of chordal graphs.

%\medskip
Some of our results are obtained by establishing a relationship between the recognition problems considered in the above-mentioned classes of graphs 
and certain problems related to the Helly property studied in the class of circular-arc graphs.

\end{abstract}

\maketitle

\section{Introduction}

The \emph{intersection graph} $G$ of a finite family $\mathcal{F}$ of sets is an undirected graph where
each set in $\mathcal{F}$ is associated with a vertex of $G$, and each pair of vertices in $G$ are joined
by an edge if and only if the corresponding sets in $\mathcal{F}$ have a non-empty intersection.
In this paper, we consider a special kind of intersection graphs, called \emph{$H$-graphs}, introduced 
by Bir\'{o}, Hujter and Tuza~\cite{biroHT92}.
We first define $H$-graphs formally using the terminology we adapt throughout the paper.

Let $H$ be a connected graph. 
An \emph{$H$-model} of a graph $G$ is a pair $(H^{\phi}, \phi)$, where $H^{\phi}$ is a subdivision of $H$
and $\phi$ is a mapping from $V(G)$ to the subsets of $V(H^{\phi})$, such that:
\begin{itemize}
\item for every $v \in V(G)$ the subgraph of $H^{\phi}$ induced by the set $\phi(v)$ is connected,
\item for every distinct $u, v \in V(G)$ we have $uv \in E(G)$ iff $\phi(u) \cap \phi(v) \neq \emptyset$.
\end{itemize}
A graph $G$ is an \emph{$H$-graph} if $G$ admits an $H$-model.

It turns out that many known geometric intersection graph classes are $H$-graphs for an appropriately chosen graph $H$ or can be shown as the union of $H$-graphs, where $H$ ranges over a simpler family of graphs.
For example:
\begin{itemize}
 \item $K_2$-graphs correspond to the class of \emph{interval graphs}, which are defined as the intersection graphs of intervals on the line,
 \item $K_3$-graphs correspond to the class of \emph{circular-arc graphs}, which are defined as the intersection graphs of arcs of a fixed circle,
 \item the union of $T$-graphs, where $T$ ranges over all trees, corresponds to the class of \emph{chordal graphs}, which are defined as graphs containing no induced cycles of size \mbox{$\geq 4$~\cite{chordalityInters}},
 \item the union of $P$-graphs, where $P$ ranges over all planar graphs, corresponds to the class of \emph{string graphs}, which are defined as the intersection graphs
 of curves in the plane. 
\end{itemize}
The recent research on $H$-graphs, initiated by Chaplick, Toepfer, Voborn{\'{\i}}k, and Zeman~\cite{CTVZ21}, 
aims to generalize efficient optimization algorithms from simple classes of graphs on 
wider families of $H$-graphs,
as well as to determine the boundary of ``polynomial tractability'' for such computational problems as
recognition or isomorphism testing.
Here we often search for efficient \emph{parameterized algorithms}, whose running time depends on the size $n$ of the input graph and the parameter $|H|$, where $|H|$ is the size of the graph~$H$.
First, we search for algorithms that work in polynomial time in $n$ and $|H|$,
then for \emph{\FPT\ algorithms} working in time $f(|H|)n^{O(1)}$ for some computable function $f$, and
finally for \emph{\XP\ algorithms} working in time $n^{f(|H|)}$ for some computable function $f$.
Various \NP-complete problems on $H$-graphs were studied in the parameterized setting and shown to admit \FPT\ and \XP\ algorithms~\cite{DBLP:journals/algorithmica/AgaogluH22, CTVZ21, Tiso, DBLP:conf/wg/ArvindNPZ22,DBLP:journals/endm/ChaplickZ17, DBLP:journals/siamdm/ChaplickFGKZ21,DBLP:conf/esa/FominGR18, DBLP:journals/algorithmica/JaffkeKT20}.

For a graph class $\mathcal{C}$, the \emph{recognition problem} for $\mathcal{C}$ is to decide whether an input graph $G$ belongs to~$\mathcal{C}$.
For a graph class $\mathcal{C}$ defined by geometry, the recognition problem of $\mathcal{C}$ usually boils down to testing whether the input graph 
has a representation appropriate for the class~$\mathcal{C}$.
There are known linear time algorithms for interval graphs~\cite{BoothLueker76} and chordal graphs~\cite{RoseTL76}.
The first polynomial time algorithm recognizing circular-arc graphs was given by Tucker~\cite{Tuck80},
currently there are known two linear-time algorithms recognizing this class of graphs~\cite{McConnel03,KapNus11}.
%On the other hand, the recognition of string graphs is \NP-hard~\cite{Kratochvil91a}.
Although the recognition of chordal graphs takes linear time, when a tree $T$ is on the input,
deciding whether a graph $G$ is a $T$-graph, is \NP-complete~\cite{KLAVIK201585}.
On the other hand, Chaplick, Toepfer, Voborn{\'{\i}}k, and Zeman~\cite{CTVZ21} showed 
that the above problem admits an \XP\ algorithm parameterized by $|T|$.
It is open whether the problem can be solved by an \FPT\ algorithm
(in~\cite{recognizingProperT} it is shown that proper $T$-graphs can be recognized in \FPT,
where a $T$-graph $G$ is called \emph{proper} if there exists a $T$-model $(T^{\phi},\phi)$ of $G$ such that for no pair $u,v \in V(G)$ we have $\phi(u) \subseteq \phi(v)$). 
Applying the above results to $H$-graphs,
$K_2$-graphs and $K_3$-graphs are recognizable in linear time, 
and for any fixed tree $T$ the class of $T$-graphs is recognizable in polynomial-time, 
where the degree of the polynomial depends on~$T$.	
On the other hand, Chaplick, Toepfer, Voborn{\'{\i}}k, and Zeman~\cite{CTVZ21} showed that recognition of $H$-graphs is \NP-complete
if $H$ contains a diamond (a cycle on four vertices with a chord) 
as a minor~\cite{CTVZ21}.
That is, recognition of $H$-graphs is \NP-complete if $H$ contains two cycles sharing an edge.

\subsection{Our results}
Our first result states the following:
\begin{theorem}
\label{thm:recog_two_cycles_hard}
For every fixed graph $H$ containing two different cycles, recognition of $H$-graphs is $\NP$-complete. 
\end{theorem}
Theorem \ref{thm:recog_two_cycles_hard} attracts our attention to $H$-graphs, where $H$ is a \emph{unicyclic} graph (connected graph containing exactly one cycle).
In particular, we are focusing on:
\begin{itemize}
\item the recognition problem for the class of $M$-graphs, where $M$ is any fixed graph containing a cycle and some trees attached to it,
\item the recognition problem for the class of \emph{medusa graphs}, where
$$\mathcal{M} = \bigcup \{ M\text{-graphs}: M \text{ is a unicyclic graph} \}.$$
\end{itemize}
By an $\mathcal{M}$-model of a graph $G$ we mean any $M$-model of $G$, where $M$ is a unicyclic graph.
That is, the medusa graphs are just the graphs that admit an $\mathcal{M}$-model.
Note that medusa graphs extend both circular-arc graphs and chordal graphs.

\begin{figure}[h]
\centering
	\begin{subfigure}[t]{0.32\linewidth}
		\centering
		\begin{tikzpicture}[scale=0.75]
			
			\node at (0,-0.5) [circle,draw, fill=black, opacity=1, color=black, inner sep=0.5mm] (a) {};
			
			\node at (0,-1.5) [circle,draw, fill=black, opacity=1, color=black, inner sep=0.5mm] (b) {};
			
			\node at (1,-1) [circle,draw, fill=black, opacity=1, color=black, inner sep=0.5mm] (c) {};
			
			\node at (2,-1) [circle,draw, fill=black, opacity=1, color=black, inner sep=0.5mm] (d) {};
			
			\node at (2.5,-2) [circle,draw, fill=black, opacity=1, color=black, inner sep=0.5mm] (e) {};
			
			\node at (3,-1) [circle,draw, fill=black, opacity=1, color=black, inner sep=0.5mm] (f) {};
			
			\node at (2.5,-3) [circle,draw, fill=black, opacity=1, color=black, inner sep=0.5mm] (g) {};
			
			\node at (4,-0.5) [circle,draw, fill=black, opacity=1, color=black, inner sep=0.5mm] (h) {};
			
			\node at (4,-1.5) [circle,draw, fill=black, opacity=1, color=black, inner sep=0.5mm] (i) {};
			
			\draw (a) -- (c) -- (b) node[midway, above]{};
			\draw (c) -- (d) node[midway, above]{};
			\draw (d) -- (e) -- (f) -- (d) node[midway, above]{};
			\draw (e) -- (g) node[midway, above]{};
			\draw (i) -- (f) -- (h) node[midway, above]{};
			
		\end{tikzpicture}
		\caption{A fixed unicyclic graph $M$}	
		\label{fig:fixedM}
	\end{subfigure}
	~
	\begin{subfigure}[t]{0.3\linewidth}
		\centering
		\begin{tikzpicture}[scale=0.75]
			
			%jazzberryjam
			\node at (0,0) [circle,draw, fill=jazzberryjam, opacity=1, color=jazzberryjam, inner sep=0.5mm] (a) {};
			
			%brightturquoise
			\node at (1,0) [circle,draw, fill=brightturquoise, opacity=1, color=brightturquoise, inner sep=0.5mm] (b) {};
			
			%lightgray
			\node at (2,0) [circle,draw, fill=lightgray, opacity=1, color=lightgray, inner sep=0.5mm] (c) {};
			
			%slateblue
			\node at (0,-1) [circle,draw, fill=slateblue, opacity=1, color=slateblue, inner sep=0.5mm] (d) {};
			
			%fuchsia
			\node at (1,-1) [circle,draw, fill=fuchsia, opacity=1, color=fuchsia, inner sep=0.5mm] (e) {};
			
			%yellow
			\node at (2,-1) [circle,draw, fill=yellow, opacity=1, color=yellow, inner sep=0.5mm] (f) {};
			
			%red
			\node at (3,-1) [circle,draw, fill=red, opacity=1, color=red, inner sep=0.5mm] (g) {};
			
			%cyan
			\node at (3,0) [circle,draw, fill=cyan, opacity=1, color=cyan, inner sep=0.5mm] (h) {};
			
			%orange
			\node at (4,-1) [circle,draw, fill=orange, opacity=1, color=orange, inner sep=0.5mm] (i) {};
			
			%teal
			\node at (4,0) [circle,draw, fill=teal, opacity=1, color=teal, inner sep=0.5mm] (j) {};
			
			%blue
			\node at (5,0) [circle,draw, fill=blue, opacity=1, color=blue, inner sep=0.5mm] (k) {};
			
			%magenta
			\node at (5,-1) [circle,draw, fill=magenta, opacity=1, color=magenta, inner sep=0.5mm] (l) {};
			
			%green
			\node at (4,-2) [circle,draw, fill=green, opacity=1, color=green, inner sep=0.5mm] (m) {};
			
			%brown
			\node at (4,-3) [circle,draw, fill=brown, opacity=1, color=brown, inner sep=0.5mm] (n) {};
			
			%gray
			\node at (6,0) [circle,draw, fill=gray, opacity=1, color=gray, inner sep=0.5mm] (o) {};
			
			%chartreuse(traditional)
			\node at (7,0) [circle,draw, fill=chartreuse(traditional), opacity=1, color=chartreuse(traditional), inner sep=0.5mm] (p) {};
			
			%pink
			\node at (6,-1) [circle,draw, fill=pink, opacity=1, color=pink, inner sep=0.5mm] (r) {};
			
			%brilliantlavender
			\node at (7,-1) [circle,draw, fill=brilliantlavender, opacity=1, color=brilliantlavender, inner sep=0.5mm] (s) {};
			
			\draw (a) -- (b) -- (c) node[midway, above]{};
			\draw (d) -- (e) --(f) node[midway, above]{};
			\draw (c) --(f) node[midway, above]{};
			\draw (c) -- (h) -- (g) --(c) node[midway, above]{};
			\draw (g) -- (i) --(h) node[midway, above]{};
			\draw (h) -- (j) --(k) node[midway, above]{};
			\draw (i) -- (l) --(k) node[midway, above]{};
			\draw (i) -- (m) --(n) node[midway, above]{};
			\draw (l) -- (m) node[midway, above]{};
			\draw (k) -- (l) --(o) -- (k) node[midway, above]{};
			\draw (o) -- (p) node[midway, above]{};
			\draw (k) -- (l) --(r) -- (k) node[midway, above]{};
			\draw (r) -- (s) node[midway, above]{};
			
		\end{tikzpicture}
		\caption{An $M$-graph $G$}	
		\label{fig:inputG}
	\end{subfigure}
	~
	\begin{subfigure}[t]{0.4\linewidth}
		\centering
		\begin{tikzpicture}[scale=0.75]
			
			%BLACK
			\tikzstyle{every path}=[dotted, thick]
			\draw (6,1) circle (0.9cm);
			
			%branches on the left 
			\foreach \i in {(2.2,1.9),(2.2,-0.1),(5,0.9)} 
			{
				\draw \i -- (3.7,0.9);
			}
			
			%branch in the middle 
			\draw (6,0.1) -- (6,-1.4);
			
			%branches on the right 
			\foreach \i in {(8.5,0),(8.5,1.8)} 
			{
				\draw \i -- (7,0.9);
			}
			
			%COLORED
			\tikzstyle{every path}=[thick]
			
			%blue
			\centerarc[blue](6,1)(-12:66:1.1cm);
			\foreach \i in {(8,0.2),(8,1.34)} 
			{
				\draw[blue] \i -- (7.06,0.78);
			}
			
			%brightturquoise
			\draw[brightturquoise] (2.51,1.85) -- (3.11,1.45);
			
			%brilliantlavender
			\draw[brilliantlavender] (8.2,0.31) -- (8.65,0.03);
			
			%brown
			\draw[brown] (5.91,-0.7) -- (5.91,-1.35);
			
			%chartreuse
			\draw[chartreuse(traditional)] (8.2,1.45) -- (8.65,1.73);
			
			%cyan
			\centerarc[cyan](6,1)(100:191:1.1cm);
			\draw[cyan] (4.9,0.8) -- (4.1,0.8);
			
			%fuchsia
			\draw[fuchsia] (3.09,0.32) -- (2.49,-0.1);
			
			%gray
			\draw[gray] (7.58,1.5) -- (8.2,1.85);
			
			%green
			\draw[green] (5.8,-0.35) -- (5.8,-1);
			
			%jazzberryjam
			\draw[jazzberryjam] (2.15,1.76) -- (2.6,1.47);
			
			%lightgray
			\draw[lightgray] (4.25,1) -- (3.33,1) -- (2.7,1.4);
			
			%magenta
			\centerarc[magenta](6,1)(-95:2:1cm);
			\draw[magenta] (5.91,0.02) -- (5.91,-0.6);
			\foreach \i in {(8,1.62),(8,0.42)} 
			{
				\draw[magenta] \i -- (7,1.02);
			}
			
			%orange
			\centerarc[orange](6.1,1)(179:270:1.1cm);
			\draw[orange] (6.1,-0.08) -- (6.1,-0.6);
			\draw[orange] (5,1) -- (4.4,1);
			
			%pink
			\draw[pink] (7.58,0.3) -- (8.2,-0.05);
			
			%red
			\draw[red] (4.65,0.7) -- (3.85,0.7);
			
			\draw[slateblue] (2.6,0.31) -- (2.15,0);
			
			%teal
			\centerarc[teal](6,1)(65:115:1cm);
			
			%yellow
			\draw[yellow] (3.71,0.8) -- (3.35,0.8) -- (2.7,0.37);
			
		\end{tikzpicture}
		\caption{An $M$-model $(M^{\phi},\phi)$ of $G$}
		\label{fig:M-model}
	\end{subfigure}
\caption{}
\label{fig:exampleM-graph} 
\end{figure}

Suppose $G$ is a medusa graph and suppose $G$ admits $\mathcal{M}$-model $(M^{\phi},\phi)$ for some unicyclic graph $M$.
A clique $C$ in $G$ satisfies the \emph{Helly property} in $(M^{\phi},\phi)$ if $\bigcap_{c \in C} \phi(c) \neq \emptyset$ 
and the model $(M^{\phi},\phi)$ of $G$ satisfies the \emph{Helly property} if 
every clique of $G$ satisfies the Helly property in $(M^{\phi},\phi)$.
A medusa graph $G$ is \emph{Helly} if $G$ admits an $\mathcal{M}$-model that satisfies the Helly property.
Figure~\ref{fig:exampleM-graph} shows a fixed unicyclic graph $M$, another graph $G$ which is an $M$-graph, and an $M$-model $(M^{\phi},\phi)$ of $G$. 
Since $(M^{\phi},\phi)$ satisfies the Helly property, $G$ is a Helly medusa graph.

Our next result states the following: 
\begin{theorem}\ 
\label{thm:recognition-medusa-graphs}
\begin{enumerate} 
\item \label{item:recognition-medusa-graphs} Recognition of medusa graphs is \NP-complete.
\item \label{item:recognition-helly-medus-graphs}Recognition of Helly medusa graphs can be solved by a poly-time algorithm.
\end{enumerate}
\end{theorem}
We also consider the recognition problem in the class of $M$-graphs, where $M$ is a fixed unicyclic graph.
Our second result concerns the class of $L$-graphs (called also as \emph{lollipop graphs}), where $L$ is a unicyclic graph that consists of a cycle and an edge attached to it.
\begin{theorem}
\label{thm:lollipop-graphs}
Recognition of $L$-graphs can be solved by a poly-time algorithm.
\end{theorem}
We do not know the status of the recognition problem in the class
of $M$-graphs, for any unicyclic graph $M$ different from $L$.
However, our research seems to confirm the thesis that it is easier to test whether $G$ is an $M$-graph in the case when $G$ has holes.
We suspect that the problem is most difficult for chordal graphs.
To support our thesis, we define the class of \emph{strongly cyclic}
graphs, which, roughly speaking, consists of graphs containing some carefully defined holes.
\begin{theorem}
\label{thm:strongly-cyclic-graphs}
For every fixed unicyclic graph $M$, recognition of $M$-graphs can be solved in polynomial time if we restrict the input to strongly cyclic graphs.
\end{theorem}
Since the definition of strongly cyclic graphs is technical and requires some tools, we postpone it to Section \ref{sec:strongly-cyclic-graphs}.

\subsection{Tools}
Some of our results are obtained by establishing a relationship between our problems and certain problems
related to the Helly property studied in the class of circular-arc graphs.
Lin and Schwarzfiter~\cite{LinSchw06} showed a linear-time algorithm recognizing Helly circular-arc graphs.
That is, their algorithm tests whether an input graph $G$ has a circular-arc model in which
all the cliques of $G$ satisfy the Helly property.
In our work, we exploit a variant of this problem in which we want to 
test whether $G$ has a circular-arc model in which some particular cliques satisfy the Helly property.

\medskip
\noindent \textbf{Helly Cliques:} \\
\begin{tabular}{rl}
\textbf{Input:}& A circular-arc graph $G$ and some of its cliques $C_1,\ldots,C_k$, \\
\textbf{Output:}& \textbf{YES} if $G$ admits a circular-arc model in which all the cliques \\
& $C_1,\ldots,C_k$ satisfy the Helly property.
\end{tabular}
\medskip

A{\u{g}}ao{\u{g}}lu {\c{C}}a{\u{g}}{\i}r{\i}c{\i} and Zeman ~\cite{AgaZem2022}, and independently, Derbisz and Krawczyk \cite{DerKra22+} (with a different proof) showed that the Helly Cliques problem is \NP-hard.
Derbisz and Krawczyk \cite{DerKra22+} considered the Helly Cliques problem from the parameterized complexity point of view, 
where the parameter is the number of cliques given in the input. 
The main results of \cite{DerKra22+} are as follows:
\begin{itemize}
 \item under ETH the Helly Cliques problem can not be solved in time $2^{o(k)}poly(n)$,
 \item the Helly Cliques problem is \FPT\ as it admits an $2^{O(k\log{k})}poly(n)$ algorithm\footnote{Actually, the algorithm of Derbisz and Krawczyk works in time $2^{k}poly(n)$ 
 except for one peculiar case where it works in time $2^{O(k\log{k})}poly(n)$.},
 \item the Helly Cliques problem admits a polynomial kernel.
\end{itemize}
In particular, to prove Theorem \ref{thm:recognition-medusa-graphs} we show that recognition of medusa graphs is polynomial time equivalent to
the Helly Cliques problem and recognition of Helly medusa graphs boils down to testing whether
some carefully chosen subgraph of the input graph is a Helly circular-arc graph. 
To prove Theorem~\ref{thm:lollipop-graphs} we use an algorithm for the Helly Cliques problem where the input is restricted to one clique ($k=1$):
in such case the \FPT-algorithm of Derbisz and Krawczyk works in polynomial time.
One of the main tools used to prove Theorem~\ref{thm:strongly-cyclic-graphs} is a polynomial time algorithm by A{\u{g}}ao{\u{g}}lu {\c{C}}a{\u{g}}{\i}r{\i}c{\i}, 
Derbisz, Gutowski, and Krawczyk \cite{ADGK22+} solving the following variant of the Helly Cliques problem:

\medskip
\noindent \textbf{Helly Cliques with Given Intersection Points:} \\
\begin{tabular}{rl}
\textbf{Input:}& A circular-arc graph $G$, cliques $C_1,\ldots,C_k$ in $G$, \\
& and points $P_1,\ldots, P_k$ on the circle $\circle$,\\
\textbf{Output:}& \textbf{YES} if $G$ admits a circular-arc model $\psi$ on the circle $\circle$ in which \\
&for every $i \in [k]$ the intersection of $\psi(C_i)$ contains the point $P_i$.
\end{tabular}

\medskip
Note that the Helly Cliques and the Helly Cliques with Given Intersection Points problems are equivalent if the input is restricted to one clique ($k=1$).

Our paper is organized as follows:
\begin{itemize}
 \item in Section \ref{sec:preliminaries} we introduce basic concepts and definitions used throughout the paper,
 \item in Section \ref{sec:medusa_graphs} we prove Theorem \ref{thm:recognition-medusa-graphs},
 \item in Section~\ref{sec:M-models} we prepare some tools to prove Theorems~\ref{thm:lollipop-graphs} and~\ref{thm:strongly-cyclic-graphs},
 \item in Section \ref{sec:strongly-cyclic-graphs} we prove Theorem~\ref{thm:strongly-cyclic-graphs},
 \item in Sections \ref{sec:lollipop_graphs}-\ref{sec:strongly_centered_lollipop_graphs} we prove Theorem~\ref{thm:lollipop-graphs},
 \item in Section \ref{sec:butterfly-NP-hard} we prove Theorem \ref{thm:recog_two_cycles_hard}.
\end{itemize}

\section{Preliminaries}
\label{sec:preliminaries}
\subsection{Graphs}
All graphs considered in this paper are \emph{simple}, that is, they have no multiedges and no loops.
For a graph $G$, by $V(G)$ and $E(G)$ we denote the vertex set of $G$ and the edge set of $G$.
For a subset $S \subseteq V(G)$, the subgraph of $G$ induced by $S$ is the graph $G[S] = (S, \{uv \mid uv \in E(G), u,v\in S\})$.
The \emph{neighborhood} of a vertex $u \in V$ is the set $N(u) = \{v \in V(G) \mid uv \in E(G)\}$.
Similarly, we write $N(U)=\bigcup_{u\in U} N(u)\setminus U$ for a set $U\subseteq V$. 
We denote a complete graph and a cycle on $n$ vertices by $K_n$ and $C_n$, respectively.
By a \emph{hole} we mean an induced cycle on at least four vertices.

A subset $I$ of $V(G)$ is called a \emph{component} of $G$ if $G[I]$ is a maximal connected (there is a path from any vertex of $G[I]$ to any other vertex of $G[I]$) induced subgraph of $G$.

A vertex $u$ of $G$ is \emph{universal} in $G$ if $N(u) = V(G) \setminus \{u\}$.
Two distinct vertices $u,v$ of $G$ are \emph{twins} in $G$ if $uv \in E(G)$ and $N(u) = N(v)$.

A \emph{unicyclic} graph is a connected graph that has exactly one cycle.
For a unicyclic graph $M$, by $M_{\circle}$ we denote the vertices belonging to the cycle of $M$. 

Suppose $M$ is a fixed unicyclic graph.
Let $(M^{\phi},\phi)$ be an $M$-model for a graph $G$.
%Note that $M^{\phi}_{\circle}$ denotes the set of all vertices from the unique circle of $M^{\phi}$. 
If the subdivision $M^{\phi}$ of $M$ is not relevant for our considerations, we denote the model $(M^{\phi},\phi)$ simply by $(M,\phi)$ or even by $\phi$ (if $M$ is clear from the context).
In this case $(M,\phi)$ can be seen as the intersection model of $G$ in which every set $\phi(v)$ 
forms an arcwise connected subset of some fixed plane drawing of the unicyclic graph $M$.
Under this assumption $M_{\circle}$ can be seen as the part of the drawing which contains the vertices and the edges from the circle of $M$.

\subsection{Partially Ordered Sets}
A \emph{partially ordered set} (shortly \emph{partial order} or \emph{poset}) is a pair $(X,{\sqsubseteq})$ that consists of a set $X$ and a reflexive, transitive, and antisymmetric relation ${\sqsubseteq}$ on $X$.
For a poset $(X,{\sqsubseteq})$, let the \emph{strict partial order} ${\sqsubset}$ be a binary relation defined on $X$ such that $x \sqsubset y$ if and only if $x \sqsubseteq y$ and $x \neq y$. 
Equivalently, $(X,{\sqsubset})$ is a strict partial order if ${\sqsubset}$ is irreflexive and transitive.
Two elements $x,y \in X$ are \emph{comparable} in $(X,{\sqsubseteq})$ if $x \sqsubseteq y$ or $y \sqsubseteq x$; otherwise, $x,y$ are \emph{incomparable} in $P$.
A \emph{chain} (an \emph{antichain}) of $P$ is a set consisting of pairwise comparable (pairwise incomparable, respectively) elements.
The \emph{width} of the poset $(X,{\sqsubseteq})$ is the maximum size of an antichain in $(X,{\sqsubseteq})$.
A \emph{linear order} $(X,{\preceq})$ is a partial order in which every two vertices $x,y \in X$ are comparable.
A \emph{strict linear order} $(X,{\prec})$ is a binary relation defined in a way that $x \prec y$ if and only if $x \preceq y$ and $x \neq y$. 
Let $(X,{\preceq})$ be a linear order and let $Y \subseteq X$.
We say $Y$ is an \emph{interval} in $(X,{\preceq})$ if 
for every $x,y,z \in X$ such that $x \prec y \prec z$ and $x,z \in Y$ we have $y \in Y$.

Given a poset $(X,{\sqsubseteq})$ and a set $Y \subseteq X$, 
by $(Y,{\sqsubseteq})$ we denote the poset induced by the set $Y$.
We use the same notation with respect to strict partial orders, 
linear orders, and strict linear orders.

Suppose $(X,{\sqsubseteq})$ is a poset. 
For an antichain $A$ in $(X,{\sqsubseteq})$ we let
$$
\begin{array}{ccl}
US(A)& = & \{x \in X: a \sqsubset x \text{ for some $a \in A$}\}, \\
DS(A)& = & \{x \in X: x \sqsubset a \text{ for some $a \in A$}\}. \\
\end{array}
$$
If $a$ is an element of $P$, then by $US(a)$ and $DS(a)$ we denote the sets $US(\{a\})$ and $DS(\{a\})$, respectively.

A \emph{co-comparability graph} of a poset $P=(X,{\sqsubseteq})$ 
has $X$ as the set of its vertices and the set including every two vertices incomparable in $(X,{\sqsubseteq})$ as the set of its edges.
A graph $G$ is a \emph{co-comparability} graph if $G$ is the co-comparability graph of some poset.
Chaplick, T{\"{o}}pfer, Voborn{\'{\i}}k, and Zeman~\cite{CTVZ21} considered \textbf{s-list coloring problem for co-comparability} graphs, defined as follows:
given a co-comparablility graph $G$, a set of colors $S$ such that $|S| \leq s$, 
and a set $L(v) \subseteq S$ for each vertex $v \in V(G)$, we want to find a proper coloring $c : V(G) \to S$ such that for every vertex $v$ we have $c(v) \in L(v)$.
In particular, it is shown in \cite{CTVZ21} that $s$-list coloring problem can be solved in time $O(n^{s^2+1}s^3)$.
Since partitioning a poset into chains is equivalent to coloring its co-comparability graph, using the same algorithm we can solve \textbf{s-list chain partitioning} problem \textbf{of posets} in which for a given poset $(X,{\sqsubseteq})$, a set of colors $S$ such that $|S| \leq s$, and a set $L(x) \subseteq S$ for every $x \in X$, 
we want to assign to each element $x \in X$ a color $c(x) \in L(x)$ such that for every $c \in S$ the elements of $X$ colored $c$ form a chain in $(X,{\sqsubseteq})$.

\subsection{Interval graphs}
A graph $H$ is an \emph{interval graph} if there is a mapping $\phi$ from $V(H)$ to the set of closed intervals in $\mathbb{R}$ such that $\psi(u) \cap \psi(v) \neq \emptyset$ iff $uv \in E(H)$.
If the above holds, the mapping $\psi$ is called an \emph{interval model} of $H$.
Clearly, the class of interval graphs coincides with the class of $K_2$-graphs.

Let $H$ be an interval graph and let $\phi$ be an interval model of $H$.
For every $x \in \mathbb{R}$ by $C(x)$ we denote the set $\{v \in V(H): x \in \phi(v) \}$.
A \emph{sector} $S$ of $\phi$ is a maximal interval of $\mathbb{R}$ such that 
$C(x)= C(y)$ holds for every $x,y \in S$.
Given a sector $S$ of $\phi$, \emph{the clique set} $C(S)$ of~$S$
is equal to $C(x)$, where $x$ is any point inside the sector $S$.
Clearly, every two sectors of $\phi$ are disjoint and the union of all sectors of $\phi$ covers $\mathbb{R}$ 
(note that $\phi$ has at least two sectors $S$ such that $C(S) = \emptyset$).
We say that a sector $S$ of $\phi$ is:
\begin{itemize}
 \item \emph{maximal} if $C(S)$ is a maximal clique in $H$,
 \item \emph{minimal} if $C(S) \subsetneq C(S')$ for any
 sector $S'$ adjacent to $S$.
\end{itemize}
Two interval models $\phi_1$ and $\phi_2$ of $H$ are \emph{equivalent} if the sequence of cliques on the consecutive sectors of $\phi_1$
is equal to the sequence of cliques on the consecutive sectors of~$\phi_2$.

An interval model $\phi$ of $H$ is \emph{normalized} if for every 
$u,v \in V$ the right (left) endpoints of $\phi(u)$ and $\phi(v)$ are equal
if $\phi(u)$ and $\phi(v)$ have the same set of intervals from $\phi(V)$ 
lying entirely to the right (to the left) of $\phi(u)$ and $\phi(v)$, respectively.
Note that in any normalized model $\phi$ of $G$ the minimal and the maximal sectors are intertwining when we go from left to right: 
we start with empty minimal sector, then we have maximal sector, minimal sector, e.t.c., and finally we have an empty minimal sector.
In particular, we can equivalently define normalized models of $H$ as the set of those models of $H$ which have exactly $2c+1$ sectors, where $c$ is the number of maximal cliques in $H$.
See Figure~\ref{fig:exampleINTNormalziedgraph} for an illustration.

Let $\mathcal{C}(H)$ be the set of all maximal cliques of an interval graph 
$H$ and let $\mathcal{C}(v) = \{C \in \mathcal{C}(H): v \in C\}$ for $v \in V$.
A linear ordering $\delta$ of $\mathcal{C}(H)$ is \emph{consecutive} if for every 
$v \in V$ the elements of $\mathcal{C}(v)$ appear consecutively in $\delta$.
It is easy to observe that there is the correspondence between consecutive orderings of $\mathcal{C}(H)$ and the normalized models of $H$.

\begin{figure}[h]
		\centering
		\begin{subfigure}[t]{0.6\linewidth}
\begin{tikzpicture}[xscale=0.55,yscale=0.7,>=latex,shorten >=-0.4pt,shorten <=-0.4pt]
  \tikzstyle{every node}=[inner sep=2pt,fill=white]  

%P1  
\draw[opacity=0.4,color=lightgray,fill=lightgray](-0.5,-0.5) rectangle (16.5,7);
\coordinate (lP1) at (0.3,6.4) {};

%P2
\draw[opacity=0.4,color=red,fill=red](-0.4,2.5) rectangle (5.4,5.5);
\coordinate (lP2) at (2.5,5.2) {};

%P3
\draw[opacity=0.4,color=teal,fill=teal](5.6,1.5) rectangle (16.3,6.8);
\coordinate (lP3) at (15.5,6.4) {};

%Q1
\draw[opacity=0.4,color=magenta,fill=magenta](5.8,2.5) rectangle (13.2,6.6);
\coordinate (lQ1) at (6.6,6.2) {};

%sectors
\draw[dotted,opacity=0.4,thick] (0,-0.7) -- (0,7);
\draw[dotted,opacity=0.4,thick] (2,-0.7) -- (2,7);
\draw[dotted,opacity=0.4,thick] (3,-0.7) -- (3,7);
\draw[dotted,opacity=0.4,thick] (5,-0.7) -- (5,7);
\draw[dotted,opacity=0.4,thick] (6,-0.7) -- (6,7);
\draw[dotted,opacity=0.4,thick] (8,-0.7) -- (8,7);
\draw[dotted,opacity=0.4,thick] (9,-0.7) -- (9,7);
\draw[dotted,opacity=0.4,thick] (10,-0.7) -- (10,7);
\draw[dotted,opacity=0.4,thick] (11,-0.7) -- (11,7);
\draw[dotted,opacity=0.4,thick] (13,-0.7) -- (13,7);
\draw[dotted,opacity=0.4,thick] (14,-0.7) -- (14,7);
\draw[dotted,opacity=0.4,thick] (16,-0.7) -- (16,7);
%\draw[opacity=0.4,color=gray,fill=gray](0,-1) rectangle (2,7);
\coordinate (lC1) at (1,6.8) {};

%P3
%\draw[opacity=0.4,color=teal,fill=teal](2.8,0.8) rectangle (7.7,-1.80);

%MAX CLIQUES
%C1
%\draw[dotted,opacity=0.4,thick] (1,-0.6) -- (1,6.6);
\coordinate (lC1) at (1,7.2) {};

%C2
%\draw[dotted,opacity=0.4,thick] (4,-0.6) -- (4,6.6);
\coordinate (lC2) at (4,7.2) {};

%C3
%\draw[dotted,opacity=0.4,thick] (7,-0.6) -- (7,6.6);
\coordinate (lC3) at (7,7.2) {};

%C4
%\draw[dotted,opacity=0.4,thick] (9.5,-0.6) -- (9.5,6.6);
\coordinate (lC4) at (9.5,7.2) {};

%C5
%\draw[dotted,opacity=0.4,thick] (12,-0.6) -- (12,6.6);
\coordinate (lC5) at (12,7.2) {};

%C6
%\draw[dotted,opacity=0.4,thick] (15,-0.6) -- (15,6.6);
\coordinate (lC6) at (15,7.2) {};

%v1
\draw[|-|,thick] (0,0) -- (16,0);
\coordinate (lv1) at (8,-0.3) {};

%v2
\draw[|-|,thick] (0,1) -- (16,1);
\coordinate (lv2) at (8,0.7) {};

%P1-node
%v3
\draw[|-|,thick] (0,3) -- (5,3);
\coordinate (lv3) at (2.5,2.7) {};
%v4
\draw[|-|,thick] (0,4) -- (2,4);
\coordinate (lv4) at (1,3.7) {};
%v5
\draw[|-|,thick] (0,5) -- (2,5);
\coordinate (lv5) at (1,4.7) {};

%v6
\draw[|-|,thick] (3,4) -- (5,4);
\coordinate (lv6) at (4,3.7) {};

%P2-node
%v7
\draw[|-|,thick] (6,2) -- (16,2);
\coordinate (lv7) at (11,1.7) {};

%Q1-node
%v8
\draw[|-|,thick] (6,3) -- (13,3);
\coordinate (lv8) at (9.5,2.7) {};
%v9
\draw[|-|,thick] (6,4) -- (10,4);
\coordinate (lv9) at (8,3.7) {};
%v11
\draw[|-|,thick] (11,4) -- (13,4);
\coordinate (lv11) at (12,3.7) {};
%v10
\draw[|-|,thick] (6,5) -- (8,5);
\coordinate (lv10) at (7,4.7) {};
%v12
\draw[|-|,thick] (9,5) -- (13,5);
\coordinate (lv12) at (11,4.7) {};
%v13
\draw[|-|,thick] (9,6) -- (13,6);
\coordinate (lv13) at (11,5.7) {};

%v14
\draw[|-|,thick] (14,3) -- (16,3);
\coordinate (lv14) at (15,2.7) {};
%v15
\draw[|-|,thick] (14,4) -- (16,4);
\coordinate (lv15) at (15,3.7) {};
%v16
\draw[|-|,thick] (14,5) -- (16,5);
\coordinate (lv16) at (15,4.7) {};

\tikzstyle{every node}=[inner sep=1pt]
\begin{scriptsize}
\node at (lv1) {$v_1$};
\node at (lv2) {$v_2$};
\node at (lv3) {$v_3$};
\node at (lv4) {$v_4$};
\node at (lv5) {$v_5$};
\node at (lv6) {$v_6$};
\node at (lv7) {$v_7$};
\node at (lv8) {$v_8$};
\node at (lv9) {$v_9$};
\node at (lv10) {$v_{10}$};
\node at (lv11) {$v_{11}$};
\node at (lv12) {$v_{12}$};
\node at (lv13) {$v_{13}$};
\node at (lv14) {$v_{14}$};
\node at (lv15) {$v_{15}$};
\node at (lv16) {$v_{16}$};

\node at (lP1) {$\mathbb{P}_1$};
\node at (lP2) {$\mathbb{P}_2$};
\node at (lP3) {$\mathbb{P}_3$};
\node at (lQ1) {$\mathbb{Q}_1$};

\node at (lC1) {$C_1$};
\node at (lC2) {$C_2$};
\node at (lC3) {$C_3$};
\node at (lC4) {$C_4$};
\node at (lC5) {$C_5$};
\node at (lC6) {$C_6$};
\end{scriptsize}

%\draw[black] (-1,-0.5)--(-1,0);
%\draw[black] (17,6.7)--(17,7);
\draw[white] (-1,-0.5)--(-1,0);
\draw[white] (17,6.7)--(17,7);

%MAX CLIQUES
%\draw[dotted,opacity=0.4] (0.5,2.2) -- (0.5,-2.05);
%\node[font=\fontsize{7}{7}\selectfont] at (0.5,2.4) {$C_1$};

%\draw[dotted,opacity=0.4] (2,2.2) -- (2,-2.05);
%\node[font=\fontsize{7}{7}\selectfont] at (2,2.4) {$C_2$};
%\draw[dotted,opacity=0.4] (3.5,2.2) -- (3.5,-2.05);
%\node[font=\fontsize{7}{7}\selectfont] at (3.5,2.4) {$C_3$};
%\draw[dotted,opacity=0.4] (4.45,2.2) -- (4.45,-2.05);
%\node[font=\fontsize{7}{7}\selectfont] at (4.45,2.4) {$C_4$};
%\draw[dotted,opacity=0.4] (5.45,2.2) -- (5.45,-2.05);
%\node[font=\fontsize{7}{7}\selectfont] at (5.45,2.4) {$C_5$};
%\draw[dotted,opacity=0.4] (6.9,2.2) -- (6.9,-2.05);
%\node[font=\fontsize{7}{7}\selectfont] at (6.9,2.4) {$C_6$};

\end{tikzpicture}
\caption{A normalized interval model $\phi$ of $H$. Sectors of $\phi$ are separated by dashed lines}
\label{fig:normalizedModelx}
\end{subfigure}
~
\begin{subfigure}[t]{0.3\linewidth}
\centering
\begin{tikzpicture}[scale=0.85,>=latex,shorten >=-0.4pt,shorten <=-0.4pt]
  \tikzstyle{every node}=[circle,minimum size=10pt,inner sep=0.5,draw];
  \begin{scriptsize}
  \node (p1) at (0.0,1) {$\mathbb{P}_1$};
  \node (p2) at (-1.5,0.0) {$\mathbb{P}_2$};
  \node (p3) at (1.5,0,0) {$\mathbb{P}_3$};
  \node (c1) at (-2,-1) {$\mathbb{L}_1$};
  \node (c2) at (-1,0.-1) {$\mathbb{L}_2$};
  \node (q1) at (0.5,-1) {$\mathbb{Q}_1$};
  \node (c6) at (2,0.-1) {$\mathbb{L}_6$};
  \node (c3) at (-0.25,-2) {$\mathbb{L}_3$};
  \node (c4) at (0.5,-2) {$\mathbb{L}_4$};
  \node (c5) at (1.25,-2) {$\mathbb{L}_5$};
  \end{scriptsize}
  \tikzstyle{every node}=[square,minimum size=15pt,inner sep=0.5,draw];
  \begin{scriptsize}
  %\node (n1) at (0.75,1.0) {$N_1$};
  \end{scriptsize}
    \path (p1) edge (p2); 
    \path (p1) edge (p3); 
    \path (p2) edge (c1); 
    \path (p2) edge (c2); 
    \path (p3) edge (q1); 
    \path (p3) edge (c6); 
    \path (q1) edge (c3); 
    \path (q1) edge (c4); 
    \path (q1) edge (c5); 

%\draw[black] (-3.2,-3.4)--(-3.2,-1.8);
%\draw[black] (2.5,1.0)--(2.5,1.5);
\draw[white] (-3.2,-3.4)--(-3.2,-1.8);
\draw[white] (2.5,1.0)--(2.5,1.5);
\end{tikzpicture}
\caption{$PQ$-tree $\mathbb{T}$ of $H$ with leaf order $\mathbb{L}_1,\mathbb{L}_2,\mathbb{L}_3,\mathbb{L}_4,\mathbb{L}_5,\mathbb{L}_6$ corresponding to consecutive clique ordering $C_1,C_2,C_3,C_4,C_5,C_6$}	
\label{fig:PQ-treex}
\end{subfigure}
		\caption{}
		\label{fig:exampleINTNormalziedgraph} 
\end{figure}

In 1976, Lueker and Booth \cite{BoothLueker76} introduced a data structure, called a \emph{$PQ$-tree}, which allows to encode all consecutive orderings of the maximal cliques of an interval graph.
A~$PQ$-tree $\mathbb{T}$ for an interval graph $H$ is a rooted, labeled tree, 
in which the elements of $\mathcal{C}(H)$ are in correspondence with the leaf nodes of $\mathbb{T}$, and each non-leaf node is labeled $P$ or $Q$. 
A~$P$-node has at least two children, and a~$Q$-node has at least three children.
For any inner node $\mathbb{N}$ of $\mathbb{T}$ a set of \emph{admissible orderings} of the children of $\mathbb{N}$ is defined: if $\mathbb{N}$ is a $P$-node, then this set contains all possible permutation of the children of $\mathbb{N}$ and 
if $\mathbb{N}$ is a $Q$-node then this set contains two permutation of the children of $\mathbb{N}$ such that one is the reverse of the other.
Lueker and Booth proved that the consecutive ordering of $\mathcal{C}(H)$ are in the correspondence with the orderings of the leaf nodes of $\mathbb{T}$
obtained by ordering the children of every inner node $\mathbb{N}$ of $\mathbb{T}$ according to an admissible order for~$\mathbb{N}$.

\begin{figure}[h]
		\centering
		\begin{subfigure}[t]{0.6\linewidth}
\begin{tikzpicture}[xscale=0.55,yscale=0.7,>=latex,shorten >=-0.4pt,shorten <=-0.4pt]
  \tikzstyle{every node}=[inner sep=2pt,fill=white]  

%P1  
\draw[opacity=0.4,color=lightgray,fill=lightgray](-0.5,-0.5) rectangle (16.5,7.5);
\coordinate (lP1) at (15.7,6.4) {};

%P2
\draw[opacity=0.4,color=red,fill=red](10.6,2.5) rectangle (16.4,5.5);
\coordinate (lP2) at (13.5,5.2) {};

%P3
\draw[opacity=0.4,color=teal,fill=teal](-0.4,1.5) rectangle (10.3,6.8);
\coordinate (lP3) at (9.5,6.4) {};

%Q1
\draw[opacity=0.4,color=magenta,fill=magenta](-0.2,2.5) rectangle (7.2,6.6);
\coordinate (lQ1) at (6.6,6.2) {};

%sectors
\draw[dotted,opacity=0.4,thick] (0,-0.7) -- (0,7);
\draw[dotted,opacity=0.4,thick] (0.5,-0.7) -- (0.5,7);
\draw[dotted,opacity=0.4,thick] (2,-0.7) -- (2,7);
\draw[dotted,opacity=0.4,thick] (3,-0.7) -- (3,7);
\draw[dotted,opacity=0.4,thick] (4,-0.7) -- (4,7);
\draw[dotted,opacity=0.4,thick] (5,-0.7) -- (5,7);
\draw[dotted,opacity=0.4,thick] (6.5,-0.7) -- (6.5,7);
\draw[dotted,opacity=0.4,thick] (7,-0.7) -- (7,7);
\draw[dotted,opacity=0.4,thick] (8,-0.7) -- (8,7);
\draw[dotted,opacity=0.4,thick] (9.5,-0.7) -- (9.5,7);
\draw[dotted,opacity=0.4,thick] (10,-0.7) -- (10,7);
\draw[dotted,opacity=0.4,thick] (11,-0.7) -- (11,7);
\draw[dotted,opacity=0.4,thick] (13,-0.7) -- (13,7);
\draw[dotted,opacity=0.4,thick] (14,-0.7) -- (14,7);
\draw[dotted,opacity=0.4,thick] (15,-0.7) -- (15,7);
\draw[dotted,opacity=0.4,thick] (15.5,-0.7) -- (15.5,7);
\draw[dotted,opacity=0.4,thick] (16,-0.7) -- (16,7);
%\draw[opacity=0.4,color=gray,fill=gray](0,-1) rectangle (2,7);
\coordinate (lC1) at (1,7.2) {};

%P3
%\draw[opacity=0.4,color=teal,fill=teal](2.8,0.8) rectangle (7.7,-1.80);

%MAX CLIQUES
%C1
%\draw[dotted,opacity=0.4,thick] (1,-0.6) -- (1,6.6);
\coordinate (lC1) at (14.5,7.2) {};

%C2
%\draw[dotted,opacity=0.4,thick] (4,-0.6) -- (4,6.6);
\coordinate (lC2) at (11.75,7.2) {};

%C3
%\draw[dotted,opacity=0.4,thick] (7,-0.6) -- (7,6.6);
\coordinate (lC3) at (6,7.2) {};

%C4
%\draw[dotted,opacity=0.4,thick] (9.5,-0.6) -- (9.5,6.6);
\coordinate (lC4) at (3.5,7.2) {};

%C5
%\draw[dotted,opacity=0.4,thick] (12,-0.6) -- (12,6.6);
\coordinate (lC5) at (1.25,7.2) {};

%C6
%\draw[dotted,opacity=0.4,thick] (15,-0.6) -- (15,6.6);
\coordinate (lC6) at (9,7.2) {};

%v1
\draw[|-|,thick] (0,0) -- (16,0);
\coordinate (lv1) at (8,-0.3) {};

%v2
\draw[|-|,thick] (0.5,1) -- (15,1);
\coordinate (lv2) at (8,0.7) {};

%P1-node
%v3
\draw[|-|,thick] (11,3) -- (15.5,3);
\coordinate (lv3) at (13.5,2.7) {};
%v4
\draw[|-|,thick] (14,4) -- (16,4);
\coordinate (lv4) at (15,3.7) {};
%v5
\draw[|-|,thick] (14,5) -- (16,5);
\coordinate (lv5) at (15,4.7) {};
%v6
\draw[|-|,thick] (11,4) -- (13,4);
\coordinate (lv6) at (12,3.7) {};

%P2-node
%v7
\draw[|-|,thick] (0,2) -- (9.5,2);
\coordinate (lv7) at (5,1.7) {};

%Q1-node
%v8
\draw[|-|,thick] (0.5,3) -- (7,3);
\coordinate (lv8) at (3.5,2.7) {};
%v9
\draw[|-|,thick] (3,4) -- (6.5,4);
\coordinate (lv9) at (5,3.7) {};
%v10
\draw[|-|,thick] (5,5) -- (7,5);
\coordinate (lv10) at (6,4.7) {};
%v11
\draw[|-|,thick] (0,4) -- (2,4);
\coordinate (lv11) at (1,3.7) {};
%v12
\draw[|-|,thick] (0.5,5) -- (4,5);
\coordinate (lv12) at (2.25,4.7) {};
%v13
\draw[|-|,thick] (0.5,6) -- (4,6);
\coordinate (lv13) at (2.25,5.7) {};

%v14
\draw[|-|,thick] (8,3) -- (10,3);
\coordinate (lv14) at (9,2.7) {};
%v15
\draw[|-|,thick] (8,4) -- (9.5,4);
\coordinate (lv15) at (8.75,3.7) {};
%v16
\draw[|-|,thick] (8,5) -- (10,5);
\coordinate (lv16) at (9,4.7) {};

\tikzstyle{every node}=[inner sep=1pt]
\begin{scriptsize}
\node at (lv1) {$v_1$};
\node at (lv2) {$v_2$};
\node at (lv3) {$v_3$};
\node at (lv4) {$v_4$};
\node at (lv5) {$v_5$};
\node at (lv6) {$v_6$};
\node at (lv7) {$v_7$};
\node at (lv8) {$v_8$};
\node at (lv9) {$v_9$};
\node at (lv10) {$v_{10}$};
\node at (lv11) {$v_{11}$};
\node at (lv12) {$v_{12}$};
\node at (lv13) {$v_{13}$};
\node at (lv14) {$v_{14}$};
\node at (lv15) {$v_{15}$};
\node at (lv16) {$v_{16}$};

\node at (lP1) {$\mathbb{P}_1$};
\node at (lP2) {$\mathbb{P}_2$};
\node at (lP3) {$\mathbb{P}_3$};
\node at (lQ1) {$\mathbb{Q}_1$};

\node at (lC1) {$C_1$};
\node at (lC2) {$C_2$};
\node at (lC3) {$C_3$};
\node at (lC4) {$C_4$};
\node at (lC5) {$C_5$};
\node at (lC6) {$C_6$};
\end{scriptsize}

%\draw[black] (-1,-0.5)--(-1,0);
%\draw[black] (17,6.7)--(17,7);
\draw[white] (-1,-0.5)--(-1,0);
\draw[white] (17,6.7)--(17,7);

%MAX CLIQUES
%\draw[dotted,opacity=0.4] (0.5,2.2) -- (0.5,-2.05);
%\node[font=\fontsize{7}{7}\selectfont] at (0.5,2.4) {$C_1$};

%\draw[dotted,opacity=0.4] (2,2.2) -- (2,-2.05);
%\node[font=\fontsize{7}{7}\selectfont] at (2,2.4) {$C_2$};
%\draw[dotted,opacity=0.4] (3.5,2.2) -- (3.5,-2.05);
%\node[font=\fontsize{7}{7}\selectfont] at (3.5,2.4) {$C_3$};
%\draw[dotted,opacity=0.4] (4.45,2.2) -- (4.45,-2.05);
%\node[font=\fontsize{7}{7}\selectfont] at (4.45,2.4) {$C_4$};
%\draw[dotted,opacity=0.4] (5.45,2.2) -- (5.45,-2.05);
%\node[font=\fontsize{7}{7}\selectfont] at (5.45,2.4) {$C_5$};
%\draw[dotted,opacity=0.4] (6.9,2.2) -- (6.9,-2.05);
%\node[font=\fontsize{7}{7}\selectfont] at (6.9,2.4) {$C_6$};

\end{tikzpicture}
\caption{An interval model $\phi$ of $H$. Sectors of $\phi$ are separated by dashed lines}
\label{fig:normalizedModelx}
\end{subfigure}
~
\begin{subfigure}[t]{0.3\linewidth}
\centering
\begin{tikzpicture}[scale=0.85,>=latex,shorten >=-0.4pt,shorten <=-0.4pt]
  \tikzstyle{every node}=[circle,minimum size=10pt,inner sep=0.5,draw];
  \begin{scriptsize}
  \node (p1) at (0.0,1) {$\mathbb{P}_1$};
  \node (p2) at (1.5,0.0) {$\mathbb{P}_2$};
  \node (p3) at (-1.5,0,0) {$\mathbb{P}_3$};
  \node (c1) at (2,-1) {$\mathbb{L}_1$};
  \node (c2) at (0.75,0.-1) {$\mathbb{L}_2$};
  \node (q1) at (-2,-1) {$\mathbb{Q}_1$};
  \node (c6) at (-0.75,0.-1) {$\mathbb{L}_6$};
  \node (c3) at (-1.25,-2) {$\mathbb{L}_3$};
  \node (c4) at (-2,-2) {$\mathbb{L}_4$};
  \node (c5) at (-2.75,-2) {$\mathbb{L}_5$};
  \end{scriptsize}
  \tikzstyle{every node}=[square,minimum size=15pt,inner sep=0.5,draw];
  \begin{scriptsize}
  %\node (n1) at (0.75,1.0) {$N_1$};
  \end{scriptsize}
    \path (p1) edge (p2); 
    \path (p1) edge (p3); 
    \path (p2) edge (c1); 
    \path (p2) edge (c2); 
    \path (p3) edge (q1); 
    \path (p3) edge (c6); 
    \path (q1) edge (c3); 
    \path (q1) edge (c4); 
    \path (q1) edge (c5); 

%\draw[black] (-3.2,-3.4)--(-3.2,-1.8);
%\draw[black] (2.5,1.0)--(2.5,1.5);
\draw[white] (-3.2,-3.4)--(-3.2,-1.8);
\draw[white] (2.5,1.0)--(2.5,1.5);
\end{tikzpicture}
\caption{$PQ$-tree $\mathbb{T}$ of $H$ with leaf order $\mathbb{L}_5,\mathbb{L}_4,\mathbb{L}_3,\mathbb{L}_6,\mathbb{L}_2,\mathbb{L}_1$ 
corresponding to clique ordering $C_5,C_4,C_3,C_6,C_2,C_1$}	
\label{fig:PQ-treex}
\end{subfigure}
		\caption{}
		\label{fig:exampleINTgraph} 
\end{figure}

Finally, observe that we can obtain any interval model of $H$ in the following way:
first we choose a normalized model $\phi$ of $H$ (which is equivalent to picking a consecutive ordering of the maximal cliques of $H$) 
and then for each maximal sector $S$ of $\phi$ we shift (by a little) the endpoints 
of the intervals of $\phi$ that lie on the borders of $S$ -- 
see Figure~\ref{fig:exampleINTgraph} for an illustration.

%Figure~\ref{fig:exampleINTNormalized} shows the sectors of the normalized interval model $\phi$ of 
%$H$ given in Figure~\ref{fig:exampleINTgraph}(b) where the maximal and minimal sectors are colored gray and light gray, respectively. 
%Note that each maximal sector of $\phi$ corresponds to a maximal clique of 
%$H$ which appear on some leaf of the $PQ$-tree $\mathbb{T}$ of $H$ given in Figure~\ref{fig:exampleINTgraph}(c).

\subsection{Circular-arc graphs}
A graph $G$ is a \emph{circular-arc} graph if there exists a mapping $\psi$ from $V(G)$ to the set of arcs of a fixed circle such that for every distinct $u,v \in V(G)$ we have $\psi(u) \cap \psi(v) \neq \emptyset$ iff $uv \in E(G)$.
If the above holds, then $\psi$ is called a \emph{circular-arc model} of $G$.
In particular, $G$ is a circular-arc graph iff $G$ has a $K_3$-model.

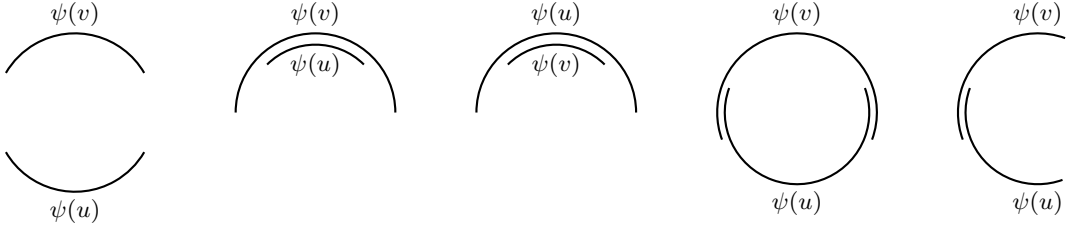
\begin{figure}[htp!]
%disjoint
\begin{tikzpicture}[scale=0.5]
\coordinate (center) at (0,0) {};
\coordinate (v) at ($(center)+(90:2cm)$) {};
\coordinate (u) at ($(center)+(270:2cm)$) {};

\coordinate (lv) at ($(center)+(90:2.6cm)$) {};
\coordinate (lu) at ($(center)+(270:2.6cm)$) {};

\tikzstyle{every node}=[inner sep=1pt]
\begin{scriptsize}
\node at (lv) {$\psi(v)$};
\node at (lu) {$\psi(u)$};
\end{scriptsize}

%u
\draw[thick] ([shift=(30:2.1cm)]0,0) arc (30:150:2.1cm);
%u
\draw[thick] ([shift=(210:2.1cm)]0,0) arc (210:330:2.1cm);

\draw[thick, white] (-3.0,-3)--(-3.0,-2.8);
\draw[thick, white] (3.0,3)--(3.0,2.8);

\end{tikzpicture}
%contains
\begin{tikzpicture}[scale=0.5]
\coordinate (center) at (0,0) {};
\coordinate (v) at ($(center)+(90:2cm)$) {};
\coordinate (u) at ($(center)+(270:2cm)$) {};

\coordinate (lv) at ($(center)+(90:2.6cm)$) {};
\coordinate (lu) at ($(center)+(90:1.25cm)$) {};

\tikzstyle{every node}=[inner sep=1pt]
\begin{scriptsize}
\node at (lv) {$\psi(v)$};
\node at (lu) {$\psi(u)$};
\end{scriptsize}

%v
\draw[thick] ([shift=(0:2.1cm)]0,0) arc (0:180:2.1cm);
%u
\draw[thick] ([shift=(45:1.8cm)]0,0) arc (45:135:1.8cm);

\draw[thick, white] (-3.0,-3)--(-3.0,-2.8);
\draw[thick, white] (3.0,3)--(3.0,2.8);

\end{tikzpicture}
%is_contained
\begin{tikzpicture}[scale=0.5]
\coordinate (center) at (0,0) {};
\coordinate (v) at ($(center)+(90:2cm)$) {};
\coordinate (u) at ($(center)+(270:2cm)$) {};

\coordinate (lu) at ($(center)+(90:2.6cm)$) {};
\coordinate (lv) at ($(center)+(90:1.25cm)$) {};

\tikzstyle{every node}=[inner sep=1pt]
\begin{scriptsize}
\node at (lv) {$\psi(v)$};
\node at (lu) {$\psi(u)$};
\end{scriptsize}

%v
\draw[thick] ([shift=(0:2.1cm)]0,0) arc (0:180:2.1cm);
%u
\draw[thick] ([shift=(45:1.8cm)]0,0) arc (45:135:1.8cm);

\draw[thick, white] (-3.0,-3)--(-3.0,-2.8);
\draw[thick, white] (3.0,3)--(3.0,2.8);

\end{tikzpicture}
%cover_the_circle
\begin{tikzpicture}[scale=0.5]
\coordinate (center) at (0,0) {};
\coordinate (v) at ($(center)+(90:2cm)$) {};
\coordinate (u) at ($(center)+(270:2cm)$) {};

\coordinate (lv) at ($(center)+(90:2.6cm)$) {};
\coordinate (lu) at ($(center)+(270:2.4cm)$) {};

\tikzstyle{every node}=[inner sep=1pt]
\begin{scriptsize}
\node at (lv) {$\psi(v)$};
\node at (lu) {$\psi(u)$};
\end{scriptsize}

%v
\draw[thick] ([shift=(-20:2.1cm)]0,0) arc (-20:200:2.1cm);
%u
\draw[thick] ([shift=(160:1.9cm)]0,0) arc (160:380:1.9cm);

\draw[thick, white] (-3.0,-3)--(-3.0,-2.8);
\draw[thick, white] (3.0,3)--(3.0,2.8);

\end{tikzpicture}
%overlap
\begin{tikzpicture}[scale=0.5]
\coordinate (center) at (0,0) {};
\coordinate (v) at ($(center)+(90:2cm)$) {};
\coordinate (u) at ($(center)+(270:2cm)$) {};

\coordinate (lv) at ($(center)+(90:2.6cm)$) {};
\coordinate (lu) at ($(center)+(270:2.4cm)$) {};

\tikzstyle{every node}=[inner sep=1pt]
\begin{scriptsize}
\node at (lv) {$\psi(v)$};
\node at (lu) {$\psi(u)$};
\end{scriptsize}
\draw[thick] ([shift=(70:2.1cm)]0,0) arc (70:200:2.1cm);
\draw[thick] ([shift=(160:1.9cm)]0,0) arc (160:290:1.9cm);

\draw[thick, white] (-3.0,-3)--(-3.0,-2.8);
\draw[thick, white] (3.0,3)--(3.0,2.8);
\end{tikzpicture}

\caption{\label{fig:mutual_arc_position} From left to right:
$\psi(v)$ and $\psi(u)$ are disjoint, $\psi(v)$ contains $\psi(u)$, $\psi(v)$ is contained in $\psi(u)$,
$\psi(v)$ and $\psi(u)$ cover the circle, and $\psi(v)$ and $\psi(u)$ overlap.
}

\end{figure}

Let $G$ be a circular-arc graph with no universal vertices and no twins and let $\psi$ be a circular-arc model of $G$ on the circle $C$.
Let $(v,u)$ be a pair of distinct vertices in $G$.
We say that:
\begin{itemize}
 \item $\psi(v)$ and $\psi(u)$ are \emph{disjoint} if $\psi(v) \cap \psi(u) = \emptyset$,
 \item $\psi(v)$ \emph{contains} $\psi(u)$ if $\psi(v) \supsetneq \psi(u)$,
 \item $\psi(v)$ \emph{is contained} in $\psi(u)$ if $\psi(v) \subsetneq \psi(u)$,
 \item $\psi(v)$ and $\psi(u)$ \emph{cover the circle} if $\psi(v) \cup \psi(u) = C$,
 \item $\psi(v)$ and $\psi(u)$ \emph{overlap}, otherwise.
\end{itemize}
See Figure~\ref{fig:mutual_arc_position} for an illustration.

In so-called \emph{normalized models}, introduced by Hsu in \cite{Hsu95}, the relative positions of the arcs reflects the neighbourhood relation between the vertices of $G$, as follows.
\begin{definition}
\label{def:normalized_model}
Let $G$ be a circular-arc graph with no universal vertices and no twins.
A circular-arc model $\psi$ of $G$ is \emph{normalized} if for every pair $(v,u)$ of distinct vertices of $G$ the following conditions are satisfied:
\begin{enumerate}
 \item \label{item:u_v_disjoint} if $uv \notin E(G)$, then $\psi(v)$ and $\psi(u)$ are disjoint,
 \item \label{item:v_contains_u} if $N[u] \subsetneq N[v]$, then $\psi(v)$ contains $\psi(u)$,
 \item \label{item:u_contains_v} if $N[v] \subsetneq N[u]$, then $\psi(v)$ is contained in $\psi(u)$,
 \item \label{item:u_v_cover_the_circle} if $uv \in E(G)$, 
 $N[v] \cup N[u] = V(G)$, $N[w] \subsetneq N[v]$ for every $w \in N[v]\setminus N[u]$, and $N[w] \subsetneq N[u]$ for every $w \in N[u]\setminus N[v]$, then $\psi(v)$ and $\psi(u)$ cover the circle, 
 \item \label{item:u_v_overlap} If none of the above condition holds, then $\psi(v)$ and $\psi(u)$ overlap.
\end{enumerate}
Furthermore, for a pair $(v,u)$ of vertices from $G$, we say that $v$ \emph{contains} $u$, $v$ \emph{is contained} in $u$, $v$ and $u$ \emph{cover the circle}, and 
$v$ and $u$ \emph{overlap} if the pair $(v,u)$ satisfies the assumption of statement \eqref{item:v_contains_u},
\eqref{item:u_contains_v}, \eqref{item:u_v_cover_the_circle}, and \eqref{item:u_v_overlap}, respectively.
\end{definition}
Hsu~\cite{Hsu95} showed that every circular-arc model $\psi$ of $G$ can be turned into a normalized model $\phi$ by 
possibly extending some arcs of $\psi$. 
%This process, called as \emph{normalization procedure for $\psi$}, goes as follows: 
%whenever we find a pair of vertices $(v,u)$ in $G$ which violate one of the conditions \eqref{item:v_contains_u}, \eqref{item:u_contains_v}, or \eqref{item:u_v_cover_the_circle},
%we extend the arc $\psi(v)$ and/or the arc $\psi(u)$ so as the condition became valid
%(conditions \eqref{item:u_v_disjoint} and \eqref{item:u_v_overlap} are satisfied in any circular-arc model $\psi$ of~$G$).
%This procedure, performed in a particular order, leads to a normalized model of~$G$ --
%we refer to \cite{Hsu95} for more details.
Note the following property of $\psi$ and $\phi$: if $C$ is a subset of $V(G)$ which satisfies the Helly property in $\psi$, 
then $C$ satisfies the Helly property in the normalized model $\phi$.
%obtained by performing normalization procedure on $\psi$.

Usually, the notion of normalized models is extended on all circular-arc graphs $G$ 
by requiring $\psi(v)=C$ for any universal vertex $v$ in $G$ and $\psi(u)=\psi(v)$ for any
pair $(u,v)$ of twin vertices in $G$.

%Suppose $G$ is a circular-arc graph.
%A circular-arc model $\psi$ of $G$ is \emph{Helly} if the set $\{\psi(v): v \in V(G)\}$ satisfies the Helly property, which means that for every clique $C$ of $G$ we have $\bigcap_{c \in C} \phi(c) \neq \emptyset$.
%A circular-arc graph is \emph{Helly} if it admits a Helly circular-arc model.
%Lin and Schwarzfiter \cite{LinSchw06} showed the following property of circular-arc graphs (see \cite{DerKra22+} for an alternative proof).
%\begin{theorem}
%\label{thm:ca-helly-graphs}
%Suppose $G$ is a circular-arc graph. Then either every normalized model of $G$ is Helly or no model of $G$ is Helly.
%\end{theorem}
%Lin and Schwarzfiter \cite{LinSchw06} devised also a linear time algorithm recognizing Helly circular-arc graphs.

\section{Medusa graphs - normalized models, recognition}
\label{sec:medusa_graphs}
In this section we prove Theorem \ref{thm:recognition-medusa-graphs}. 
As for circular-arc graphs, we show that any $\mathcal{M}$-model of a graph $G$ 
can be turned into a model satisfying certain properties; such models will be called \emph{normalized}.
In particular, to test whether a graph $G$ admits an $\mathcal{M}$-model it suffices to
check whether $G$ admits a normalized $\mathcal{M}$-model.
To define normalized models for medusa graphs we need some preparation first.

For a graph $G$ (not necessarily a medusa graph) we define a partition of its vertex set $V(G)$ into a \emph{circle part} $V_C$ and a \emph{tree part} $V_T$ as follows.
We start by setting $V_C = \emptyset$.
We repeat one of the following operations as long as we can extend the set $V_C$:
\begin{itemize}
 \item if $C$ is a hole in $G$, add the vertices of $C$ to $V_C$,
 \item if $P$ is an induced path in $G$ joining two non-adjacent vertices from $V_C$, add the vertices of $P$ to $V_C$.
\end{itemize}
Finally, we set $V_T = V \setminus V_C$.

Let $T_1,\ldots,T_k$ be a partition of $V_T$ into connected components of $G[V_T]$.
Furthermore, let $N_C(T_i)$ be the neighbourhood of $T_i$ in the cycle part of $G$, that is:
$$N_C(T_i) = N(T_i) \cap V_C.$$

\begin{lemma}
\label{lem:tree_parts_are_chordal}
For every $i \in [k]$, $G[T_i \cup N_C(T_i)]$ is a chordal graph. 
Moreover, $N_C(T_i)$ is a clique in $G[V_C]$.
\end{lemma}
\begin{proof}
First, we prove that $N_C(T_i)$ is a clique in $G[V_C]$. 
Suppose for the sake of contradiction that there exist non-adjacent vertices $u,v\in N_C(T_i)$. 
Since $u,v\in N_C(T_i)$, there exist $p,q\in T_i$ that are neighbours of $u$ and $v$, respectively. 
Since $T_i$ is connected, there exists a path $P\subseteq T_i$ joining $p$ and $q$. 
Pick $p,q$ and $P$ such that $P$ is the shortest possible among all choices of $p,q,P$. 
Note that $P$ extended by $u$ and $v$ is an induced path in $G$. 
Therefore, $p$ is on an induced path joining two non-adjacent vertices from $V_C$, which implies $p$ would have been added to $V_C$. 
This is a contradiction since $p\in T_i$ and $T_i \cap V_C = \emptyset$.

Now we prove that for every $i \in [k]$, $G[T_i \cup N_C(T_i)]$ is a chordal graph. Suppose for the sake of contradiction that there is a hole $H$ in $G[T_i \cup N_C(T_i)]$. 
Since $H$ is a hole, we have $H\subseteq V_C$. 
Since $T_i\cap V_C=\emptyset$, we have $H\subseteq N_C(T_i)$. 
But $N_C(T_i)$ is a clique, so it cannot contain any holes, which leads to a contradiction.
\end{proof}

The next lemma captures the role of the sets $V_C$ and $V_T$ in description of $\mathcal{M}$-models of medusa graphs.
\begin{lemma}
\label{lem:medusa_graphs_normalized_model_existence}
Let $G$ be a medusa graph. Then:
\begin{enumerate}
\item In every $\mathcal{M}$-model $(M,\psi)$ of $G$ we have $\psi(v) \cap M_{\circle} \neq \emptyset$ for every $v \in V_C$.
\item There is an $\mathcal{M}$-model $(M',\phi)$ of $G$ such that $\phi(v) \cap M'_{\circle} = \emptyset$ for every $v \in V_T$.
\end{enumerate}
\end{lemma}
\begin{proof}
Let $(M,\psi)$ be an $\mathcal{M}$-model of $G$. 
First, we prove $\psi(v) \cap M_{\circle} \neq \emptyset$ for every $v \in V_C$. 
Suppose for the sake of contradiction that there is $v \in V_C$ such that $\psi(v) \cap M_O = \emptyset$. 
Consider two possible cases:
\begin{itemize}
\item vertex $v$ was added to $V_C$ because there is a hole $H$ in $G$ such that $v \in H$.
But then there is no way to model vertices of $H$ without introducing a chord, which is a contradiction.
\item vertex $v$ was added to $V_C$ because there is an induced path $P$ joining two non-adjacent vertices from $V_C$ such that $v\in P$. 
But then there is no way to model vertices of $P$ without introducing an additional edge, which is a contradiction.
\end{itemize}

Now we prove that there is an $M'$-model $\phi$ of $G$, where $M'$ is a unicyclic connected graph, 
such that $\phi(v) \cap M'_{\circle} = \emptyset$ for every $v \in V_T$. 
First, we show that there is a circular-arc model $\phi_C$ of $G[V_C]$ such that for every 
$i \in [k]$ the set $N_C(T_i)$ satisfies the Helly property.
Since $G$ is a medusa graph, $G$ has an $M$-model $\psi$ for some unicyclic graph $M$. 
We will denote $\psi(X)=\bigcup_{x\in X} \psi(x)$ for a set of vertices $X\subseteq V$. 
Note that:
\begin{itemize}
 \item for every $i \in [k]$ such that $\psi(T_i) \cap M_{\circle} = \emptyset$ 
 all the arcs from the set $\{\psi(v) \cap M_{\circle}: v \in N_C(T_i)\}$ cover the vertex from $M_{\circle}$ adjacent to the branch of $M$ containing the representation of $T_i$,
 \item the sets $\{\psi(T_i) \cap M_{\circle}: i \in [k]\}$ are pairwise disjoint as $T_1,\ldots,T_k$ are components of $G[V_T]$. 
 Moreover, if $\psi(T_i) \cap M_{\circle} \neq \emptyset$ then $\psi(v) \cap (\psi(T_i) \cap M_{\circle}) \neq \emptyset$ for every $v \in N_C(T_i)$.
\end{itemize}
Now, for every $v \in V_C$ we set
$$\phi_C(v) = \big{(}\psi(v) \cap M_{\circle}\big{)} \cup \bigcup \big{\{} \psi(T_i) \cap M_{\circle}: v \in N_C(T_i)\big{\}}.$$
Given the above properties one can easily check that $\phi_C$ satisfies the desired properties.
For $i \in [k]$ denote by $y_i$ the vertex of $M_{\circle}$ such that 
$y_i \in \bigcap_{v \in N_C(T_i)} \phi_C(v)$.
Since for every $i \in [k]$, $G[T_i \cup N_C(T_i)]$ is a chordal graph, 
there is a $F_i$-model $\psi_i$ of $G[T_i \cup N_C(T_i)]$ for some tree $F_i$. 
Due to the Helly property of subtrees of a tree, there is a vertex $x_i\in F_i$ such that $x_i\in \bigcap_{v\in N_C(T_i)} \psi_i(v)$ since $N_C(T_i)$ is a clique. 
Now, note that
$$\phi(v) = \left\{
\begin{array}{ll}
 \phi_C(v) &\text{if } v \in V_C \setminus \bigcup_{i=1}^k N_C(T_i), \\
 \phi_C(v) \cup \bigcup \big{\{} \psi_i(v): v \in N_C(T_i)\big{\}} &\text{if } v \in \bigcup_{i=1}^k N_C(T_i), \\
 \psi_i(v) &\text{if } v \in T_i \text{ for some $i \in [k]$.}\\
\end{array}
\right.
$$
is an $M'$-model of $G$, where $M'$ is a unicyclic graph that arises from $M_{\circle}$ by joining $x_i \in F_i$ with $y_i \in M_{\circle}$ with an edge $x_iy_i$, for $i \in [k]$.
One can easily check that $\phi$ satisfies the desired properties.
\end{proof}

In particular, Lemma~\ref{lem:medusa_graphs_normalized_model_existence} asserts that every medusa graph $G$ admits an~$\mathcal{M}$-model $(M,\phi)$ that satisfies:
\begin{itemize}
 \item $\phi(v) \cap M_O \neq \emptyset$ for every $v \in V_C$,
 \item $\phi(v) \cap M_O = \emptyset$ for every $v \in V_T$,
 \item $\{\phi(v) \cap M_O: v \in V_C\}$ is a normalized model of 
 circular-arc graph $G[V_C]$.
\end{itemize}
We call such an $\mathcal{M}$-model of $G$ \emph{normalized}.
Now we are ready to prove the theorem characterizing medusa graphs.
\begin{theorem}
\label{thm:medusa_graphs_characterization}
Let $G$ be a graph, let $V(G) = V_C \cup V_T$ be a partition of $V(G)$ into the circle part $V_C$ and the tree part $V_T$ of $G$, and let $T_1,\ldots,T_k$ be a partition of $V_T$ into connected components of $G[V_T]$.
Then: 
\begin{enumerate}
\item \label{item:medusa_graph_characterization} $G$ is a medusa graph if and only if $G[V_C]$ is a circular arc graph admitting a normalized circular-arc model $\phi$ in which every set $\bigcap_{v \in N_C[T_i]} \phi(v)$ is non-empty.
\item \label{item:helly_medusa_graph_characterization} $G$ is a Helly medusa graph if and only if $G[V_C]$ is a Helly circular arc graph.
\end{enumerate}
\end{theorem}
\begin{proof}
Given a (Helly) normalized $\mathcal{M}$-model $(M,\phi)$ of $G$, the model $\phi_C$ of $G[V_C]$ defined by 
$\phi_C(v) = \phi(v) \cap M_{\circle}$ satisfies all the properties required by statement \eqref{item:medusa_graph_characterization} (statement~\eqref{item:helly_medusa_graph_characterization}, respectively). 
On the other hand, given a (Helly) normalized circular-arc model $\phi_C$
of $G[V_C]$, the $\mathcal{M}$-model $(M',\phi)$ of $G$ constructed as in the proof of Lemma ~\ref{lem:medusa_graphs_normalized_model_existence} proves that $G$ is a (Helly) medusa graph.
\end{proof}

\noindent The above theorem focuses our attention to the Helly Cliques problem.
Recall that in this problem, for a circular-arc graph $G$ and cliques $C_1,\ldots,C_k$ of $G$
we want to test if $G$ admits a circular-arc model in which all the cliques
$C_1,\ldots,C_k$ satisfy the Helly property.
Note that a yes-instance $G,C_1,\ldots,C_k$ to the Helly Cliques problem 
can be always witnessed by a normalized model.

\begin{lemma}
\label{lem:medusa-graphs-rec-and-helly-cliques-are-equivalent}
Recognition of medusa graphs is poly-time equivalent to the Helly Cliques problem.
\end{lemma}
\begin{proof}
First, we show that if there is a polynomial algorithm solving the Helly Cliques problem, 
then there is a polynomial algorithm solving recognition of medusa graphs. 
Let $G$ be the graph on the input. 
Let $V(G)=V_C \cup V_T$ be the partition into the circle part $V_C$ and the tree part $V_T$, 
and let $T_1,\ldots,T_k$ be a partition of $V_T$ into connected components of $G[V_T]$. 
Due to Theorem~\ref{thm:medusa_graphs_characterization}.\eqref{item:medusa_graph_characterization}, 
there is a normalized circular-arc model of $G[V_C]$ in which the cliques $N_C(T_i)$ have the Helly property for every $i\in [k]$ if and only if $G$ admits an $\mathcal{M}$-model (i.e.\ $G$ is a medusa graph).
Due to the above, we conclude that $G$ is a medusa graph if and only if $G[V_C]$ and $N_C(T_1), \ldots, N_C(T_k)$ is a yes-instance of the Helly Cliques problem. 
Note that computing $G[V_C]$ and the cliques $N_C(T_1), \ldots, N_C(T_k)$ can easily be done in polynomial time.
 
Now we show that if there is a polynomial algorithm solving recognition of medusa graphs, 
then there is a polynomial algorithm solving the Helly Cliques problem. 
Let $G$ be the graph on the input and let $C_1, \ldots, C_k$ be the cliques on the input. 
We assume $G$ has no universal vertices and no twins as every instance 
of the Helly Cliques problem can be easily transformed to an equivalent one which satisfies these conditions.
We define a graph $G'$ from $G$ the following way:
 \begin{enumerate}
    \item For every $i\in [k]$ add a vertex $c_i$ to $V(G')$.
    \item For every $v\in V$ add four vertices $v, v_1, v_2, v_3$ to $V(G')$.
    Let $X_v = \{v,v_3\}$ and let $Y_v = \{v_1,v_2\}$.
    \item For every $i\in [k]$ and for each $v\in C_i$ add an edge $vc_i$ to $E(G')$.
    \item For every $v\in V$ add four edges $vv_1, v_1v_2, v_2v_3, v_3v_4$ to $E(G')$.
    \item For every $v\in V$ and for every $u\in V$ such that $v$ and $u$ overlap
    join with an edge in $G'$ any vertex in $X_u \cup Y_u$ with any vertex in $X_v \cup Y_v$.
    \item For every $v\in V$ and for every $u\in V$ such that $u$ and $v$ are disjoint 
    join with an edge in $G'$ any vertex in $X_u$ with any vertex in $Y_v$ 
    and any vertex in $Y_u$ and any vertex in $X_v \cup Y_v$.
    \item For every $v\in V$ and for every $u\in V$ such that $u$ and $v$ cover the circle
    join with an edge in $G'$ any vertex in $X_u$ with any vertex in $X_v \cup Y_v$ 
    and any vertex in $Y_u$ with any vertex in $X_v$.
    \item For every $v\in V$ and for every $u\in V$ such that $u$ is contained in $v$ 
    join with an edge in $G'$ any vertex in $X_u$ with any vertex in $X_v$ 
    and any vertex in $Y_u$ with any vertex in $X_v \cup Y_v$.
 \end{enumerate}
 Note that the construction of $G'$ can be easily done in polynomial time. 
 We claim that $G'$ is a medusa graph if and only if there exists a circular-arc model of $G$ in which the cliques $C_1, \ldots, C_k$ have the Helly property. 
 
 In one direction, suppose $G'$ is a medusa graph. 
 Let $V'_C$ be the circle part of $G'$ and let $V'_T$ be the tree part of $G'$, and
 let $(M,\phi)$ be a normalized $\mathcal{M}$-model of $G'$.
 Note that for every $v\in V$, $v,v_1,v_2,v_3$ form an induced cycle in $G'$ and therefore $v\in V'_C$. 
 Note that for every $i\in [k]$ we have $N(c_i)=C_i$ and, since $C_i$ is a clique, 
 we have $c_i \in V'_T$. 
 Since $(M,\phi)$ is a normalized $\mathcal{M}$-model we conclude 
 that $\phi$ restricted to $M_{\circle}$ is a circular-arc model of $G$ 
 in which for every $i\in [k]$ the clique $C_i$ has the Helly property.
 
 In the other direction, suppose there is a normalized circular-arc model $\phi$ of $G$ in which for every $i\in [k]$ the clique $C_i$ has the Helly property. 
 Note that $V'_C = \bigcup_{v\in V} \{ v, v_1, v_2, v_3\}$ and $V'_T=\{c_1, \ldots , c_k\}$. 
 Note that we can obtain a circular-arc model $\phi'$ of $G'[V'_C]$ from $\phi$ 
 by placing the endpoints of the arcs of $v_1, v_2, v_3$ next to the endpoints of $\phi(v)$ -- see Figure \ref{fig:reduction} for an illustration. 
 Therefore, due to Lemma~\ref{lem:medusa_graphs_normalized_model_existence} 
 we conclude that $G'$ admits an $\mathcal{M}$-model (i.e.\ $G'$ is a medusa graph).
\end{proof}

\begin{figure}[htp!]
\begin{tikzpicture}[scale=0.5]
\coordinate (center) at (0,0) {};
\coordinate (v) at ($(center)+(90:2cm)$) {};
\coordinate (u) at ($(center)+(270:2cm)$) {};

\coordinate (lv) at ($(center)+(180:2.6cm)$) {};
\coordinate (lv3) at ($(center)+(180:1.6cm)$) {};
\coordinate (lv1) at ($(center)+(0:1.6cm)$) {};
\coordinate (lv2) at ($(center)+(0:2.65cm)$) {};

\tikzstyle{every node}=[inner sep=1pt]
\begin{scriptsize}
\node at (lv) {$v$};
\node at (lv1) {$v_1$};
\node at (lv2) {$v_2$};
\node at (lv3) {$v_3$};
\end{scriptsize}

%v
\draw[thick] ([shift=(96:2.3cm)]0,0) arc (96:272:2.3cm);
%v3
\draw[thick] ([shift=(88:2.0cm)]0,0) arc (88:264:2.0cm);

%v1
\draw[thick,red] ([shift=(276:2.3cm)]0,0) arc (276:452:2.3cm);
%v2
\draw[thick,red] ([shift=(268:2.0cm)]0,0) arc (268:444:2.0cm);

\draw[thick, white] (-3.0,-3)--(-3.0,-2.8);
\draw[thick, white] (3.0,3)--(3.0,2.8);

\end{tikzpicture}
%overlap
\begin{tikzpicture}[scale=0.5]
\coordinate (center) at (0,0) {};
\coordinate (v) at ($(center)+(90:2cm)$) {};
\coordinate (u) at ($(center)+(270:2cm)$) {};

\coordinate (Yv) at ($(center)+(90:2.7cm)$) {};
\coordinate (Xv) at ($(center)+(270:2.7cm)$) {};
\coordinate (Xu) at ($(center)+(90:1.6cm)$) {};
\coordinate (Yu) at ($(center)+(270:1.6cm)$) {};

\tikzstyle{every node}=[inner sep=1pt]
\begin{scriptsize}
\node at (Xu) {$X_u$};
\node at (Xv) {$X_v$};
\node at (Yu) {$Y_u$};
\node at (Yv) {$Y_v$};
\end{scriptsize}
\draw[very thick] ([shift=(47:2.0cm)]0,0) arc (47:223:2.0cm);
\draw[very thick,red] ([shift=(225:2.0cm)]0,0) arc (225:405:2.0cm);
\draw[very thick] ([shift=(137:2.3cm)]0,0) arc (137:313:2.3cm);
\draw[very thick,red] ([shift=(315:2.3cm)]0,0) arc (315:495:2.3cm);

\draw[thick, white] (-3.0,-3)--(-3.0,-2.8);
\draw[thick, white] (3.0,3)--(3.0,2.8);
\end{tikzpicture}
%disjoint
\begin{tikzpicture}[scale=0.5]
\coordinate (center) at (0,0) {};
\coordinate (v) at ($(center)+(90:2cm)$) {};
\coordinate (u) at ($(center)+(270:2cm)$) {};

\coordinate (Xu) at ($(center)+(270:2.7cm)$) {};
\coordinate (Yu) at ($(center)+(90:2.7cm)$) {};
\coordinate (Xv) at ($(center)+(90:1.6cm)$) {};
\coordinate (Yv) at ($(center)+(270:1.6cm)$) {};

\tikzstyle{every node}=[inner sep=1pt]
\begin{scriptsize}
\node at (Xu) {$X_u$};
\node at (Xv) {$X_v$};
\node at (Yu) {$Y_u$};
\node at (Yv) {$Y_v$};
\end{scriptsize}

%u
\draw[very thick] ([shift=(32:2.0cm)]0,0) arc (32:148:2.0cm);
\draw[very thick,red] ([shift=(150:2.0cm)]0,0) arc (150:390:2.0cm);
%u
\draw[very thick] ([shift=(212:2.3cm)]0,0) arc (212:328:2.3cm);
\draw[very thick,red] ([shift=(330:2.3cm)]0,0) arc (330:570:2.3cm);

\draw[thick, white] (-3.0,-3)--(-3.0,-2.8);
\draw[thick, white] (3.0,3)--(3.0,2.8);

\end{tikzpicture}
%cover the circle
\begin{tikzpicture}[scale=0.5]
\coordinate (center) at (0,0) {};
\coordinate (v) at ($(center)+(90:2cm)$) {};
\coordinate (u) at ($(center)+(270:2cm)$) {};

\coordinate (Yu) at ($(center)+(270:2.7cm)$) {};
\coordinate (Xu) at ($(center)+(90:2.7cm)$) {};
\coordinate (Yv) at ($(center)+(90:1.6cm)$) {};
\coordinate (Xv) at ($(center)+(270:1.6cm)$) {};

\tikzstyle{every node}=[inner sep=1pt]
\begin{scriptsize}
\node at (Xu) {$X_u$};
\node at (Xv) {$X_v$};
\node at (Yu) {$Y_u$};
\node at (Yv) {$Y_v$};
\end{scriptsize}

%u
\draw[very thick,red] ([shift=(32:2.0cm)]0,0) arc (32:148:2.0cm);
\draw[very thick] ([shift=(150:2.0cm)]0,0) arc (150:390:2.0cm);
%u
\draw[very thick,red] ([shift=(212:2.3cm)]0,0) arc (212:328:2.3cm);
\draw[very thick] ([shift=(330:2.3cm)]0,0) arc (330:570:2.3cm);

\draw[thick, white] (-3.0,-3)--(-3.0,-2.8);
\draw[thick, white] (3.0,3)--(3.0,2.8);

\end{tikzpicture}
%contains
\begin{tikzpicture}[scale=0.5]
\coordinate (center) at (0,0) {};
\coordinate (v) at ($(center)+(90:2cm)$) {};
\coordinate (u) at ($(center)+(270:2cm)$) {};

\coordinate (Xv) at ($(center)+(90:2.7cm)$) {};
\coordinate (Yv) at ($(center)+(270:2.7cm)$) {};
\coordinate (Xu) at ($(center)+(90:1.6cm)$) {};
\coordinate (Yu) at ($(center)+(270:1.6cm)$) {};

\tikzstyle{every node}=[inner sep=1pt]
\begin{scriptsize}
\node at (Xv) {$X_u$};
\node at (Xu) {$X_v$};
\node at (Yv) {$Y_u$};
\node at (Yu) {$Y_v$};
\end{scriptsize}

%v
\draw[very thick] ([shift=(2:2.0cm)]0,0) arc (2:178:2.0cm);
\draw[very thick,red] ([shift=(180:2.0cm)]0,0) arc (180:360:2.0cm);
%u
\draw[very thick] ([shift=(47:2.3cm)]0,0) arc (47:133:2.3cm);
\draw[very thick,red] ([shift=(135:2.3cm)]0,0) arc (135:405:2.3cm);

\draw[thick, white] (-3.0,-3)--(-3.0,-2.8);
\draw[thick, white] (3.0,3)--(3.0,2.8);

\end{tikzpicture}

\caption{\label{fig:reduction}}

\end{figure}

We can summarize the section with the following theorem (which extends Theorem~\ref{thm:recognition-medusa-graphs}).
\begin{theorem} \
\begin{enumerate}
\item Recognition of medusa graphs is \NP-complete. 
\item Recognition of medusa graphs parameterized by the number of components in the tree part of the input graph is \FPT.
\item Recognition of Helly medusa graphs is poly-time solvable. 
\end{enumerate}
\end{theorem}
\begin{proof}
The statments of the theorem follow from Lemma~\ref{lem:medusa-graphs-rec-and-helly-cliques-are-equivalent},
from the fact that the Helly Cliques problem is \NP-complete \cite{AgaZem2022} and can be solved in time $2^{O(k \log k )}poly(n)$ \cite{DerKra22+}, 
and from the fact that Helly circular-arc graphs recognition can be solved in linear-time \cite{LinSchw06}.
\end{proof}

\section{Recognition of $M$-graphs for a fixed unicyclic graph $M$}
\label{sec:M-models}
In the rest of this paper we deal with the recognition problem
of $M$-graphs, where $M$ is a fixed unicyclic graph.

\subsection{Saturated $M$-graphs and saturated $M$-models}
Let $M$ be a unicyclic graph.
We say that a graph $G$ is \emph{saturated $M$-graph} 
if $G$ has an $M$-model and has no $M^*$-model for any minor $M^* \neq M$ of $M$.
\begin{observation}
Suppose $M^*$ is a minor of $M$. 
If $G$ has an $M^*$-model, then $G$ has an $M$-model.
\end{observation}
Note that $G$ is an $M$-graph iff $G$ is an $M^*$-graph for some minor $M^*$ of 
$M$ or $G$ is a saturated $M$-graph.
Hence, to show Theorems~\ref{thm:lollipop-graphs} and~\ref{thm:strongly-cyclic-graphs},
we can restrict our attention to the recognition problems for saturated $L$-graphs 
and saturated strongly-cyclic $M$-graphs for any fixed unicyclic graph $M$, respectively.

Suppose $M$ is a unicyclic graph and $G$ is an $M$-saturated graph.
Suppose $V_C$, $V_T$ are as defined in the previous section, and let
$\mathcal{T}$ be the set of all maximal cliques from all chordal graphs $G[T \cup N_C(T)]$, where $T$ runs 
over all connected components of $G[V_T]$.  
Since $G$ has no $M^*$-model for any minor $M^*$ of $M$, we can
transform an $M$-model $\phi$ of $G$ into so-called \emph{saturated} $M$-model of $G$.

Let $\mathcal{E}$ be the set of all edges adjacent to exactly one vertex of $M_{\circle}$.
Let $(M^{\phi},\phi)$ be an $M$-model of $G$ and let $P(e)$ for every $e \in \mathcal{E}$ 
be the vertex from $e^{\phi} \setminus M^{\phi}_{\circle}$ closest to $M^{\phi}_{\circle}$, where $e^\phi$
contains the vertices of $M^{\phi}$ resulting from the subdivision of $e$ together with the ends of $e$.
Model $(M^{\phi},\phi)$ is \emph{saturated} if for every $e \in \mathcal{E}$ the set $$C(e) = \{v \in V(G): P(e) \in \phi(v)\}$$ 
satisfies the following conditions:
\begin{itemize}
\item $C(e)$ contains a vertex $v_e \in V(G)$ such that $\phi(v_e) \cap M^{\phi}_{\circle} = \emptyset$,
\item $C(e)$ is a maximal clique from the set $\mathcal{T}$.
\end{itemize}
The clique $C(e)$ satisfying the above properties is called \emph{the clique closest to $M_{\circle}$ on the edge~$e$ in the model~$(M^{\phi},\phi)$}.
Note that for every $e \in \mathcal{E}$ the vertex $v_e$ asserted by the first condition belongs to $V_T$ as $\phi(v_e) \cap M^{\phi}_{\circle} = \emptyset$.

\begin{claim}
\label{claim:saturated_model_existence}
Let $M$ be a unicyclic graph. 
Every saturated $M$-graph $G$ admits a saturated $M$-model.
\end{claim}
\begin{proof}
Suppose $\phi = (M^{\phi}, \phi)$ is an $M$-model of $G$.
Let $e \in \mathcal{E}$ and let $Q(e),R(e)$ be the vertices of $M$ such that $e = Q(e)R(e)$ 
and $Q(e) \in M_{\circle}$.
Since $G$ is $M$-saturated, there is a vertex $v \in V$ such that
$\phi(v) \cap e^{\phi} \neq \emptyset$ and $Q(e) \notin \phi(v)$.
Indeed, assuming otherwise, we can show that $G$ has an $M^*$-model, 
where $M^*$ is a minor of $M$ arisen from $M$
by contracting the edge $e$ into the vertex $Q(e)R(e)$.
To see this, we first extend every set $\phi(v)$ such that $Q(e) \in \phi(v)$ 
by the set $e^{\phi}$. 
Our assumption asserts that $\phi$ is still an $M$-model of $G$ and 
for every $v \in V$ we have either $e^{\phi} \subseteq \phi(v)$ or $e^{\phi} \cap \phi(v) = \emptyset$.
Then, we transform $\phi$ into an $M^*$-model by 
replacing every set $\phi(v)$ containing $e^{\phi}$ by the set 
$(\phi(v)  \setminus e^{\phi}) \cup \{P(e)Q(e)\}$.
Now, let $P(e)$ be a vertex from $e^{\phi}$ and $v_e$ be a vertex from $V_T$ such that 
$P(e) \in \phi(v)$, $\phi(v) \cap M^{\phi}_{\circle} = \emptyset$, and such that $P(e)$ is the closest vertex to $M^{\phi}_{\circle}$ among all choices of 
$P(e)$ and $v_e$ which satisfy these conditions.
Suppose $v_e \in T$ for a connected component $T$ in $G[V_T]$.
Next, we modify $(M^{\phi},\phi)$ so as the set $C(e) = \{u \in V: P(e) \in \phi(u)\}$ 
is a maximal clique in the graph $G[T \cup N_C(T)]$ (which means $C(e) \in \mathcal{T}$).
Note that we have $v_e \in C(e)$.
Suppose $C(e)$ is not maximal in $G[T \cup N_C(T)]$.
It means that the set
$$D = \Big{\{}v' \in V(G): P(e) \notin \phi(v') \text{ and } \{v'\} \cup C(e) \text{ is a clique in $G[T \cup N_C(T)]$}
\Big{\}}$$
is non-empty.
By the choice of $P(e)$ and by $v(e) \in C(e)$, we have that $D \subseteq T$ and that $\bigcup \phi(D)$ is entirely contained in the tree branch of $M^{\phi}$ rooted at~$P(e)$.
Then, we extend every set $\phi(v')$ for $v' \in D$ (in any order) 
towards the point $P(e)$ as long as we keep $\phi$ a model of $G$.
Clearly, we obtain $P(e) \in \phi(v')$ for some $v' \in D$, thus extending the clique $C(e)$ by the vertex $v'$.
Applying the above step possibly many times, we finally get a model $\phi$ 
such that $C(e)$ is a maximal clique in $G[T \cup N_C(T)]$.

We proceed analogously for every edge $e \in \mathcal{E}$, 
thus defining $P(e)$, $v_e$, and $C(e)$ for every edge $e \in \mathcal{E}$.
Now, let $v$ be any vertex from $V(G) \setminus \bigcup \{ C(e): e \in \mathcal{E}\}$ 
such that $\phi(v) \cap M_{\circle} \neq \emptyset$.
By the choice of $P(e)$ we can replace $\phi(v)$ by the set $\phi(v) \cap M_{\circle}$,
keeping $\phi$ an $M$-model of $G$.
Finally, we can modify $(M^{\phi},\phi)$ (by contracting all the vertices between $Q(e)$ and $P(e)$ in $M^{\phi}$) 
to get the model $(M^{\phi},\phi)$ in which $P(e)$ is the vertex closest to $M^\phi_{\circle}$ in the set $e^{\phi} \setminus M^{\phi}_{\circle}$.
\end{proof}
The next claim follows easily from the definition of saturated $M$-models.
Recall that $C(e)$ denotes the clique closest to $M_{\circle}$ on the edge $e \in \mathcal{E}$.
\begin{claim}
\label{claim:saturated_models_components}
Let $\phi$ be a saturated $M$-model of $G$.
Then, every connected component $F$ of the graph $G[V \setminus \bigcup \{C(e): e \in \mathcal{E}\}]$
satisfies $$\text{either}\quad \bigcup \phi(F) \subseteq M_{\circle} \quad \text{or} \quad \bigcup \phi(F) \cap M_{\circle} = \emptyset.$$
\end{claim}

\section{Recognizing strongly cyclic $M$-graphs}
\label{sec:strongly-cyclic-graphs}

In this section we prove Theorem~\ref{thm:strongly-cyclic-graphs}, that is, 
for every fixed unicyclic graph $M$ we show a poly-time algorithm testing whether 
an input strongly cyclic graph $G$ is an $M$-graph.
By remarks made in the previous section, to complete the task it suffices to show a poly-time algorithm testing whether $G$ is a saturated $M$-graph.

Let $G$ be a graph, let $V_C,V_T,\mathcal{T}$ be as defined as in the previous section, and let  $N_C(V_T) = V_C \cap N(V_T)$.
Graph $G$ is said to be \emph{strongly cyclic} if
$G\big{[}V \setminus \big{(}V_T \cup N_C(V_T)\big{)}\big{]}$
contains a hole.

\begin{proof}[Proof of Theorem \ref{thm:strongly-cyclic-graphs}]
Suppose $G$ is a strongly-cyclic graph given on the input.
By the results of Section~\ref{sec:M-models}, 
we need to check whether $G$ is a saturated $M$-graph.
Let $\mathcal{E}$ be the set of all tree edges of $M$ attached 
to $M_{\circle}$ and let $P_1,\ldots,P_k$ be the circular order of all vertices of degree $\geq 3$  
from $M_{\circle}$.
Finally, let $M_{T}$ be a forest that arises by subdividing every edge 
$e \in \mathcal{E}$ by a vertex $Q(e)$ and then by deleting the vertices from $M_{\circle}$.

Suppose for a while that $G$ admits a saturated $M$-model $\phi$.
For every $e \in \mathcal{E}$ let $C(e)$ be the maximal clique from $\mathcal{T}$ 
which is closest to $M_{\circle}$ on the edge $e$ in $\phi$.
Clearly, since $C(e)$ in $\mathcal{T}$, we have
$$\bigcup \left\{ C(e): e \in \mathcal{E}\right\} \subseteq V_T \cup N_C(V_T).$$
Consider the connected components of the graph $G[V \setminus \bigcup \left\{ C(e): e \in \mathcal{E}\right\}]$.
Since $G$ is strongly cyclic, the inclusion above asserts  $G\left[V \setminus \bigcup \left\{ C(e): e \in \mathcal{E} \right\}\right]$ has a component
containing a hole.
Since any hole must be represented by the sets that cover $M_{\circle}$, 
we must have exactly one such a component, say $CA$.
Since $\phi$ is $M$-saturated, Claim~\ref{claim:saturated_models_components} yields
$\bigcup \phi(CA) \subseteq M_{\circle}$.
By Claim~\ref{claim:saturated_models_components}, for every node $P_i$ and every edge $e \in \mathcal{E}$ adjacent to $P_i$ we have
$$\{v \in C(e): \phi(v) \text{ covers } P_i \} = \{v \in C(e): v \text{ has a neighbour in } CA \}.$$
For every $e \in \mathcal{E}$ let $C'(e) = \{v \in C(e): v \text{ has a neighbour in } CA \}$ 
and for every $i \in [k]$ let 
$C'_i = \bigcup \{C'(e): e \in \mathcal{E} \text{ is adjacent to $P_i$}\}$.
Note that:
\begin{itemize}
 \item for every $v \in V$, $\phi(v)$ intersects $M_{\circle}$ iff $v \in (CA \cup \bigcup_{i=1}^k C'_i)$,
 \item for every $i \in [k]$, $C'_i$ is a clique and $P_i \in \bigcap \phi(C'_i)$.
\end{itemize}
In particular, $G\big{[}CA \cup \bigcup_{i=1}^k C'_i\big{]}$ admits a circular-arc model on $M_{\circle}$ such that for every $i \in [k]$
the clique $C'_i$ is represented by arcs covering the node $P_i$.
Note also that the graph $G[V \setminus CA]$
has an $M_T$-model such that for every $e \in \mathcal{E}$ 
we have $Q(e) \in \bigcap \phi(C(e))$.

Given the above description, we may easily prove that
Algorithm~\ref{alg:saturated-M-model} correctly tests whether a strongly cyclic graph $G$ is $M$-saturated.
Roughly speaking, the algorithm iterates over all possible assignments of 
the maximal cliques from $\mathcal{T}$ to the edges of~$\mathcal{E}$ and tests whether any such assignment $C$ can be extended to a saturated $M$-model of $G$ 
in which for every $e \in \mathcal{E}$ the clique closest to $M_{\circle}$ 
on the edge $e$ is $C(e)$.
See Algorithm~\ref{alg:saturated-M-model} for the details.

\begin{algorithm}
\caption{Testing whether a strongly cyclic graph $G$ admits a saturated $M$-model}
\label{alg:saturated-M-model}
\begin{algorithmic}[1]
\State $\mathcal{F} = 
\begin{array}{c}
\text{the set of all injective functions from $\mathcal{E}$ to $\mathcal{T}$}
\end{array}$
\For {\text{\textbf{every} $C$ in $\mathcal{F}$}}
    \If{$G\left[V \setminus \bigcup \left\{C(e): e \in \mathcal{E}\right\}\right]$ \text{has exactly one non-chordal component $CA$}}
    \For {$e \in \mathcal{E}$}
        \State {$C'(e) = \{v \in C(e): v \text{ has a neighbour in }CA\}$}
    \EndFor
    \For {$i \in [k]$}
        \State {$C_i = \bigcup \{ C'(e): e \in \mathcal{E} \text{ is adjacent to }P_i\}$}
    \EndFor
    \smallskip
    \If{ \label{cond:saturated-M-model} $\left\{\begin{array}{c}
                \text{$G[CA \cup \bigcup_{i=1}^k C_i]$ admits a circular-arc model $\phi$ on $M_{\circle}$} \\
                \text{such that for every $i \in [k]$ we have $P_i \in \bigcap \phi(C_i)$} \\
            \text{\ \ \textbf{and}} \\
                \text{$G[V \setminus CA]$ has an $M_{T}$-model $\phi$ such that for every $e \in \mathcal{E}$} \\
                \text{we have $Q(e) \in \bigcap \phi(C(e))$}
        \end{array}\right.$}
        \State{\textbf{accept} $G$}
    \EndIf
	\EndIf
\EndFor
\State{\textbf{reject} $G$}
\end{algorithmic}
\end{algorithm}

Clearly, to test the first condition from Line~\ref{cond:saturated-M-model} of 
Algorithm~\ref{alg:saturated-M-model} we use poly-time algorithm of A{\u{g}}ao{\u{g}}lu {\c{C}}a{\u{g}}{\i}r{\i}c{\i}, 
Derbisz, Gutowski, and Krawczyk \cite{ADGK22+} for the Helly Cliques with Given Intersection Points problem, 
and to test the second condition we can use poly-time algorithm of Chaplick, T{\"{o}}pfer, Voborn{\'{\i}}k, and Zeman recognizing $T$-graphs~\cite{CTVZ21}.
\end{proof}

\section{Recognizing lollipop graphs}
\label{sec:lollipop_graphs}

This section develops a poly-time algorithm for recognizing saturated lollipop graphs.
Let $L$ be a lollipop graph, $P$ be the only node that has degree $3$ in $L$,
let $L_\circle \subseteq V(L)$ be the circle part of $L$
and $L_\stick \subseteq V(L)$ be the stick part of $L$.
We assume $P \in L_{\circle}$ and $P \in L_{\stick}$.

Suppose $G$ is an input graph.
For a maximal clique $C$ of $G$, an $L$-model $\phi$ of $G$ is called \emph{$C$-centered} if $C = \{v \in V(G): P \in \phi(v)\}$.
To test whether $G$ is a saturated $L$-graph we use as a black-box an algorithm solving the following problem: 
\medskip

\noindent \textbf{Centered Lollipop Graph Recognition:} \\
\begin{tabular}{rl}
\textbf{Input:}& A graph $G$ and a maximal clique $C \subseteq V(G)$, \\
\textbf{Output:}& \textbf{YES} if $G$ admits a $C$-centered $L$-model.
\end{tabular}

\medskip
\noindent A poly-time algorithm for the above problem is shown in the next section.

%The first step is to output candidates for the maximal cliques $C$
%	that lies on the node $P$, the only onde
%
%\ \\
%
%In this section we show a poly-time algorithm recognizing lollipop graphs.
%Our algorithm uses as a black-box a poly-time algorithm testing whether 
%an input graph $G$ admits a \emph{$C$-centered $L$-model}, that is, 
%an $L$-model $\phi$ of $G$ such that $C = \{v \in V(G): P \in \phi(v)\}$, where 
%$P$ is the only node of degree $3$ in $L$. 
%Figure~\ref{fig:example-C-centered-L-model} shows an example of a $C$-centered $L$-model $\phi$ of some $L$-graph $G$, 
%where $C = \{a,b,c\}$ and $\{I_1,\dots,I_7\}$ are the connected components of $G[V \setminus C]$.

%\subsection{Determining Candidates for a Centered Model}

%This section gives an algorithm
%	that outputs polynomial many candidates $C$
%	for the test whether $G$ is $C$-centered $L$-graph.
%
%\begin{lemma}
%\label{lemma:poly:candidates:C}
%There is a poly-time algorithm,
%	that given a graph $G$,
%	that outputs polynomial many candidates $C \subseteq V(G)$
%	such that $G$ is an $L$-graph if and only if $G$ is a $C$-centered $L$-graph for one output candidate~$C$.
%\end{lemma}

\medskip

Suppose $G$ is a saturated lollipop graph and $\phi$ is a saturated $L$-model of $G$.
Let $C$ be the maximal clique from $\mathcal{T}$ closest
to~$L_{\circle}$ on the edge $L_{\stick}$ in $\phi$.
Let $\mathcal{I}$ denote the set of all connected components of $G[V \setminus C]$.
By Claim~\ref{claim:saturated_models_components} every component $I$ from $\mathcal{I}$ 
satisfies either $\bigcup \phi(I) \subseteq L_{\circle}$ or 
$\bigcup \phi(I) \cap L_{\circle} = \emptyset$.
Clearly, if for every $I \in \mathcal{I}$ the node $P$ is not covered by $\bigcup \phi(I)$,
then $\phi$ can be easily turned into $C$-centered model.
So, let $I \in \mathcal{I}$ be such that $P \in \bigcup \phi(I)$.
Let $C' = \{v \in C: v \text{ has a neighbour in }I\}$.
Note that $C' = \{v \in C: P \in \phi(v) \}$ by $P \in \bigcup \phi(I)$ and by $\bigcup \phi(I) \subseteq L_{\circle}$.
Since $G$ is saturated, we fall into two cases:
\begin{itemize}
 \item $I$ is the only component from $\mathcal{I}$ such that $\bigcup \phi(I) \subseteq L_{\circle}$.
 In this case $G[I]$ can induce a circular-arc graph in $G$.
 \item There is some other component $I'$ in $\mathcal{I}$ such that $\bigcup \phi(I') \subseteq L_{\circle}$.
 In this case all the components from $\mathcal{I}$ induce interval graphs in $G$. 
\end{itemize}
Suppose the first case holds.
Note that:
\begin{itemize}
 \item the graph $G[I \cup C']$ admits a circular-arc model on $L_{\circle}$
 such that $P \in \bigcap_{c \in C'} \phi(c)$,
 \item the graph $G[V \setminus I]$ admits an interval model 
 with $C$ as the leftmost maximal clique.
\end{itemize}

Consider the second case.
Note that $C' \neq \emptyset$ as otherwise $\phi$ can be easily turned into $C$-centered $L$-model. 
Next, for every maximal clique $D$ in $G[I]$ let
$$C(D) = C' \cup \{v \in D: v \text{ is adjacent to every vertex in $C'$}\}.$$
We show that there is a maximal clique $D$ in $G[I]$ such that either $\phi$ is $C(D)$-centered or $\phi$ can be can be transformed into $C(D)$-centered model.
Consider the following cases:
\begin{itemize}
 \item $P$ is covered by a maximal sector $S$ of $\phi(I)$.
 Then, $\phi$ is $C(D)$-centered, where $D$ is the maximal clique of $G[I]$ 
 corresponding to the sector $S$.
 \item $P$ is between two maximal sectors $S_1$ and $S_2$ of $\phi(I)$.
 Suppose $S_1$ and $S_2$ correspond to maximal cliques $D_1$ and $D_2$ of $G[I]$.
 In particular, $P \in \phi(v)$ for every $v \in D_1 \cap D_2$.
 Consider the sets $D_1 \setminus D_2$ and $D_2 \setminus D_1$ and note that 
 $P \notin \bigcup \phi(D_1 \setminus D_2)$ or $P \notin \bigcup \phi(D_2 \setminus D_1)$.
 Suppose that $P \notin \bigcup \phi(D_2 \setminus D_1)$.
 Then, we can transform $\phi$ so as $\phi$ is $C(D_1)$-centered.
 For this purpose, for every $v \in C(D_1) \setminus C'$ such that $P \notin \phi(v)$
 we extend $\phi(v)$ towards $P$ in order to get $P \in \phi(v)$.
 The property $P \notin \bigcup \phi(D_2 \setminus D_1)$ asserts $\phi$ keeps to be a model of $G$.
 The second case is symmetric.
 \item $P$ is to the left (to the right) of all maximal sectors in $\phi(I)$.
 Suppose $D$ is a maximal clique of $G[I]$ whose sector is closest to the node $P$.
 Clearly, we can transform $\phi$ (using the same ideas as above) 
 to $C(D)$-centered model of $G$.
\end{itemize}
Finally, note that $C(D)$ is a maximal clique in $G$ as $C(D)$ can not be extended by a vertex from $C \setminus C'$ 
as well as by a vertex from $I \setminus C(D)$.

Given the above description one can easily check that Algorithm~\ref{alg:recognizing-lollipop-graphs} properly tests whether $G$ admits a saturated $L$-model.
Also, one can easily verify Algorithm~\ref{alg:recognizing-lollipop-graphs} works in polynomial time.
\begin{algorithm}
\caption{Testing whether $G$ admits a saturated $L$-model}
\label{alg:recognizing-lollipop-graphs}
\begin{algorithmic}[1]
\For {\text{\textbf{every} $C$ in $\mathcal{T}$}}
    \If {$G$ has a $C$-centered $L$-model}
        \State{ \textbf{accept} $G$}
    \EndIf
    \For {\text{\textbf{every} connected component $I$ from $G[V \setminus C]$}}
        \State {$C' = \{v \in C: v \text{ has a neighbour in $I$}\}$}
        \If{$ \left\{ \begin{array}{c}
                \text{$G[I \cup C']$ admits a circular-arc model $\phi$} \\
                \text{such that $\bigcap \phi(C')$ is non-empty} \\
            \text{\ \ \textbf{and}} \\
                \text{$G[V \setminus I]$ has an interval model with} \\
                \text{$C$ as the leftmost maximal clique}
        \end{array} \right.$ }
        \State{\textbf{accept} $G$}
        \EndIf
        \For {\text{\textbf{every} maximal clique $D$ of $G[I]$}}
           \State{$C(D) = C' \cup \{v \in D: v \text{ is a neighbour of every vertex in $C'$}\}$} 
           \If{$G$ has a $C(D)$-centered $L$-model}
                \State{\textbf{accept} $G$}
            
           \EndIf
           
        \EndFor

    \EndFor
    
\EndFor
\State{\textbf{reject} $G$}
\end{algorithmic}
\end{algorithm}

\section{Recognizing Centered Lollipop Graphs}
\label{sec:centered_lollipop_graphs}
The remaining task is to recognize $C$-centered $L$-graphs for some given maximal clique~$C$.

Consider an input consisting of a graph $G$ and a maximal clique $C \subseteq V(G)$.
Let $\mathcal{I}$ be the set of connected components of $G[V \setminus C]$.
If $G$ has a $C$-centered model $\phi$, then every $I \in \mathcal{I}$ induces an interval graph and $\{\phi(v) \mid v \in I\}$ is an interval model for $G[I]$.
So, for the remainder we assume that the instance $G,C$ satisfies the following condition:
\begin{equation}
\label{eq:input_interval_components}
\text{For every $I \in \mathcal{I}$ the graph $G[I]$ is an interval graph.}
\tag{I1}
\end{equation}

\begin{figure}[h]
\centering
	\begin{subfigure}[t]{1\linewidth}
		\centering
		\begin{tikzpicture}[scale=-1.25]
			
			%BLACK
			\tikzstyle{every path}=[dotted, thick]
			\draw (6,1) circle (0.88cm);
			
			\draw (1.5,1) -- (5.12,1);
			
			%COLORED
			\tikzstyle{every path}=[thick]
			
			%blue
			\centerarc[blue](6,1)(14:205:1cm);
			
			\foreach \i in {0.96,0.98,1.02,1.04} 
			{
				\centerarc[blue](6,1)(14:15:\i cm);
				\centerarc[blue](6,1)(204:205:\i cm);
			}
		
			\draw[blue] (3.5,1.1) -- (5,1.1);
			\node[font=\fontsize{4}{4}\selectfont] at (4.4,1.25) {\textcolor{blue}{$\phi(c)$}};
			
			\draw[blue] (3.5,1.03) -- (3.5,1.17);
	
			%teal
			\centerarc[teal](6,1)(155:405:1.2cm);
			\foreach \i in {1.16,1.18,1.22,1.24} 
			{
				\centerarc[teal](6,1)(155:156:\i cm);
				\centerarc[teal](6,1)(404:405:\i cm);
			}
		
			\draw[teal] (3.1,0.9) -- (4.8,0.9);
			\draw[teal] (3.1,0.83) -- (3.1,0.97);
			
			\node[font=\fontsize{4}{4}\selectfont] at (4.4,0.77) {\textcolor{teal}{$\phi(a)$}};
		
			%red
			\centerarc[red](6,1)(85:277:1.3cm);
			\foreach \i in {1.26,1.28,1.32,1.34} 
			{
				\centerarc[red](6,1)(85:86:\i cm);
				\centerarc[red](6,1)(276:277:\i cm);
			}
		
			\draw[red] (2.15,0.65) -- (4.75,0.65);
			\draw[red] (2.15,0.58) -- (2.15,0.72);
			\node[font=\fontsize{4}{4}\selectfont] at (4.4,0.5) {\textcolor{red}{$\phi(b)$}};

			%COMPONENTS
			
			%I1
			\centerarc[black](6,1)(140:160:1.1cm);
			\foreach \i in {1.06,1.08,1.12,1.14} 
			{
				\centerarc[black](6,1)(140:141:\i cm);
				\centerarc[black](6,1)(159:160:\i cm);
			}
		
			\centerarc[black](6,1)(130:150:1.2cm);
			\foreach \i in {1.16,1.18,1.22,1.24} 
			{
				\centerarc[black](6,1)(130:131:\i cm);
				\centerarc[black](6,1)(149:150:\i cm);
			}
		
			\node[font=\fontsize{7}{7}\selectfont] at (5.4,1.98) {$I_1$};
		
			%I2
			\centerarc[black](6,1)(70:90:1.1cm);
			\foreach \i in {1.06,1.08,1.12,1.14} 
			{
				\centerarc[black](6,1)(70:71:\i cm);
				\centerarc[black](6,1)(89:90:\i cm);
			}
		
			\centerarc[black](6,1)(40:60:1.1cm);
			\foreach \i in {1.06,1.08,1.12,1.14} 
			{
				\centerarc[black](6,1)(40:41:\i cm);
				\centerarc[black](6,1)(59:60:\i cm);
			}
		
			\centerarc[black](6,1)(50:80:1.2cm);
			\foreach \i in {1.16,1.18,1.22,1.24} 
			{
				\centerarc[black](6,1)(50:51:\i cm);
				\centerarc[black](6,1)(79:80:\i cm);
			}
		
			\node[font=\fontsize{7}{7}\selectfont] at (6.55,2.25) {$I_2$};
			
			%I3
			\centerarc[black](6,1)(-12:8:1cm);
			\foreach \i in {0.96,0.98,1.02,1.04} 
			{
				\centerarc[black](6,1)(-12:-11:\i cm);
				\centerarc[black](6,1)(7:8:\i cm);
			}
			
			\centerarc[black](6,1)(2:22:1.1cm);
			\foreach \i in {1.06,1.08,1.12,1.14} 
			{
				\centerarc[black](6,1)(2:3:\i cm);
				\centerarc[black](6,1)(21:22:\i cm);
			}
		
			\node[font=\fontsize{7}{7}\selectfont] at (6.97,0.6) {$I_3$};
			
			%I4
			\centerarc[black](6,1)(280:300:1cm);
			\foreach \i in {0.96,0.98,1.02,1.04} 
			{
				\centerarc[black](6,1)(280:281:\i cm);
				\centerarc[black](6,1)(299:300:\i cm);
			}
		
			\centerarc[black](6,1)(270:290:1.1cm);
			\foreach \i in {1.06,1.08,1.12,1.14} 
			{
				\centerarc[black](6,1)(270:271:\i cm);
				\centerarc[black](6,1)(289:290:\i cm);
			}
		
			\node[font=\fontsize{7}{7}\selectfont] at (6.645,0.18) {$I_4$};
			
			%I5
			\centerarc[black](6,1)(210:230:1cm);
			\foreach \i in {0.96,0.98,1.02,1.04} 
			{
				\centerarc[black](6,1)(210:211:\i cm);
				\centerarc[black](6,1)(229:230:\i cm);
			}
		
			\node[font=\fontsize{7}{7}\selectfont] at (5.49,0.11) {$I_5$};
			
			%I6
			\draw[black] (3.2,1.25) -- (3.7,1.25);
			\draw[black] (3.2,1.18) -- (3.2,1.32);
			\draw[black] (3.7,1.18) -- (3.7,1.32);
			
			\draw[black] (2.85,1.1) -- (3.35,1.1);
			\draw[black] (2.85,1.03) -- (2.85,1.17);
			\draw[black] (3.35,1.03) -- (3.35,1.17);
			
			\draw[black] (2.5,1.25) -- (3,1.25);
			\draw[black] (2.5,1.18) -- (2.5,1.32);
			\draw[black] (3,1.18) -- (3,1.32);
			
			\node[font=\fontsize{4}{4}\selectfont] at (3.1,1.45) {$I_6$};
			
			%I7
			\draw[black] (2.3,0.8) -- (1.8,0.8);
			\draw[black] (2.3,0.73) -- (2.3,0.87);
			\draw[black] (1.8,0.73) -- (1.8,0.87);
			
			\draw[black] (2,0.65) -- (1.5,0.65);
			\draw[black] (2,0.58) -- (2,0.72);
			\draw[black] (1.5,0.58) -- (1.5,0.72);
			
			\node[font=\fontsize{4}{4}\selectfont] at (1.9,0.5) {$I_7$};
			
		\end{tikzpicture}
		\caption{$C$-centered $L$-model $\phi$ of $G$.}
		\label{fig:C-centered-L-model-phi}
	\end{subfigure}
	
\begin{subfigure}[t]{1\linewidth}
\begin{tikzpicture}[xscale=0.36,yscale=0.7,>=latex,shorten >=-0.4pt,shorten <=-0.4pt]
  \tikzstyle{every node}=[inner sep=2pt,fill=white]

%I1
\draw[thick] (0,3) -- (2,3);
\draw[thick] (1,3.4) -- (3,3.4);
\coordinate (lI1) at (1.5,3.8) {};
\coordinate (I1) at (1.5,-0.2) {};

%I2
\draw[thick] (4,3) -- (6,3);
\draw[thick] (5,3.4) -- (8,3.4);
\draw[thick] (7,3) -- (9,3);
\coordinate (lI2) at (6.5,3.8) {};
\coordinate (I2) at (6.5,0.3) {};

%I3
\draw[thick] (10,3) -- (12,3);
\draw[thick] (11,3.4) -- (13,3.4);
\coordinate (lI3) at (11.5,3.8) {};
\coordinate (I3) at (11.5,0.3) {};

%I4
\draw[thick] (14,3) -- (16,3);
\draw[thick] (15,3.4) -- (17,3.4);
\coordinate (lI4) at (15.5,3.8) {};
\coordinate (I4) at (15.5,0.3) {};

%I5
\draw[thick] (18,3.2) -- (20,3.2);
\coordinate (lI5) at (19,3.8) {};
\coordinate (I5) at (19,-0.2) {};

%c
\draw[very thick,blue,-|] (-3,2.5) -- (10.8,2.5);
\draw[very thick,blue,|-] (20.2,2.5) -- (23.3,2.5);

\coordinate (lcleft) at (-1.6,2.75) {};
\coordinate (lcright) at (22,2.75) {};

%\draw[very thick, blue,|-|] (10,-0.5) -- (20,-0.5);

%b
\draw[very thick,red,-|] (-3,2) -- (4.8,2);
\draw[very thick,red,|-] (16.2,2) -- (23.3,2);

\coordinate (lbleft) at (-1.6,2.25) {};
\coordinate (lbright) at (22,2.25) {};

%\draw[very thick, red,|-|] (4,0) -- (17,0);

%a
\draw[very thick,teal,-|] (-3,1.5) -- (0.8,1.5);
\draw[very thick,teal,|-] (8.2,1.5) -- (23.3,1.5);
%\draw[very thick, teal,|-|] (0,-0.5) -- (9,-0.5);

\coordinate (laleft) at (-1.6,1.75) {};
\coordinate (laright) at (22,1.75) {};

%\tikzstyle{every node}=[circle,minimum size=5pt,inner sep=0pt,draw,fill]
%\node at (I1) {};
%\node at (I2) {};
%\node at (I3) {};
%\node at (I4) {};
%\node at (I5) {};

\tikzstyle{every node}=[inner sep=1pt]
\begin{tiny}
\node at (lI1) {$I_1$};
\node at (lI2) {$I_2$};
\node at (lI3) {$I_3$};
\node at (lI4) {$I_4$};
\node at (lI5) {$I_5$};

%\node at ($(I1) + (0.6,0)$) {$I_1$};
%\node at ($(I2) + (0.6,0)$) {$I_2$};
%\node at ($(I3) + (0.6,0)$) {$I_3$};
%\node at ($(I4) + (0.6,0)$) {$I_4$};
%\node at ($(I5) + (0.6,0)$) {$I_5$};

\node at (lbleft) {$\phi_{left}(b)$};
\node at (lcleft) {$\phi_{left}(c)$};
\node at (laleft) {$\phi_{left}(a)$};

\node at (lbright) {$\phi_{right}(b)$};
\node at (lcright) {$\phi_{right}(c)$};
\node at (laright) {$\phi_{right}(a)$};
\end{tiny}

%\draw[black] (-3.5,1) -- (-3.5,1.5);
%\draw[black] (24,3) -- (24,4.5);

\draw[white] (-3.5,1) -- (-3.5,1.5);
\draw[white] (24,3) -- (24,4.5);

\begin{scope}[shift={(30,0)}]
%I6
\draw[thick] (0,3) -- (2,3);
\draw[thick] (1,3.4) -- (4,3.4);
\draw[thick] (3,3) -- (5,3);
\coordinate (lI6) at (1.5,3.8) {};
\coordinate (I6) at (1.5,-0.2) {};

%I7
\draw[thick] (6,3) -- (8,3);
\draw[thick] (7,3.4) -- (9,3.4);
\coordinate (lI7) at (7.5,3.8) {};
\coordinate (I7) at (7.5,0.8) {};

%c
\draw[very thick, blue,-|] (-3,2.5) -- (0.8,2.5);
\coordinate (lcstick) at (-1.6,2.75) {};

%\draw[very thick, blue,|-|] (0,-0.5) -- (9,-0.5);

%b
\draw[very thick,red,-|] (-3,2) -- (6.8,2);
\coordinate (lbstick) at (-1.6,2.25) {};

%\draw[very thick, red,|-|] (6,0) -- (9,0);

%a
\draw[very thick,teal,-|] (-3,1.5) -- (2.8,1.5);
\coordinate (lastick) at (-1.6,1.75) {};
%\draw[very thick, teal,|-|] (6,0.5) -- (9,0.5);

\tikzstyle{every node}=[circle,minimum size=5pt,inner sep=0pt,draw,fill]
%\node at (I6) {};
%\node at (I7) {};

\tikzstyle{every node}=[inner sep=1pt]
\begin{tiny}
\node at (lI6) {$I_6$};
\node at (lI7) {$I_7$};
%\node at ($(I6) + (0.6,0)$) {$I_6$};
%\node at ($(I7) + (0.6,0)$) {$I_7$};
\end{tiny}

\tikzstyle{every node}=[inner sep=1pt]
\begin{tiny}
\node at (lbstick) {$\phi_{stick}(b)$};
\node at (lcstick) {$\phi_{stick}(c)$};
\node at (lastick) {$\phi_{stick}(a)$};
\end{tiny}

%\draw[black] (-3.5,1) -- (-3.5,1.5);
%\draw[black] (9,3) -- (9,4.5);

\draw[white] (-3.5,1) -- (-3.5,1.5);
\draw[white] (9,3) -- (9,4.5);
\end{scope}
\end{tikzpicture}
\caption{equivalent representation of $\phi$}
\label{fig:interval-representation-phi}
\end{subfigure}
	
\caption{}
\label{fig:example-C-centered-L-model} 
\end{figure}

Recall $L_\circle \subseteq V(L)$ is the circle part of $L$
and $L_\stick \subseteq V(L)$ is the stick part of $L$.
We define sets of $I \in \mathcal{I}$ according to the intersection with $L_\circle$ and $L_\stick$:
%$\{\mathcal{I}^{\phi}_{\circle},\mathcal{I}^{\phi}_{\stick}\}$, where:
$$\mathcal{I}^{\phi}_{\circle} = \left\{I \in \mathcal{I}: \bigcup \phi(I) \subseteq L_{\circle} \right\} \quad \text{and} \quad \mathcal{I}^{\phi}_{\stick} = \left\{I \in \mathcal{I}: \bigcup \phi(I) \subseteq L_{\stick}\right\}.$$

If $G$ has a $C$-centered $L$-model $\phi$,
then $\{\mathcal{I}^{\phi}_{\circle},\mathcal{I}^{\phi}_{\stick}\}$ partitions $\mathcal{I}$ --
see Figure \ref{fig:example-C-centered-L-model} for an example.
In the rest of the paper we use the following convention: given 
a $C$-centered model $\phi$ of $G$ and a set $\mathcal{L} \subseteq \mathcal{I}$, by $\mathcal{L}^{\phi}_{\circle}$ and $\mathcal{L}^{\phi}_{\stick}$
we denote the sets $\mathcal{L} \cap \mathcal{I}^{\phi}_{\circle}$ and $\mathcal{L} \cap \mathcal{I}^{\phi}_{\stick}$, respectively.

The main idea behind the algorithm testing whether $G$ has a $C$-centered model
is to search for a partition $\{\mathcal{J},\mathcal{J}'\}$ of the components of $\mathcal{I}$ which can be used to construct a $C$-centered $L$-model $\phi$ of $G$ such that $\mathcal{J} = \mathcal{I}^{\phi}_{\circle}$ and $\mathcal{J}' = \mathcal{I}^{\phi}_{\stick}$.

First, we classify the components of $\mathcal{I}$ 
depending on the properties of their interval models with 
respect to the clique $C$.
For this purpose we let:
$$
\begin{array}{ccl}
 N_C(I) = \{c \in C: c \text{ is adjacent to some vertex in }I\}& \text{and} &C^s(I)= C \setminus N_C(I),\\
 N^s_C(I) = \{c \in C: c \text{ is adjacent to every vertex in }I\} & \text{and} &C(I)= C \setminus N^s_C(I).
\end{array}
$$
Note that
$$
\begin{array}{c}
 C^s(I) = \{c \in C: c \text{ is not adjacent to every vertex in }I\}, \\
 C(I) =\{c \in C: c \text{ is not adjacent to some vertex in }I\}. \\
\end{array}
$$
Note also that $N^s_C(I) \subseteq N_C(I)$, $C^s(I) \subseteq C(I)$, 
and the sets $C(I) \setminus C^s(I)$ and $N_C(I) \setminus N_C^s(I)$ are equal and contain the vertices
from $C$ which are adjacent to some but not all vertices in $I$.
Now, we are ready to classify the components, as follows:
\begin{itemize}
\item a component $I \in \mathcal{I}$ is \emph{ambiguous} if 
$G[C \cup I]$ has an interval model with $C$ as its left-most clique; otherwise 
$I$ is a \emph{circle component},
\item an ambiguous component $I \in \mathcal{I}_{a}$ is \emph{simple} if we have $N_{C}(I)=N^{s}_{C}(I)$, otherwise $I \in \mathcal{I}_a$ is \emph{non-simple}.
\end{itemize}
We denote the sets of circle, ambiguous, ambiguous simple, and ambiguous non-simple components by $\mathcal{I}_c$, $\mathcal{I}_a$, $\mathcal{I}_{as}$, and $\mathcal{I}_{ans}$, respectively. Clearly, $\{\mathcal{I}_a, \mathcal{I}_c\}$ is a partition of $\mathcal{I}$ and $\{\mathcal{I}_{as}, \mathcal{I}_{ans}\}$ is a partition of $\mathcal{I}_a$.
Table~\ref{tab:component_types_and_subseteqT_poset} shows the types of the components of the graph whose $C$-centered model is shown in Figure \ref{fig:example-C-centered-L-model}.

\begin{table}[h!]
		\centering

\begin{tabular}{c | c | c | c | c |c|}
\textcolor{red}{$I_i \in \mathcal{I}$} & \textcolor{red}{$N_C(I_i)$} & \textcolor{red}{$N_C^s(I_i)$} & \textcolor{red}{$C(I_i)$} & \textcolor{red}{$C^s(I_i)$} & \textcolor{red}{type}\\
\hline
$I_1$ & $abc$ & $bc$ & $a$ & $\emptyset$& \scriptsize{ambiguous non-simple}\\
\hline

$I_2$ & $abc$ & $c$ & $ab$ & $\emptyset$& \scriptsize{circle}\\
\hline

$I_3$ & $ac$ & $a$ & $bc$ & $b$& \scriptsize{ambiguous non-simple}\\
\hline

$I_4$ & $ab$ & $a$ & $bc$ & $c$& \scriptsize{ambiguous non-simple}\\
\hline

$I_5$ & $ab$ & $ab$ & $c$ & $c$& \scriptsize{ambiguous simple}\\
\hline

$I_6$ & $abc$ & $b$ & $ac$ & $\emptyset$& \scriptsize{ambiguous non-simple}\\
\hline

$I_7$ & $c$ & $\emptyset$ & $abc$ & $ab$& \scriptsize{ambiguous non-simple}\\
\hline

\hline

\end{tabular}
\smallskip

		\caption{The types of the components from $\mathcal{I}$ in graph $G$ whose $\{a,b,c\}$-centered $L$-model is shown in Figure~\ref{fig:example-C-centered-L-model}}
		\label{tab:component_types_and_subseteqT_poset}
\end{table}

Clearly, in any $C$-centered model $\phi$ of $G$, a circle component must be represented on the circle, but an ambiguous component can be represented either on the stick or on the circle.
\begin{observation}
$\mathcal{I}_c \subseteq \mathcal{I}^{\phi}_{\circle}$ for any $C$-centered model $\phi$ of $G$.
\end{observation}

\begin{observation}
Suppose $I_1,I_2$ are two simple components such that $C(I_1) = C(I_2)$.
Then $G$ has a $C$-centered model iff $G[V \setminus I_1]$ has a $C$-centered model.
\end{observation}
\begin{proof}
Suppose $G[V \setminus I_1]$ has a $C$-centered model $\phi$.
Since $I_1,I_2$ are two simple components such that $C(I_1) = C(I_2)$,
the neighbourhoods of every vertex from $I_1 \cup I_2$ in the set $C$ are the same.
In particular, we can simply extend $\phi$ to be a model of $G$:
for this purpose, for every $c \in N_{C}(I_2)$ we extend $\phi(c)$ 
so as it contains the interval $\bigcup \phi(I_2)$ and then we place an interval model of $I_1$ just next to $\bigcup \phi(I_2)$.
\end{proof}
The above observation allows us to assume
	that the input instance $G,C$ satisfies the condition:
\begin{equation}
\label{eq:input_simple_components}
 \text{For every two simple components $I_1,I_2$ we have $C(I_1) \neq C(I_2)$.}
 \tag{I2} 
\end{equation}

Next, we introduce the left-right relation on the points from $L_{\circle} \setminus P$: given two different points $Q,R \in L_{\circle} \setminus P$ we say $Q$ is \emph{to the left} of $R$ (or $R$ is \emph{to the right} of $Q$) if we encounter $Q$ before $R$ when we traverse $L_{\circle}$ in the clockwise order starting from ~$P$.
Similarly, we introduce left-right ordering on the points from $L_{\stick} \setminus P$:
given two different points $Q$ and $R$ from $L_{\stick} \setminus P$ we say $Q$ is \emph{to the left} of $R$ if $Q$ is closer to $P$ than $R$.
We naturally extend the left-right relation on the subsets of $L_{\circle} \setminus P$ and $L_{\stick} \setminus P$, respectively.
Given a $C$-centered model $\phi$ of $G$,
by $(\mathcal{I}^{\phi}_{\circle},{\prec_{\phi}})$ we denote strict linear orders of the components from 
$\mathcal{I}_{\circle}^{\phi}$ defined $I_1 \prec_{\phi} I_2$ if 
the interval $\bigcup \phi(I_1)$ is to the left of the interval $\bigcup \phi(I_2)$.
For the model $\phi$ shown in Figure~\ref{fig:example-C-centered-L-model}, $(\mathcal{I}^{\phi}_{\circle},{\prec_{\phi}})$ is equal to
$I_1 \prec_{\phi} \dots \prec_{\phi} I_5$.
Eventually, 
\begin{itemize}
 \item for every $c \in C$ such that $(L_\circle \setminus P ) \setminus \phi(c) \neq \emptyset$, let $\phi_{left}(c)$ and $\phi_{right}(c)$ be 
 the unique intervals in $L_{\circle} \setminus P$ such that 
 $\phi_{left}(c) \cup \phi_{right}(c) = \phi(c) \cap (L_\circle \setminus P)$ and $\phi_{left}(c)$ is to the left of $\phi_{right}(c)$.
 For $c \in C$ such that $(L_{\circle} \setminus P) \subseteq \phi(c)$ let $\phi_{left}(c) = \phi_{right}(c) = L_{\circle} \setminus P$.
 \item for every $c \in C$ let $\phi_{stick}(c) = L_{\stick} \cap \phi(c)$.
\end{itemize}
This way we can equivalently represent $\phi$ in such a way that the set $\phi(c)$ for every $c$ is represented by means of the intervals $\phi_{left}(c)$, $\phi_{right}(c)$, and $\phi_{stick}(c)$
-- see Figure~\ref{fig:interval-representation-phi} for an illustration.

Suppose that $G$ has a $C$-centered model $\phi$.
First, we extend the sets $\phi(c)$ for $c \in C$ as much as possible, keeping $\phi$ an $L$-model of $G$ and keeping $\phi(I)$ for $I \in \mathcal{I}$ unchanged.
Model $\phi$ obtained this way is called \emph{$C$-expanded}. 
It satisfies the following properties:
\begin{itemize}
 \item for every $I \in \mathcal{I}$ and every $c \in C$ we have $\bigcup \phi(I) \subseteq \phi(c)$ iff $c \in N^s_C(I)$,
 \item for every $c \in C$ we have $L_{\circle} \subseteq \phi(c)$ iff $c \in N^s_C(I)$ for every $I \in \mathcal{I}^{\phi}_{\circle}$ (or equivalently, $c \notin C(I)$ for every $I \in \mathcal{I}^{\phi}_{\circle}$).
\end{itemize}
For example, model $\phi$ shown in Figure~\ref{fig:example-C-centered-L-model} is $C$-expanded.

Next, observe that for every $c \in C$ the set $\{I \in \mathcal{I}^{\phi}_{\circle}: c \in C(I)\}$ forms an interval in $(\mathcal{I}^{\phi}_{\circle},{\prec_\phi})$.
Let $C' = \bigcup \{C(I): I \in \mathcal{I}^{\phi}_{\circle}\}$ and let $\phi^*(c)$ be the shortest interval in $L_{\circle} \setminus P$
that contains the interval $\bigcup \phi(I)$ for every $I \in \mathcal{I}^{\phi}_{\circle}$ such that $c \in C(I)$.
Note that the family $\{\phi^*(c): c \in C'\}$ forms an interval model of a graph $H^{\phi}$, where:
$$
\begin{array}{ccl}
V(H^{\phi}) &=& C' \ = \ \bigcup \{C(I): I \in \mathcal{I}^{\phi}_{\circle}\}, \\ 
E(H^{\phi}) &=& \big{\{} \{c,c'\}: \text{there is $I$ in $\mathcal{I}^{\phi}_{\circle}$ such that $c,c' \in C(I)$}\big{\}}.
\end{array}
$$
Observe that the interval model $\phi^*$ of $H^\phi$ satisfies the following properties:
\begin{itemize}
\item every interval $\bigcup \phi(I)$ is contained in some sector $S$ of $\phi^*$ (and then we say \emph{$I$ is contained in sector $S$} of $\phi^*$),
\item every non-minimal sector of $\phi^*$ contains some component from $\mathcal{I}^{\phi}_{\circle}$,
\item if $\phi$ is $C$-expanded, then for every $c \in C'$ the set $\{\phi_{left}(c),\phi^*(c),\phi_{right}(c)\}$ covers $L_{\circle} \setminus P$.
In particular, if a sector $S$ of $\phi^*$ is not covered by $\phi^*(c)$, then $S$ is covered by $\phi_{left}(c)$ or $\phi_{right}(c)$.
\end{itemize}
The first property allows us to denote by $S^{\phi^*}(I)$ the unique sector of $\phi^*$ containing~$I$
(usually we omit superscript in $S^{\phi^*}(I)$ if $\phi^*$ is clear from the context). 
Note that $C(S^{\phi^*}(I))=C(I)$, where $C(S^{\phi^*}(I))$ is the clique of the sector $S^{\phi^*}(I)$.
Figure~\ref{fig:phi-star-schematic-view} shows the model $\phi^*$ of $H^{\phi}$ (and its schematic view) obtained 
from $C$-centered model $\phi$ shown in Figure~\ref{fig:example-C-centered-L-model}.
In particular, we have $V(H^{\phi}) = \{a,b,c\}$, and $E(H^{\phi}) = \{ab,bc\}$.

\begin{figure}[h]
\centering
\begin{subfigure}[t]{0.6\linewidth}
\begin{tikzpicture}[xscale=0.36,yscale=0.7,>=latex,shorten >=-0.4pt,shorten <=-0.4pt]
  \tikzstyle{every node}=[inner sep=2pt,fill=white]

%I1
\draw[thick] (0,3) -- (2,3);
\draw[thick] (1,3.4) -- (3,3.4);
\coordinate (lI1) at (1.5,3.8) {};
\coordinate (I1) at (1.5,-0.2) {};

%I2
\draw[thick] (4,3) -- (6,3);
\draw[thick] (5,3.4) -- (8,3.4);
\draw[thick] (7,3) -- (9,3);
\coordinate (lI2) at (6.5,3.8) {};
\coordinate (I2) at (6.5,0.3) {};

%I3
\draw[thick] (10,3) -- (12,3);
\draw[thick] (11,3.4) -- (13,3.4);
\coordinate (lI3) at (11.5,3.8) {};
\coordinate (I3) at (11.5,0.3) {};

%I4
\draw[thick] (14,3) -- (16,3);
\draw[thick] (15,3.4) -- (17,3.4);
\coordinate (lI4) at (15.5,3.8) {};
\coordinate (I4) at (15.5,0.3) {};

%I5
\draw[thick] (18,3.2) -- (20,3.2);
\coordinate (lI5) at (19,3.8) {};
\coordinate (I5) at (18.6,-0.2) {};

%c
\draw[blue,-|] (-3,2.5) -- (10.8,2.5);
\draw[blue,|-] (20.2,2.5) -- (23.3,2.5);

\coordinate (lcleft) at (-1.6,2.75) {};
\coordinate (lcright) at (22,2.75) {};
\coordinate (lc*) at (21.2,-0.5) {};

\draw[very thick, blue,|-|] (10,-0.5) -- (20,-0.5);

%b
\draw[red,-|] (-3,2) -- (4.8,2);
\draw[red,|-] (16.2,2) -- (23.3,2);

\coordinate (lbleft) at (-1.6,2.25) {};
\coordinate (lbright) at (22,2.25) {};

\draw[very thick, red,|-|] (4,0) -- (17,0);
\coordinate (lb*) at (3.0,0.3) {};

%a
\draw[teal,-|] (-3,1.5) -- (0.8,1.5);
\draw[teal,|-] (8.2,1.5) -- (23.3,1.5);
\draw[very thick, teal,|-|] (0,-0.5) -- (9,-0.5);

\coordinate (laleft) at (-1.6,1.75) {};
\coordinate (laright) at (22,1.75) {};
\coordinate (la*) at (-1,-0.5) {};

\tikzstyle{every node}=[circle,minimum size=5pt,inner sep=0pt,draw,fill]
\node at (I1) {};
\node at (I2) {};
\node at (I3) {};
\node at (I4) {};
\node at (I5) {};

\tikzstyle{every node}=[inner sep=1pt]
\begin{tiny}
\node at (lI1) {$I_1$};
\node at (lI2) {$I_2$};
\node at (lI3) {$I_3$};
\node at (lI4) {$I_4$};
\node at (lI5) {$I_5$};
\node at ($(I1) + (0.6,0)$) {$I_1$};
\node at ($(I2) + (0.6,0)$) {$I_2$};
\node at ($(I3) + (0.6,0)$) {$I_3$};
\node at ($(I4) + (0.6,0)$) {$I_4$};
\node at ($(I5) + (0.6,0)$) {$I_5$};

\node at (lbleft) {$\phi_{left}(b)$};
\node at (lcleft) {$\phi_{left}(c)$};
\node at (laleft) {$\phi_{left}(a)$};

\node at (lbright) {$\phi_{right}(b)$};
\node at (lcright) {$\phi_{right}(c)$};
\node at (laright) {$\phi_{right}(a)$};

\node at (la*) {$\phi^*(a)$};
\node at (lb*) {$\phi^*(b)$};
\node at (lc*) {$\phi^*(c)$};

\end{tiny}

%\draw[black] (-3.5,-1) -- (-3.5,0);
%\draw[black] (24,3) -- (24,4);

\draw[white] (-3.5,-1) -- (-3.5,0);
\draw[white] (24,3) -- (24,4);

\end{tikzpicture}
\caption{schematic view of $\phi^*$}
\label{fig:phi-star-schematic-view}
\end{subfigure}
~
\begin{subfigure}[t]{0.35\linewidth}
\centering
\begin{tikzpicture}[xscale=0.4,yscale=0.7,>=latex,shorten >=-0.4pt,shorten <=-0.4pt]
%I6
\draw[thick] (0,3) -- (2,3);
\draw[thick] (1,3.4) -- (4,3.4);
\draw[thick] (3,3) -- (5,3);
\coordinate (lI6) at (2,3.8) {};
\coordinate (I6) at (2,0.3) {};

%I7
\draw[thick] (6,3) -- (8,3);
\draw[thick] (7,3.4) -- (9,3.4);
\coordinate (lI7) at (7.5,3.8) {};
\coordinate (I7) at (7.5,0.8) {};

%c
\draw[ blue,-|] (-3,2.5) -- (0.8,2.5);
\coordinate (lcstick) at (-1.6,2.75) {};
\coordinate (lc**) at (10.3,0) {};

\draw[very thick, blue,|-|] (0,0) -- (9,0);

%b
\draw[red,-|] (-3,2) -- (6.8,2);
\coordinate (lbstick) at (-1.6,2.25) {};
\coordinate (lb**) at (10.3,0.5) {};

\draw[very thick, red,|-|] (6,0.5) -- (9,0.5);

%a
\draw[teal,-|] (-3,1.5) -- (2.8,1.5);
\coordinate (lastick) at (-1.6,1.75) {};
\draw[very thick, teal,|-|] (0,-0.5) -- (9,-0.5);
\coordinate (la**) at (10.3,-0.5) {};

\tikzstyle{every node}=[circle,minimum size=5pt,inner sep=0pt,draw,fill]
\node at (I6) {};
\node at (I7) {};

\tikzstyle{every node}=[inner sep=1pt]
\begin{tiny}
\node at (lI6) {$I_6$};
\node at (lI7) {$I_7$};
\node at ($(I6) + (0.6,0)$) {$I_6$};
\node at ($(I7) + (0.6,0)$) {$I_7$};
\end{tiny}

\tikzstyle{every node}=[inner sep=1pt]
\begin{tiny}
\node at (lbstick) {$\phi_{stick}(b)$};
\node at (lcstick) {$\phi_{stick}(c)$};
\node at (lastick) {$\phi_{stick}(a)$};

\node at (la**) {$\phi^{**}(a)$};
\node at (lb**) {$\phi^{**}(b)$};
\node at (lc**) {$\phi^{**}(c)$};
\end{tiny}

%\draw[black] (-3.5,-1) -- (-3.5,0);
%\draw[black] (9,3) -- (9,4);

\draw[white] (-3.5,-1) -- (-3.5,0);
\draw[white] (9,3) -- (9,4);

\end{tikzpicture}
\caption{schematic view of $\phi^{**}$}
\label{fig:phi-star-star-schematic-view}
\end{subfigure}
\caption{}
\label{fig:phi-star-phi-star-star-schematic-view} 
\end{figure}

Now we describe some technical properties satisfied by the components of the strict linear order
$(\mathcal{I}^{\phi}_{\circle},{\prec_{\phi}})$ -- see Figure~\ref{fig:technical-lemma-fig} for an illustration.
In what follows, for a component $I \in \mathcal{I}$ and cliques $C_L, C_R \subseteq C(I)$, by
$G^{ext}(C_L,I,C_R)$ we denote the graph obtained from $G[C \cup I]$ by adding two extra vertices $v_L$
and $v_R$ adjacent to $C \setminus C_L$ and $C \setminus C_R$, respectively.
\begin{lemma} 
\label{lemma:phi-star-technical-lemma}
Let $\phi$ be a $C$-centered model of $G$, $I$ be the component from $\mathcal{I}^{\phi}_{\circle}$, and
$S_L$ and $S_R$ be the slots of $\phi^*$ adjacent to $S(I)$ from the left and the right side, respectively.
\begin{enumerate}
\item \label{item:phi-star-technical-lemma-ambiguous-left} If there is $J \in \mathcal{I}^{\phi}_{\circle}$ such that $C(I) \subseteq C(J)$ and $I \prec_{\phi} J$, 
then $I$ is an ambiguous component.
Moreover, if $I$ is ambiguous non-simple, then $I$ is the leftmost component in $S(I)$ and $C(S_L) \subseteq C^s(I)$.
See Figure~\ref{fig:technical-lemma-a} for an illustration.
\item \label{item:phi-star-technical-lemma-ambiguous-right} If there is $J \in \mathcal{I}^{\phi}_{\circle}$ such that $C(I) \subseteq C(J)$ and $J \prec_{\phi} I$, then
$I$ is an ambiguous component.
Moreover, if $I$ is ambiguous non-simple, then $I$ is the rightmost component in $S(I)$ and $C(S_R) \subseteq C^s(I)$.
\item \label{item:phi-star-technical-lemma-maximal} Suppose $I$ is such that $S(I)$ is a maximal sector of $\phi^*$ and $I$ is the only component of $S(I)$.
Then:
  \begin{itemize}
   \item $N_C(I) \setminus N^s_C(I) \subseteq \big{(}C \setminus C(S_L)\big{)} \cup \big{(}C \setminus C(S_R)\big{)}$,
   \item the graph $G^{ext}\big{(}C(S_L),I,C(S_R)\big{)}$ admits a circular-arc model.
  \end{itemize}
See Figure~\ref{fig:technical-lemma-b} for an illustration.
\end{enumerate}
\end{lemma}
\begin{figure}[h]
\centering
\begin{subfigure}[t]{0.5\linewidth}
\begin{tikzpicture}[xscale=0.38,yscale=0.5,>=latex,shorten >=-0.4pt,shorten <=-0.4pt]
  \tikzstyle{every node}=[inner sep=2pt,fill=white]  

%I1
\draw[thick] (2,5.5) -- (3.75,5.5);
\draw[thick] (3.5,5) -- (5,5);
\coordinate (lI1) at (3.5,6) {};

%I2
\draw[thick] (6,5.5) -- (7.75,5.5);
\draw[thick] (7.5,5) -- (9,5);
\coordinate (lI2) at (7.5,6) {};

\draw[dotted,opacity=0.4,thick] (2,-5) -- (2,6.5);
\coordinate (lSL) at (4,-4.5) {};
\draw[dotted,opacity=0.4,thick] (6,-5) -- (6,6.5);
\coordinate (lSI) at (10,-4.5) {};
\draw[dotted,opacity=0.4,thick] (14,-5) -- (14,6.5);

%I3
\draw[thick] (10.5,5.5) -- (12,5.5);
\coordinate (lI3) at (11.25,6) {};

%J
\draw[thick] (14,5) -- (15.75,5);
\draw[thick] (15.25,5.5) -- (17,5.5);
\coordinate (lJ) at (15.5,6) {};
\draw[dotted,opacity=0.4,thick] (17,-5) -- (17,6.5);

%a
\draw[thick, red,-|] (0,4) -- (3,4);
\coordinate (laleft) at (1.5,4.3) {};
\draw[thick, red,|-] (18,4) -- (21.5,4);
\coordinate (laright) at (20,4.3) {};

\draw[very thick, red,|-] (2,-4) -- (17.5,-4);
\coordinate (la) at (18.5,-4) {};

%\coordinate (lbright) at (22,2.25) {};
\coordinate (I1) at (3.5,-3.5) {};

%b
\draw[thick, teal,-|] (0,3) -- (7,3);
\coordinate (lbleft) at (1.5,3.3) {};
\draw[thick, teal,|-] (16,3) -- (21.5,3);
\coordinate (lbright) at (20,3.3) {};

\draw[very thick, teal,|-|] (6,-3) -- (17,-3);
\coordinate (lb) at (18.5,-3) {};

%c
\draw[thick, green,-|] (0,2) -- (7,2);
\coordinate (lcleft) at (1.5,2.3) {};
\draw[thick, green,|-] (18,2) -- (21.5,2);
\coordinate (lcright) at (20,2.3) {};

\draw[very thick, green,|-] (6,-2) -- (17.5,-2);
\coordinate (lc) at (18.5,-2) {};

\coordinate (I2) at (7.5,-1.5) {};
\coordinate (I3) at (11,-1.5) {};
%d
\draw[thick, blue,-|] (0,1) -- (13.75,1);
\coordinate (ldleft) at (1.5,1.3) {};
\draw[thick, blue,|-] (16,1) -- (21.5,1);
\coordinate (ldright) at (20,1.3) {};

\draw[very thick, blue,|-|] (14,-1) -- (17,-1);
\coordinate (ld) at (18.5,-1) {};
\coordinate (J) at (15.5,-0.5) {};

\tikzstyle{every node}=[circle,minimum size=5pt,inner sep=0pt,draw,fill]
\node at (I1) {};
\node at (I2) {};
\node at (I3) {};
\node at (J) {};

\tikzstyle{every node}=[inner sep=1pt]
\begin{tiny}
%\node at (lI1) {$I_1$};
\node at (lI2) {$I$};
%\node at (lI3) {$I_3$};
\node at (lJ) {$J$};

%\node at ($(I1) + (0.6,0)$) {$I_1$};
\node at ($(I2) + (0.6,0)$) {$I$};
%\node at ($(I3) + (0.6,0)$) {$I_3$};
\node at ($(J) + (0.6,0)$) {$J$};

\node at (la) {$\phi^*(a)$};
\node at (lb) {$\phi^*(b)$};
\node at (lc) {$\phi^*(c)$};
\node at (ld) {$\phi^*(d)$};

\node at (lSL) {$S_L$};
\node at (lSI) {$S(I)$};

\node at (laleft) {$\phi_{left}(a)$};
\node at (lbleft) {$\phi_{left}(b)$};
\node at (lcleft) {$\phi_{left}(c)$};
\node at (ldleft) {$\phi_{left}(d)$};

\node at (laright) {$\phi_{right}(a)$};
\node at (lbright) {$\phi_{right}(b)$};
\node at (lcright) {$\phi_{right}(c)$};
\node at (ldright) {$\phi_{right}(d)$};
\end{tiny}

%\draw[black] (0,-5) -- (0,-4);
%\draw[black] (21.5,6) -- (21.5,7);

\draw[white] (0,-5) -- (0,-4);
\draw[white] (21.5,6) -- (21.5,7);
\end{tikzpicture}
\caption{$N_C(I) \setminus N^s_C(I) = \{b,c\}$,
$C(I) = \{a,b,c\}$, and $C(J) = \{a,b,c,d\}$.}
\label{fig:technical-lemma-a}
\end{subfigure}
\hspace{0.5cm}
\begin{subfigure}[t]{0.4\linewidth}
\begin{tikzpicture}[xscale=0.34,yscale=0.5,>=latex,shorten >=-0.4pt,shorten <=-0.4pt]
  \tikzstyle{every node}=[inner sep=2pt,fill=white]

%I

\draw[very thick,-] (1,6) -- (1.5,6);
\coordinate (lIL) at (1.25,6.5) {};

\draw[very thick,-] (10.5,6) -- (11,6);
\coordinate (lIR) at (10.75,6.5) {};

\draw[thick] (2,6) -- (4,6);
\draw[thick] (5,6) -- (10,6);
\draw[thick] (2,5.5) -- (7,5.5);
\draw[thick] (8,5.5) -- (10,5.5);
\draw[thick] (5,5) -- (7,5);

\coordinate (lI) at (6,6.5) {};

\draw[dotted,opacity=0.4,thick] (0,-5) -- (0,6.5);
\draw[dotted,opacity=0.4,thick] (2,-5) -- (2,6.5);
\draw[dotted,opacity=0.4,thick] (10,-5) -- (10,6.5);
\draw[dotted,opacity=0.4,thick] (12,-5) -- (12,6.5);

\coordinate (lSI) at (6,-4.5) {};
\coordinate (lSL) at (1,-4.5) {};
\coordinate (lSR) at (11,-4.5) {};

%a red
\draw[thick, red,-|] (-3,4) -- (3.5,4);
\coordinate (laleft) at (-1.5,4.3) {};
\draw[thick, red,|-] (10.2,4) -- (15.5,4);
\coordinate (laright) at (14,4.3) {};

\draw[very thick, red,|-|] (2,-4) -- (10,-4);
\coordinate (la) at (8.8,-3.6) {};

%\coordinate (lbright) at (22,2.25) {};
\coordinate (I) at (6,-1.5) {};

%b teal
\draw[thick, teal,-|] (-3,3) -- (0.0,3);
\coordinate (lbleft) at (-1.5,3.3) {};
\draw[thick, teal,|-] (8.5,3) -- (15.5,3);
\coordinate (lbright) at (14,3.3) {};

\draw[very thick, teal,-|] (0,-3) -- (10,-3);
\coordinate (lb) at (8.8,-2.6) {};

% c green
\draw[thick, green,-|] (-3,2) -- (3.5,2);
\coordinate (lcleft) at (-1.5,2.3) {};
\draw[thick, green,|-] (12,2) -- (15.5,2);
\coordinate (lcright) at (14,2.3) {};

\draw[very thick, green,|-] (2,-2) -- (12,-2);
\coordinate (lc) at (8.8,-1.6) {};

%d blue
\draw[thick, blue,-] (-3,1) -- (15.5,1);
\coordinate (ldleft) at (6,1.3) {};
%\draw[thick, blue,|-] (8.5,1) -- (14.5,1);
\coordinate (ldright) at (6,1.3) {};
%\draw[very thick, blue,|-|] (2,-1) -- (10,-1);
%\coordinate (ld) at (8.8,-0.6) {};

\tikzstyle{every node}=[circle,minimum size=5pt,inner sep=0pt,draw,fill]
\node at (I) {};

\tikzstyle{every node}=[inner sep=1pt]
\begin{tiny}
\node at (lI) {$I$};

\node at (lIL) {$I_L$};
\node at (lIR) {$I_R$};

\node at (lSI) {$S(I)$};
\node at (lSL) {$S_L$};
\node at (lSR) {$S_R$};

\node at ($(I) + (0.6,0)$) {$I$};

\node at (la) {$\phi^*(a)$};
\node at (lb) {$\phi^*(b)$};
\node at (lc) {$\phi^*(c)$};
%\node at (ld) {$\phi^*(d)$};

\node at (laleft) {$\phi_{left}(a)$};
\node at (lbleft) {$\phi_{left}(b)$};
\node at (lcleft) {$\phi_{left}(c)$};
\node at (ldleft) {$\phi_{left}(d) \cup \phi_{right}(d)$};

\node at (laright) {$\phi_{right}(a)$};
\node at (lbright) {$\phi_{right}(b)$};
\node at (lcright) {$\phi_{right}(c)$};
%\node at (ldright) {$\phi_{right}(d)$};
\end{tiny}

%\draw[black] (-2.5,-5) -- (-2.5,-4);
%\draw[black] (14.5,6) -- (14.5,7);

\draw[white] (-2.5,-5) -- (-2.5,-4);
\draw[white] (14.5,6) -- (14.5,7);
\end{tikzpicture}
\caption{$N^s_C(I) \setminus N_C(I) = \{a,b,c\}$,
$a,c,d \in C \setminus C(S_L)$ and $a,b,d \in C \setminus C(S_R)$.}
\label{fig:technical-lemma-b}
\end{subfigure}
\caption{}
\label{fig:technical-lemma-fig} 
\end{figure}
\begin{proof}
Suppose $J \in \mathcal{I}^{\phi}_{\circle}$ is a component such that
$I \prec_{\phi} J$ and $C(I) \subseteq C(J)$ -- see Figure~\ref{fig:technical-lemma-a} for an illustration.
If $N_C(I) \setminus N^s_C(I) = \emptyset$, then 
$I$ is ambiguous simple.
Assume that $I$ is ambiguous non-simple, that is, assume that 
$N_{C}(I) \setminus N^s_{C}(I) \neq \emptyset$.
Let $c \in N_{C}(I) \setminus N^s_{C}(I)$.
Since $N_{C}(I) \setminus N^s_{C}(I) = C(I) \setminus C^s(I) \subseteq C(J)$, 
we have $c \in C'$ and there is a vertex in $J$ which is not adjacent to $c$.
Thus, the left endpoint of $\phi_{right}(c)$ is contained in the interval $\bigcup \phi(J)$ or is to the right of this interval.
In particular, $\phi_{right}(c)$ is disjoint with $\bigcup \phi(I)$.
Since $\bigcup \phi(I)$ is contained in $S(I)$, we have that:
\begin{itemize}
\item the interval $\phi_{left}(c)$ overlaps the slot 
$S(I)$ and the left side of $\phi^*(c)$ coincides with the left side of $S(I)$.
\end{itemize}
See Figure~\ref{fig:technical-lemma-a} for an illustration.
In particular, $I$ is the leftmost component in $S(I)$ and we have $C(S_L) \subseteq C^s(I)$.
Now, for every $v \in C \cup I$ we set
$$\psi(v) = \left\{
\begin{array}{lcl}
\phi_{left}(v) & \text{if} & v \in N_C(I) \setminus N^s_C(I), \\
L_{\circle} \setminus P & \text{if} & v \in N^s_C(I), \\
\phi(v) & \text{if} & v \in I. \\
\end{array}
\right.
$$
The observation made above assert $\psi$ forms an interval model of $G[C \cup I]$ with $C$ as the leftmost clique.
This proves statement~\eqref{item:phi-star-technical-lemma-ambiguous-left}.

Statement~\eqref{item:phi-star-technical-lemma-ambiguous-right} is proven analogously, however, in this case we show that for every $c \in N_C(I) \setminus N^s_C(I)$:
\begin{itemize}
\item the interval $\phi_{right}(c)$ overlaps the slot $S(I)$ and the right side of $\phi^*(c)$ coincides with the right side of $S(I)$.
\end{itemize}

To show statement \eqref{item:phi-star-technical-lemma-maximal} let $I$ be such that $S(I)$ is a maximal sector of $\phi^*$ and 
$I$ is the only component in $S(I)$. 
Note that for every $c \in N_C(I) \setminus N^s_C(I)$: 
\begin{itemize}
\item if the interval $\phi_{left}(c)$ overlaps the slot $S(I)$,  then the left side of $\phi^*(c)$ coincides with the left side of $S(I)$ and $c \in C \setminus C(S_L)$, 
\item if the interval $\phi_{right}(c)$ overlaps the slot $S(I)$, 
 then the right side of $\phi^*(c)$ coincides with the right side of $S(I)$ and $c \in C \setminus C(S_R)$.
\end{itemize}
Let $c \in N^s_C(I) \setminus N_C(I)$.
Clearly, we have $\phi_{left}(c) \cap \bigcup \phi(I) \neq \emptyset$ or 
$\phi_{right}(c) \cap \bigcup \phi(I) \neq \emptyset$.
If $\phi_{left}(c) \cap \bigcup \phi(I) \neq \emptyset$, 
then $\phi_{left}(c)$ overlaps $S(I)$, and hence $c \in C \setminus C(S_L)$. 
If $\phi_{right}(c) \cap \bigcup \phi(I) \neq \emptyset$, 
then $\phi_{right}(c)$ overlaps $S(I)$, 
and hence $c \in C \setminus C(S_R)$.
This shows the first statement of \eqref{item:phi-star-technical-lemma-maximal}.
Suppose that $\phi$ is $C$-expanded. 
Then for every $c \in C$ the set $\{\phi_{left}(c), \phi^*(c), \phi_{right}(c)\}$ covers $L_\circle \setminus P$.
Hence, for every $c \in C \setminus C(S_L)$ the set $\phi_{left}(c) \cup \phi_{right}(c)$ covers the slot $S_L$
and for every $c \in C \setminus C(S_R)$ the set $\phi_{left}(c) \cup \phi_{right}(c)$ covers the slot $S_R$.
Now, let $I_L$ be a tiny interval next to the right endpoint of $S_L$, $I_R$ be a tiny interval next to the left endpoint of $S_R$, and $\psi$ be a mapping on $C \cup I \cup \{v_L, v_R\}$ defined as follows:
$$\psi(v) = 
\left\{
\begin{array}{lcl}
\phi_{left}(c) \cup \phi_{right}(c) \cup P & \text{if} & v \in C, \\
\phi(v) & \text{if} & v \in I, \\
I_L & \text{if} & v = v_L, \\
I_R & \text{if} & v = v_R. \\
\end{array}
\right.
$$
See Figure \ref{fig:technical-lemma-b} for an illustration.
One can easily check that $\psi$ is a circular arc model of $G^{ext}\big{(}C(S_L),I_S,C(S_R)\big{)}$.
\end{proof}
A similar lemma can be stated for the components lying on the stick in $\phi$.
Here, for every $c \in C$ by $\phi^{**}(c)$ we denote the shortest 
interval in $L_S \setminus P$ such that $\phi^{**}(c)$ contains the interval $\bigcup \phi(I)$ for every $I \in \mathcal{I}^{\phi}_{\stick}$ such that $c \in C(I)$.
We use the same notation for the sectors of $\phi^{**}$ as for the sectors of $\phi^*$.
\begin{lemma}
\label{lemma:phi-star-star-technical-lemma}
Let $\phi$ be a $C$-centered model of $G$, $I$ be a component from $\mathcal{I}^{\phi}_{\stick}$, and $S_L$ be the sector of $\phi^{**}$ adjacent to $S(I)$.
Then $I$ is ambiguous and $C(S_L) \subseteq C^s(I)$. 
Moreover, if $I$ is ambiguous non-simple, then $I$ is the leftmost component in the sector $S(I)$.
\end{lemma}
\begin{proof}
Note that for every $c \in N_C(I) \setminus N^s_C(I)$:
\begin{itemize} 
\item the interval $\phi_{stick}(c)$ overlaps the slot $S(I)$ and 
 the left side of $\phi^{**}(c)$ coincides with the left side of $S(I)$.
\end{itemize}
In particular, we have $C(S_L) \subseteq C^s(I)$. 
The rest of the lemma is proved the same as Lemma \ref{lemma:phi-star-technical-lemma}.
\end{proof}

The above lemmas allow to characterize the types of components in particular sectors of $\phi^*$ and $\phi^{**}$.
We partition the maximal sectors of $\phi^*$ into two groups:
\emph{one-component maximal sectors} of $\phi^*$ and \emph{multi-component maximal sectors} of $\phi^*$, 
which contain, respectively, exactly one or at least two components from $\mathcal{I}^{\phi}_{\circle}$. 
Note that:
\begin{itemize}
 \item $I$ satisfies the assumption of Lemma \ref{lemma:phi-star-star-technical-lemma}.\eqref{item:phi-star-technical-lemma-ambiguous-left} if and only if
 $I$ is contained in a minimal sector, or $I$ is contained in a multi-component maximal sector and $I$ is not the rightmost component of this sector, 
 or $I$ is contained in a sector $S$ that satisfies $S_L \subsetneq S \subsetneq S_R$, where $S_L$ and $S_R$ are sectors adjacent to $S$ from the left and the right side, respectively,
 \item $I$ satisfies the assumption of Lemma \ref{lemma:phi-star-star-technical-lemma}.\eqref{item:phi-star-technical-lemma-ambiguous-right} if and only if
 $I$ is contained in a minimal sector, or $I$ is contained in a multi-component maximal sector and $I$ is not the leftmost component of this sector, 
 or $I$ is contained in a sector $S$ that satisfies $S_R \subsetneq S \subsetneq S_L$, where $S_L$ and $S_R$ are sectors adjacent to $S$ from the left and the right side, respectively.
\end{itemize}
Taking into account property ~\eqref{eq:input_simple_components} which asserts no sector of $\phi^*$ and $\phi^{**}$ can contain two ambiguous simple 
components, we obtain the following lemma:
\begin{lemma}
\label{lemma:sectors-phi-star-phi-star-star}
Let $\phi$ be a $C$-centered model of $G$.
Let $S$ be a sector of $\phi^*$ and let $S_L$ and $S_R$ be the sectors adjacent to $S$.
\begin{enumerate}
 \item \label{item:multi-component-maximal-sector-phi-star} If $S$ is a multi-component maximal sector, then $S$ contains at most $3$ components and all of them are ambiguous.
 \begin{itemize}
  \item If $S$ contains three components $I_L \prec_{\phi} I_M \prec_{\phi} I_R$, then 
  $I_L$ and $I_R$ are ambiguous non-simple and $I_M$ is ambiguous simple.
  \item If $S$ contains two components $I_L \prec_{\phi} I_R$, then at most one of them is simple.
 \end{itemize}
 \item \label{item:one-component-maximal-sector-phi-star} If $S$ is a one-component maximal sector, then $S$ contains any kind component. 
 \item \label{item:left-zone-sector-phi-star} If $S$ is such that $C(S_L) \subsetneq C(S) \subsetneq C(S_R)$, 
 then $S$ contains at most two ambiguous components, at most one simple and at most one non-simple.
 If $S$ contains non-simple component, it is the leftmost in $S$.
 \item \label{item:right-zone-sector-phi-star} If $S$ is such that $C(S_R) \subsetneq C(S) \subsetneq C(S_L)$, 
 then $S$ contains at most two ambiguous components, at most one simple and at most one non-simple.
 If $S$ contains non-simple component, it is the rightmost in $S$.
 \item \label{item:minimal-sector-phi-star} If $S$ is a minimal sector of $\phi^*$, then $S$ contains at most one component which is ambiguous simple. 
\end{enumerate}
Let $S$ be a sector of $\phi^{**}$.
\begin{enumerate}[resume]
 \item \label{item:left-zone-sector-phi-star-star} $S$ contains at most two ambiguous components, 
 at most one simple and at most one non-simple. 
 If $S$ contains non-simple component, it is the leftmost in $S$.
\end{enumerate}
\end{lemma}

The above lemma asserts that every circle component $I$ occupies a one-component maximal 
sector of~$\phi^*$.  
In particular, if $G$ has a $C$-centered model, then:
\begin{equation}
\label{eq:input_circle_components}
\begin{array}{c}
 \text{There are no two distinct circle components $I,J \in \mathcal{I}$}\\
 \text{such that $C(I) \subseteq C(J)$.}
 \end{array}
 \tag{I3}
\end{equation}
\begin{equation}
\label{eq:deletion_rule_upset}
\begin{array}{c}
\text{If distinct $I,J \in \mathcal{I}$ are such that $I$ is circle and  $C(I) \subseteq C(J)$,} \\ 
\text{then $J \in \mathcal{I}^{\phi}_{\stick}$ for every $C$-centered model $\phi$ of $G$.}
\end{array}
 \tag{D1}
\end{equation}

Our algorithm testing whether $G$ has a $C$-centered model,
instead of finding a valid model $\phi$ directly,
first determines all candidates for its `fingerprint' $H^\phi$.
A $C$-centered $L$-model of $G$ is \emph{$(C,H)$-centered} if
if it has a $C$-centered model $\phi$ such that $H^\phi$ equals $H$.

\begin{theorem}
\label{thm:H-family-has-poly-size}
There is a polynomial time algorithm that, given a graph $G$ and a maximal clique $C \subseteq V(G)$,
outputs a polynomial-size set of interval graphs $\mathcal{H}$
such that, if $G$ is a $C$-centered $L$- model $\phi$,
then $H^{\phi}=H$  for some $H \in \mathcal{H}$.
\end{theorem}
Given the above theorem, the remaining task, solved in the next section, is to find a poly-time algorithm for the following problem.

\medskip
\noindent \textbf{Strongly Centered Lollipop Graph Recognition:} \\
\begin{tabular}{rl}
\textbf{Input:}& A graph $G$, a maximal clique $C \subseteq V(G)$, 
and an interval graph $H$ on $C' \subseteq C$. \\
\textbf{Output:}& \textbf{YES} if $G$ admits a $(C,H)$-centered model.
\end{tabular}
\medskip

In the rest of this section we prove Theorem~\ref{thm:H-family-has-poly-size}.
For this purpose, we study the properties of a strict partial order ${\subset_t}$, defined in the set $\mathcal{I}$ as follows:
$$
\begin{array}{ccl}
I \subset_t J & \text{if} & \big{(}C(I) \subsetneq C(J)\big{)} \quad \text{or} \\
&&\big{(}C(I) = C(J) \text{ and $I \in \mathcal{I}_c$ and $J \in \mathcal{I}_a$}\big{)}\quad \text{or} \\
&& \big{(}C(I) = C(J) \text{ and $I \in \mathcal{I}_{ans}$ and $J \in \mathcal{I}_{as}$}\big{)}.
\end{array}
$$
That is, ${\subset_t}$ extends the strict order ${\subsetneq}$ between the sets $\{C(I): I \in \mathcal{I}\}$ such that 
in every group of components $I \in \mathcal{I}$ sharing the same set $C(I)$ the relation ${\subset_t}$ gives the priority to ambiguous components over circle components and to ambiguous simple components over ambiguous non-simple components.
One can easily check that $(\mathcal{I},{\subset_t})$ is indeed a strict partial order. 
Hence, $(\mathcal{I},{\subseteq_t})$ is a poset, where ${\subseteq_t}$ is a reflexive closure of ${\subset_t}$.

\begin{figure}[h]
\centering
\begin{subfigure}[t]{1\linewidth}
\centering
\begin{tikzpicture}[xscale=1,yscale=1,>=latex,shorten >=-0.4pt,shorten <=-0.4pt]
\coordinate (I7) at (0,0) {};
\coordinate (I2) at (-1.5,-1) {};
\coordinate (I6) at (-0.5,-1) {};
\coordinate (I3) at (0.5,-1) {};
\coordinate (I4) at (1.5,-1) {};
\coordinate (I1) at (-1,-2) {};
\coordinate (I5) at (1,-2) {};

\tikzstyle{every node}=[circle,minimum size=5pt,inner sep=0pt,draw,fill]
\node at (I1) {};
\node at (I2) {};
\node at (I3) {};
\node at (I4) {};
\node at (I5) {};
\node at (I6) {};
\node at (I7) {};

\draw[->, shorten >= 3pt] (I2) -- (I7) {};					
\draw[->, shorten >= 3pt] (I6) -- (I7) {};					
\draw[->, shorten >= 3pt] (I3) -- (I7) {};					
\draw[->, shorten >= 3pt] (I4) -- (I7) {};					

\draw[->, shorten >= 3pt] (I1) -- (I2) {};					
\draw[->, shorten >= 3pt] (I1) -- (I6) {};					
\draw[->, shorten >= 3pt] (I5) -- (I3) {};					
\draw[->, shorten >= 3pt] (I5) -- (I4) {};

\tikzstyle{every node}=[inner sep=1pt]
\begin{tiny}
\node at ($(I7) + (0,0.25)$) {$I_7$};
\node at ($(I2) + (-0.25,0)$) {$I_2$};
\node at ($(I6) + (0.25,0)$) {$I_6$};
\node at ($(I3) + (-0.25,0)$) {$I_3$};
\node at ($(I4) + (0.25,0)$) {$I_4$};
\node at ($(I1) + (0,-0.25)$) {$I_1$};
\node at ($(I5) + (0,-0.25)$) {$I_5$};
\end{tiny}
\end{tikzpicture}
\caption{The poset $(\mathcal{I},{\subseteq}_t)$ for the components shown in Table~\ref{tab:component_types_and_subseteqT_poset}.}
\label{fig:poset-subseteqt-example}
\end{subfigure}
\caption{}
\label{fig:posetSubseteqTExampple} 
\end{figure}

\begin{lemma}
\label{lem:poset_on_components}
Suppose $\phi$ is a $C$-centered model of $G$.
Then:
\begin{enumerate}
 \item \label{item:poset_on_components_stick_components_form_chain} $\mathcal{I}^{\phi}_{\stick}$ forms a chain in $(\mathcal{I},{\subseteq_t})$.
 \item \label{item:poset_on_components_circle_components_has_width_2} The downset of every component $I \in \mathcal{I}^{\phi}_{\circle}$ has the width at most $2$ in $(\mathcal{I}^{\phi}_{\circle},{\subseteq_t})$.
 \item \label{item:poset_on_components_downset_has_width_3} The downset of every component $I \in \mathcal{I}^{\phi}_{\circle}$ has the width at most $3$ in $(\mathcal{I},{\subseteq_t})$.
\end{enumerate}
\end{lemma}
\begin{proof}
By Lemma \ref{lemma:phi-star-star-technical-lemma}, for every two slots $S,S'$ of $\phi^{**}$ such that $S$ is to the left of $S'$ we have $C(S) \subsetneq C(S')$.
Hence the components of $\mathcal{I}^{\phi}_{\stick}$ contained in $S$ are $\subseteq_t$--below the components contained in $S'$.
Additionally, Lemma~\ref{lemma:sectors-phi-star-phi-star-star}.\eqref{item:left-zone-sector-phi-star-star} 
shows that at most two components might occupy a single sector of $\phi^{**}$ and, if this is the case, 
the left one is ambiguous non-simple and the right one is ambiguous simple. 
In particular, they are $\subseteq_t$--comparable.
This shows statement~\eqref{item:poset_on_components_stick_components_form_chain}. 

To show~\eqref{item:poset_on_components_circle_components_has_width_2} 
it suffices to prove that for every $I \in \mathcal{I}^{\phi}_{\circle}$ the sets 
$$\{J \in \mathcal{I}^{\phi}_{\circle}: J \subseteq_t I \text{ and } J \prec_{\phi}I\} \quad \text{and} \quad \{J \in \mathcal{I}^{\phi}_{\circle}: J \subseteq_t I \text{ and } I \prec_{\phi} J\}$$
are chains in $(\mathcal{I},{\subseteq_t})$.
Suppose $J_1,J_2 \in \{J \in \mathcal{I}^{\phi}_{\circle}: J \subseteq_t I \text{ and } J \prec_{\phi}I\}$ and $J_1 \prec_{\phi} J_2$.
Clearly, $C(J_1) \subseteq C(J_2) \subseteq C(I)$ as $C(J_1),C(J_2) \subseteq C(I)$, $J_1 \prec_{\phi} J_2 \prec_{\phi} I$, and $\phi^*$ is an interval model of $H^{\phi}$.
If $C(J_1) \subsetneq C(J_2)$, then $J_1 \subseteq_t J_2$.
If $C(J_1) = C(J_2)$, then $J_1$ and $J_2$ share the same sector of $\phi^*$, 
$J_1$ is ambiguous non-simple and $J_2$ is ambiguous simple, as asserted by Lemma~\ref{lemma:sectors-phi-star-phi-star-star}.
Hence, $J_1 \subseteq_t J_2$.
This completes the proof of~\eqref{item:poset_on_components_circle_components_has_width_2}.

Statement~\eqref{item:poset_on_components_downset_has_width_3} follows from \eqref{item:poset_on_components_stick_components_form_chain} and 
\eqref{item:poset_on_components_circle_components_has_width_2} and from the fact that the sets $\mathcal{I}_{\circle}^{\phi},\mathcal{I}_{\stick}^{\phi}$ form a partition of $\mathcal{I}$.
\end{proof}

Lemma~\ref{lem:poset_on_components}.\eqref{item:poset_on_components_downset_has_width_3} allows us to identify another set of components from $\mathcal{I}$ which need to be represented on the stick in any $C$-centered model of $G$.
\begin{equation}
\label{eq:deletion_rule_width}
\begin{array}{c}
\text{If $J \in \mathcal{I}$ is such that the downset of $J$ has the width at least $4$ in $(\mathcal{I},{\subseteq_t})$,} \\ 
\text{then $J \in \mathcal{I}^{\phi}_{\stick}$ for every $C$-centered model $\phi$ of $G$.}
\end{array}
 \tag{D2}
\end{equation}

Let $\mathcal{A}^{\phi}$ be the set of all maximal components from $(\mathcal{I}^{\phi}_{\circle},{\subseteq_t})$.
Clearly, every component from $\mathcal{A}^{\phi}$ is contained in a maximal sector of $\phi^*$.
One can also show, using Lemma \ref{lemma:sectors-phi-star-phi-star-star}, that every maximal sector of $\phi^*$ is occupied by at least one component from $\mathcal{A}^{\phi}$.
In particular, the set of all maximal cliques of $H^{\phi}$, 
and hence the entire graph $H^{\phi}$, is determined by the antichain~$\mathcal{A}^{\phi}$.
More precisely, we have
$$H^{\phi} = H(\mathcal{A}^{\phi}),$$
where for an antichain $\mathcal{A}$ of $(\mathcal{I},{\subseteq_t})$ the graph $H(\mathcal{A})$ is defined such that:
$$V\big{(}H(\mathcal{A})\big{)} = \bigcup \{ C(I) : I \in \mathcal{A} \} \quad \text{and} \quad 
E\big{(}H(\mathcal{A})\big{)} = \big{\{}cc': c,c' \in C(I) \text{ for some } I \in \mathcal{A}\big{\}}.$$

Now, we refer to Algorithm~\ref{alg:recognizing_centered_graphs} that computes a family of interval graphs $\mathcal{H}$ fulfilling the claims of Theorem~\ref{thm:H-family-has-poly-size}.
From the considerations made above, we may assume an input instance 
$G,C$ to Algorithm \ref{alg:recognizing_centered_graphs} satisfies conditions~\eqref{eq:input_interval_components}-\eqref{eq:input_circle_components}.
Let $\mathcal{D}$ denote the set of components $J \in \mathcal{I}$ such that $J$ satisfies~\eqref{eq:deletion_rule_upset} or~\eqref{eq:deletion_rule_width}.
We assume $\mathcal{D}$ is a chain in $(\mathcal{I},{\subseteq}_t)$. 
Otherwise, since for any $C$-centered $L$-model $\phi$ of $G$ we have $\mathcal{D} \subseteq \mathcal{I}^{\phi}_{\stick}$
and $\mathcal{I}^{\phi}_{\stick}$ is a chain in $(\mathcal{I},{\subseteq}_t)$, we could conclude $G$ 
has no $C$-centered $L$-model.
To prove that Algorithm~\ref{alg:recognizing_centered_graphs} is correct
it suffices to show $H(\mathcal{A}^{\phi}) \in \mathcal{H}$ 
for any $C$-centered model $\phi$ of~$G$.
To the rest of this section, 
the upset $US$ and downset $DS$ operators are referring to the poset $(\mathcal{I},{\subseteq_t})$.  

Suppose $\phi$ is a $C$-centered model of $G$.
Clearly, we have $\mathcal{I}_c \subseteq \mathcal{A}^{\phi}$ and $US(\mathcal{I}_c) \subseteq \mathcal{D}$ by property \eqref{eq:deletion_rule_upset}.

\begin{figure}[h]
\centering
\begin{subfigure}[t]{1\linewidth}
\centering
\begin{tikzpicture}[xscale=0.3,yscale=0.3,>=latex]

%IC
\coordinate (lIc) at (-2,0) {};

\coordinate (lUSIc) at (5,6) {};
\coordinate (lDSIc) at (5,-6) {};

\coordinate (Ic1) at (0,0) {};
\coordinate (Ic2) at (2,0) {};
\coordinate (Ic3) at (4,0) {};
\coordinate (Ic4) at (6,0) {};
\coordinate (Ic5) at (8,0) {};
\coordinate (Ic6) at (10,0) {};

%J1

\coordinate (A21) at (14,0) {};
\coordinate (A22) at (16,0) {};
\coordinate (A23) at (18,0) {};
\coordinate (A24) at (20,0) {};
\coordinate (A25) at (22,0) {};
\coordinate (A12) at (24,0) {};
\coordinate (A13) at (26,0) {};
\coordinate (A14) at (28,0) {};
\coordinate (A15) at (30,0) {};
\coordinate (A16) at (32,0) {};

\coordinate (lA1) at (34,0) {};
\coordinate (lA2) at (12,0) {};

%\coordinate (lJ1) at (31,0) {};

\coordinate (lDSJ) at (23,-6) {};

%A2

\coordinate (lI) at (18.5,0.3) {};

%D
\coordinate (D1) at (5,11) {};
\coordinate (D2) at (6,9) {};
\coordinate (D3) at (9,7) {};
\coordinate (D4) at (18,5) {};
\coordinate (D5) at (14,-9) {};
\coordinate (D6) at (14,-11) {};

%(D1) at (5,10)

\coordinate (B1L) at (23,0) {};
\coordinate (B1R) at (13,0) {};

%\coordinate (B2R) at (21,-9) {};
%\coordinate (B2L) at (15,-9) {};

%US(I_C)
\draw[fill=red!30] (-0.5,1)--(-1.5,12)--(11.5,12)--(10.5,1)--cycle;
%I_C
\draw[fill=red!30,opacity=0.7,rounded corners=6pt] (-1,-1)--(11,-1)--(11,1)--(-1,1)--cycle;

%DS(I_C)
\draw[fill=red!30] (-0.5,-1)--(-1.5,-12)--(11.5,-12)--(10.5,-1)--cycle;

%J
\draw[fill=teal!30,opacity=0.7,rounded corners=6pt] (12.8,-1)--(33.2,-1)--(33.2,1)--(12.8,1)--cycle;
\draw[fill=teal!30] (13.5,-1)--(12.5,-12)--(33.5,-12)--(32.5,-1)--cycle;

\draw[fill=green!60,opacity=0.7,rounded corners=6pt] (12.8,-1)--(22.8,-1)--(22.8,1)--(12.8,1)--cycle;
\draw[fill=red!60,opacity=0.7,rounded corners=6pt] (23.2,-1)--(33.2,-1)--(33.2,1)--(23.2,1)--cycle;

\draw[fill=gray!50] (A23)--(15,-3)--(15,-12)--(21,-12)--(21,-3)--cycle;

\draw[gray,very thick] (A23)--(15,-3);
\draw[gray,very thick] (A23)--(21,-3);

\draw[gray,very thick] (15,-12)--(15,-3);
\draw[gray,very thick] (21,-12)--(21,-3);

\draw[gray,very thick] (17,-1.5)--(17,-10);
\draw[gray,very thick] (18,-0.5)--(18,-10);
\draw[gray,very thick] (19,-1.5)--(19,-10);

\coordinate (A31) at (19,-2.5) {};
\coordinate (A32) at (18,-1.5) {};
\coordinate (A33) at (17,-2.4) {};

\coordinate (lA3) at (18,-2.5) {};

\coordinate (lDSI) at (16.5,-4) {};

%chainD
\draw[red,very thick] (D1) -- (D2);
\draw[red,very thick] (D2) -- (D3);
\draw[red,very thick] (D3) -- (D4);
%\draw[red,very thick] (D4) -- (D5);
\draw[red,very thick] (D5) -- (D6);

%incD4 (18,4)
%\draw[red,very thick,dashed] (D4) -- (25,12);
\draw[red,very thick,dashed] (D4) -- (35,-12);

\draw[red,very thick,red] (D4) -- (B1L);
\draw[red,very thick,red] (D4) -- (B1R);
\draw[red,very thick,red] (D5) -- (B1L);
\draw[red,very thick,red] (D5) -- (B1R);

%incD5 (18,-6)
\draw[red,very thick,dashed] (D5) -- (35,12);
%\draw[red,very thick,dashed] (D5) -- (25,-12);

\tikzstyle{every node}=[circle,minimum size=5pt,inner sep=0pt,draw,fill]
\node at (Ic1) {};
\node at (Ic2) {};
\node at (Ic3) {};
\node at (Ic4) {};
\node at (Ic5) {};
\node at (Ic6) {};

\node at (A12) {};
\node at (A13) {};
\node at (A14) {};
\node at (A15) {};

\node at (A21) {};
\node at (A22) {};
\node at (A24) {};
\node at (A25) {};

\node at (A31) {};
\node at (A32) {};
\node at (A33) {};

\node[red] at (D1) {};
\node[red] at (D2) {};
\node[red] at (D3) {};
\node[red] at (D4) {};
\node[red] at (D5) {};
\node[red] at (D6) {};

\tikzstyle{every node}=[circle,minimum size=5pt,inner sep=0pt,draw, fill=white]
\node at (A23) {};

\tikzstyle{every node}=[inner sep=1pt]
\begin{tiny}
\node at (lA1) {$\mathcal{A}_1$};
\node at (lA2) {$\mathcal{A}_2$};
\node at (lA3) {$\mathcal{A}_3$};
\node at (lIc) {$\mathcal{I}_c$};
%\node at (lJ1) {$\mathcal{J}$};
\node at (lDSIc) {$DS(\mathcal{I}_c)$};
\node at (lUSIc) {$US(\mathcal{I}_c)$};
\node at (lI) {$I$};
\node at (lDSI) {$DS(I)$};
\node at (lDSJ) {$\mathcal{J}$};
%\node at (lJ22) {$\mathcal{J}$};
\end{tiny}

\end{tikzpicture}
\end{subfigure}
\caption{Schematic view of the poset $(\mathcal{I},{\subseteq_t})$. The elements of $\mathcal{A}^{\phi}$ are marked with dots filled in black.
Chain $\mathcal{D}$ is shown in red.}
\label{fig:poset-subseteqt} 
\end{figure}
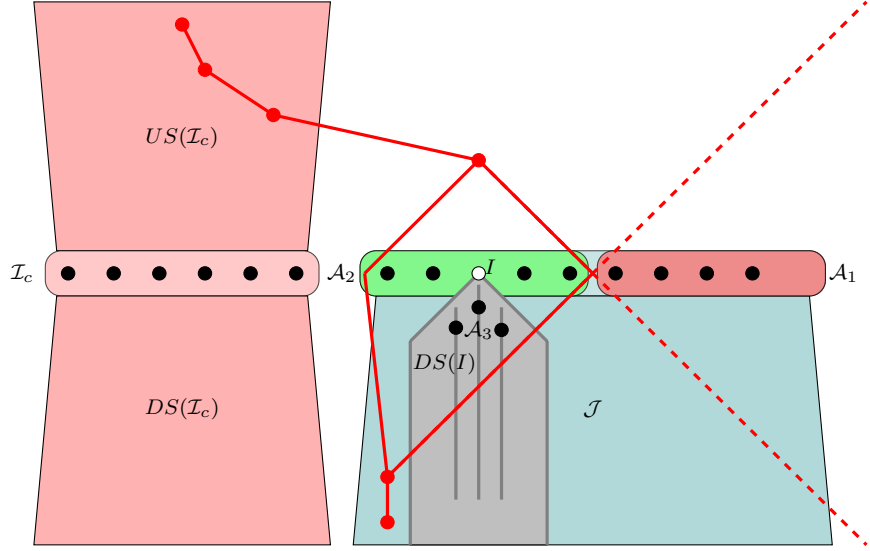

Let $\mathcal{J}_c$, $\mathcal{J}$, $\mathcal{A}$, $\mathcal{A}_1$ and $\mathcal{A}_2$ 
be as defined by Algorithm~\ref{alg:recognizing_centered_graphs} -- see Figure \ref{fig:poset-subseteqt} for an illustration.
Since $\mathcal{I}_c \subseteq \mathcal{I}^{\phi}_{\circle}$, $\mathcal{D} \subseteq \mathcal{I}^\phi_{\stick}$, and $\mathcal{I}^\phi_{\stick}$ is a chain in $(\mathcal{I}, {\subseteq_t})$, 
we have $\mathcal{I}^{\phi}_{\stick} \cap (\mathcal{I}_c \cup \mathcal{A}_1) = \emptyset$.
Because $\mathcal{I}^\phi_{\stick}$ is a chain and $\mathcal{A}$ is an antichain, $|\mathcal{I}^{\phi}_{\stick} \cap \mathcal{A}_2| \leq 1$.
Since $\mathcal{I}^{\phi}_{\circle} \subseteq DS(\mathcal{A}^{\phi}) \cup \mathcal{A}^{\phi}$ and since any element from $\mathcal{A}$ is not ${\subset_t}$-below
any element from~$\mathcal{A}^{\phi}$, we conclude $(\mathcal{I}_{c} \cup \mathcal{A}_1) \subseteq \mathcal{A}^{\phi}$
and either $(\mathcal{I}_{c} \cup \mathcal{A}_1 \cup \mathcal{A}_2) = \mathcal{A}^{\phi}$ or 
there is $I \in \mathcal{A}_2$ such that $\big{(}(\mathcal{I}_{c} \cup \mathcal{A}_1 \cup \mathcal{A}_2) \setminus \{I\} \big{)} \subseteq \mathcal{A}^{\phi}$. 
If the second case holds, we must have
$\mathcal{A}^{\phi} =  \mathcal{A}_I \cup \mathcal{A}_3,$
where $\mathcal{A}_I = \big{(}(\mathcal{I}_c \cup \mathcal{A}_1 \cup \mathcal{A}_2) \setminus \{I\} \big{)}$ and 
$\mathcal{A}_3$ is an antichain contained in $DS(I)$ (possibly empty).
In any case, we have $H(\mathcal{A}^{\phi}) \in \mathcal{H}$.
Since $I \notin \mathcal{D}$, $(DS(I),{\subseteq_t})$ has the width at most $3$, and thus the set $\mathcal{H}$ contains polynomially many interval graphs.
This completes the proof of Theorem~\ref{thm:H-family-has-poly-size}.
\begin{algorithm}
\caption{Computing the family $\mathcal{H}$ for graph $G$ and clique $C$ of $G$}
\label{alg:recognizing_centered_graphs}
\begin{algorithmic}[1]
\State {$\mathcal{H} = \emptyset$}
\State {$\mathcal{J}_c = US(\mathcal{I}_c) \cup \mathcal{I}_c \cup DS(\mathcal{I}_c)$}
\State {$\mathcal{J} = \mathcal{I} \setminus (\mathcal{J}_c \cup \mathcal{D})$
}
%\thcomment{$US(\mathcal{I}_c) \subseteq \mathcal{D}$, right?
%Then I don't see why it is added to $\mathcal{J}_c$.}
%\tkcomment{We have $US(\mathcal{I}_c) \subseteq \mathcal{D}$. However, $\mathcal{D}$ might contain components which are not in $US(\mathcal{I}_c)$: see the smallest three components in red in Figure 6.5}
\State {$\mathcal{A} = \text{maximal components in $(\mathcal{J},{\subseteq_t})$}$}
\State {$\mathcal{A}_1 = \{I \in \mathcal{A}: A \text{ is incomparable to some component in $\mathcal{D}$}\}$}
\State {$\mathcal{A}_2 = \mathcal{A} \setminus \mathcal{A}_1$}
\If{$H(\mathcal{I}_c \cup \mathcal{A}_1 \cup \mathcal{A}_2)$ is an interval graph}
\State {add $H(\mathcal{I}_c \cup \mathcal{A}_1 \cup \mathcal{A}_2)$ to $\mathcal{H}$}
\EndIf
\For {$I \in \mathcal{A}_2$}
    \State{$\mathcal{A}_I = (\mathcal{I}_c \cup \mathcal{A}_1 \cup \mathcal{A}_2) \setminus \{I\}$}
    \For{every antichain $\mathcal{A}_3$ in $DS(I)$}
        \If{$(\mathcal{A}_I \cup \mathcal{A}_3)$ is an antichain in $(\mathcal{I},{\subseteq_t})$}
                \If{$H(\mathcal{A}_{I} \cup \mathcal{A}_3)$ is an interval graph}
		  \State{Add $H(\mathcal{A}_{I} \cup \mathcal{A}_3)$ to $\mathcal{H}$}
		\EndIf
        \EndIf
    \EndFor
\EndFor
\end{algorithmic}
\end{algorithm}

% !TeX spellcheck = en_US
\section{Recognition of strongly centered lollipop graphs}
\label{sec:strongly_centered_lollipop_graphs}

Suppose $G,C,H$ is an input to the recognition problem of strongly centered graphs.

Suppose $\mathbb{T}$ is an unordered $PQ$-tree of $H$ and $\mathbb{R}$ is the root of $\mathbb{T}$.
Let $\mathbb{N}$ be a node and let $\mathcal{L}(\mathbb{N})$ denote 
the set of all leaves of $\mathbb{T}$ which descend $\mathbb{N}$ in $\mathbb{T}$.
We set $\mathcal{L} = \mathcal{L}(\mathbb{R})$.
A~linear ordering $\mathbb{L}_1,\ldots,\mathbb{L}_n$ of $\mathcal{L}(\mathbb{N})$ is \emph{admissible} for $\mathbb{N}$ if 
$\mathbb{L}_1,\ldots,\mathbb{L}_n$ is left-right order obtained by ordering the children of every descendant $\mathbb{M}$ (including $\mathbb{N}$) of 
$\mathbb{N}$ according to an admissible order for $\mathbb{M}$.  

Consider a vertex $v \in V(H)$.
We define $\Cl(v) = \{ \LLL \in \LL \mid v \in C(\LLL) \}$,
hence the set of leaves of $\LL$ where $v$ occurs in $C(\LLL)$.
Note that $\Cl(v) \neq \emptyset$ for every $v \in V(H)$.
For a node $\NNN$ of $\TTT$, we define subsets $U(\mathbb{N})$, $V(\mathbb{N})$,
$I(\mathbb{N})$, $E(\mathbb{N})$ of $V(H)$, as follows:
$$ V(\NNN) = \{ v \in V(H) \mid \Cl(v) \subseteq \LL(\NNN) \} ,$$
$$ U(\NNN) = \{v \in V(H) \mid \Cl(v) = \LL(\NNN)  \} ,$$
$$ I(\NNN) = \{v \in V(H) \mid \Cl(v) \subsetneq \LL(\NNN) \} ,$$
$$ E(\NNN) = \{v \in V(H) \mid  \LL(\NNN) \subsetneq \Cl(v) \}.$$
In particular, we have $V(\mathbb{N}) = U(\mathbb{N}) \cup I(\mathbb{N})$.
In Figures~\ref{fig:exampleINTNormalziedgraph} and \ref{fig:exampleINTgraph} the intervals representing the vertices 
from the sets $V(\mathbb{P}_1)$, $V(\mathbb{P}_2)$, $V(\mathbb{P}_3)$ and $V(\mathbb{Q}_1)$ are contained in the boxes marked by
colors gray, red, blue, and purple, respectively.
For example, for the node $\mathbb{Q}_1$ we have: $V(\mathbb{Q}_1) = \{v_8,v_9,v_{10},v_{11},v_{12},v_{13}\}$, 
$I(\mathbb{Q}_1) = \{v_9,v_{10},v_{11},v_{12},v_{13}\}$, $U(\mathbb{Q}_1) = \{v_8\}$, and $E(\mathbb{Q}_1) = \{v_1,v_2,v_7\}$.

Suppose $\phi$ is a $(C,H)$-centered model of $G$.
By Lemma~\ref{lemma:phi-star-technical-lemma}, if $I$ is a component contained in a minimal sector $S$ of $\phi^*$, then $I$ is ambiguous simple and $C(S)=C(I)$.
Since the set $\{C(S): S \text{ is a minimal sector of $\phi^*$}\}$ is independent on an interval model $\phi^*$ of $H$, 
any model of $H$ contains a sector where the component $I$ can be placed.
The above observation allows us to assume the following property of the instance $G,C,H$:
\begin{equation}
\label{eq:input_no_simple_cliques_on_minimal_sectors}
\begin{array}{c}
\text{There is no ambiguous simple component $I \in \mathcal{I}$ such that $C(I)=C(S)$} \\
\text{for some minimal sector $S$ of some interval model of $H$.}
\end{array}
\tag{I4}
\end{equation}

In order to describe some properties of the linear orders $(\mathcal{I}^{\phi}_{\circle}, \prec_{\phi})$ for $(C,H)$-centered models $\phi$ of~$G$,
we introduce a binary relation $\sqsubset$ defined in the set $\mathcal{I}_a$ as follows:
$$
I \sqsubset J \quad \text{if} \quad C(I) \subseteq C^s(J), \quad \text{for distinct $I,J \in \mathcal{I}_a$}.
$$
One can easily check that $(\mathcal{I}_{a},{\sqsubset})$ is a strict partial order and
hence $(\mathcal{I}_{a},{\sqsubseteq})$ is a poset, 
where ${\sqsubseteq}$ is the reflexive closure of ${\sqsubset}$.
Let us denote a clique $B$ in $H$ as a \emph{border}.
For a chain $\mathcal{A}$ in $(\mathcal{I}_{a}, {\sqsubseteq})$ and a border $B$,
we say $\mathcal{A}$ \emph{respects the border $B$} if $B \subseteq C^s(I)$ for every $I \in \mathcal{A}$ 
(or equivalently, if $B \subseteq C^s(I)$ holds for the minimal component $I$ in $(\mathcal{A}, {\sqsubseteq})$).
Lemma~\ref{lemma:phi-star-star-technical-lemma} proves that:
\begin{lemma}
\label{lemma:D_forms_a_sqsubseteq_chain}
For every $(C,H)$-centered model $\phi$ of $G$, the set $\mathcal{I}^{\phi}_{\stick}$ forms a chain in $(\mathcal{I}_a,{\sqsubseteq})$.
\end{lemma}
As in the previous section, let $\mathcal{D}$ denote the set of all components $I$ from $\mathcal{I}_a$ 
which satisfy \eqref{eq:deletion_rule_upset} or \eqref{eq:deletion_rule_width}.
Since $\mathcal{D} \subseteq \mathcal{I}^{\phi}_{\stick}$ for any $(C,H)$-centered model $\phi$ of $G$, 
we can assume the instance $G,C,H$ is such that:
\begin{equation}
\label{eq:input_deleted_components_form_chain}
\text{The set $\mathcal{D}$ forms a chain in $(\mathcal{I}_a,{\sqsubseteq})$.}
\tag{I5}
\end{equation}
We denote also by $\mathcal{I'}$ the set $\mathcal{I} \setminus \mathcal{D}$.

To test whether $G$ admits a $(C,H)$-centered model we check whether there exists
a partition $(\mathcal{J}, \mathcal{J}')$ of the set $\mathcal{I}'$ and a strict linear order ${\prec}$ of $\mathcal{J}$ such that
$(\mathcal{J},\mathcal{J}',{\prec})$ can be turned into a $(C,H)$-centered model $\phi$ of $G$ that satisfies
$(\mathcal{J},{\prec}) = (\mathcal{I}^{\phi}_{\circle}, {\prec_{\phi}})$ and 
$\mathcal{J}' \cup \mathcal{D} = \mathcal{I}^{\phi}_{\stick}$.
The triples $(\mathcal{J},\mathcal{J}',{\prec})$ possessing this property will be called \emph{good} for the set~$\mathcal{I}'$ -- see Definition \ref{def:good_triple}.
First, we characterize the strict linear orders $(\mathcal{J},{\prec})$ which can occur in good triples for $\mathcal{I}'$ as
\emph{good orders for the root $\mathbb{R}$ of $\mathbb{T}$} -- see Definition~\ref{def:good-order}.
In particular, for every $(C,H)$-model $G$ the pair $(\mathcal{I}^{\phi}_{\circle}, {\prec_{\phi}})$ is a good order for the root $\mathbb{R}$.
On the other hand, we define good orders in a broader context, for all nodes in $\mathbb{T}$. 
It turns out that good orders for the nodes of $\mathbb{T}$ appear as intervals in good orders for $\mathbb{R}$.
Moreover, the structure of a good order for a node $\mathbb{N}$ of $\mathbb{T}$ can be described 
by means of the structure of good orders for the children of $\mathbb{N}$.
This allows to construct good orders for $\mathbb{R}$ in a bottom-up manner along the tree $\mathbb{T}$.

First we need some more definitions.
For a leaf $\mathbb{L} \in \mathcal{L}$, and a subset $\mathcal{K} \in \mathcal{I}'$, let
$$inner(\mathbb{L}) = \{I \in \mathcal{I}': C(I) = C(\mathbb{L})\} \quad \text{and} \quad inner^{\mathcal{K}}(\mathbb{L}) = inner(\mathbb{L}) \cap \mathcal{K}.$$

Consider a node $\mathbb{N}$ of $\mathbb{T}$ and a subset $\mathcal{K} \in \mathcal{I}'$.
Let ${\prec}$ be a strict linear order on $\mathcal{K}$
such that for every leaf $\mathbb{L} \in \mathcal{L}(\mathbb{N})$ the set
$inner^{\mathcal{K}}(\mathbb{L})$ is a non-empty interval in $(\mathcal{K},\prec)$.
Hence this defines an order $\mathbb{L}_1,\ldots, \mathbb{L}_{n}$ of $\mathcal{L}(\mathbb{N})$ given by 
$inner^{\mathcal{K}}(\mathbb{L}_1) \prec \dots \prec inner^{\mathcal{K}}(\mathbb{L}_n)$.
If additionally $\mathbb{L}_1,\ldots, \mathbb{L}_{n}$ is admissible for $\mathbb{N}$, then we denote $(\mathcal{K},{\prec})$ as \emph{nice to $\mathbb{N}$} and 
$\mathbb{L}_1,\ldots,\mathbb{L}_n$ as \emph{being induced by $(\mathcal{K},{\prec})$}.
%For example, $(\mathcal{I}^\phi_\circle, {\prec_\phi})$ is nice to $\mathbb{R}$.
Additionally, we define the following subsets of $\mathcal{K}$:
$$
\begin{array}{rcl} \medskip
zone_{L}^{\mathcal{K},\prec}(\mathbb{L}_i) &=& \{I \in \mathcal{K}: I \prec inner^{\mathcal{K}}(\mathbb{L}_i) \text{ and } C(I) \cap (C(\mathbb{L}_i) \setminus C(\mathbb{L}_{i-1})) \neq \emptyset\}, \\ \medskip
zone_{R}^{\mathcal{K},\prec}(\mathbb{L}_i) &=& \{I \in \mathcal{K}: inner^{\mathcal{K}}(\mathbb{L}_i) \prec I \text{ and } C(I) \cap (C(\mathbb{L}_i) \setminus C(\mathbb{L}_{i+1})) \neq \emptyset\}, \\
zone^{\mathcal{K},\prec}(\mathbb{L}_i) & = & zone_L^{\mathcal{K},\prec}(\mathbb{L}_i) \cup 
inner^{\mathcal{K}}(\mathbb{L}_i) \cup zone_R^{\mathcal{K},\prec}(\mathbb{L}_i).
\end{array}
$$
Here we conveniently assumed $C(\mathbb{L}_0) = \emptyset$ and $C(\mathbb{L}_{n+1}) = \emptyset$.
For a non-leaf node $\mathbb{N}$, let $zone^{\mathcal{K},\prec}(\mathbb{N}) = \bigcup_{\mathbb{L} \in \mathcal{L}(\mathbb{N})} zone^{\mathcal{K},\prec}(\mathbb{L})$.
We may drop the superscript `$\mathcal{K},\prec$' if it is clear from the context.

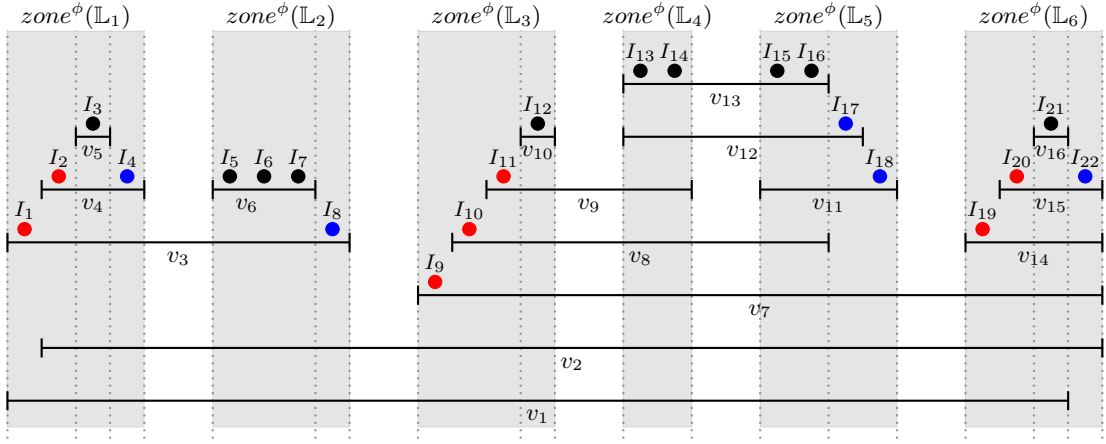
\begin{figure}[h]
\centering
\begin{subfigure}[t]{1\linewidth}
\begin{tikzpicture}[xscale=0.9,yscale=0.7,>=latex,shorten >=-0.4pt,shorten <=-0.4pt]
  \tikzstyle{every node}=[inner sep=2pt,fill=white]  

%P1  
%\draw[opacity=0.4,color=lightgray,fill=lightgray](-0.5,-0.5) rectangle (16.5,7);
\coordinate (lP1) at (0,6.4) {};

%P2
%\draw[opacity=0.4,color=red,fill=red](-0.4,-0.3) rectangle (5.4,6);
\coordinate (lP2) at (2.5,5.6) {};

%P3
%\draw[opacity=0.4,color=teal,fill=teal](5.6,-0.3) rectangle (16.3,6.8);
\coordinate (lP3) at (16,6.4) {};

%Q1
%\draw[opacity=0.4,color=magenta,fill=magenta](5.8,-0.1) rectangle (13.2,6.6);
\coordinate (lQ1) at (6.2,6.2) {};

%sectors
%L1
\draw[dotted,opacity=0.4,thick] (0,-0.7) -- (0,7);
\draw[dotted,opacity=0.4,thick] (1,-0.7) -- (1,7);
%\draw[opacity=0.2,color=red,fill=red](0,-0.5) rectangle (1,7);

\draw[dotted,opacity=0.4,thick] (1.5,-0.7) -- (1.5,7);
%\draw[opacity=0.2,color=gray,fill=gray](1,-0.5) rectangle (1.5,7);

\draw[dotted,opacity=0.4,thick] (2,-0.7) -- (2,7);
%\draw[opacity=0.2,color=blue,fill=blue](1.5,-0.5) rectangle (2,7);

\draw[opacity=0.2,color=gray,fill=gray](0,-0.5) rectangle (2,7);
\coordinate (lC1) at (1,7.3) {};

%L2
\draw[dotted,opacity=0.4,thick] (3,-0.7) -- (3,7);
\draw[dotted,opacity=0.4,thick] (4.5,-0.7) -- (4.5,7);
%\draw[opacity=0.2,color=gray,fill=gray](3,-0.5) rectangle (4,7);
\draw[dotted,opacity=0.4,thick] (5,-0.7) -- (5,7);
%\draw[opacity=0.2,color=blue,fill=blue](4,-0.5) rectangle (5,7);

\draw[opacity=0.2,color=gray,fill=gray](3,-0.5) rectangle (5,7);
\coordinate (lC2) at (4,7.3) {};

%L3
\draw[dotted,opacity=0.4,thick] (6,-0.7) -- (6,7);
\draw[dotted,opacity=0.4,thick] (7.5,-0.7) -- (7.5,7);
%\draw[opacity=0.2,color=red,fill=red](6,-0.5) rectangle (7.5,7);

\draw[dotted,opacity=0.4,thick] (8,-0.7) -- (8,7);
%\draw[opacity=0.2,fill=gray](7.5,-0.5) rectangle (8,7);
\draw[opacity=0.2,color=gray,fill=gray](6,-0.5) rectangle (8,7);
\coordinate (lC3) at (7,7.3) {};

%L4
\draw[dotted,opacity=0.4,thick] (9,-0.7) -- (9,7);
\draw[dotted,opacity=0.4,thick] (10,-0.7) -- (10,7);
%\draw[opacity=0.2,fill=gray](9,-0.5) rectangle (10,7);
\draw[opacity=0.2,color=gray,fill=gray](9,-0.5) rectangle (10,7);
\coordinate (lC4) at (9.5,7.3) {};

%L5
\draw[dotted,opacity=0.4,thick] (11,-0.7) -- (11,7);
\draw[dotted,opacity=0.4,thick] (12,-0.7) -- (12,7);
%\draw[opacity=0.2,fill=gray](11,-0.5) rectangle (12,7);

\draw[dotted,opacity=0.4,thick] (13,-0.7) -- (13,7);
%\draw[opacity=0.2,fill=blue](12,-0.5) rectangle (13,7);
\draw[opacity=0.2,color=gray,fill=gray](11,-0.5) rectangle (13,7);
\coordinate (lC5) at (12,7.3) {};

%L6
\draw[dotted,opacity=0.4,thick] (14,-0.7) -- (14,7);
\draw[dotted,opacity=0.4,thick] (15,-0.7) -- (15,7);
%\draw[opacity=0.2,fill=red](14,-0.5) rectangle (15,7);

\draw[dotted,opacity=0.4,thick] (15.5,-0.7) -- (15.5,7);
%\draw[opacity=0.2,fill=gray](15,-0.5) rectangle (15.5,7);

\draw[dotted,opacity=0.4,thick] (16,-0.7) -- (16,7);
%\draw[opacity=0.2,fill=blue](15.5,-0.5) rectangle (16,7);
\draw[opacity=0.2,color=gray,fill=gray](14,-0.5) rectangle (16,7);
\coordinate (lC6) at (15,7.3) {};

%v1
\draw[|-|,thick] (0,0) -- (15.5,0);
\coordinate (lv1) at (7.75,-0.3) {};

%v2
\draw[|-|,thick] (0.5,1) -- (16,1);
\coordinate (lv2) at (8.25,0.7) {};

%P1-node
%v3
\draw[|-|,thick] (0,3) -- (5,3);
\coordinate (lv3) at (2.5,2.7) {};

\coordinate (I1) at (0.25,3.25) {};
\coordinate (I8) at (4.75,3.25) {};

%v4
\draw[|-|,thick] (0.5,4) -- (2,4);
\coordinate (lv4) at (1.25,3.7) {};
\coordinate (I2) at (0.75,4.25) {};
\coordinate (I4) at (1.75,4.25) {};

%v5
\draw[|-|,thick] (1,5) -- (1.5,5);
\coordinate (lv5) at (1.25,4.7) {};
\coordinate (I3) at (1.25,5.25) {};

%v6
\draw[|-|,thick] (3,4) -- (4.5,4);
\coordinate (lv6) at (3.5,3.7) {};
\coordinate (I5) at (3.25,4.25) {};
\coordinate (I6) at (3.75,4.25) {};
\coordinate (I7) at (4.25,4.25) {};

%P2-node
%v7
\draw[|-|,thick] (6,2) -- (16,2);
\coordinate (lv7) at (11,1.7) {};
\coordinate (I9) at (6.25,2.25) {};

%Q1-node
%v8
\draw[|-|,thick] (6.5,3) -- (12,3);
\coordinate (lv8) at (9.25,2.7) {};
\coordinate (I10) at (6.75,3.25) {};
%v9
\draw[|-|,thick] (7,4) -- (10,4);
\coordinate (lv9) at (8.5,3.7) {};
\coordinate (I11) at (7.25,4.25) {};

%v11
\draw[|-|,thick] (11,4) -- (13,4);
\coordinate (lv11) at (12,3.7) {};
\coordinate (I18) at (12.75,4.25) {};
%v10
\draw[|-|,thick] (7.5,5) -- (8,5);
\coordinate (lv10) at (7.75,4.7) {};
\coordinate (I12) at (7.75,5.25) {};

%v12
\draw[|-|,thick] (9,5) -- (12.5,5);
\coordinate (lv12) at (10.75,4.7) {};
\coordinate (I17) at (12.25,5.25) {};
%v13
\draw[|-|,thick] (9,6) -- (12,6);
\coordinate (lv13) at (10.5,5.7) {};
\coordinate (I13) at (9.25,6.25) {};
\coordinate (I14) at (9.75,6.25) {};

\coordinate (I15) at (11.25,6.25) {};
\coordinate (I16) at (11.75,6.25) {};

%v14
\draw[|-|,thick] (14,3) -- (16,3);
\coordinate (lv14) at (15,2.7) {};
\coordinate (I19) at (14.25,3.25) {};
%v15
\draw[|-|,thick] (14.5,4) -- (16,4);
\coordinate (lv15) at (15.25,3.7) {};
\coordinate (I20) at (14.75,4.25) {};
\coordinate (I22) at (15.75,4.25) {};

%v16
\draw[|-|,thick] (15,5) -- (15.5,5);
\coordinate (lv16) at (15.25,4.7) {};
\coordinate (I21) at (15.25,5.25) {};

\tikzstyle{every node}=[inner sep=1pt]
\begin{scriptsize}
\node at (lv1) {$v_1$};
\node at (lv2) {$v_2$};
\node at (lv3) {$v_3$};
\node at (lv4) {$v_4$};
\node at (lv5) {$v_5$};
\node at (lv6) {$v_6$};
\node at (lv7) {$v_7$};
\node at (lv8) {$v_8$};
\node at (lv9) {$v_9$};
\node at (lv10) {$v_{10}$};
\node at (lv11) {$v_{11}$};
\node at (lv12) {$v_{12}$};
\node at (lv13) {$v_{13}$};
\node at (lv14) {$v_{14}$};
\node at (lv15) {$v_{15}$};
\node at (lv16) {$v_{16}$};

%\node at (lP1) {$\mathbb{P}_1$};
%\node at (lP2) {$\mathbb{P}_2$};
%\node at (lP3) {$\mathbb{P}_3$};
%\node at (lQ1) {$\mathbb{Q}_1$};

\node at (lC1) {$zone^{\phi}(\mathbb{L}_1)$};
\node at (lC2) {$zone^{\phi}(\mathbb{L}_2)$};
\node at (lC3) {$zone^{\phi}(\mathbb{L}_3)$};
\node at (lC4) {$zone^{\phi}(\mathbb{L}_4)$};
\node at (lC5) {$zone^{\phi}(\mathbb{L}_5)$};
\node at (lC6) {$zone^{\phi}(\mathbb{L}_6)$};
\end{scriptsize}

\tikzstyle{every node}=[inner sep=1pt]
\begin{tiny}
\node at ($(I1) + (0,0.35)$) {$I_1$};
\node at ($(I2) + (0,0.35)$) {$I_2$};
\node at ($(I3) + (0,0.35)$) {$I_3$};
\node at ($(I4) + (0,0.35)$) {$I_4$};
\node at ($(I5) + (0,0.35)$) {$I_5$};
\node at ($(I6) + (0,0.35)$) {$I_6$};
\node at ($(I7) + (0,0.35)$) {$I_7$};
\node at ($(I8) + (0,0.35)$) {$I_8$};
\node at ($(I9) + (0,0.35)$) {$I_9$};

\node at ($(I10) + (0,0.35)$) {$I_{10}$};
\node at ($(I11) + (0,0.35)$) {$I_{11}$};
\node at ($(I12) + (0,0.35)$) {$I_{12}$};
\node at ($(I13) + (0,0.35)$) {$I_{13}$};
\node at ($(I14) + (0,0.35)$) {$I_{14}$};
\node at ($(I15) + (0,0.35)$) {$I_{15}$};
\node at ($(I16) + (0,0.35)$) {$I_{16}$};
\node at ($(I17) + (0,0.35)$) {$I_{17}$};
\node at ($(I18) + (0,0.35)$) {$I_{18}$};
\node at ($(I19) + (0,0.35)$) {$I_{19}$};
\node at ($(I20) + (0,0.35)$) {$I_{20}$};
\node at ($(I21) + (0,0.35)$) {$I_{21}$};
\node at ($(I22) + (0,0.35)$) {$I_{22}$};
\end{tiny}

\tikzstyle{every node}=[circle,minimum size=5pt,inner sep=0pt,draw,fill]
\node[red] at (I1) {};
\node[red] at (I2) {};
\node at (I3) {};
\node[blue] at (I4) {};
\node at (I5) {};
\node at (I6) {};
\node at (I7) {};
\node[blue] at (I8) {};
\node[red] at (I9) {};
\node[red] at (I10) {};
\node[red] at (I11) {};
\node at (I12) {};
\node at (I13) {};
\node at (I14) {};
\node at (I15) {};
\node at (I16) {};
\node[blue] at (I17) {};
\node[blue] at (I18) {};
\node[red] at (I19) {};
\node[red] at (I20) {};
\node at (I21) {};
\node[blue] at (I22) {};

%\draw[black] (-1,-0.5)--(-1,0);
%\draw[black] (17,6.7)--(17,7);
\draw[white] (-1,-0.5)--(-1,0);
\draw[white] (17,6.7)--(17,7);

\end{tikzpicture}
\end{subfigure}
\caption{Zones in $\phi^*$. Components from the sets $zone^{\phi}_L(\mathbb{L}_i)$, $inner^{\phi}_{\circle}(\mathbb{L}_i)$, and $zone^{\phi}_R(\mathbb{L}_i)$  
are depicted as red, black, and blue dots, respectively}
\label{fig:zones_of_phi*}
\end{figure}

Suppose $\phi$ is $(C,H)$-model of $G$.
For simplicity, let $zone^{\phi}(\mathbb{L}_i) = zone^{\mathcal{I}^{\phi}_{\circle},{\prec_\phi}}(\mathbb{L}_i)$ and let $zone_L^{\phi}(\mathbb{L}_i), zone_R^{\phi}(\mathbb{L}_i)$ be defined analogously.
To follow our convention we denote the set $inner^{\mathcal{I}^\phi_{\circle}}(\mathbb{L}_i)$ 
by $inner^{\phi}_{\circle}(\mathbb{L}_i)$.
Clearly, $(\mathcal{I}^{\phi}_{\circle},{\prec_{\phi}})$ is nice for $\mathbb{R}$ and $(zone^{\phi}(\mathbb{N}),{\prec_{\phi}})$ is nice for every node $\mathbb{N}$ in $\mathbb{T}$.
Figure~\ref{fig:zones_of_phi*} shows the zones for some example model $\phi^*$.
Components from the sets $zone^{\phi}_L(\mathbb{L}_i)$, $inner^{\phi}_{\circle}(\mathbb{L}_i)$, and $zone^{\phi}_R(\mathbb{L}_i)$  
are illustrated as red, black, and blue dots, respectively.
We have, for example, $zone^{\phi}_L(\mathbb{L}_3) = \{I_9,I_{10},I_{11}\}$, $zone^{\phi}_R(\mathbb{L}_2) = \{I_8\}$, and $inner^{\phi}_{\circle}(\mathbb{L}_2)=\{I_5,I_6,I_7\}$.
We have $zone^{\phi}(\mathbb{Q}_1) = \{I_9,\ldots,I_{18}\}$ as $\mathcal{L}(\mathbb{Q}_1) = \{\mathbb{L}_3,\mathbb{L}_4,\mathbb{L}_5\}$.

\begin{definition}
\label{def:good-order}
Let ${\prec}$ be a strict linear ordering of $\mathcal{K} \subseteq \mathcal{I'}$
that is nice to a node $\mathbb{N}$ in~$\mathbb{T}$.
Let $\mathbb{L}_1,\ldots,\mathbb{L}_n$ be an admissible order of $\mathcal{L}(\mathbb{N})$ induced by $(\mathcal{K},{\prec})$ and let $B_L$ and $B_R$ be two cliques in $H$.
Then $(\mathcal{K},{\prec})$ is \emph{good for the node $\mathbb{N}$ and the borders $(B_L,B_R)$} if the following conditions hold (below we conveniently assume $C(\mathbb{L}_0) = B_L$ and $C(\mathbb{L}_{n+1})=B_R$):
\begin{enumerate}
\item \label{item:def_good_order_partition} The set 
$\{zone(\mathbb{L}_i): i \in [n]\}$ forms a partition of $\mathcal{K}$ and we have
$$zone(\mathbb{L}_1) \prec \dots \prec zone(\mathbb{L}_n).$$
\item \label{item:def_good_order_zones} For every $i \in [n]$:
\begin{itemize}
\item the set $zone_L(\mathbb{L}_i)$ is a chain in $(\mathcal{I}_a,{\sqsubseteq})$ that respects the border $C(\mathbb{L}_{i-1}) \cap C(\mathbb{L}_i)$
and $\big{(}zone_L(\mathbb{L}_i),{\sqsubset}\big{)}$ and $\big{(}zone_L(\mathbb{L}_i),{\prec}\big{)}$ are equal. 
\item the set $zone_R(\mathbb{L}_i)$ is a chain in $(\mathcal{I}_a,{\sqsubseteq})$ that respects the border $C(\mathbb{L}_{i}) \cap C(\mathbb{L}_{i+1})$ 
and $\big{(}zone_R(\mathbb{L}_i), {\sqsubset}\big{)}$ equals to the reverse of 
$\big{(}zone_R(\mathbb{L}_i),{\prec}\big{)}$. 
\end{itemize}
\item \label{item:def-good-order-multi-component-maximal-sectors}
Suppose $i\in [n]$ is such that $|inner^{\mathcal{K}}(\mathbb{L}_i)| \geq 2$.
Let $I_L$ and $I_R$ be the leftmost and the rightmost component in $(inner^{\mathcal{K}},{\prec})$.
Then:
\begin{itemize}
 \item $inner^{\mathcal{K}}(\mathbb{L}_i) \subseteq \mathcal{I}_a$ and the set $inner^{\mathcal{K}}(\mathbb{L}_i) \setminus \{I_L,I_R\}$, if non-empty,
 contains a single ambiguous simple component,
 \item $zone_L(\mathbb{L}_i) \sqsubset I_L$ and $\big{(}zone_L(\mathbb{L}_i) \cup \{I_L\},{\sqsubseteq}\big{)}$ 
 respects the border $C(\mathbb{L}_{i-1}) \cap C(\mathbb{L}_i)$,
 \item $zone_R(\mathbb{L}_i) \sqsubset I_R$ and $\big{(}zone_R(\mathbb{L}_i) \cup \{I_R\},{\sqsubseteq}\big{)}$ 
 respects the border $C(\mathbb{L}_{i}) \cap C(\mathbb{L}_{i+1})$.
\end{itemize}
\item \label{item:def-good-ordering-one-component-maximal-sectors} Suppose $i \in [n]$ is such that $|inner^{\mathcal{K}}(\mathbb{L}_i)| = 1$.
Let $inner^{\mathcal{K}}(\mathbb{L}_i) = \{I_S\}$ and let $C_L$ and $C_R$ be the cliques of $H$ defined as
$$C_L = 
\left\{ 
\begin{array}{ll}
C(I') & \text{where $I'$ is the maximum component in $(zone_L(\mathbb{L}_i),{\sqsubseteq})$}, \\
C(\mathbb{L}_{i-1}) \cap C(\mathbb{L}_{i})& \text{if $zone_L(\mathbb{L}_i) = \emptyset$},
\end{array}
\right.
$$
and 
$$C_R = 
\left\{ 
\begin{array}{cl}
C(I') & \text{where $I'$ is the maximum component in $(zone_R(\mathbb{L}_i),{\sqsubseteq})$}, \\
C(\mathbb{L}_{i}) \cap C(\mathbb{L}_{i+1}) & \text{if $zone_R(\mathbb{L}_i) = \emptyset$.}
\end{array}
\right.
$$
Then:
\begin{itemize}
 \item $N_C(I_S) \setminus N^s_C(I_S) \subseteq (C \setminus C_L) \cup (C \setminus C_R)$,
 \item the graph $G^{ext}(C_L, I_S,C_R)$ has a circular-arc model.
\end{itemize}
\end{enumerate}
If $(B_L,B_R) = (\emptyset,\emptyset)$, then we simply say $(\mathcal{K},{\prec})$ is a \emph{good order} for $\mathbb{N}$.
\end{definition}

\begin{lemma}
\label{lemma:prec_phi_is_good_order_for_R}
Let $\phi$ be a $(C,H)$-centered model for $G$. 
Then $(\mathcal{I}^{\phi}_{\circle},{\prec_{\phi}})$ is a good order for~$\mathbb{R}$.  
\end{lemma}
\begin{proof}
Let $\mathbb{L}_1,\ldots, \mathbb{L}_n$ be an admissible order of $\mathcal{L}$ induced by $(\mathcal{I}^{\phi}_{\circle},{\prec_{\phi}})$, let
$C(\mathbb{L}_0)=C(\mathbb{L}_{n+1})=\emptyset$.

For every $i \in [n]$, definition of the sets $zone^{\phi}_L(\mathbb{L}_i)$ and $zone^{\phi}_R(\mathbb{L}_i)$ asserts we have 
$zone^{\phi}_L(\mathbb{L}_i) \prec_{\phi} inner^{\phi}_{\circle}(\mathbb{L}_i) \prec_{\phi} zone^{\phi}_R(\mathbb{L}_i)$,
for every $I \in zone^{\phi}_L(\mathbb{L}_i)$ there is a vertex $v_I \in C(\mathbb{L}_i) \setminus C(\mathbb{L}_{i-1})$ such that $I$ is contained in $\phi^*(v_I)$, and 
for every $I \in zone^{\phi}_R(\mathbb{L}_{i})$ there is a vertex $v_{I} \in C(\mathbb{L}_i) \setminus C(\mathbb{L}_{i+1})$ such that $I$ is contained in $\phi^*(v_{I})$.
This shows $inner^{\phi}_{\circle}(\mathbb{L}_{i-1}) \prec_{\phi} zone^{\phi}_L(\mathbb{L}_i)$ for every $i \in [2,n]$ and $zone^{\phi}_R(\mathbb{L}_i) \prec_{\phi} inner^{\phi}_{\circle}(\mathbb{L}_{i+1})$ for every $i \in [n-1]$.
Suppose $I \in zone_R(\mathbb{L}_i)$ and $I' \in zone_L(\mathbb{L}_{i+1})$.
Thus, there are $v_I \in C(\mathbb{L}_i) \setminus C(\mathbb{L}_{i+1})$ and
and $v_{I'} \in C(\mathbb{L}_{i+1}) \setminus C(\mathbb{L}_{i})$ such that 
$I$ is contained in $\phi^*(v_I)$ and $I'$ is contained in $\phi^*(v_{I'})$.
Since $\phi^*(v_I)$ is to the left of $\phi^*(v_{I'})$, we have $I \prec_{\phi} I'$.
This shows $zone^{\phi}(\mathbb{L}_i) \prec_\phi zone^{\phi}(\mathbb{L}_{i+1})$.
Now, let $I \in \mathcal{I}^{\phi}_{\circle}$ be such that $ inner^{\phi}_{\circle}(\mathbb{L}_{i})\prec_{\phi} I \prec_{\phi} inner^{\phi}_{\circle}(\mathbb{L}_{i+1})$ for some $i \in [n-1]$.
Condition \eqref{eq:input_no_simple_cliques_on_minimal_sectors} asserts that
either there is $v_I \in C(\mathbb{L}_{i}) \setminus C(\mathbb{L}_{i+1})$ such that $I$ is contained in $\phi^*(v_I)$ or there is $v'_I \in C(\mathbb{L}_{i+1}) \setminus C(\mathbb{L}_{i})$ 
such that $I$ is contained in $\phi^*(v'_I)$.
In the first case $I \in zone^{\phi}_R(\mathbb{L}_i)$ and in the second case 
$I \in zone^{\phi}_R(\mathbb{L}_{i+1})$.
This proves the property~\eqref{item:def_good_order_partition}.

Property~\eqref{item:def_good_order_zones} follows by Lemma~\ref{lemma:phi-star-technical-lemma}.\eqref{item:left-zone-sector-phi-star}-\eqref{item:right-zone-sector-phi-star}.
Property~\eqref{item:def-good-order-multi-component-maximal-sectors} follows 
by Lemma~\ref{lemma:phi-star-technical-lemma} and by Lemma \ref{lemma:sectors-phi-star-phi-star-star}.\eqref{item:multi-component-maximal-sector-phi-star}.

Finally, let $I_S$ be a component in a one-component maximal sector $S$ of $\phi^*$.
Let $C_L$ and $C_R$ be as defined in Definition~\ref{def:good-order}.\eqref{item:def-good-ordering-one-component-maximal-sectors}.
Note that $C_L$ and $C_R$ denote the clique set of the sectors adjacent to $S$ from the left and the right side, respectively.
Then, Lemma~\ref{lemma:phi-star-technical-lemma}.\eqref{item:phi-star-technical-lemma-maximal} asserts that:
\begin{itemize}
 \item $N^s_C(I_S) \setminus N_C(I) \subseteq (C \setminus C_L) \cup (C \setminus C_R)$,
 \item $G^{ext}(C_L,I,C_R)$ is a circular-arc graph.
\end{itemize}
This shows that $(\mathcal{I}^{\phi}_{\circle},{\prec_{\phi}})$ satisfies property \eqref{item:def-good-ordering-one-component-maximal-sectors} 
of Definition~\ref{def:good-order}.
\end{proof}

\begin{definition}
\label{def:good_triple}
Let $(\mathcal{J},\mathcal{J}')$ be a partition of the set $\mathcal{I}'$ and let ${\prec}$ be a strict linear ordering on $\mathcal{J}$. 
A triple $(\mathcal{J}, \mathcal{J}',{\prec})$ is \emph{good} for the set $\mathcal{I}'$ if 
\begin{itemize}
\item $(\mathcal{J},{\prec})$ is a good order for the root $\mathbb{R}$.
\item $\mathcal{J'} \cup \mathcal{D}$ forms a chain in $(\mathcal{I}_{a},{\sqsubseteq})$,
\end{itemize}
\end{definition}
The next lemma shows that the good triples for the set $\mathcal{I}'$ are in the correspondence with $(C,H)$-centered models of $G$.
\begin{lemma}
\label{lemma:models_good_triples}
Suppose $G$ is a graph.
\begin{enumerate}
 \item For every $(C,H)$-centered model $\phi$ of $G$ the triple
 $(\mathcal{I}^{\phi}_{\circle},\mathcal{I}^{\phi}_{\stick} \setminus \mathcal{D},{\prec_{\phi}})$ is good for the set~$\mathcal{I}'$.
 \item For every good triple $(\mathcal{J},\mathcal{J}',{\prec})$ for the set $\mathcal{I}'$ there is a $(C,H)$-centered model $\phi$ of $G$ such that
$(\mathcal{J},{\prec}) = (\mathcal{I}^{\phi}_{\circle},{\prec_{\phi}})$ and
$\mathcal{J}' \cup \mathcal{D} = \mathcal{I}^{\phi}_{\stick}$.
\end{enumerate}
% A graph $G$ admits a $(C,H)$-centered model if and only if
%there is a good partition $(\mathcal{J},\mathcal{J}')$ of the set $%\mathcal{I'}$ witnessed by a good ordering $(\mathcal{J})$ for the root of $\mathbb{T}$ and for the borders $(\emptyset, \emptyset)$.
\end{lemma}
\begin{proof}
The first statement follows by Lemmas~\ref{lemma:D_forms_a_sqsubseteq_chain} and~\ref{lemma:prec_phi_is_good_order_for_R}.

Let $(\mathcal{J}, \mathcal{J}',{\prec})$ be a good triple for the set $\mathcal{I}'$.
Suppose $(\mathcal{J},{\prec})$ induces an admissible ordering $\mathbb{L}_1,\ldots,\mathbb{L}_n$ of the set $\mathcal{L}$.
First, we show that:
\begin{itemize}
\item for every vertex $v \in C'$ the set $J(v) = \{I \in \mathcal{J}: v \in C(I)\}$ forms a non-empty interval in $(\mathcal{J},{\prec})$,
\item the sets $\{J(v): v \in C'\}$ form an interval model of $H$, that is, for every $u,v \in C'$ we have $uv \in E(H)$ iff $J(u) \cap J(v) \neq \emptyset$.
\end{itemize}
Since $\mathbb{L}_1,\ldots,\mathbb{L}_n$ is admissible for $\mathbb{T}$, 
for every $c \in C'$ the set $\Cl(v) = \{i \in [n]: v \in C(\mathbb{L}_i)\}$ forms a non-empty interval in $[n]$.
Fix $v \in C'$ and assume $\Cl(v) = [l,k]$ for some $l \leq k$.
Clearly, $inner^{\mathcal{J}}(\mathbb{L}_i) \subseteq J(v)$ for every $i \in [l,k]$. 
Since for every $i \in [l,k-1]$ the chains $(zone_R(\mathbb{L}_i), \sqsubseteq)$ and $(zone_L(\mathbb{L}_{i+1}), \sqsubseteq)$ 
respect the borders $C(\mathbb{L}_i) \cap C(\mathbb{L}_{i+1})$,
every component $J$ from $zone_R(\mathbb{L}_i) \cup zone_L(\mathbb{L}_{i+1})$ satisfies $v \in C(\mathbb{L}_i) \cap C(\mathbb{L}_{i+1}) \subseteq C(J)$.
This shows $zone_R(\mathbb{L}_i) \cup zone_L(\mathbb{L}_{i+1}) \subseteq J(v)$
for every $i \in [l,k-1]$.
Finally, since the strict linear orders $(zone_L(\mathbb{L}_l),{\sqsubset})$ and $(zone_L(\mathbb{L}_l),{\prec})$ are equal, 
the set $\{J \in zone_L(\mathbb{L}_l): v \in C(J)\}$ forms an upset in $(zone_L(\mathbb{L}_l), {\prec})$.
Similarly, since $(zone_R(\mathbb{L}_k),{\sqsubset})$ equals to the reverse of $(zone_R(\mathbb{L}_k),{\prec})$, the set $\{J \in zone_R(\mathbb{L}_k): v \in C(J)\}$ 
forms a downset in $(zone_R(\mathbb{L}_k), {\prec})$.
This shows that $J(v)$ forms a non-empty interval in $(\mathcal{J},{\prec})$.
Now, note that $zone_R(\mathbb{L}_{l-1}) \prec zone_L(\mathbb{L}_{l})$ and $zone_R(\mathbb{L}_{k}) \prec zone_L(\mathbb{L}_{k+1})$, which 
follows by $zone(\mathbb{L}_1) \prec \ldots \prec zone(\mathbb{L}_n)$.
So, we have $J(v) \cap J(w_L) = \emptyset$ for every $w_L \in C(\mathbb{L}_{l-1}) \setminus C(L_{l})$ 
and $J(v) \cap J(w_R) = \emptyset$ for every $w_R \in C(\mathbb{L}_{k+1}) \setminus C(L_{k})$. 
This shows $\{J(v): v \in C'\}$ is an interval model of $H$.

Next, we represent every component $I$ in $\mathcal{J}$ as a tiny interval $\phi^*(I)$ in $L_{\circle} \setminus P$ such that 
$\phi^*(I)$ is to the left of $\phi^*(I')$ if and only if $I \prec I'$.
For every $c \in C'$ let $\phi^*(c)$ be the shortest interval in $L_{\circle} \setminus P$ 
such that $\phi^*(c)$ contains $\phi^*(I)$ for every $I \in J(v)$.
Clearly, for every $c \in C'$ the set $(L_{\circle} \setminus P) \setminus \phi^*(c)$ consists of two disjoint intervals,
$\phi_{left}(c)$ and $\phi_{right}(c)$, where $\phi_{left}(c)$ is to the left of $\phi_{right}(c)$.
For every $c \in C \setminus C'$ we let $\phi_{left}(c) = \phi_{right}(c) = L_{\circle} \setminus P$.
Similarly, we represent every component $I$ in $\mathcal{J}'$ as a tiny interval $\phi^{**}(I)$ in $L_{\stick} \setminus P$ such that 
$\phi^{**}(I)$ is to the left of $\phi^{**}(I')$ if and only if $I \sqsubset I'$.
For every $c \in C$ let $\phi^{**}(c)$ be the shortest interval in $L_{\stick} \setminus P$ 
containing $\phi^{**}(I)$ for every $I \in J(v)$ and let $\phi_{stick}(c) = L_{\stick} \setminus \phi^{**}(c)$.
Note the following properties of the sectors of $\phi^*$ and $\phi^{**}$:
\begin{itemize}
 \item For every $I \in \mathcal{J}$ and every $c \in C$ we have $c \in N^s_C(I)$ if and only if the sector $S^{\phi^*}(I)$ is covered by $\phi_{left}(c) \cup \phi_{right}(c)$,   
 \item For every $I \in \mathcal{J}'$ and every $c \in C$ we have $c \in N^s_C(I)$ if and only if the sector $S^{\phi^{**}}(I)$ is covered by $\phi_{stick}(c)$.   
\end{itemize}
Further, we partition ambiguous components from $\mathcal{I}$ that are not contained in one-component maximal sectors of $\phi^*$ into three sets: $\mathcal{I}_L, \mathcal{I}_R, \mathcal{I}_{S}$, as follows:
\begin{itemize}
\item $\mathcal{I}_L$ is the set of all ambiguous non-simple components $I$ from $\mathcal{J}$ such that $C(I) \subseteq C(J)$ and $I \prec J$ for 
some component $J$ from $inner^{\mathcal{J}}(\mathbb{L}_i)$ for some  $i \in [n]$, 
\item $\mathcal{I}_R$ is the set of all ambiguous non-simple components $I$ from $\mathcal{J}$ such that $C(I) \subseteq C(J)$ and $J \prec I$ for 
some component $J$ from $inner^{\mathcal{J}}(\mathbb{L}_i)$ for some  $i \in [n]$, 
\item $\mathcal{I}_S$ is the set of all ambiguous non-simple components from the set $\mathcal{J}'$.
\end{itemize}

Let $I \in \mathcal{I}_L$ and let $J$ be a component from $\mathcal{J}$ chosen such that $C(I) \subseteq C(J)$, $I \prec J$, $J \in inner^{\mathcal{J}}(\mathbb{L}_i)$, 
and $i \in [n]$ is as small as possible.
Note that $I \in zone_L(\mathbb{L}_i)$ if $S^{\phi^*}(J)$ is one-component maximal sector 
and $I \in zone_L(\mathbb{L}_i) \cup \{I_L\}$ if $S^{\phi^*}(J)$ is multi-component maximal sector, 
where $I_L$ is the leftmost component in $S^{\phi^*}(J)$.
Suppose the second case holds (if the first case holds, we proceed similarly).
Suppose $I$ is contained in sector $S$ of $\phi^*$.  
Since the strict linear orderings $\big{(}zone_L(\mathbb{L}_i) \cup \{I_L\},{\prec}\big{)}$ 
and $\big{(}zone_L(\mathbb{L}_i) \cup \{I_L\},{\sqsubset}\big{)}$ are equal,
$\big{(}zone_L(\mathbb{L}_i) \cup \{I_L\},{\sqsubseteq}\big{)}$ respects the border 
$C(\mathbb{L}_{i-1}) \cap C(\mathbb{L}_i)$, and $I$ is ambiguous non-simple, 
we have
$$C(\mathbb{L}_{i-1}) \cap C(\mathbb{L}_i) \subseteq C^s(I) \subsetneq C(I) = C(S).$$
Suppose $I'$ is a component preceding $I$ in $(\mathcal{J},{\prec})$ and suppose $I' \in zone_L(\mathbb{L}_i)$. 
Then $I' \sqsubset I$ and hence $C(I') \subseteq C^s(I) \subsetneq C(I) = C(S)$.
We conclude $I$ is the leftmost component in $S$ and the sector $S_L$ adjacent to $S$ from the left satisfies $C(S_L) \subseteq C^s(I)$.  
Since $N_C(I) \setminus N^s_C(I) = C(I) \setminus C^{s}(I)$, we have $$N_C(I) \setminus N^s_C(I) \subseteq C(S) \setminus C(S_L),$$
which shows that for every $c \in N_C(I_L) \setminus N^s_C(I_L)$ the interval $\phi_{left}(c)$ 
touches the left side of $S$ (that is, $\phi_{left}(c)$ and $S$ are disjoint and $\phi_{left}(c)$
covers $S_L$). 
Thus, we have shown the following property of the components from the set $\mathcal{I}_L$:
\begin{itemize}
\item Let $I \in \mathcal{I}_L$. 
The component $I$ is the leftmost component in the slot $S^{\phi^*}(I_L)$.
Moreover, for every $c \in N_C(I) \setminus N^s_C(I)$ the interval $\phi_{left}(c)$ touches the left side of the slot $S^{\phi^*}(I)$.
\end{itemize}
Using similar arguments we prove the following:
\begin{itemize}
\item Let $I \in \mathcal{I}_R$.
The component $I$ is the rightmost component in the slot $S^{\phi^*}(I)$.
Moreover, for every $c \in N_C(I) \setminus N^s_C(I)$ the interval $\phi_{right}(c)$ touches the right side of the slot $S^{\phi^*}(I_R)$.
\item Let $I \in \mathcal{I}_S$.
The component $I$ is the leftmost component in the slot $S^{\phi^{**}}(I)$.
Moreover, for every $c \in N_C(I) \setminus N^s_C(I)$ the interval $\phi_{stick}(c)$ touches the left side of the slot $S^{\phi^{**}}(I_S)$.
\end{itemize}

Finally, suppose $I_S$ is a component contained in a one-component maximal sector $S$ of~$\phi^*$.
Suppose $C(S) =C(\mathbb{L}_i)$ for some $i \in [n]$.
Let $S_L$ and $S_R$ be the slots of $\phi^*$ adjacent to $S$ from the left and the right side, respectively.
Note that
$$C(S_L) = \left\{
\begin{array}{ll}
C(I')&\text{where $I'$ is the maximum component in $(zone_L(\mathbb{L}_i),{\sqsubseteq}),$} \\
C(\mathbb{L}_{i-1}) \cap C(\mathbb{L}_i)& \text{if $zone_L(\mathbb{L}_i) = \emptyset$,}
\end{array}
\right.$$
and 
$$C(S_R) = \left\{
\begin{array}{ll}
C(I')&\text{where $I'$ is the maximum component in $(zone_R(\mathbb{L}_i),{\subseteq})$,} \\
C(\mathbb{L}_{i}) \cap C(\mathbb{L}_{i+1})& \text{if $zone_R(\mathbb{L}_i) = \emptyset$.}
\end{array}
\right.$$
Since $N_C(I_S) \setminus N^s_C(I_S) \subseteq (C \setminus C(S_L)) \cup (C \setminus C(S_R))$, $I_S$ and $\phi^*$ satisfy the following: 
\begin{itemize}
\item For every $c \in N_C(I_S) \setminus N^s_C(I_S)$ we have $c \in C \setminus C(S_L)$ or $c \in C \setminus C(S_R)$,
\item For every $c \in N_C(I_S) \setminus N^s_C(I_S)$, if $c \in C \setminus C(S_L)$ then $\phi_{left}(c)$ touches the left side of $S^{\phi^*}(I_S)$ 
and if $c \in C \setminus C(S_R)$ then $\phi_{right}(c)$ touches the right side of $S^{\phi^*}(I_S)$.
\end{itemize}

Suppose $I \in \mathcal{I}_L$.
Assume $I$ is contained in a sector $S$. 
Since $I$ is ambiguous non-simple, $G[C \cup I]$ admits an interval model $\psi$ with $C$ as the left-most clique.
Clearly, the properties of $I$ and $\phi^*$ shown above assert 
that for every $c \in N_C(I) \setminus N^s_C(I)$ the interval $\phi_{left}(c)$ touches the left side of $S$. 
We shrink the model $\psi|I$ of $G[I]$ and we paste it into the interval $\phi^*(I)$.
Then, for every $c \in N_C(I) \setminus N^s_C(I)$ we extend $\phi_{left}(c)$ to the right 
such that we have $\phi_{left}(c) \cap \psi(v)$ iff $cv \in E(G)$ for every $v \in I_L$.

We perform an analogous procedure for every ambiguous non-simple component $I$ from the set $\mathcal{I}_R$ 
(this time we extend the intervals $\phi_{right}(c)$ for $c \in N_C(I_L) \setminus N^s_C(I_L)$)
and for every ambiguous non-simple component $I$ from the set $\mathcal{I}_S$ (this time we extend the intervals $\phi_{stick}(c)$ for $c \in N_C(I_L) \setminus N^s_C(I_L)$).
Also, for every ambiguous simple component $I \in \mathcal{J}$ ($I \in \mathcal{J}'$) 
we paste any interval model $\psi$ of $G[I]$ into $\phi^*(I)$ (into $\phi^{**}(I)$, respectively).

Finally, let $I_S$ be a component contained in a one-component maximal sector of~$\phi^*$. 
Property~\eqref{item:def-good-ordering-one-component-maximal-sectors} of Definition \ref{def:good-order} 
asserts that the graph $G^{ext}(C(S_L), I_S, C(S_R))$ has a circular-arc model $\psi$.
Note that the arcs $\psi(v_L)$, $\bigcup \psi(I)$, $\psi(v_R)$ are disjoint.
By possibly mirroring the model $\psi$ we may assume we encounter these arcs in the order $\psi(v_L)$, $\bigcup \psi(I)$, $\psi(v_R)$ 
when we traverse the circle in the clockwise direction starting from $\psi(v_L)$.
Let $c \in N_C(I_S) \setminus N^s_C(I_S)$.
Since $N_C(I_S) \setminus N^s_C(I_S) \subseteq (C \setminus C(S_L)) \cup (C \setminus C(S_R))$, we have 
\begin{itemize}
\item $\psi(c) \cap \psi(v_L) \neq \emptyset$ or $\psi(c) \cap \psi(v_R) \neq \emptyset$ (note that both of these conditions might be satisfied).
\end{itemize}
Also, since $\psi$ is a circular-arc model of $G^{ext}(C(S_L), I_S, C(S_R))$, we have 
\begin{itemize}
\item $\psi(c) \cap \psi(v_L) \neq \emptyset$ iff $c \in C \setminus C(S_L)$,
\item $\psi(c) \cap \psi(v_R) \neq \emptyset$ iff $c \in C \setminus C(S_R)$.
\end{itemize}
If $c \in C \setminus C(S_L)$, then the set $\psi(c) \setminus \psi(v_L)$ consists of at most two arcs
and at most one of them intersects $\bigcup \psi(I_S)$ and is disjoint from $\psi(v_R)$ 
-- if such the arc exists, we denote it by $\psi_{left}(c)$.
Similarly, if $c \in C \setminus C(S_R)$, then the set $\psi(c) \setminus \psi(v_R)$ consists of at most two arcs
and at most one of them intersects $\bigcup \psi(I_S)$ and is disjoint from $\psi(v_L)$ -- if such the arc exists, we denote it by $\psi_{right}(c)$.
Now, we paste the model $\psi|I_S$ of $G[I_S]$ into the interval $\phi^*(I_S)$.
Let $c \in N(S) \setminus N^s_C(I)$.
If $c \in C \setminus C(S_L)$ and $\psi_{left}(c)$ is defined, we extend $\phi_{left}(c)$ to the right so as 
for every $v \in I_S$ we have $\psi_{left}(c) \cap \psi(v) \neq \emptyset$ iff $\phi_{left}(c) \cap \psi(v) \neq \emptyset$.
If $c \in C \setminus C(S_R)$ and $\psi_{right}(c)$ is defined, we extend $\phi_{right}(c)$ to the left so as 
for every $v \in I_S$ we have $\psi_{right}(c) \cap \psi(v) \neq \emptyset$ iff $\phi_{right}(c) \cap \psi(v) \neq \emptyset$.

Given the above construction, we can easily verify that $\phi$ given by
$$\phi(v) = \left\{ 
\begin{array}{ccl}
\phi_{left}(v) \cup \phi_{right}(v) \cup \phi_{stick}(v) \cup \{P\} & \text{if}& v \in C \\
\psi(v)& \text{if}& v \in \bigcup \{I:I \in \mathcal{I}\}
\end{array}
\right.$$
is a $(C,H)$-centered model of $G$.
\end{proof}

In order to describe the structure of a good triple for $\mathcal{I}'$, for each node $\mathbb{N}$ of $\mathbb{T}$ we define four sets:
$inner(\mathbb{N}), comp(\mathbb{N}) \subseteq \mathcal{I}'$ and $middle(\mathbb{N}), outer(\mathbb{N}) \subseteq \mathcal{I}' \cap \mathcal{I}_a$.
These sets will be defined so that they satisfy the following conditions:
\begin{itemize}
 \item $\mathcal{I}' = comp(\mathbb{R})$,
 \item for every node $\mathbb{N}$ of $\mathbb{T}$ the set $\{inner(\mathbb{N}), middle(\mathbb{N}), outer(\mathbb{N})\}$ forms a partition of the set $comp(\mathbb{N})$,
 \item for every inner node $\mathbb{N}$ of $\mathbb{T}$ the set
 $\{comp(\mathbb{N}_1),\ldots, comp(\mathbb{N}_k)\}$
 forms a partition of $inner(\mathbb{N})$, where $\mathbb{N}_1,\ldots,\mathbb{N}_k$ are the children of $\mathbb{N}$ in $\mathbb{T}$.
\end{itemize}
For every $(C,H)$-centered model $\phi$ of $G$ and every node $\mathbb{N}$ of $\mathbb{T}$, 
by $comp^{\phi}_{\circle}(\mathbb{N})$ and $comp^{\phi}_{\stick}(\mathbb{N})$ 
we denote the sets $comp(\mathbb{N}) \cap \mathcal{I}^{\phi}_\circle$ and $comp(\mathbb{N}) \cap \mathcal{I}^{\phi}_\stick$, respectively.
We define the sets $inner^\phi_\circle(\mathbb{N})$, $inner^\phi_\stick(\mathbb{N})$,
$middle^{\phi}_\circle(\mathbb{N})$, and $middle^{\phi}_\stick(\mathbb{N})$ analogously.

Our algorithm needs to test whether there exists a good triple for the set $\mathcal{I}' = comp(\mathbb{R})$.
For this purpose, we extend the notion of a good triple to the sets $comp(\mathbb{N})$ and $inner(\mathbb{N})$ and introduce 
the notion of a \emph{three-chain-partition} of the sets $inner(\mathbb{N})$ and $outer(\mathbb{N})$, for any node $\mathbb{N}$ in $\mathbb{T}$.
Then we show that any good triple for the set $comp(\mathbb{R})$ is, in some particular way, composed of these smaller components,
where the components for a node $\mathbb{N}$ are composed from suitable components for the children of $\mathbb{N}$ in $\mathbb{T}$.
We start with some definitions.

\begin{definition}
\label{def:good:for:comp:inner}
Let $\mathbb{N}$ be a node of $\mathbb{T}$.
A triple $(\mathcal{J},\mathcal{J}',{\prec})$ is called \emph{good} for the set $comp(\mathbb{N})$ (for the set $inner(\mathbb{N})$, respectively)
and the borders $(B_L,B_R)$, if:
\begin{itemize}
 \item $\{\mathcal{J},\mathcal{J}'\}$ is a partition of the set $comp(\mathbb{N})$ (of the set $inner(\mathbb{N})$, respectively),
 \item $(\mathcal{J},{\prec})$ is a good order for the node $\mathbb{N}$ and the borders $(B_L,B_R)$,
 \item $\mathcal{J}' \cup \mathcal{D}$ is a chain in $(\mathcal{I}_{a},\sqsubseteq)$.
\end{itemize}
If $(B_L,B_R) = (\emptyset,\emptyset)$, we simply say $(\mathcal{J},\mathcal{J}',{\prec})$ is \emph{good} for the set $comp(\mathbb{N})$ (for the set $inner(\mathbb{N})$, respectively).
\end{definition}
Note that the notion of a good triple for the set $\mathcal{I}'$, as defined by Definition~\ref{def:good_triple},
and the notion of a good triple for the set $comp(\mathbb{R})$, are equivalent. 
\begin{definition}
Let $\mathcal{A}$ be a subset of $\mathcal{I}_a \setminus \mathcal{D}$. 
A tuple $(\mathcal{C}_L, \mathcal{C}_R, \mathcal{C}_D)$ is called a \emph{three-chain-partition} of the set $\mathcal{A}$ 
if $\{\mathcal{C}_L,\mathcal{C}_R,\mathcal{C}_D \cup \mathcal{D}\}$ partitions the poset $(\mathcal{A} \cup \mathcal{D},{\sqsubseteq})$ into three chains.
\end{definition}

\begin{figure}[h]
\centering
\input{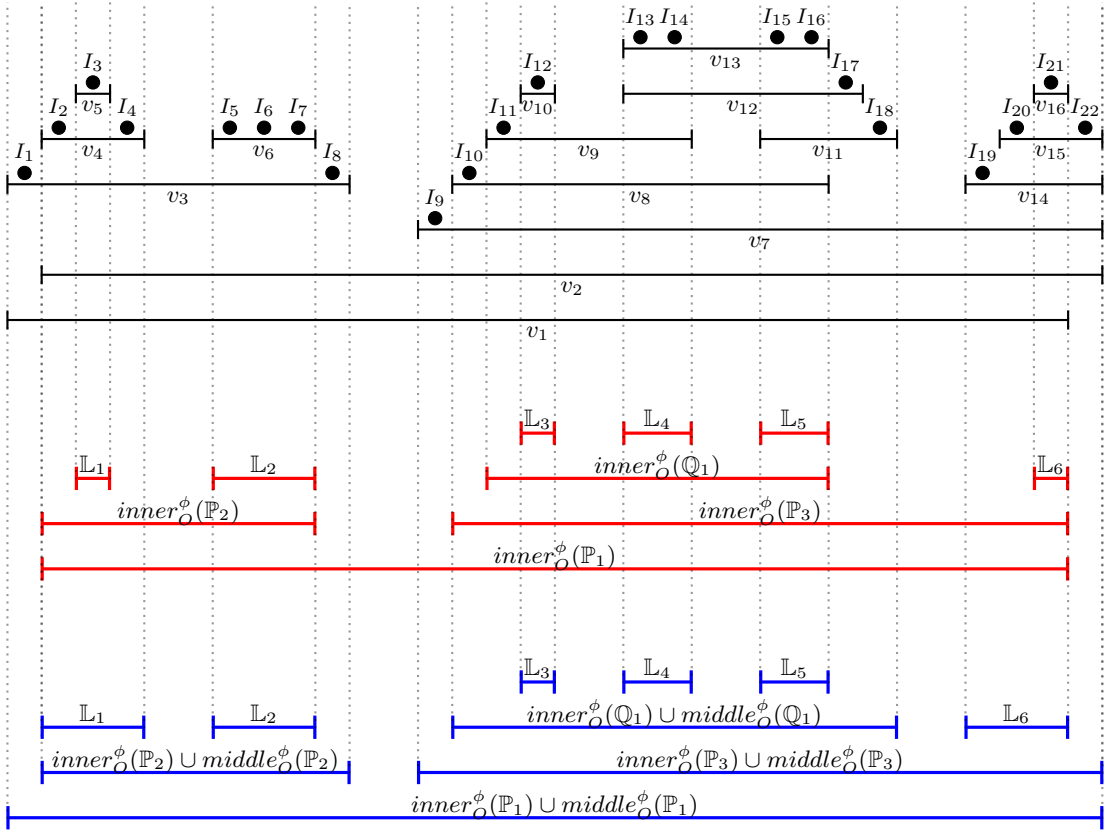}
\caption{The intervals $inner^{\phi}_{\circle}$ for the nodes of $\mathbb{T}$ are drawn in red.
The intervals $inner^{\phi}_{\circle} \cup middle^{\phi}_{\circle}$ for the nodes of $\mathbb{T}$ are drawn in blue.}
\label{fig:inner-middle-sets} 
\end{figure}

We start by defining the sets $inner(\mathbb{N})$ for all nodes of $\mathbb{T}$.
Recall that for every leaf node $\mathbb{L}$ of $\mathbb{T}$ we have defined $$inner(\mathbb{L}) = \{I \in \mathcal{I}': C(I) = C(\mathbb{L})\}.$$
For any inner node $\mathbb{N}$ of $\mathbb{T}$ we set
$$
inner(\mathbb{N}) = \bigl\{  I \in \mathcal{I'}: 
C(I) \cap I(\mathbb{N}) \neq \emptyset \quad \text{and} \quad U(\mathbb{N}) \cup E(\mathbb{N}) \subsetneq C(I)
\bigr\}.
$$
See Figure~\ref{fig:inner-middle-sets} for an illustration.
We have $I_1 \notin inner(\mathbb{P}_1)$ as $U(\mathbb{P}_1) \nsubseteq C(I_1)$ as witnessed by vertex $v_2$,
and $I_{17} \notin inner(\mathbb{Q}_1)$ as $U(\mathbb{Q}_1) \nsubseteq C(I_{17})$ as witnessed by vertex $v_{8}$,
and $I_8 \notin inner(\mathbb{P}_2)$ as $C(I_8) \cap I(\mathbb{P}_2) = \emptyset$.

\begin{lemma}
\label{lem:inner_components}
The following statements hold:
\begin{enumerate}
 \item \label{item:inner_components_properties}
For every two nodes $\mathbb{N} \neq \mathbb{M}$ of $\mathbb{T}$ we have:
\begin{itemize}
 \item $inner(\mathbb{N}) \subseteq inner(\mathbb{M})$ if $\mathbb{N}$ is a descendant of $\mathbb{M}$,
 \item $inner(\mathbb{N}) \cap inner(\mathbb{M}) = \emptyset$ 
 if $\mathbb{N}$ and $\mathbb{M}$ are incomparable in $\mathbb{T}$.
 \end{itemize}
\item \label{item:inner_components_model} For any $(C,H)$-centered model $\phi$ and any inner node $\mathbb{N}$ of $\mathbb{T}$:
\begin{itemize}
\item the set $inner^{\phi}_{\circle}(\mathbb{N})$ forms a non-empty interval in 
$(\mathcal{I}^{\phi}_{\circle}, {\prec_{\phi}})$ contained in the set $zone^{\phi}(\mathbb{N})$.
\item the triple $(inner^{\phi}_{\circle}(\mathbb{N}), inner^{\phi}_{\stick}(\mathbb{N}),{\prec_{\phi}})$ 
is good for the set $inner(\mathbb{N})$ and the set $inner^{\phi}_{\stick}(\mathbb{N})$, if non-empty, forms an upset in the chain $(comp^{\phi}_{\stick}(\mathbb{R}),{\sqsubseteq})$.
\end{itemize}
\item \label{item:inner_components_model_stick} All nodes $\mathbb{N}$ from $\mathbb{T}$ which satisfy $inner^{\phi}_{\stick}(\mathbb{N}) \neq \emptyset$
are contained in a single root to leaf path in $\mathbb{T}$. 
\end{enumerate}
\end{lemma}
\begin{proof}
Statement~\eqref{item:inner_components_properties} is obvious when both $\mathbb{N}$ and $\mathbb{M}$ are leaves in $\mathbb{T}$.
Now, suppose $\mathbb{M}$ is an inner node in $\mathbb{T}$, $\mathbb{N}$ is a leaf in $\mathbb{T}$, and $I \in inner(\mathbb{N})$.
Hence $C(I) = C(\mathbb{N})$.
Suppose $\mathbb{N}$ is a descendant of $\mathbb{M}$, that is, suppose $\mathbb{N} \in \mathcal{L}(\mathbb{M})$.
Then, we have $E(\mathbb{M}) \cup U(\mathbb{M}) \subsetneq C(\mathbb{N})$, which means there is $v \in C(\mathbb{N})$ such that $v \in I(\mathbb{M})$.
Thus, $E(\mathbb{M}) \cup U(\mathbb{M}) \subsetneq C(I)$ and $C(I) \cap I(\mathbb{M}) \neq \emptyset$, which proves $I \in inner(\mathbb{M})$.
Suppose $\mathbb{N} \notin \mathcal{L}(\mathbb{M})$.
Since any $J \in inner(\mathbb{M})$ satisfies $C(J) \cap I(\mathbb{M}) \neq \emptyset$, and since we have $I(\mathbb{M}) \cap C(\mathbb{N}) = \emptyset$, we can not have $C(J) = C(\mathbb{N})$. 
So, we have $J \notin inner(\mathbb{N})$.
This proves statement \eqref{item:inner_components_properties} for the case when $\mathbb{N}$ is a leaf and $\mathbb{M}$ is an inner node.
Now, suppose both $\mathbb{N}, \mathbb{M}$ are inner nodes in $\mathbb{T}$.
Suppose $I \in inner(\mathbb{N})$, that is, suppose $C(I) \cap I(\mathbb{N}) \neq \emptyset$
and $U(\mathbb{N}) \cup E(\mathbb{N}) \subsetneq C(I)$.
Suppose $\mathbb{N}$ is a descendant of $\mathbb{M}$.
Then $E(\mathbb{M}) \cup U(\mathbb{M}) \subseteq E(\mathbb{N})$ and 
$I(\mathbb{N}) \subseteq I(\mathbb{M})$.
In particular, $C(I) \cap I(\mathbb{M}) \neq \emptyset$ and $E(\mathbb{M}) \cup U(\mathbb{M}) \subsetneq C(I)$.
Thus $I \in inner(\mathbb{M})$.
Suppose $\mathbb{N}, \mathbb{M}$ are incomparable in $\mathbb{T}$.
Since there is no edge in $H$ between the sets $I(\mathbb{N})$ and $I(\mathbb{M})$, we must have $inner(\mathbb{N}) \cap inner(\mathbb{M}) = \emptyset$.
This proves~\eqref{item:inner_components_properties}.

Suppose $\mathbb{N}$ is an inner node of $\mathbb{T}$.
Suppose $I_1,I, I_2$ are the components from $\mathcal{I}^{\phi}_{\circle}$ such that $I_1 \prec_{\phi} I \prec_{\phi} I_2$ and $I_1,I_2 \in inner(\mathbb{N})$.
Suppose $v \in E(\mathbb{N}) \cup U(\mathbb{N})$.
Since $I_1,I_2 \in inner^\phi_{\circle}(\mathbb{N})$, we have $I_1,I_2 \in \phi^*(v)$.
Since $\phi^*(v)$ is an interval and $I_1 \prec_{\phi} I \prec_{\phi} I_2$, we also have $I \in \phi^*(v)$.

Suppose $\mathbb{N}$ is a $Q$-node in $\mathbb{T}$.
Since $\mathbb{N}$ is a $Q$-node, the set $\bigcup \{\phi^*(v) : v \in I(\mathbb{N})\}$ is an interval in $L_{\circle} \setminus P$. 
Since $I_1, I_2 \in inner^{\phi}_{\circle}(\mathbb{N})$, 
we have $I_1,I_2 \in \bigcup \{\phi^*(v) : v \in I(\mathbb{N})\}$.
Since $I_1 \prec_{\phi} I \prec_{\phi} I_2$, we also have $I \in \bigcup \{\phi^*(v) : v \in I(\mathbb{N})\}$.
It means $C(I) \cap I(\mathbb{N}) \neq \emptyset$ and hence $I \in inner^{\phi}_{\circle}(\mathbb{N})$.

Suppose $\mathbb{N}$ is a $P$-node and suppose $\mathbb{N}_1, \ldots,\mathbb{N}_k$ are the children of $\mathbb{N}$.
Since $\mathbb{N}$ is a $P$-node, 
the set $\bigcup \phi^*(I(\mathbb{N}))$ consists of $k$ intervals $\bigcup \phi^*(V(\mathbb{N}_i))$ for $i \in [k]$.
Suppose the children of $\mathbb{N}$ are enumerated such that $\bigcup \phi^*(V(\mathbb{N}_i))$ is to the left of $\bigcup \phi^*(V(\mathbb{N}_j))$ for $i < j$.
Since $I_1 \prec_{\phi} I \prec_{\phi} I_2$ and $I_1,I_2 \in \bigcup \phi^*(I(\mathbb{N}))$, 
either $I$ is contained in the interval $\bigcup \phi^*(V(\mathbb{N}_i))$ for some $i \in [k]$ or 
$I$ is between the intervals $\bigcup \phi^*(V(\mathbb{N}_i))$ and $\bigcup \phi^*(V(\mathbb{N}_{i+1}))$ for some $i \in [k-1]$.
In the first case $I \in \bigcup \phi^*(I(\mathbb{N}))$, which shows $C(I) \cap I(\mathbb{N})\neq \emptyset$ and proves $I \in inner(\mathbb{N})$.
In the second case, we have $C(I) = E(\mathbb{N}) \cup U(\mathbb{N})$, which means $I$ is contained in a minimal sector of $\phi^*$, contradicting~\eqref{eq:input_no_simple_cliques_on_minimal_sectors}.

The inclusion $inner^{\phi}_{\circle}(\mathbb{N}) \subseteq zone^{\phi}(\mathbb{N})$ is obvious when $\mathbb{N}$ is a leaf.
Suppose $\mathbb{N}$ is an inner node and suppose $I \in inner^{\phi}_{\circle}(\mathbb{N})$.
It means that $C(I) \cap I(\mathbb{N}) \neq \emptyset$.
Note that for every vertex $v \in I(\mathbb{N})$ all the components from $\mathcal{I}^{\phi}_{\circle}$ contained in $\phi^*(v)$ belong to $zone^{\phi}(\mathbb{N})$.
In particular, we have $I \in zone^{\phi}(\mathbb{N})$.
This shows the first part of statement~\eqref{item:inner_components_model}.
Since $(zone^{\phi}(\mathbb{N}),{\prec_{\phi}})$ is a good order for $\mathbb{N}$ and since
$I \sqsubset J$ and $I \in inner(\mathbb{N})$ yields $J \in inner(\mathbb{N})$, 
the second part of statement~\eqref{item:inner_components_model} follows.

Finally, statement \eqref{item:inner_components_model_stick} follows from statements~\eqref{item:inner_components_properties} and \eqref{item:inner_components_model}.
\end{proof}
Next, we define the sets $middle(\mathbb{N})$ for every node $\mathbb{N}$ of $\mathbb{T}$.
For the root $\mathbb{R}$ of $\mathbb{T}$ we set
$$middle(\mathbb{R}) = \mathcal{I}' \setminus inner(\mathbb{R}).$$
Clearly, $\{middle(\mathbb{R}),inner(\mathbb{R})\}$ forms a partition of $\mathcal{I}'$ and we have
$$middle(\mathbb{R}) = \{I \in \mathcal{I}': U(\mathbb{R}) \setminus C(I) \neq \emptyset \quad \text{or} \quad C(I) = U(I)\}.$$
Now, supposing $\mathbb{N}$ is a node of $\mathbb{T}$ and $\mathbb{M}$ is the parent of $\mathbb{N}$ in $\mathbb{T}$, we set 
$$middle(\mathbb{N}) = \{I \in inner(\mathbb{M}) \setminus inner(\mathbb{N}): C(I) \cap V(\mathbb{N}) \neq \emptyset\}.$$
In particular, note that
$$middle(\mathbb{N}) \cup inner(\mathbb{N}) = 
\left\{
\begin{array}{ccl}
\{I \in inner(\mathbb{M}): C(I) \cap V(\mathbb{N}) \neq \emptyset\} \quad & \text{if} & V(\mathbb{N}) \neq \emptyset, \\
inner(\mathbb{N}) &\text{if} & V(\mathbb{N}) = \emptyset.
\end{array}
\right.
$$
Notice that $V(\mathbb{N}) = \emptyset$ might happen only when $\mathbb{N}$ is a leaf of $\mathbb{T}$ and the parent of $\mathbb{N}$ is a $Q$-node.
	(In figure \ref{fig:inner-middle-sets}, an example is $V(\mathbb{L}_4)=\emptyset$.)
Note also that the set $middle(\mathbb{N})$ can be equivalently set such that
$$middle(\mathbb{N}) = \left\{I \in inner(\mathbb{M}): 
\begin{array}{c}
C(I) \cap V(\mathbb{N}) \neq \emptyset \quad \text{and} \\
C(I) \cap I(\mathbb{N}) = \emptyset \quad \text{or} \quad \big{(}U(\mathbb{N}) \cup E(\mathbb{N})\big{)} \setminus C(I) \neq \emptyset
\end{array}
\right\}.
$$
See Figure~\ref{fig:inner-middle-sets} for an illustration.
\begin{lemma}
\label{lem:middle_components}
The following statements hold:
\begin{enumerate}
\item \label{item:middle_components_property} Suppose $\mathbb{N}$ is a non-leaf node
 and suppose $\mathbb{N}_1,\ldots,\mathbb{N}_k$ are the children of $\mathbb{N}$ in $\mathbb{T}$.  
Then for every $i \neq j$ in $[k]$ we have
$$\big{(}middle(\mathbb{N}_i) \cup inner(\mathbb{N}_i)\big{)} \cap \big{(}middle(\mathbb{N}_j) \cup inner(\mathbb{N}_j)\big{)} = \emptyset.$$
\item \label{item:middle_components_model} For any $(C,H)$-centered model $\phi$ of $G$ and any node $\mathbb{N}$ of $\mathbb{T}$ 
the set $\big{(}middle^{\phi}_{\circle}(\mathbb{N}) \cup inner^{\phi}_{\circle}(\mathbb{N})\big{)}$ forms an interval in $(\mathcal{I}^{\phi}_{\circle}, {\prec_{\phi}})$ contained in the set $zone^{\phi}(\mathbb{N})$.
\item \label{item:middle_components_comparability} If $I \in inner(\mathbb{N})$ and $J \in middle(\mathbb{N})$ are $\sqsubseteq$-comparable, then $J \sqsubset I$.
\end{enumerate}
\end{lemma}
\begin{proof}
Let $i$ be such that $V(\mathbb{N}_i) \neq \emptyset$.
It means we have 
$$
\big{(}middle(\mathbb{N}_i) \cup inner(\mathbb{N}_i)\big{)} = \{I \in inner(\mathbb{M}): C(I) \cap V(\mathbb{N}_i) \neq \emptyset\}.
$$

Suppose $j$ is such that $V(\mathbb{N}_j) \neq \emptyset$.
Similarly, we have
$$\big{(}middle(\mathbb{N}_j) \cup inner(\mathbb{N}_j)\big{)} = \{I \in inner(\mathbb{M}): C(I) \cap V(\mathbb{N}_j) \neq \emptyset\}.
$$
Now, statement~\eqref{item:middle_components_property} follows by the fact that there is no edge in $H$ between the sets $V(\mathbb{N}_i)$ and $V(\mathbb{N}_j)$.

Suppose $j$ is such that $V(\mathbb{N}_j) = \emptyset$.
It means $\mathbb{N}_j$ is a leaf of $\mathbb{T}$ and we have 
$$middle(\mathbb{N}_j) \cup inner(\mathbb{N}_j) = inner(\mathbb{N}_j) = \{I \in \mathcal{I}': C(I) = C(\mathbb{N}_j)\}.$$
In this case statement~\eqref{item:middle_components_property} follows by the fact that there is a vertex $v \in C(\mathbb{N}_j)$ such that there is no edge in $H$ between $v$ and the set $V(\mathbb{N}_i)$.

Suppose $i,j$ are such that $V(\mathbb{N}_i) = V(\mathbb{N}_j) = \emptyset$.
Then, statement~\eqref{item:middle_components_property} follows by Lemma~\ref{lem:inner_components}.\eqref{item:inner_components_properties}. 

Statement~\eqref{item:middle_components_model} is trivially satisfied when $\mathbb{N}$ is the root of $\mathbb{T}$ or when $V(\mathbb{N}) = \emptyset$.
So, suppose $\mathbb{N}$ has the parent $\mathbb{M}$ in $\mathbb{T}$ and $V(\mathbb{N}) \neq \emptyset$.

Suppose $\mathbb{N}$ is a $Q$-node or $\mathbb{N}$ is a leaf node such that $V(\mathbb{N}) \neq \emptyset$.
In this case the set $V(\mathbb{N})$ is non-empty and the set $\bigcup_{v \in V(\mathbb{N})} \phi^*(v)$ forms an interval in $L_{\circle} \setminus P$.
Then, using the same arguments as in Lemma~\ref{lem:inner_components}.\eqref{item:inner_components_model}, we deduce the set $\big{(}middle^{\phi}_{\circle}(\mathbb{N}) \cup inner^{\phi}_{\circle}(\mathbb{N})\big{)}$ forms an interval in $(\mathcal{I}^{\phi}_{\circle},{\prec_{\phi}})$.

Suppose $\mathbb{N}$ is a $P$-node.
If $U(\mathbb{N}) \neq \emptyset$, then  $\bigcup_{v \in V(\mathbb{N})} \phi^*(v)$ is an interval, otherwise  $\bigcup_{v \in V(\mathbb{N})} \phi^*(v)$ consists of pairwise disjoint intervals, each corresponding to a child of $\mathbb{N}$.
However, in the second case condition~\eqref{eq:input_no_simple_cliques_on_minimal_sectors} asserts that no component from $\mathcal{I}^{\phi}_{\circle}$ can be contained between two such intervals.
In particular, it means that the set 
$\big{(}middle^{\phi}_{\circle}(\mathbb{N}) \cup inner^{\phi}_{\circle}(\mathbb{N})\big{)}$ forms an interval in $(\mathcal{I}^{\phi}_{\circle},{\prec_{\phi}})$.

Now we show $\big{(}middle^{\phi}_{\circle}(\mathbb{N}) \cup inner^{\phi}_{\circle}(\mathbb{N})\big{)} \subseteq zone^{\phi}(\mathbb{N})$ for every node $\mathbb{N}$ of $\mathbb{T}$.
If $V(\mathbb{N}) = \emptyset$ then the inclusion $\big{(}middle^{\phi}_{\circle}(\mathbb{N}) \cup inner^{\phi}_{\circle}(\mathbb{N})\big{)} \subseteq zone^{\phi}(\mathbb{N})$ follows by Lemma \ref{lem:inner_components}.\eqref{item:inner_components_model}.
Otherwise, note that every component from $\mathcal{I}^{\phi}_{\circle}$ contained in $\phi^*(v)$ for $v \in V(\mathbb{N})$ must belong to $zone^{\phi}(\mathbb{N})$.
In particular, it proves $\big{(}middle^{\phi}_{\circle}(\mathbb{N}) \cup inner^{\phi}_{\circle}(\mathbb{N})\big{)} \subseteq zone^{\phi}(\mathbb{N})$ in the case $V(\mathbb{N}) \neq \emptyset$.

To show statement \eqref{item:middle_components_comparability},
suppose $I \in inner(\mathbb{N})$ and $J \in middle(\mathbb{N})$ are ${\sqsubseteq}$-comparable. 
To show $J \sqsubset I$ it is enough to prove we can not have  $C(I) \subseteq C(J)$.
Suppose $\mathbb{N}$ is a leaf of $\mathbb{T}$.
In this case we can not have $C(I) \subseteq C(J)$ as otherwise we would have $C(\mathbb{N}) = C(I) = C(J)$, which would yield $J \in inner(\mathbb{N})$.
Suppose $\mathbb{N}$ is an inner node of $\mathbb{T}$.
In this case $I \in inner(\mathbb{N})$ is equivalent to $C(I) \cap I(\mathbb{N}) \neq \emptyset$ and $U(\mathbb{N}) \cup E(\mathbb{N}) \subsetneq C(I)$.
In particular, $C(I) \subseteq C(J)$ would yield $J \in inner(\mathbb{N})$, which is not the case.
\end{proof} 

Eventually, we define the sets $outer(\mathbb{N})$ for every node $\mathbb{N}$ of $\mathbb{T}$.
Note that this way we will also define the sets $comp(\mathbb{N})$ as we have
$comp(\mathbb{N}) = outer(\mathbb{N}) \cup middle(N) \cup inner(\mathbb{N})$ for every node $\mathbb{N}$ in $\mathbb{T}$.

\begin{figure}[h]
\begin{center}
\begin{subfigure}[t]{1\linewidth}
\centering
\begin{tikzpicture}[xscale=1.2,yscale=0.6,>=latex,shorten >=-0.4pt,shorten <=-0.4pt]
  \tikzstyle{every node}=[inner sep=2pt,fill=white]

\draw[|-|,very thick,black] (1.52,-1.5) -- (3.48,-1.5);
\coordinate (linR) at (2.5,-1.1) {};

\draw[|-|,very thick,red] (0,-2) -- (1.48,-2);
\coordinate (lmidL) at (0.75,-1.7) {};

\draw[|-|,very thick,blue] (3.52,-2) -- (6,-2);
\coordinate (lmidR) at (4.75,-1.7) {};

%v1

\draw[dotted,opacity=0.4,thick] (0,-1.5) -- (0,6.5);
\draw[dotted,opacity=0.4,thick] (1.5,-1.5) -- (1.5,6.5);
\draw[dotted,opacity=0.4,thick] (3.5,-1.5) -- (3.5,6.5);
\draw[dotted,opacity=0.4,thick] (6,-1.5) -- (6,6.5);

\draw[|-|,thick] (0,0) -- (6,0);
\coordinate (lv1) at (3,-0.3) {};
\coordinate (I1) at (0.25,0.25) {};
\coordinate (I12) at (5.75,0.25) {};

%v2
\draw[|-|,thick] (0.5,1) -- (3.5,1);
\coordinate (lv2) at (1.75,0.7) {};
\coordinate (I2) at (0.75,1.25) {};

%v3
\draw[|-|,thick] (1,2) -- (5.5,2);
\coordinate (lv3) at (3.5,1.7) {};
\coordinate (I10) at (4.75,2.25) {};
\coordinate (I11) at (5.25,2.25) {};

%v4
%\draw[|-|,thick,red] (1,3) -- (2.5,3);
\coordinate (lv4) at (1.75,2.7) {};
\coordinate (I3) at (1.25,2.25) {};

%IR
\draw[thick,opacity=0.3,fill=red, red] (1.5,2.5) rectangle (3.5,5.5);
\coordinate (lIR) at (2.5,5.0) {};
\coordinate (I4) at (1.75,5.75) {};
\coordinate (I5) at (2.25,5.75) {};
\coordinate (I6) at (2.75,5.75) {};
\coordinate (I7) at (3.25,5.75) {};

%v5
\draw[|-|,thick,red] (2.5,3.5) -- (4.5,3.5);
\coordinate (lv5) at (3.5,3.2) {};
\coordinate (I8) at (3.75,3.75) {};
\coordinate (I9) at (4.25,3.75) {};

\tikzstyle{every node}=[inner sep=1pt]
\begin{scriptsize}
\node at (lv1) {$u_1$};
\node at (lv2) {$u_2$};
\node at (lv3) {$u_3$};
%\node at (lv4) {$w_1$};
\node at (lv5) {$w_1$};
\node at (lIR) {$I(\mathbb{R})$};

\node at (linR) {$inner^{\phi}_{\circle}(\mathbb{R})$};
\node at (lmidL) {$mid_L$};
\node at (lmidR) {$mid_R$};
\end{scriptsize}

\tikzstyle{every node}=[inner sep=1pt]
\begin{tiny}
\node at ($(I1) + (0,0.45)$) {$I_1$};
\node at ($(I2) + (0,0.45)$) {$I_2$};
\node at ($(I3) + (0,0.45)$) {$I_3$};
\node at ($(I4) + (0,0.45)$) {$I_4$};
\node at ($(I5) + (0,0.45)$) {$I_5$};
\node at ($(I6) + (0,0.45)$) {$I_6$};
\node at ($(I7) + (0,0.45)$) {$I_7$};
\node at ($(I8) + (0,0.45)$) {$I_8$};
\node at ($(I9) + (0,0.45)$) {$I_9$};
\node at ($(I10) + (0,0.45)$) {$I_{10}$};
\node at ($(I11) + (0,0.45)$) {$I_{11}$};
\node at ($(I12) + (0,0.45)$) {$I_{12}$};

\end{tiny}

\tikzstyle{every node}=[circle,minimum size=5pt,inner sep=0pt,draw,fill]
\node[red] at (I1) {};
\node[red] at (I2) {};
\node[red] at (I3) {};
\node at (I4) {};
\node at (I5) {};
\node at (I6) {};
\node at (I7) {};
\node[blue] at (I8) {};
\node[blue] at (I9) {};
\node[blue] at (I10) {};
\node[blue] at (I11) {};
\node[blue] at (I12) {};

%\draw[black] (-1,-1.5)--(-1,0);
%\draw[black] (7,5.5)--(7,6.5);
\draw[white] (-1,-1.5)--(-1,0);
\draw[white] (7,5.5)--(7,6.5);

\end{tikzpicture}
\end{subfigure}
\end{center}
\caption{The structure of $(comp^{\phi}_{\circle}(\mathbb{R}),{\prec_{\phi}})$. 
We have $U(\mathbb{R}) = \{u_1,u_2,u_3\}$, $w_1 \in I(\mathbb{R})$. 
The chains $(mid_L,{\sqsubseteq})$ and $(mid_R,{\sqsubseteq})$ are depicted in red and blue, respectively.
We have $max(mid_L,{\sqsubseteq})=I_3$ and $max(mid_R,{\sqsubseteq})=I_8$.} 
\label{fig:root-node-structure} 
\end{figure}
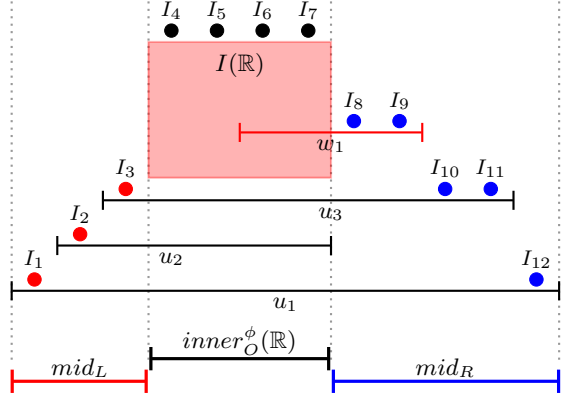

We set $outer(\mathbb{R}) = \emptyset$ for the root node $\mathbb{R}$.
Now we are ready to describe the structure of any good triple for the set $comp(\mathbb{R})$.
Lemma~\ref{lemma:models_good_triples} asserts that for any such triple $(\mathcal{J},\mathcal{J}',{\prec})$ there is a
$(C,H)$-centered model $\phi$ of $G$ which satisfies $(\mathcal{J},\mathcal{J}',{\prec}) = (comp^{\phi}_{\circle}(\mathbb{R}),comp^{\phi}_{\stick}(\mathbb{R}), {\prec_{\phi}})$. 
The next lemma describes $(comp^{\phi}_{\circle}(\mathbb{R}),comp^{\phi}_{\stick}(\mathbb{R}),{\prec_{\phi}})$
as a certain composition of good triple $(inner^{\phi}_{\circle}(\mathbb{R}),inner^{\phi}_{\stick}(\mathbb{R}),{\prec_{\phi}})$ 
for the set $inner(\mathbb{R})$ and a three-chain-partition $(mid_L,mid_R,mid_D)$ of the set $middle(\mathbb{R})$. 
\begin{lemma}[root node structure -- see Figure \ref{fig:root-node-structure}]
\label{lemma:root-node-structure}
Let $(comp^{\phi}_{\circle}(\mathbb{R}),comp^{\phi}_{\stick}(\mathbb{R}),{\prec_{\phi}})$ be a good triple for $comp(\mathbb{R})$ corresponding to a $(C,H)$-model $\phi$ of $G$.
Then there exists a three-chain-partition $(mid_L,mid_R,mid_D)$ of the set $middle(\mathbb{R})$ such that:
\begin{itemize}
 \item $comp^{\phi}_{\circle}(\mathbb{R}) = mid_L \cup mid_R \cup inner^{\phi}_{\circle}(\mathbb{R})$ and $mid_L \prec_{\phi} inner^{\phi}_{\circle}(\mathbb{R}) 
 \prec_{\phi} mid_R$,
 \item $comp^{\phi}_{\stick}(\mathbb{R}) = mid_D \cup inner^{\phi}_{\stick}(\mathbb{R})$ and $mid_D \sqsubset inner^{\phi}_{\stick}(\mathbb{R})$,
 \item $(mid_L,{\prec_{\phi}})$ equals to $(mid_L,{\sqsubset})$ and $(mid_R,{\prec_{\phi}})$ equals to the reverse of $(mid_R,{\sqsubset})$,
 \item $\big{(}inner^{\phi}_{\circle}(\mathbb{R}),inner^{\phi}_{\stick}(\mathbb{R}), {\prec_{\phi}})$ is good for the set $inner(\mathbb{R})$ and the borders $(C_L,C_R)$, where:
 $$
 C_L  = \left\{
 \begin{array}{ll}
 C\big{(}max(mid_L,{\sqsubseteq})\big{)} & \text{if $mid_L \neq \emptyset$,} \\
 \emptyset & \text{otherwise,}
 \end{array}
 \right.
 \quad
 C_R  = \left\{
 \begin{array}{ll}
 C\big{(}max(mid_R,{\sqsubseteq})\big{)} & \text{if $mid_R \neq \emptyset$,} \\
 \emptyset & \text{otherwise.}
 \end{array}
 \right.
 $$
 \end{itemize}

\end{lemma}
\begin{proof}
The statements of the lemma follow easily from Lemmas~\ref{lem:inner_components} and~\ref{lem:middle_components}. 
\end{proof}

Let $\mathbb{P}$ be a $P$-node in $\mathbb{T}$ and
let $\mathbb{N}_1,\ldots, \mathbb{N}_k$ be the children of 
$\mathbb{P}$.
We set $outer(\mathbb{N}_i) = \emptyset$ for every $i \in [k]$.
Since $\mathbb{P}$ is a $P$-node in $\mathbb{T}$, the set $\{V(\mathbb{N}_i): i \in [k]\}$ forms a partition of the set $I(\mathbb{P})$. 
Since $C(I) \cap I(\mathbb{P}) \neq \emptyset$ for every $I \in inner(\mathbb{P})$, there is a unique $i \in [k]$ such that $C(I) \cap V(\mathbb{N}_i) \neq \emptyset$, which means $I \in middle(\mathbb{N}_i) \cup inner(\mathbb{N}_i)$.
So, the set $\{comp(\mathbb{N}_i): i \in [k]\}$ forms a partition of the set
$inner(\mathbb{\mathbb{P}})$.
Finally, let $\mathcal{C}$ be a chain in the poset $(inner(\mathbb{P}) \cap \mathcal{I}_{a}, {\sqsubseteq})$.
The \emph{trace} of $\mathcal{C}$ is a tuple that lists all the sets $comp(\mathbb{N}_i)$ intersected by $\mathcal{C}$ when we traverse $C$ upward.
If $\mathcal{C} = \emptyset$, then the trace of $\mathcal{C}$ is an empty tuple.
Otherwise, since for every $i \neq j$ the sets $comp(\mathbb{N}_i)$ and 
$comp(\mathbb{N}_j)$ are $\sqsubseteq$-incomparable, 
there exists a unique $d \in [k]$ such that 
$\mathcal{C} \cap comp(\mathbb{N}_d) \neq \emptyset$.
In this case the trace of $C$ is just $(d)$.
Summing up, we have proved that:
\begin{lemma}
\label{lem:P-node-chain-structure}
Let $\mathbb{P}$ be a $\mathbb{P}$-node in $\mathbb{T}$ and let $\mathbb{N}_1,\ldots,\mathbb{N}_k$ be the children of $\mathbb{P}$.
\begin{enumerate}
 \item The set $\{comp(\mathbb{N}_i): i \in [k]\}$ forms a partition of $inner(\mathbb{P})$.
 \item The trace of every chain $\mathcal{C}$ in $(\mathcal{I}_a \cap inner(\mathbb{P}),{\sqsubseteq})$ consists of at most one entry.
\end{enumerate}
\end{lemma}

\begin{figure}[h]
\begin{center}
\begin{subfigure}[t]{1\linewidth}
\centering
\begin{tikzpicture}[xscale=1,yscale=0.6,>=latex,shorten >=-0.4pt,shorten <=-0.4pt]
  \tikzstyle{every node}=[inner sep=2pt,fill=white]  

\draw[|-|,very thick,black] (0.5,-2) -- (15,-2);
\coordinate (linP) at (7.5,-1.6) {};

\draw[dotted,opacity=0.4,very thick] (0.5,-2) -- (0.5,7.5);
\draw[dotted,opacity=0.4,very thick] (15.0,-2) -- (15.0,7.5);

\draw[|-|,very thick,black] (0.5,-3.25) -- (4.5,-3.25);
\coordinate (lcompN1) at (2.5,-2.85) {};

\draw[|-|,very thick,black] (1.5,-4.5) -- (3.5,-4.5);
\coordinate (linN1) at (2.5,-4.1) {};
\draw[|-|,very thick,red] (0.5,-5) -- (1.5,-5);
\coordinate (lmid1L) at (1,-4.6) {};
\draw[|-|,very thick,red] (3.5,-5) -- (4.5,-5);
\coordinate (lmid1R) at (4,-4.6) {};

\draw[dotted,opacity=0.4,thick] (0.5,-5) -- (0.5,7.5);
\draw[dotted,opacity=0.4,thick] (1.5,-5) -- (1.5,7.5);
\draw[dotted,opacity=0.4,thick] (3.5,-5) -- (3.5,7.5);
\draw[dotted,opacity=0.4,thick] (4.5,-5) -- (4.5,7.5);

\draw[|-|,very thick,black] (5.5,-3.25) -- (9.5,-3.25);
\coordinate (lcompN2) at (7.5,-2.85) {};

\draw[|-|,very thick,black] (6.5,-4.5) -- (8.5,-4.5);
\coordinate (linN2) at (7.5,-4.1) {};
\draw[|-|,very thick,red] (5.5,-5) -- (6.5,-5);
\coordinate (lmid2L) at (6,-4.6) {};
\draw[|-|,very thick,red] (8.5,-5) -- (9.5,-5);
\coordinate (lmid2R) at (9,-4.6) {};

\draw[dotted,opacity=0.4,thick] (5.5,-5) -- (5.5,7.5);
\draw[dotted,opacity=0.4,thick] (6.5,-5) -- (6.5,7.5);
\draw[dotted,opacity=0.4,thick] (8.5,-5) -- (8.5,7.5);
\draw[dotted,opacity=0.4,thick] (9.5,-5) -- (9.5,7.5);

\draw[|-|,very thick,black] (10.5,-3.25) -- (15,-3.25);
\coordinate (lcompN3) at (12.5,-2.85) {};

\draw[|-|,very thick,black] (11.5,-4.5) -- (13.5,-4.5);
\coordinate (linN3) at (12.5,-4.1) {};
\draw[|-|,very thick,red] (10.5,-5) -- (11.5,-5);
\coordinate (lmid3L) at (11,-4.6) {};
\draw[|-|,very thick,red] (13.5,-5) -- (15,-5);
\coordinate (lmid3R) at (14.25,-4.6) {};

\draw[dotted,opacity=0.4,thick] (10.5,-5) -- (10.5,7.5);
\draw[dotted,opacity=0.4,thick] (11.5,-5) -- (11.5,7.5);
\draw[dotted,opacity=0.4,thick] (13.5,-5) -- (13.5,7.5);
\draw[dotted,opacity=0.4,thick] (15,-5) -- (15,7.5);

%v1
\draw[-,thick] (-0.5,0) -- (15.5,0);
\coordinate (lv1) at (7.5,-0.3) {};

%v2
\draw[-,thick] (-0.5,1) -- (15.5,1);
\coordinate (lv2) at (7.5,0.7) {};
%v3
\draw[-,thick] (-0.5,2) -- (15.5,2);
\coordinate (lv3) at (7.5,1.7) {};
\coordinate (I0) at (0.25,2.25) {};

%u1
\draw[|-|,thick] (0.5,3) -- (3.5,3);
\coordinate (lu1) at (2.5,2.7) {};
\coordinate (I1) at (0.75,3.25) {};

%u2
\draw[|-|,thick] (1,4) -- (4.5,4);
\coordinate (lu2) at (2.5,3.7) {};
\coordinate (I2) at (1.25,4.25) {};
\coordinate (I8) at (4.25,4.25) {};

%N1
\draw[thick,opacity=0.3,fill=red, red] (1.5,4.5) rectangle (3.5,6.5);
%w1
\draw[|-|,thick,red] (3,5) -- (4,5);
\coordinate (lw1) at (3.75,4.7) {};
\coordinate (ln1) at (2.5,6) {};
\coordinate (I3) at (1.75,6.75) {};
\coordinate (I4) at (2.25,6.75) {};
\coordinate (I5) at (2.75,6.75) {};
\coordinate (I6) at (3.25,6.75) {};
\coordinate (I7) at (3.75,5.25) {};

%u3
\draw[|-|,thick] (5.5,3) -- (9.5,3);
\coordinate (lu3) at (7.5,2.7) {};
\coordinate (I9) at (5.75,3.25) {};
\coordinate (I16) at (9.25,3.25) {};

%u4
\draw[|-|,thick] (6,4) -- (9,4);
\coordinate (lu4) at (7.5,3.7) {};
\coordinate (I10) at (6.25,4.25) {};
\coordinate (I15) at (8.75,4.25) {};

%N2
\draw[thick,opacity=0.3,fill=red, red] (6.5,4.5) rectangle (8.5,6.5);
\coordinate (ln2) at (7.5,6) {};
\coordinate (I11) at (6.75,6.75) {};
\coordinate (I12) at (7.25,6.75) {};
\coordinate (I13) at (7.75,6.75) {};
\coordinate (I14) at (8.25,6.75) {};

%u5
\draw[|-|,thick] (10.5,3) -- (14.5,3);
\coordinate (lu5) at (12.5,2.7) {};

%u6
\draw[|-|,thick] (10.5,4) -- (13.5,4);
\coordinate (lu6) at (12.5,3.7) {};
\coordinate (I17) at (10.75,4.25) {};
\coordinate (I18) at (11.25,4.25) {};

%N3
\draw[thick,opacity=0.3,fill=red, red] (11.5,4.5) rectangle (13.5,6.5);
\coordinate (ln3) at (12.5,6) {};
\draw[|-|,thick,red] (13,5) -- (15,5);
\coordinate (lw2) at (14,4.7) {};
\coordinate (I25) at (14.75,5.25) {};
\draw[|-|,thick,red] (13,6) -- (14.5,6);
\coordinate (lw3) at (13.75,5.7) {};
\coordinate (I23) at (13.75,6.25) {};
\coordinate (I24) at (14.25,6.25) {};

\coordinate (I19) at (11.75,6.75) {};
\coordinate (I20) at (12.25,6.75) {};
\coordinate (I21) at (12.75,6.75) {};
\coordinate (I22) at (13.25,6.75) {};

\tikzstyle{every node}=[inner sep=1pt]
\begin{scriptsize}
\node at (lv1) {$v_1$};
\node at (lv2) {$v_2$};
\node at (lv3) {$v_3$};

\node at (lu1) {$u_1$};
\node at (lu2) {$u_2$};
\node at (lu3) {$u_3$};
\node at (lu4) {$u_4$};
\node at (lu5) {$u_5$};
\node at (lu6) {$u_6$};

\node at (lw1) {$w_1$};
\node at (lw2) {$w_2$};
\node at (lw3) {$w_3$};

\node at (ln1) {$I(\mathbb{N}_1)$};
\node at (ln2) {$I(\mathbb{N}_2)$};
\node at (ln3) {$I(\mathbb{N}_3)$};

\node at (linP) {$inner^{\phi}_{\circle}(\mathbb{P})$};
\node at (linN1) {$inner^{\phi}_{\circle}(\mathbb{N}_1)$};
\node at (linN2) {$inner^{\phi}_{\circle}(\mathbb{N}_2)$};
\node at (linN3) {$inner^{\phi}_{\circle}(\mathbb{N}_3)$};

\node at (lcompN1) {$comp^{\phi}_{\circle}(\mathbb{N}_1)$};
\node at (lcompN2) {$comp^{\phi}_{\circle}(\mathbb{N}_2)$};
\node at (lcompN3) {$comp^{\phi}_{\circle}(\mathbb{N}_3)$};

\node at (lmid1L) {$mid^1_L$};
\node at (lmid2L) {$mid^2_L$};
\node at (lmid3L) {$mid^3_L$};
\node at (lmid1R) {$mid^1_R$};
\node at (lmid2R) {$mid^2_R$};
\node at (lmid3R) {$mid^3_R$};
\end{scriptsize}

\tikzstyle{every node}=[inner sep=1pt]
\begin{tiny}
%\node at ($(I1) + (0,0.45)$) {$I_1$};
%\node at ($(I2) + (0,0.45)$) {$I_2$};
%\node at ($(I3) + (0,0.45)$) {$I_3$};
\end{tiny}

\tikzstyle{every node}=[circle,minimum size=5pt,inner sep=0pt,draw,fill]
%\node at (I0) {};
\node[red] at (I1) {};
\node[red] at (I2) {};
\node at (I3) {};
\node at (I4) {};
\node at (I5) {};
\node at (I6) {};
\node[red] at (I7) {};
\node[red] at (I8) {};

\node[red] at (I9) {};
\node[red] at (I10) {};
\node at (I11) {};
\node at (I12) {};
\node at (I13) {};
\node at (I14) {};
\node[red] at (I15) {};
\node[red] at (I16) {};

\node[red] at (I17) {};
\node[red] at (I18) {};
\node at (I19) {};
\node at (I20) {};
\node at (I21) {};
\node at (I22) {};
\node[red] at (I23) {};
\node[red] at (I24) {};
\node[red] at (I25) {};

%\draw[black] (-1,-4.5)--(-1,-3.5);
%\draw[black] (16,5.5)--(16,7.5);
\draw[white] (-1,-4.5)--(-1,-3.5);
\draw[white] (16,5.5)--(16,7.5);

\end{tikzpicture}
\end{subfigure}
\end{center}
\caption{The structure of $(inner^{\phi}_{\circle}(\mathbb{P}),{\prec_\phi})$, where $\mathbb{P}$ is a $P$-node with children $\mathbb{N}_1,\mathbb{N}_2,\mathbb{N}_3$.
We have $U(\mathbb{P}) \cup E(\mathbb{P}) = \{v_1,v_2,v_3\}$, $U(\mathbb{N}_1) = \{u_1,u_2\}$, $U(\mathbb{N}_2) = \{u_3,u_4\}$, $U(\mathbb{N}_3) = \{u_5,u_6\}$,
$w_1 \in I(\mathbb{N}_1)$, $w_2,w_3 \in I(\mathbb{N}_3)$.
The components from $mid^{i}_L$ and $mid^{i}_R$ are depicted in red.
} 
\label{fig:P-node-structure} 
\end{figure}
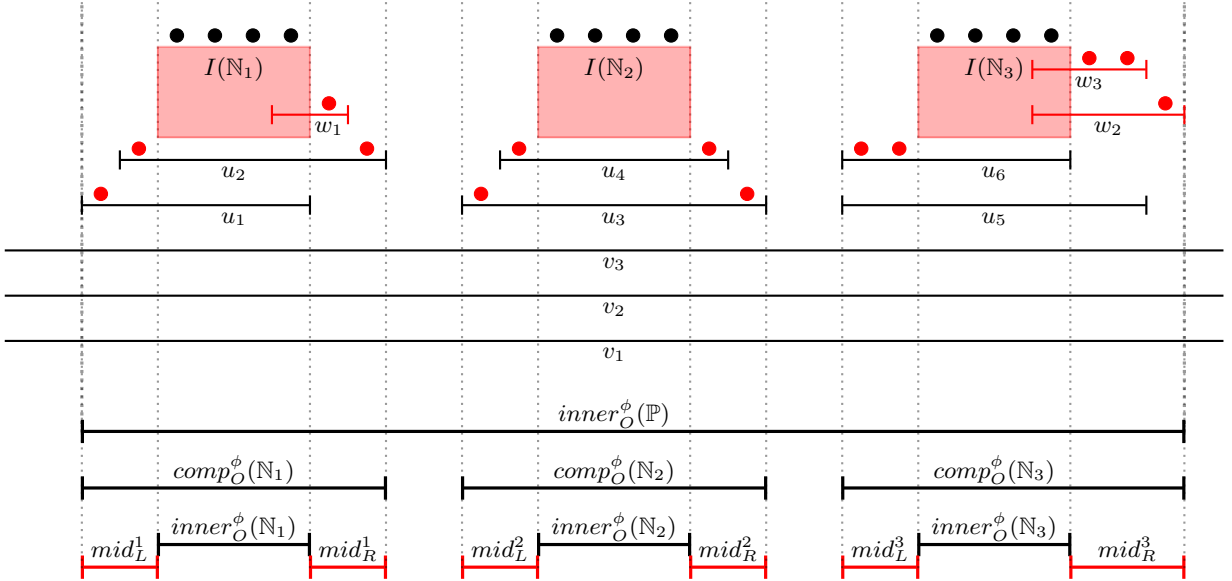

The next lemma describes good triple $(inner^{\phi}_{\circle}(\mathbb{P}),inner^{\phi}_{\stick}(\mathbb{P}),{\prec_{\phi}})$ for the set $inner(\mathbb{P})$
as a certain composition of good triples 
$$(comp^{\phi}_{\circle}(\mathbb{N}_1),comp^{\phi}_{\stick}(\mathbb{N}_1),{\prec_{\phi}}), \ldots, (comp^{\phi}_{\circle}(\mathbb{N}_k),comp^{\phi}_{\stick}(\mathbb{N}_k),{\prec_{\phi}})$$ 
for the sets $comp(\mathbb{N}_1), \ldots, comp(\mathbb{N}_k)$, respectively.
where $(comp^{\phi}_{\circle}(\mathbb{N}_i),comp^{\phi}_{\stick}(\mathbb{N}_i),{\prec_{\phi}})$ is a composition of good triple 
$(inner^{\phi}_{\circle}(\mathbb{N}_i),inner^{\phi}_{\stick}(\mathbb{N}_i),{\prec_{\phi}})$ for the set $inner(\mathbb{N}_i)$ and a three-chain-partition
$(mid^{i}_L,mid^{i}_R,mid^{i}_D)$ of the set $middle(\mathbb{N}_i)$. 
\begin{lemma}[$P$-node structure -- see Figure~\ref{fig:P-node-structure}]
\label{lem:P-node-model-structure}
Let $\big{(}inner^{\phi}_{\circle}(\mathbb{P}),inner^{\phi}_{\stick}(\mathbb{P}),{\prec_{\phi}}\big{)}$ be a good triple for the set $inner(\mathbb{P})$, where $\phi$ is a $(C,H)$-model of $G$.
Let $\mathbb{N}_1,\ldots,\mathbb{N}_k$ be an admissible order of the children of $\mathbb{P}$ induced by $(inner^{\phi}_{\circle}(\mathbb{P}),{\prec_{\phi}})$.
Then:
\begin{itemize}
\item for every $i \in [k]$, $\big{(}comp^{\phi}_{\circle}(\mathbb{N}_{i}),comp^{\phi}_{\stick}(\mathbb{N}_{i}), {\prec_{\phi}} \big{)}$ 
is good for the set $comp(\mathbb{N}_i)$ and the borders $(B^i_L,B^i_R)$,
where: 
$$
(B^{i}_L, B^i_R) = \left\{
\begin{array}{ll}
\big{(}\emptyset,U(\mathbb{P}) \cup E(\mathbb{P})\big{)}& \text{if $i=1$,} \\
\big{(}U(\mathbb{P}) \cup E(\mathbb{P}),U(\mathbb{P}) \cup E(\mathbb{P})\big{)}& \text{if $i \in [2,k-1]$,} \\
\big{(}U(\mathbb{P}) \cup E(\mathbb{P}), \emptyset \big{)}& \text{if $i=k$,}
\end{array}
\right.
$$
\item $inner^{\phi}_{\circle}(\mathbb{P}) = \bigcup_{i=1}^k comp^{\phi}_{\circle}(\mathbb{N}_{i})$ and $comp^{\phi}_{\circle}(\mathbb{N}_{1}) \prec_{\phi} \dots \prec_{\phi} comp^{\phi}_{\circle}(\mathbb{N}_{k})$,
\item if $inner^{\phi}_\stick(\mathbb{P})$ is non-empty, then $inner^{\phi}_\stick(\mathbb{P}) = comp^{\phi}_{\stick}(\mathbb{N}_d)$, 
where $(d)$ is the trace of the chain $inner^{\phi}_\stick(\mathbb{P})$ in the poset $(inner(\mathbb{P}) \cap \mathcal{I}_a, {\sqsubseteq})$
(which yields $comp^{\phi}_{\stick}(\mathbb{N}_i) = \emptyset$ for every $i \in [k] \setminus \{d\}$). 
\end{itemize}
Moreover, there are three-chain-partitions $(mid^i_L,mid^i_R,mid^i_D)$ of the sets $middle(\mathbb{N}_i)$ for $i \in [k]$ such that:
\begin{itemize}
\item $comp^{\phi}_{\circle}(\mathbb{N}_i) = inner^{\phi}_{\circle}(\mathbb{N}_i) \cup mid^{i}_L \cup mid^{i}_R$ and $mid^{i}_L \prec_{\phi} inner^{\phi}_{\circle}(\mathbb{N}_i) \prec_{\phi} mid^{i}_R$,
\item $comp^{\phi}_{\stick}(\mathbb{N}_i) = mid^i_D \cup inner^{\phi}_{\stick}(\mathbb{N}_i)$ and $mid^i_D \sqsubset inner^{\phi}_{\stick}(\mathbb{N}_i)$,
\item $(mid^{i}_L,{\prec_{\phi}})$ equals to $(mid^{i}_L,{\sqsubset})$ and $(mid^{i}_R,{\prec_{\phi}})$ equals to the reverse of $(mid^{i}_R,{\sqsubset})$,
\item $\big{(}inner^{\phi}_{\circle}(\mathbb{N}_i),inner^{\phi}_{\stick}(\mathbb{N}_i), {\prec_{\phi}})$ is good for the set $inner(\mathbb{N}_i)$ and the borders $(C^i_L,C^i_R)$, where:
$$
 C^i_L  = \left\{
 \begin{array}{ll}
 C\big{(}max(mid^{i}_L,{\sqsubseteq})\big{)} & \text{if $mid^{i}_L \neq \emptyset$,} \\
 B^i_L & \text{otherwise,}
 \end{array}
 \right.
 \quad
 C^i_R  = \left\{
 \begin{array}{ll}
 C\big{(}max(mid^{i}_R,{\sqsubseteq})\big{)} & \text{if $mid^{i}_R \neq \emptyset$,} \\
 B^i_R & \text{otherwise.}
 \end{array}
 \right.
 $$
\end{itemize}
\end{lemma}

Finally, let $\mathbb{Q}$ be a $Q$-node in $\mathbb{T}$ and let $\mathbb{N}_1,\ldots,\mathbb{N}_k$ be an admissible order of the children of $\mathbb{Q}$.
Let $I'(\mathbb{Q}) = I(\mathbb{Q}) \setminus \bigcup_{i \in [k]} V(\mathbb{N}_i)$.
Note that $I'(\mathbb{Q}) \neq \emptyset$.
In Figure \ref{fig:inner-middle-sets}, $\mathbb{Q}_1$ is a $Q$-node with children $\mathbb{L}_3,\mathbb{L}_4,\mathbb{L}_5$ and  we have $I'(\mathbb{Q}_1) = \{v_9,v_{12},v_{13}\}$.
Recall that for every $v \in V(H)$ we have defined $\Cl(v) = \{\mathbb{L} \in \mathcal{L} \mid v \in C(\mathbb{L})\}$.
Clearly, for every $v \in I'(\mathbb{Q})$ and every $i \in [k]$ we have either $\mathcal{L}(\mathbb{N}_i) \subseteq \Cl(v)$ or
$\mathcal{L}(\mathbb{N}_i) \cap \Cl(v) = \emptyset$.
So, for every $v \in I'(\mathbb{Q})$ we let
$$J(v) = \{j \in [k]: \mathcal{L}(\mathbb{N}_j) \subseteq \Cl(v) \}$$
and we note that for every $v \in I'(\mathbb{Q})$ the set $J(v)$ is a strict subinterval of $[k]$.
Also, we observe the following property of a $Q$-node:
\begin{equation}
\label{eq:Q-node-main-property}
\begin{array}{c}
 \text{For every strict subinterval $[j_1,j_2]$ of $[k]$ there is  a vertex $v \in I'(\mathbb{Q})$ such that} \\
 \text{either $j_1 \notin J(v)$ and $j_2,j_{2}+1 \in J(v)$ or $j_1-1,j_1 \in J(v)$ and $j_2 \notin J(v)$.}
\end{array}
\tag{Q}
\end{equation}
If this property does not hold, we can reverse the children $\mathbb{N}_{j_1}, \ldots ,\mathbb{N}_{j_2}$ of $\mathbb{Q}$ and we obtain 
a consecutive ordering of the leaves of $\mathbb{T}$ in which the children of $\mathbb{Q}$ occur in order different 
than $\mathbb{N}_1,\ldots,\mathbb{N}_k$ and than $\mathbb{N}_k,\ldots,\mathbb{N}_1$, which can not be the case for a Q-node.

Note that in the case of $Q$-nodes, in contrast to $P$-nodes, 
the set $$rest(\mathbb{Q})  = inner(\mathbb{Q}) \setminus \bigcup_{i \in [k]}  \big{(}middle(\mathbb{N}_i) \cup inner(\mathbb{N}_i)\big{)}$$ might be non-empty.
Indeed, notice that for every $I \in inner(\mathbb{Q})$ we have $I \in rest(\mathbb{Q})$ if and only if 
$C(I) \cap \big{(}\bigcup_{i \in [k]} V(\mathbb{N}_i)\big{)} = \emptyset$ and $C(I) \neq C(\mathbb{L})$ for every $\mathbb{L} \in \mathcal{L}(\mathbb{Q})$.
In Figure~\ref{fig:inner-middle-sets}, $I_{11} \in rest(\mathbb{Q}_1)$ as $C(I_1) \cap \big{(}\bigcup_{i \in [3,5]} V(\mathbb{L}_i)\big{)} \neq \emptyset$, $C(I) \neq C(\mathbb{L})$ 
for every $\mathbb{L} \in \mathcal{L}(\mathbb{Q}_1) = \{\mathbb{L}_3,\mathbb{L}_4,\mathbb{L}_5\}$.

Let $I \in rest(\mathbb{Q})$.
Observe that $$\big{(}E(\mathbb{Q}) \cup U(\mathbb{Q})\big{)} \quad \subsetneq  \quad C(I) \quad \subsetneq \quad \big{(} E(\mathbb{Q}) \cup U(\mathbb{Q}) \cup I'(\mathbb{Q})\big{)}.$$
Let
$$J(I) = \bigcap\{J(v): v \in C(I) \cap I'(\mathbb{Q})\}.$$
Clearly, $J(I)$ is a strict subinterval of $[k]$.
Suppose $J = [j_1,j_2]$ for some $j_1 \leq j_2$.
Notice that:
\begin{itemize}
 \item there is a vertex $v^{I}_L \in C(I) \cap I'(\mathbb{Q})$ such that $j_1 - 1 \notin J(v^{I}_L)$. 
 \item there is a vertex $v^{I}_R \in C(I) \cap I'(\mathbb{Q})$ such that $j_2 + 1 \notin J(v^{I}_R)$.
\end{itemize}
By definition of $J(I)$ we have $J(I) \subseteq J(v^{I}_L)$ and $J(I) \subseteq J(v^{I}_R)$.
Note that we might have $v^{I}_L  = v^{I}_R$.
Below we will prove the following property:
\begin{equation}
\label{eq:deletion_rule_Q_node}
 \begin{array}{c}
    \text{If there is $v \in I'(\mathbb{Q}) \setminus C(I)$ such that $\{j_1-1,j_1\} \subseteq J(v)$ and } \\
    \text{there is $v \in I'(\mathbb{Q}) \setminus C(I)$ such that $\{j_2,j_2+1\} \subseteq J(v)$, then} \\
    \text{$I \in \mathcal{I}^{\phi}_{\stick}$ in any $(C,H)$-centered model $\phi$ of $G$.}
    \tag{D3}
 \end{array}
\end{equation}
We extend $\mathcal{D}$ by $I$ whenever $I$ satisfies~\eqref{eq:deletion_rule_Q_node}.
Suppose still $I \notin \mathcal{D}$.
If $j_1=j_2$, then we add $I$ to the set $outer(\mathbb{Q}_j)$.
If $j_1 < j_2$, then:
\begin{itemize}
\item We add $I$ to $outer(\mathbb{Q}_{j_1})$ if there 
is no $v \in I'(\mathbb{Q}) \setminus C(I)$ such that $\{j_1-1,j_1\} \subseteq J(v)$.
\item We add $I$ to $outer(\mathbb{Q}_{j_2})$ if there 
is no $v \in I'(\mathbb{Q}) \setminus C(I)$ such that $\{j_2,j_2+1\} \subseteq J(v)$.
\end{itemize}
Since $\mathbb{Q}$ is a $Q$-node and $[j_1,j_2]$ is a strict subinterval of $[k]$, 
property~\eqref{eq:Q-node-main-property}  asserts 
there is $v \in I'(\mathbb{Q})$ such that either $j_1 \notin J(v)$ and $\{j_2,j_2+1\} \in J(v)$ or $j_2 \notin J(v)$ and $\{j_1-1,j_1\} \in J(v)$.
In particular, $v \notin C(I)$ and hence $I$ is added to exactly one of the sets
$outer(\mathbb{Q}_{j_1})$ and $outer(\mathbb{Q}_{j_2})$.

Clearly, the procedure described above shows the set
$\{comp(\mathbb{N}_i): i \in [k]\}$ forms a partition of $inner(\mathbb{Q})$. 
Also, note that the partition $\{comp(\mathbb{N}_i): i \in [k]\}$ 
is independent on the choice of an admissible ordering of the children of $\mathbb{Q}$.
We also note that every component $I \in outer(\mathbb{N}_i)$ satisfies the following property:
\begin{equation}
\label{eq:outer_component_property}
\begin{array}{c}
\text{there is $v \in I'(\mathbb{Q}) \cap C(I)$ such that $i-1 \notin J(v)$ and $i \in J(v)$ and} \\
\text{$\big{\{}w \in I'(\mathbb{Q}): \{i-1,i\} \subseteq J(w)\big{\}} \subsetneq C(I)$} \\
\text{or}\\
\text{there is $v \in I'(\mathbb{Q}) \cap C(I)$ such that $i+1 \notin J(v)$ and $i \in J(v)$ and} \\
\text{$\big{\{}w \in I'(\mathbb{Q}): \{i,i+1\} \subseteq J(w)\big{\}} \subsetneq C(I)$.}
\tag{*}
\end{array}
\end{equation}

Next, we show that for every $(C,H)$-centered model $\phi$ of $G$ and every $i \in [k]$ we have $comp^\phi(\mathbb{N}_1) \prec_{\phi} \dots \prec_{\phi}comp^\phi(\mathbb{N}_k)$ if the children of $\mathbb{Q}$ occur in $\phi^*$ in the order $\mathbb{N}_1,\ldots,\mathbb{N}_k$ or we have $comp^\phi(\mathbb{N}_k) \prec_{\phi} \dots \prec_{\phi} comp^\phi(\mathbb{N}_1)$ if the children of $\mathbb{Q}$ occur in $\phi^*$ in the order $\mathbb{N}_k,\ldots,\mathbb{N}_1$.
We also prove that property \eqref{eq:deletion_rule_Q_node} is correct.

Suppose $\phi$ is a $(C,H)$-centered model $\phi$ of $G$.
Suppose the children of $\mathbb{Q}$ appear in $\phi^*$ in 
the order $\mathbb{N}_1,\dots,\mathbb{N}_k$.
In particular, we have
$$zone^{\phi}(\mathbb{N}_1) \prec_{\phi}  zone^{\phi}(\mathbb{N}_2) \prec_{\phi} \dots \prec_{\phi} zone^{\phi}(\mathbb{N}_k).$$
We claim $comp^{\phi}_{\circle}(\mathbb{N}_i) \subseteq zone^{\phi}(\mathbb{N}_i)$ holds for $i \in [k]$, which proves
$$comp^{\phi}_{\circle}(\mathbb{N}_1) \prec_{\phi}  comp^{\phi}_{\circle}(\mathbb{N}_2) \prec_{\phi} \dots \prec_{\phi} comp^{\phi}_{\circle}(\mathbb{N}_k).$$

Suppose $I \in comp^{\phi}_{\circle}(\mathbb{N}_{j_1})$.
If $I \in middle(\mathbb{N}_{j_1}) \cup inner(\mathbb{N}_{j_1})$, then $C(I) \cap V(\mathbb{N}_{j_1}) \neq \emptyset$ or $\mathbb{N}_{j_1}$ is a leaf in $\mathbb{T}$ and $C(I) = C(\mathbb{N}_{j_1})$, 
and hence $I \in zone^{\phi}(\mathbb{N}_{j_1})$.
Suppose $I \in outer(\mathbb{N}_{j_1})$.
Clearly, $I$ is contained in the set $zone^{\phi}(\mathbb{N}_j)$ for some $j \in J(I)$.
So, if $J(I) = \{j_1\}$, then we have $I \in zone^{\phi}(\mathbb{N}_{j_1})$, which proves our claim.
Now, suppose $J(I) = [j_1,j_2]$ for some $j_1 < j_2$ (the other case is analogous).
First, note that $I$ can not occupy a maximal sector of $\phi^*$ as $I \in outer(\mathbb{N}_{j_1})$.
Moreover, $I$ can not occupy a sector between two maximal sectors $S^{\phi^*}(\mathbb{L})$
and $S^{\phi^*}(\mathbb{L}')$ for some $\mathbb{L}, \mathbb{L}'$ from 
$\bigcup_{j \in J(I)} \mathcal{L}(\mathbb{N}_j)$. 
Otherwise, since $C(I) \subseteq C(\mathbb{L}) \cap C(\mathbb{L}')$,
$I$ must be ambiguous simple and contained in the minimal sector between $S^{\phi^*}(\mathbb{L})$ and $S^{\phi^*}(\mathbb{L}')$,
contradicting property~\eqref{eq:input_no_simple_cliques_on_minimal_sectors}.
So, either $I \in zone^{\phi}(\mathbb{N}_{j_1})$ and $I$ is to the left of 
$S^{\phi^*}(\mathbb{N}_{j_1})$ or $I \in zone^{\phi}(\mathbb{N}_{j_2})$ and $I$ is to the right of $S^{\phi^*}(\mathbb{N}_{j_2})$.
Since $J(I) = [j_1,j_2]$ and $I \in outer(\mathbb{N}_{j_1})$, 
property \eqref{eq:Q-node-main-property} asserts we have a vertex $v \in I'(\mathbb{Q}) \setminus C(I)$ such that $\{j_2,j_2+1\} \subseteq J(v)$.
Note that $\phi^*(v)$ covers all the maximal sectors of $S^{\phi{^*}}(\mathbb{L})$ for $\mathbb{L} \in \mathcal{L}(\mathbb{N}_{j_2}) \cup \mathcal{L}(\mathbb{N}_{j_2+1})$.
In particular, $I$ can not be contained in $zone^{\phi}(\mathbb{N}_{j_2})$.
We conclude $I \in zone^{\phi}(\mathbb{N}_{j_1})$.

Now, we show property~\eqref{eq:deletion_rule_Q_node}.
If $v \in I'(v) \setminus C(I)$ is such that 
$\{j_2,j_2+1\} \subseteq J(v)$, then the interval $\phi^*(v)$ covers the maximal sectors $S^{\phi^*}(\mathbb{L})$ for $\mathbb{L} \in \mathcal{L}(\mathbb{N}_{j_2}) \cup \mathcal{L}(\mathbb{N}_{j_2+1})$.
Similarly, if there is $v \in I'(v) \setminus C(I)$ such that 
$\{j_1,j_1-1\} \subseteq J(v)$, then $\phi^*(v)$ covers the maximal sectors $S^{\phi^*}(\mathbb{L})$ for $\mathbb{L} \in \mathcal{L}(\mathbb{N}_{j_1-1}) \cup \mathcal{L}(\mathbb{N}_{j_1})$.
Using the same argument as above, we show that $I$ can not be contained between the maximal sectors from the set $S^{\phi^*}(\mathbb{L})$ for 
$\mathbb{L} \in \bigcup_{j \in J(I)} \mathcal{L}(\mathbb{N}_{j}))$.
It means $I \notin zone^{\phi}(\mathbb{N}_j)$ for every $j \in J(I)$.
So, we must have $I \in \mathcal{I}^{\phi}_{\stick}$, which proves~\eqref{eq:deletion_rule_Q_node}.

Finally, we describe some properties of a chain $\mathcal{C}$ in the poset $(inner(\mathbb{Q}) \cap \mathcal{I}_a, {\sqsubseteq})$.
Contrary to the case for P-nodes, $\mathcal{C}$ might intersect many sets from the family $\{comp(\mathbb{N}_i): i \in [k]\}$.
In order to describe precisely how such chains behave, we introduce ${\sqsubseteq_Q}$ relation on the children of $\mathbb{Q}$.
For every $i < j$ in $[n]$ we set:
$$
\begin{array}{rl}
\mathbb{N}_i \sqsubseteq_{Q} \mathbb{N}_j 
&\text{if there is $v \in I'(\mathbb{Q})$ such that $i \notin J(v)$ and $\{j,j+1\} \subseteq J(v)$,} \\
\mathbb{N}_j \sqsubseteq_{Q} \mathbb{N}_i
&\text{if there is $v \in I'(\mathbb{Q})$ such that $\{i,i-1\} \subseteq J(v)$ and $j \notin J(v)$.}
\end{array}
$$ 
\begin{claim}
\label{lem:Q_node_chain_structure_preparation}
The following statements hold:
\begin{enumerate}
\item \label{item:chain_Q_node_sqsubset_Q_prop} For every $i \in [k-1]$, we have $\mathbb{N}_i \not \sqsubseteq_{Q} \mathbb{N}_k$ and for every $i \in [2,k]$ we have $\mathbb{N}_i \not \sqsubseteq_{Q} \mathbb{N}_1$. 
However, for every $i < j$ from $[k]$ such that $(i,j) \neq (1,k)$ we have $\mathbb{N}_i \sqsubseteq_{Q} \mathbb{N}_j$ or $\mathbb{N}_j \sqsubseteq_{Q} \mathbb{N}_i$.
 \item \label{item:chain_Q_node_sqsubset_Q_consistent_with_sqsubset} For every $i \neq j$ from $[k]$, 
 if $I_1 \in comp(\mathbb{N}_i)$ and $I_2 \in comp(N_j)$ are such that $I_1 \sqsubset I_2$, 
 then $\mathbb{N}_i \sqsubseteq_{Q} \mathbb{N}_j$, $\mathbb{N}_j \not \sqsubseteq_{\mathbb{Q}} \mathbb{N}_i$, and $I_1 \in outer(\mathbb{N}_i)$.
 \item \label{item:chain_Q_node_sqsubset_Q_consistent_with_sqsubset_cons}
 For every $i \neq j$ from $[k]$, if $\mathbb{N}_i \sqsubseteq_{Q} \mathbb{N}_j$ and 
 $\mathbb{N}_j \sqsubseteq_{Q} \mathbb{N}_i$, then $comp(\mathbb{N}_i)$ and $comp(\mathbb{N}_j)$ are ${\sqsubseteq}$-incomparable.
\item \label{item:chain_Q_node_zigzag_prop} For every $i_1 \neq i_2$ and $i_3$ from $[k]$, if $I_1 \in comp(\mathbb{N}_{i_1})$, $I_2 \in comp(\mathbb{N}_{i_2})$, and $I_3 \in comp(\mathbb{N}_{i_3})$ are such that $I_1 \sqsubset I_2 \sqsubset I_3$, then:
\begin{itemize}
 \item if $i_1 < i_2$, then $i_1 < i_3$.
 \item if $i_1 > i_2$, then $i_1 > i_3$.
\end{itemize}
\end{enumerate}
\end{claim}
\begin{proof}
The first part of~\eqref{item:chain_Q_node_sqsubset_Q_prop} is obvious.
The second part of~\eqref{item:chain_Q_node_sqsubset_Q_prop} follows from property~\eqref{eq:Q-node-main-property}.

To show statement~\eqref{item:chain_Q_node_sqsubset_Q_consistent_with_sqsubset} let $I \in comp(\mathbb{N}_i)$ and $J \in comp(\mathbb{N}_j)$ be such that $I\sqsubset J$.
Let $i < j$.
Clearly, we must have $I \in outer(\mathbb{N}_i)$ as otherwise we would have 
$C(I) \cap V(\mathbb{N}_i) \neq \emptyset$ or $C(I)=C(\mathbb{L})$ for some $\mathbb{L} \in \mathcal{L}(\mathbb{N}_i)$, 
which contradicts $C(I) \subseteq C(J)$ and $J \in comp(\mathbb{N}_j)$.
First, note that $(i,j) \neq (1,k)$.
Suppose otherwise.
Since $I_1 \in outer(\mathbb{N}_1)$, property~\eqref{eq:outer_component_property} asserts 
the set $C(I_1)$ contains a vertex $v \in I'(\mathbb{Q})$ such that $1 \in J(v)$.
In particular, $k \notin J(v)$, which shows $v \notin C(I_2)$.
So, we can not have $C(I_1) \subseteq C(I_2)$, which contradicts $I_1 \sqsubset I_2$.
Now, to show statement~\eqref{item:chain_Q_node_sqsubset_Q_consistent_with_sqsubset} 
it is enough to prove
$\mathbb{N}_j \not \sqsubseteq_Q \mathbb{N}_i$.
Suppose $\mathbb{N}_j \sqsubseteq_Q \mathbb{N}_i$.
It means that there is $v \in I'(\mathbb{Q})$ such that $j \notin J(v)$ and $\{i-1,i\} \subseteq J(v)$.
We can not have $v \in C(I_1)$, as otherwise $C(I_1) \subseteq C(I_2)$ does not hold, contradicting $I_1 \sqsubset I_2$.
Then, \eqref{eq:outer_component_property}~property asserts there is $w \in I'(\mathbb{Q})$ such that $w \in C(I)$, $i \in J(w)$, and $i+1 \notin J(w)$.
In particular, it means that $C(I) \subseteq C(J)$ does not hold,  contradicting $I_1 \sqsubset I_2$.
This proves statement~\eqref{item:chain_Q_node_sqsubset_Q_consistent_with_sqsubset}.

Statement~\eqref{item:chain_Q_node_sqsubset_Q_consistent_with_sqsubset_cons} follows from~\eqref{item:chain_Q_node_sqsubset_Q_consistent_with_sqsubset}.

Suppose $I_1 \in comp(\mathbb{N}_{i_1})$, $I_2 \in comp(\mathbb{N}_{i_2})$, and 
$I_3 \in comp(\mathbb{N}_{i_3})$ are such that $i_1 < i_2$ and $I_1 \sqsubset I_2 \sqsubset I_3$.
Statement~\eqref{item:chain_Q_node_sqsubset_Q_consistent_with_sqsubset} shows 
$\mathbb{N}_{i_1} \sqsubset_{Q} \mathbb{N}_{i_2}$.
Again, using \eqref{eq:outer_component_property} property we can show, in the same way as above, 
that there is $v \in I'(\mathbb{Q}) \cap C(I_2)$ such that $i_1 \notin J(v)$ and $i_2 \in J(v)$.
Since $v \in C(I_2) \subseteq C(I_3)$, we conclude $i_1 < i_3$.
This shows statement~\eqref{item:chain_Q_node_zigzag_prop}.
\end{proof}

A tuple $(d_1,\ldots,d_z)$ of integers from $[k]$ forms a \emph{zigzag} if for every $i < j < k$ in $[z]$
we have:
\begin{itemize}
\item if $d_i < d_j$ then $d_i < d_k$,  
\item if $d_i > d_j$ then $d_i < d_k$.
\end{itemize}
Additionally, a zigzag $(d_1,\ldots,d_z)$ is \emph{$\sqsubseteq_{Q}$-consistent} if for every $i < j$ in $[z]$ we have
$\mathbb{N}_{d_i} \sqsubseteq_Q \mathbb{N}_{d_j}$ and $\mathbb{N}_{d_j} \not \sqsubseteq_Q \mathbb{N}_{d_i}$.

\begin{figure}[h]
\begin{center}
\begin{subfigure}[t]{1\linewidth}
\centering
\begin{tikzpicture}[xscale=0.34,yscale=0.4,>=latex,shorten >=-0.4pt,shorten <=-0.4pt]
 \tikzstyle{every node}=[inner sep=2pt,fill=white]  

\draw[dotted,opacity=1,thick] (1.5,-0.5) -- (1.5,12.5);
\coordinate (lN1) at (1.5,13) {};
\draw[dotted,opacity=1,thick] (4.5,-0.5) -- (4.5,12.5);
\coordinate (lN2) at (4.5,13) {};
\draw[dotted,opacity=1,thick] (6.5,-0.5) -- (6.5,12.5);
\coordinate (lN3) at (6.5,13) {};
\draw[dotted,opacity=1,thick] (8.5,-0.5) -- (8.5,12.5);
\coordinate (lN4) at (8.5,13) {};
\draw[dotted,opacity=1,thick] (10.5,-0.5) -- (10.5,12.5);
\coordinate (lN5) at (10.5,13) {};
\draw[dotted,opacity=1,thick] (12.5,-0.5) -- (12.5,12.5);
\coordinate (lN6) at (12.5,13) {};
\draw[dotted,opacity=1,thick] (15.5,-0.5) -- (15.5,12.5);
\coordinate (lN7) at (15.5,13) {};
\draw[dotted,opacity=1,thick] (18.5,-0.5) -- (18.5,12.5);
\coordinate (lN8) at (18.5,13) {};
\draw[dotted,opacity=1,thick] (20.5,-0.5) -- (20.5,12.5);
\coordinate (lN9) at (20.5,13) {};

\draw[dotted,opacity=1,thick] (23.5,-0.5) -- (23.5,12.5);
\coordinate (lN10) at (23.5,13) {};
\draw[dotted,opacity=1,thick] (25.5,-0.5) -- (25.5,12.5);
\coordinate (lN11) at (25.5,13) {};
\draw[dotted,opacity=1,thick] (27.5,-0.5) -- (27.5,12.5);
\coordinate (lN12) at (27.5,13) {};
\draw[dotted,opacity=1,thick] (29.5,-0.5) -- (29.5,12.5);
\coordinate (lN13) at (29.5,13) {};
\draw[dotted,opacity=1,thick] (31.5,-0.5) -- (31.5,12.5);
\coordinate (lN14) at (31.5,13) {};
\draw[dotted,opacity=1,thick] (34.5,-0.5) -- (34.5,12.5);
\coordinate (lN15) at (34.5,13) {};
\draw[dotted,opacity=1,thick] (37.5,-0.5) -- (37.5,12.5);
\coordinate (lN16) at (37.5,13) {};
\draw[dotted,opacity=1,thick] (39.5,-0.5) -- (39.5,12.5);
\coordinate (lN17) at (39.5,13) {};
\draw[dotted,opacity=1,thick] (41.5,-0.5) -- (41.5,12.5);
\coordinate (lN18) at (41.5,13) {};
\draw[dotted,opacity=1,thick] (43.5,-0.5) -- (43.5,12.5);
\coordinate (lN19) at (43.5,13) {};

\draw[|-|,thick,black] (0,0) -- (42,0);
\draw[|-|,thick,black] (43,0) -- (44,0);
\coordinate (lv1) at (21,-0.5) {};

\coordinate (I1) at (0.3,1) {};
\draw[|-|,thick,black] (1,1) -- (2,1);
\draw[|-|,thick,black] (3,1) -- (40,1);
\coordinate (lv2) at (21,0.5) {};
\draw[|-|,thick,black] (41,1) -- (44,1);

\coordinate (I2) at (3.3,2) {};
\draw[|-|,thick,black] (4,2) -- (5,2);
\draw[|-|,thick,black] (6,2) -- (38,2);
\coordinate (lv3) at (21,1.5) {};
\draw[|-|,thick,black] (39,2) -- (44,2);

\draw[|-|,thick,black] (6,3) -- (7,3);
\draw[|-|,thick,black] (8,3) -- (36,3);
\draw[|-|,thick,black] (37,3) -- (44,3);
\coordinate (lv4) at (21,2.5) {};

\draw[|-|,thick,black] (6,4) -- (9,4);
\draw[|-|,thick,black] (10,4) -- (33,4);
\draw[|-|,thick,black] (34,4) -- (35,4);
\coordinate (I3) at (35.7,4) {};

\draw[|-|,thick,black] (6,5) -- (11,5);
\draw[|-|,thick,black] (12,5) -- (30,5);
\draw[|-|,thick,black] (31,5) -- (32,5);
\coordinate (I4) at (32.7,5) {};

\draw[|-|,thick,black] (6,6) -- (13,6);
\draw[|-|,thick,black] (14,6) -- (28,6);
\draw[|-|,thick,black] (29,6) -- (30,6);

\coordinate (I5) at (14.3,7) {};
\draw[|-|,thick,black] (15,7) -- (16,7);
\draw[|-|,thick,black] (17,7) -- (26,7);
\draw[|-|,thick,black] (27,7) -- (30,7);

\coordinate (I6) at (17.3,8) {};
\draw[|-|,thick,black] (18,8) -- (19,8);
\draw[|-|,thick,black] (20,8) -- (24,8);
\draw[|-|,thick,black] (25,8) -- (30,8);

\draw[|-|,thick,black] (23,9) -- (30,9);
\draw[|-|,thick,black] (20,9) -- (22,9);

\draw[thick,opacity=0.3,fill=red, red] (20,10) rectangle (21,12);
\coordinate (I7) at (21.7,9.7) {};
\coordinate (I8) at (20.5,11) {};

\draw[-,blue,dotted, very thick] (I1) -- (I2)--(I3)--(I4)--(I5)--(I6)--(I7)--(I8);

\tikzstyle{every node}=[circle,minimum size=5pt,inner sep=0pt,draw,fill]
%\node at (I0) {};
\node[blue] at (I1) {};
\node[blue] at (I2) {};
\node[blue] at (I3) {};
\node[blue] at (I4) {};
\node[blue] at (I5) {};
\node[blue] at (I6) {};
\node[blue] at (I7) {};
\node[blue] at (I8) {};

\tikzstyle{every node}=[inner sep=1pt]
\begin{scriptsize}
\node at (lN1) {$\mathbb{N}_1$};
\node at (lN2) {$\mathbb{N}_2$};
\node at (lN3) {$\mathbb{N}_3$};
\node at (lN4) {$\mathbb{N}_4$};
\node at (lN5) {$\mathbb{N}_5$};
\node at (lN6) {$\mathbb{N}_6$};
\node at (lN7) {$\mathbb{N}_7$};
\node at (lN8) {$\mathbb{N}_8$};
\node at (lN9) {$\mathbb{N}_9$};
\node at (lN10) {$\mathbb{N}_{10}$};
\node at (lN11) {$\mathbb{N}_{11}$};
\node at (lN12) {$\mathbb{N}_{12}$};
\node at (lN13) {$\mathbb{N}_{13}$};
\node at (lN14) {$\mathbb{N}_{14}$};
\node at (lN15) {$\mathbb{N}_{15}$};
\node at (lN16) {$\mathbb{N}_{16}$};
\node at (lN17) {$\mathbb{N}_{17}$};
\node at (lN18) {$\mathbb{N}_{18}$};
\node at (lN19) {$\mathbb{N}_{19}$};

\node at (lv1) {$v_1$};
\node at (lv2) {$v_2$};
\node at (lv3) {$v_3$};
\node at (lv4) {$v_4$};

\end{scriptsize}

%\draw[black] (0,-1)--(0,0);
%\draw[black] (45,12)--(45,13.5);

\draw[white] (0,-1)--(0,0);
\draw[white] (45,12)--(45,13.5);

\end{tikzpicture}
\end{subfigure}
\end{center}
\caption{The trace of a blue chain is $T = (1,2,15,14,7,8,9)$. $T$ forms a zigzag respecting $\sqsubseteq_{Q}$-relation. 
We have $\mathbb{N}_1 \sqsubset_{Q} \mathbb{N}_2$ as witnessed by $v_2$, we have $\mathbb{N}_2 \sqsubset_{Q} \mathbb{N}_{15}$ as witnessed by $v_3$ or $v_4$.} 
\label{fig:zigzag} 
\end{figure}
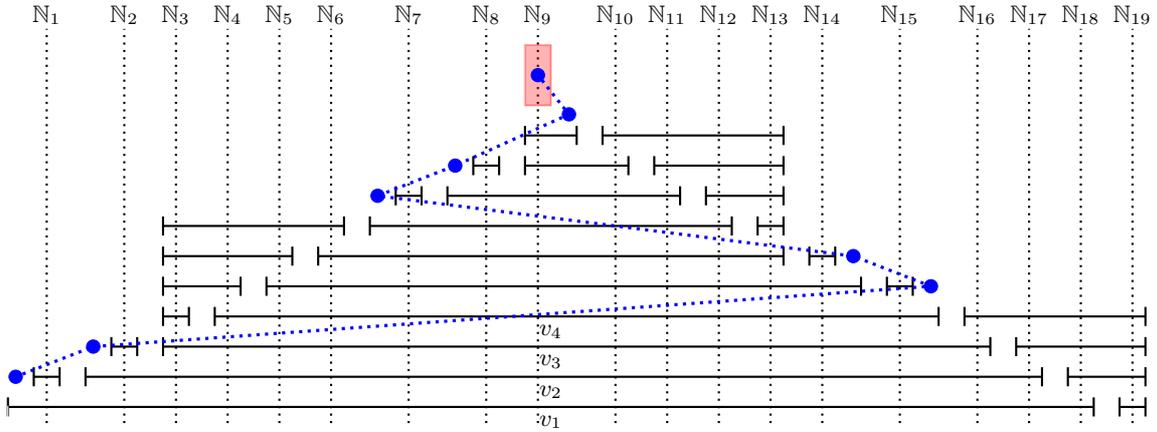

Suppose $C$ is a chain in the poset $(inner(\mathbb{Q}) \cap \mathcal{I}_{a},{\sqsubseteq})$.
As for $P$-nodes, the \emph{trace} of $\mathcal{C}$ in $(inner(\mathbb{Q}) \cap \mathcal{I}_{a}, {\sqsubseteq})$ 
is a tuple listing all the sets $comp(\mathbb{N}_i)$ intersected by $\mathcal{C}$ when we traverse $C$ upward.
Since, by Lemma \ref{lem:Q_node_chain_structure_preparation}.\eqref{item:chain_Q_node_zigzag_prop}, 
for every $i \in [k]$ the set $C \cap comp(\mathbb{N}_i)$ forms an interval in $(C,{\sqsubseteq})$, 
every entry in the trace of $C$ appears exactly once.
See Figure~\ref{fig:zigzag} for an illustration.
 
We can summarize this section with the following lemmas.
\begin{lemma}
Suppose $\mathbb{Q}$ is a $Q$-node of $\mathbb{T}$ with an admissible order of the children $\mathbb{N}_1,\ldots,\mathbb{N}_k$.
Then:
\begin{enumerate}
 \item The set $\{comp(\mathbb{N}_i) \mid i \in [k]\}$ forms a partition of $inner(\mathbb{Q})$,
 \item If $I \in outer(\mathbb{N}_i)$ and $J \in middle(\mathbb{N}_i) \cup inner(\mathbb{N}_i)$ are $\sqsubseteq$-comparable, then $I \sqsubset J$.
 \item Let $C$ be a chain in the poset $(inner(\mathbb{Q}) \cap \mathcal{I}_a, \subseteq)$.
 Then the trace $(d_1,\ldots,d_z)$ of $C$ forms a zigzag consistent with ${\sqsubseteq}_Q$-relation and 
 for every $j \in [z-1]$ we have $C \cap comp(\mathbb{N}_{d_j}) \subseteq outer(\mathbb{N}_{d_j})$.
\end{enumerate}
\end{lemma}

The next lemma describes good triple $(inner^{\phi}_{\circle}(\mathbb{Q}),inner^{\phi}_{\stick}(\mathbb{Q}),{\prec_{\phi}})$ for the set $inner(\mathbb{Q})$
as a certain composition of good triples 
$$(comp^{\phi}_{\circle}(\mathbb{N}_1),comp^{\phi}_{\stick}(\mathbb{N}_1),{\prec_{\phi}}), \ldots, (comp^{\phi}_{\circle}(\mathbb{N}_k),comp^{\phi}_{\stick}(\mathbb{N}_k),{\prec_{\phi}})$$ 
for the sets $comp(\mathbb{N}_1), \ldots, comp(\mathbb{N}_k)$, respectively, 
where $(comp^{\phi}_{\circle}(\mathbb{N}_i),comp^{\phi}_{\stick}(\mathbb{N}_i),{\prec_{\phi}})$ is a composition of good triple 
$(inner^{\phi}_{\circle}(\mathbb{N}_i),inner^{\phi}_{\stick}(\mathbb{N}_i),{\prec_{\phi}})$ for the set $inner(\mathbb{N}_i)$ and a three-chain-partition
$(midout^{i}_L,midout^{i}_R,midout^{i}_D)$ of the set $middle(\mathbb{N}_i) \cup outer(\mathbb{N}_i)$. 
\begin{lemma}[$Q$-node structure -- see Figure~\ref{fig:Q-node-structure}]
\label{lem:Q-node-model-structure}
Let $\big{(}inner^{\phi}_{\circle}(\mathbb{Q}),inner^{\phi}_{\stick}(\mathbb{Q}),{\prec_{\phi}}\big{)}$ 
be a good triple for the set $inner(\mathbb{Q})$, where $\phi$ is a $(C,H)$-model of $G$.
Let $\mathbb{N}_1,\ldots,\mathbb{N}_k$ be an admissible order of the children of $\mathbb{Q}$ induced by $(inner^{\phi}_{\circle}(\mathbb{Q}),{\prec_{\phi}})$.
Then:
\begin{itemize}
\item $\big{(}comp^{\phi}_{\circle}(\mathbb{N}_{i}),comp^{\phi}_{\stick}(\mathbb{N}_{i}), {\prec_{\phi}} \big{)}$ 
is good for the set $comp(\mathbb{N}_i)$ and the borders $(B^i_L,B^i_R)$, where: 
 $$(B^i_L,B^i_R) = 
 \left\{
 \begin{array}{ll}
  \big{(}\emptyset, E(\mathbb{N}_1) \cap E(\mathbb{N}_2)\big{)} & \text{if $i=1$,} \\
  \big{(}E(\mathbb{N}_{i-1}) \cap E(\mathbb{N}_{i}), E(\mathbb{N}_i) \cap E(\mathbb{N}_{i+1})\big{)} & \text{if $i \in [2,k-1]$,} \\
  \big{(}E(\mathbb{N}_{k-1}) \cap E(\mathbb{N}_{k}), \emptyset \big{)} & \text{if $i = k$,} \\
 \end{array}
 \right.
$$
\item $inner^{\phi}_{\circle}(\mathbb{Q}) = \bigcup_{i=1}^k comp^{\phi}_{\circle}(\mathbb{N}_{i})$ and $comp^{\phi}_{\circle}(\mathbb{N}_{1}) \prec_{\phi} \dots \prec_{\phi} comp^{\phi}_{\circle}(\mathbb{N}_{k})$,
\item $inner^{\phi}_{\stick}(\mathbb{Q}) = \bigcup_{i \in T} comp^{\phi}_{\stick}(\mathbb{N}_i)$, 
$comp^{\phi}_{\stick}(\mathbb{N}_{d_1}) \sqsubset \dots \sqsubset comp^{\phi}_{\stick}(\mathbb{N}_{d_z})$, and $comp^{\phi}_{\stick}(\mathbb{N}_{d_i}) \subseteq outer(\mathbb{N}_{d_i})$ for $i \in [z-1]$,
where $T = (d_1,\ldots,d_z)$ is the trace of $inner^{\phi}_{\stick}(\mathbb{Q})$.
\end{itemize}
Moreover, there are three-chain-partitions $(midout^i_L,midout^i_R,midout^i_D)$ of the sets $middle(\mathbb{N}_i) \cup outer(\mathbb{N}_i)$ for $i \in [k]$ such that:
\begin{itemize}
\item $comp^{\phi}_{\circle}(\mathbb{N}_i) = inner^{\phi}_{\circle}(\mathbb{N}_i) \cup mid^{i}_L \cup mid^{i}_R$ and $mid^{i}_L \prec_{\phi} inner^{\phi}_{\circle}(\mathbb{N}_i) \prec_{\phi} mid^{i}_R$,
\item $comp^{\phi}_{\stick}(\mathbb{N}_i) = midout^i_D \cup inner^{\phi}_{\stick}(\mathbb{N}_i)$ and $midout^i_D \sqsubset inner^{\phi}_{\stick}(\mathbb{N}_i)$,
\item $(midout^{i}_L,{\prec_{\phi}})$ equals to $(midout^{i}_L,{\sqsubset})$ and $(midout^{i}_R,{\prec_{\phi}})$ equals to the reverse of $(midout^{i}_R,{\sqsubset})$,
\item $\big{(}inner^{\phi}_{\circle}(\mathbb{N}_i),inner^{\phi}_{\stick}(\mathbb{N}_i), {\prec_{\phi}})$ is good for the set $inner(\mathbb{N}_i)$ and the borders $(C^i_L,C^i_R)$, where:
$$
 C^i_L  = \left\{
 \begin{array}{ll}
 C\big{(}max(midout^{i}_L,{\sqsubseteq})\big{)} & \text{if $midout^{i}_L \neq \emptyset$,} \\
 B^i_L & \text{otherwise,}
 \end{array}
 \right.
$$
$$
 C^i_R  = \left\{
 \begin{array}{ll}
 C\big{(}max(midout^{i}_R,{\sqsubseteq})\big{)} & \text{if $midout^{i}_R \neq \emptyset$,} \\
 B^i_R & \text{otherwise.}
 \end{array}
 \right.
$$
\end{itemize}
\end{lemma}
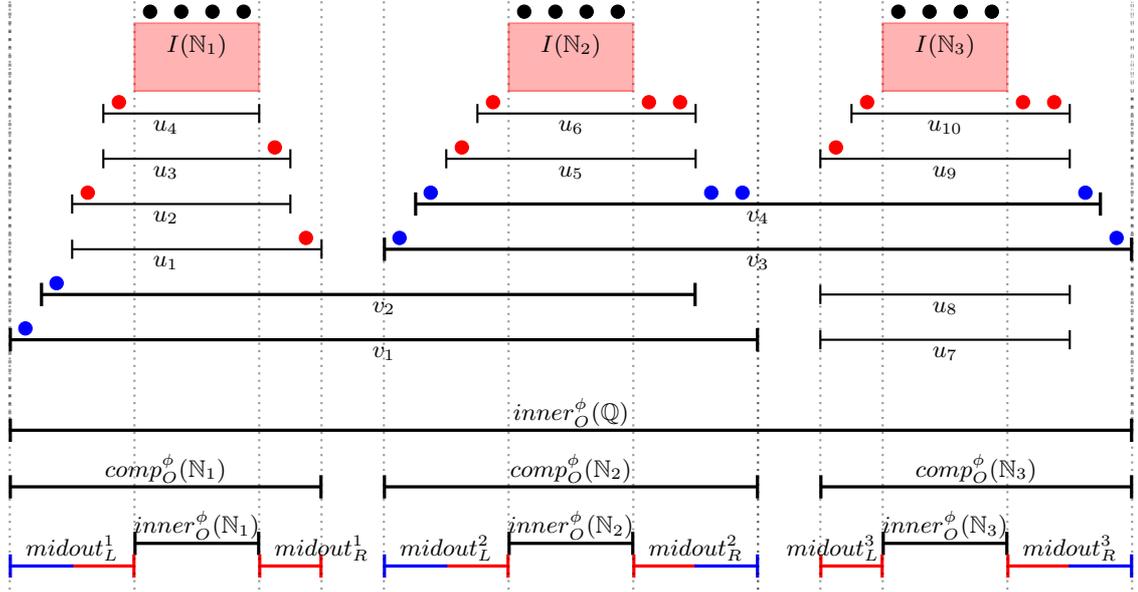
\begin{figure}[h]
\begin{center}
\begin{subfigure}[t]{1\linewidth}
\centering
\begin{tikzpicture}[xscale=0.82,yscale=0.6,>=latex,shorten >=-0.4pt,shorten <=-0.4pt]
  \tikzstyle{every node}=[inner sep=2pt,fill=white]  

\draw[|-|,very thick,black] (0,-2) -- (18,-2);
\coordinate (linQ) at (9,-1.6) {};

\draw[dotted,opacity=0.4,very thick] (0,-2) -- (0,7.5);
\draw[dotted,opacity=0.4,very thick] (18,-2) -- (18,7.5);

\draw[|-|,very thick,black] (0,-3.25) -- (5,-3.25);
\coordinate (lcompN1) at (2.5,-2.85) {};
\draw[|-|,very thick,black] (2,-4.5) -- (4,-4.5);
\coordinate (linN1) at (3,-4.1) {};

\draw[-|,very thick,red] (1,-5) -- (2,-5);
\coordinate (lmid1L) at (1,-4.6) {};
\draw[|-,very thick,blue] (0,-5.0) -- (1,-5.0);
\coordinate (lout1L) at (0.8,-4.6) {};

\draw[|-|,very thick,red] (4,-5) -- (5,-5);
\coordinate (lmid1R) at (5,-4.6) {};

\draw[dotted,opacity=0.4,thick] (0,-5.5) -- (0,7.5);
%\draw[dotted,opacity=0.4,thick] (1,-5.5) -- (1,7.5);
\draw[dotted,opacity=0.4,thick] (2,-5.5) -- (2,7.5);
\draw[dotted,opacity=0.4,thick] (4,-5.5) -- (4,7.5);
\draw[dotted,opacity=0.4,thick] (5,-5.5) -- (5,7.5);

\draw[|-|,very thick,black] (6,-3.25) -- (12,-3.25);
\coordinate (lcompN2) at (9,-2.85) {};
\draw[|-|,very thick,black] (8,-4.5) -- (10,-4.5);
\coordinate (linN2) at (9,-4.1) {};

\draw[-|,very thick,red] (7,-5) -- (8,-5);
\coordinate (lmid2L) at (7,-4.6) {};
\draw[|-,very thick,blue] (6,-5) -- (7,-5);
\coordinate (lout2L) at (7,-4.6) {};

\draw[|-,very thick,red] (10,-5) -- (11,-5);
\coordinate (lmid2R) at (11,-4.6) {};
\draw[-|,very thick,blue] (11,-5) -- (12,-5);
\coordinate (lout2R) at (11.5,-4.6) {};

\draw[dotted,opacity=0.4,thick] (6,-5.5) -- (6,7.5);
%\draw[dotted,opacity=0.4,thick] (7,-5.5) -- (7,7.5);
\draw[dotted,opacity=0.4,thick] (8,-5.5) -- (8,7.5);
\draw[dotted,opacity=0.4,thick] (10,-5.5) -- (10,7.5);
%\draw[dotted,opacity=0.4,thick] (11,-5.5) -- (11,7.5);
\draw[dotted,opacity=0.4,thick] (12,-5.5) -- (12,7.5);

\draw[|-|,very thick,black] (13,-3.25) -- (18,-3.25);
\coordinate (lcompN3) at (15.5,-2.85) {};
\draw[|-|,very thick,black] (14,-4.5) -- (16,-4.5);
\coordinate (linN3) at (15,-4.1) {};

\draw[|-|,very thick,red] (13,-5) -- (14,-5);
\coordinate (lmid3L) at (13.2,-4.6) {};

\draw[|-,very thick,red] (16,-5) -- (17,-5);
\coordinate (lmid3R) at (17,-4.6) {};
\draw[-|,very thick,blue] (17,-5) -- (18,-5);
\coordinate (lout3R) at (17,-4.6) {};

\draw[dotted,opacity=0.4,thick] (12,-5.5) -- (12,7.5);
\draw[dotted,opacity=0.4,thick] (13,-5.5) -- (13,7.5);
\draw[dotted,opacity=0.4,thick] (14,-5.5) -- (14,7.5);
\draw[dotted,opacity=0.4,thick] (16,-5.5) -- (16,7.5);
\draw[dotted,opacity=0.4,thick] (18,-5.5) -- (18,7.5);

%v1
\draw[|-|,very thick] (0,0) -- (12,0);
\coordinate (lv1) at (6,-0.3) {};
\coordinate (I1) at (0.25,0.25) {};

\coordinate (I21) at (11.25,3.25) {};
\coordinate (I22) at (11.75,3.25) {};

%v2
\draw[|-|,very thick] (0.5,1) -- (11,1);
\coordinate (lv2) at (6,0.7) {};
\coordinate (I2) at (0.75,1.25) {};

%v3
\draw[|-|,very thick] (6,2) -- (18,2);
\coordinate (lv3) at (12,1.7) {};
\coordinate (I11) at (6.25,2.25) {};
\coordinate (I32) at (17.75,2.25) {};

%v4
\draw[|-|,very thick] (6.5,3) -- (17.5,3);
\coordinate (lv4) at (12,2.7) {};
\coordinate (I12) at (6.75,3.25) {};
\coordinate (I31) at (17.25,3.25) {};

%u1
\draw[|-|,thick] (1,2) -- (5,2);
\coordinate (lu1) at (2.5,1.7) {};
\coordinate (I10) at (4.75,2.25) {};

%u2
\draw[|-|,thick] (1,3) -- (4.5,3);
\coordinate (lu2) at (2.5,2.7) {};
\coordinate (I3) at (1.25,3.25) {};

%u3
\draw[|-|,thick] (1.5,4) -- (4.5,4);
\coordinate (lu3) at (2.5,3.7) {};
\coordinate (I9) at (4.25,4.25) {};

%u4
\draw[|-|,thick] (1.5,5) -- (4,5);
\coordinate (lu4) at (2.5,4.7) {};
\coordinate (I4) at (1.75,5.25) {};

%N1
\draw[thick,opacity=0.3,fill=red, red] (2,5.5) rectangle (4,7);
\coordinate (ln1) at (3,6.5) {};
\coordinate (I5) at (2.25,7.25) {};
\coordinate (I6) at (2.75,7.25) {};
\coordinate (I7) at (3.25,7.25) {};
\coordinate (I8) at (3.75,7.25) {};

%u5
\draw[|-|,thick] (7,4) -- (11,4);
\coordinate (lu5) at (9,3.7) {};
\coordinate (I13) at (7.25,4.25) {};

%u6
\draw[|-|,thick] (7.5,5) -- (11,5);
\coordinate (lu6) at (9,4.7) {};
\coordinate (I14) at (7.75,5.25) {};
\coordinate (I19) at (10.25,5.25) {};
\coordinate (I20) at (10.75,5.25) {};

%N2
\draw[thick,opacity=0.3,fill=red, red] (8,5.5) rectangle (10,7);
\coordinate (ln2) at (9,6.5) {};
\coordinate (I15) at (8.25,7.25) {};
\coordinate (I16) at (8.75,7.25) {};
\coordinate (I17) at (9.25,7.25) {};
\coordinate (I18) at (9.75,7.25) {};

%u7
\draw[|-|,thick] (13,0) -- (17,0);
\coordinate (lu7) at (15,-0.3) {};
%u8
\draw[|-|,thick] (13,1) -- (17,1);
\coordinate (lu8) at (15,0.7) {};

%u9
\draw[|-|,thick] (13,4) -- (17,4);
\coordinate (lu9) at (15,3.7) {};
\coordinate (I23) at (13.25,4.25) {};

%u10
\draw[|-|,thick] (13.5,5) -- (17,5);
\coordinate (lu10) at (15,4.7) {};
\coordinate (I24) at (13.75,5.25) {};
\coordinate (I29) at (16.25,5.25) {};
\coordinate (I30) at (16.75,5.25) {};

%N3
\draw[thick,opacity=0.3,fill=red, red] (14,5.5) rectangle (16,7);
\coordinate (ln3) at (15,6.5) {};
\coordinate (I25) at (14.25,7.25) {};
\coordinate (I26) at (14.75,7.25) {};
\coordinate (I27) at (15.25,7.25) {};
\coordinate (I28) at (15.75,7.25) {};

\tikzstyle{every node}=[inner sep=1pt]
\begin{scriptsize}
\node at (lv1) {$v_1$};
\node at (lv2) {$v_2$};
\node at (lv3) {$v_3$};
\node at (lv4) {$v_4$};

\node at (lu1) {$u_1$};
\node at (lu2) {$u_2$};
\node at (lu3) {$u_3$};
\node at (lu4) {$u_4$};
\node at (lu5) {$u_5$};
\node at (lu6) {$u_6$};
\node at (lu7) {$u_7$};
\node at (lu8) {$u_8$};
\node at (lu9) {$u_9$};
\node at (lu10) {$u_{10}$};

\node at (ln1) {$I(\mathbb{N}_1)$};
\node at (ln2) {$I(\mathbb{N}_2)$};
\node at (ln3) {$I(\mathbb{N}_3)$};

\node at (linQ) {$inner^{\phi}_{\circle}(\mathbb{Q})$};
\node at (linN1) {$inner^{\phi}_{\circle}(\mathbb{N}_1)$};
\node at (linN2) {$inner^{\phi}_{\circle}(\mathbb{N}_2)$};
\node at (linN3) {$inner^{\phi}_{\circle}(\mathbb{N}_3)$};

\node at (lcompN1) {$comp^{\phi}_{\circle}(\mathbb{N}_1)$};
\node at (lcompN2) {$comp^{\phi}_{\circle}(\mathbb{N}_2)$};
\node at (lcompN3) {$comp^{\phi}_{\circle}(\mathbb{N}_3)$};

\node at (lmid1L) {$midout^1_L$};
\node at (lmid1R) {$midout^1_R$};
%\node at (lout1L) {$midout^1_L$};
\node at (lmid2L) {$midout^2_L$};
\node at (lmid2R) {$midout^2_R$};
%\node at (lout2L) {$midout^2_L$};
%\node at (lout2R) {$midout^2_R$};
\node at (lmid3L) {$midout^3_L$};
\node at (lmid3R) {$midout^3_R$};
%\node at (lout3R) {$midout^3_R$};
\end{scriptsize}

\tikzstyle{every node}=[inner sep=1pt]
\begin{tiny}
%\node at ($(I1) + (0,0.45)$) {$I_1$};
%\node at ($(I2) + (0,0.45)$) {$I_2$};
%\node at ($(I3) + (0,0.45)$) {$I_3$};
\end{tiny}

\tikzstyle{every node}=[circle,minimum size=5pt,inner sep=0pt,draw,fill]
%\node at (I0) {};
\node[blue] at (I1) {};
\node[blue] at (I2) {};
\node[red] at (I3) {};
\node[red] at (I4) {};
\node at (I5) {};
\node at (I6) {};
\node at (I7) {};
\node at (I8) {};
\node[red] at (I9) {};
\node[red] at (I10) {};

\node[blue] at (I11) {};
\node[blue] at (I12) {};
\node[red] at (I13) {};
\node[red] at (I14) {};
\node at (I15) {};
\node at (I16) {};
\node at (I17) {};
\node at (I18) {};
\node[red] at (I19) {};
\node[red] at (I20) {};

\node[blue] at (I21) {};
\node[blue] at (I22) {};

\node[red] at (I23) {};
\node[red] at (I24) {};

\node at (I25) {};
\node at (I26) {};
\node at (I27) {};
\node at (I28) {};
\node[red] at (I29) {};
\node[red] at (I30) {};
\node[blue] at (I31) {};
\node[blue] at (I32) {};

%\draw[black] (-1,-6)--(-1,-3.5);
%\draw[black] (18,5.5)--(18,7.5);
\draw[white] (-1,-6)--(-1,-3.5);
\draw[white] (18,5.5)--(18,7.5);

\end{tikzpicture}
\end{subfigure}
\end{center}
\caption{The structure of $(inner^{\phi}_{\circle}(\mathbb{Q}),{\prec_\phi})$, where $\mathbb{Q}$ is a $Q$-node with children $\mathbb{N}_1,\mathbb{N}_2,\mathbb{N}_3$.
We have $I'(\mathbb{Q}) = \{v_1,v_2,v_3,v_4\}$, $U(\mathbb{N}_1) = \{u_1,u_2,u_3,u_4\}$, $U(\mathbb{N}_2) = \{u_5,u_6\}$, $U(\mathbb{N}_3) = \{u_7,u_8,u_9,u_{10}\}$.
The components from the sets $midddle(\mathbb{N}_i)$ and $outer(\mathbb{N}_i)$ are depicted in red and blue, respectively.
} 
\label{fig:Q-node-structure} 
\end{figure}

In the rest of this section we show a poly-time algorithm that tests whether the set of all good triples
for the set $\mathcal{I}' = comp(\mathbb{R})$ is non-empty (which is equivalent to testing whether $G$ has a $(C,H)$-centered model).
Roughly speaking, the algorithm traverses the tree $\mathbb{T}$ bottom-up and for every inner node $\mathbb{N}$ 
it computes the set of all good triples for the set $inner(\mathbb{N})$ and the borders $(\emptyset,\emptyset)$.
One can show that any good triple for $inner(\mathbb{N})$ follows the decomposition scheme given by Lemma~\ref{lem:P-node-model-structure} and 
Lemma~\ref{lem:Q-node-model-structure}, depending on whether $\mathbb{N}$ is $P$ or a $Q$-node.
However, there is a certain problem with this approach as the set of all good triples for the set $inner(\mathbb{N})$ might have an exponential size in the case
$\mathbb{N}$ is an inner node of $\mathbb{T}$.
Therefore, for such nodes $\mathbb{N}$ the algorithm computes the set $vbt_{inner}(\mathbb{N})$ of \emph{valid boundary tuples} for the set $inner(\mathbb{N})$, 
which keeps `fingerprints' of all good triples for the set $inner(\mathbb{N})$.
The set $vbt_{inner}(\mathbb{N})$ has a polynomial size and satisfies the following properties:
\begin{itemize}
\item it allows to compute the sets $vbt_{inner}(\mathbb{N})$ in a bottom-up order along the tree $\mathbb{T}$,
\item it allows to test whether the set of good triples for the set $comp(\mathbb{R})$ is non-empty (and hence to test whether $G$ admits a $(C,H)$-centered model).
\end{itemize}
Formally, if $\mathbb{N}$ is an inner node and $(\mathcal{J}, \mathcal{J'},{\prec})$ is a good triple for the set $inner(\mathbb{N})$, 
where $(\mathcal{J},{\prec})$ has the form $(I_F,I_{SF},\ldots,I_{SL},I_L)$ 
(that is, $I_F,I_{SF},I_{SL},I_L$ denote, respectively, the first, the second, the last but one, and the last component in the linear order $(\mathcal{J},{\prec})$, respectively),
$(\mathcal{J}, \mathcal{J'},{\prec})$ is represented in the set $vbt_{inner}(\mathbb{N})$ 
by its \emph{fingerprint} $(I_F, C_R, I_L, C_L, MinDel)$, where:
$$
\begin{array}{rcl}
 \smallskip
    C_R &=& \left\{
        \begin{array}{cl}
        C(I_F) \cap C(I_{SF})& \text{if $inner(\mathbb{L}) \cap \mathcal{J} = \{I_F\}$ \text{for some }$\mathbb{L} \in \mathcal{L}(\mathbb{N})$}, \\
        null& \text{otherwise,}
        \end{array}
    \right. \\
    \smallskip
    C_L &=& \left\{
        \begin{array}{cl}
        C(I_L) \cap C(I_{SL})& \text{if $inner(\mathbb{L}) \cap \mathcal{J} = \{I_L\}$ \text{for some }$\mathbb{L} \in \mathcal{L}(\mathbb{N})$}, \\
        null& \text{otherwise,}
        \end{array}
    \right.
    \\
    \smallskip
    MinDel &=& \left\{
        \begin{array}{cl}
        min(\mathcal{J}',{\sqsubseteq})& \text{if $\mathcal{J'} \neq \emptyset$}, \\
        null& \text{otherwise.}
        \end{array}
    \right.
  \end{array}
$$
In particular, if $\phi$ is a $(C,H)$-centered model of $G$,
good triple $\big{(}inner^{\phi}_{\circle}(\mathbb{N}), inner^{\phi}_{\stick}(\mathbb{N}), {\prec_{\phi}}\big{)}$ 
for the set $inner(\mathbb{N})$ is represented in the set $vbt_{inner}(\mathbb{N})$ by
$\big{(}I_F, C_R, I_L, C_L, minDel\big{)}$, where:
\begin{itemize}
 \item $I_F$ and $I_L$ are, respectively, the first and the last component in $(inner^{\phi}_{\circle}(\mathbb{N}),{\prec_{\phi}})$,
 \item if $I_F$ is contained in a one-component maximal sector $S$ of $\phi^*$, then $C_R$ is the clique of the sector of $\phi^*$ adjacent to $S$ from the right; otherwise $C_R = null$,
 \item if $I_L$ is contained in a one-component maximal sector $S$ of $\phi^*$, then $C_L$ is the clique of the sector of $\phi^*$ adjacent to $S$ from the left; otherwise $C_L = null$,
 \item $minDel$ is the minimal component in $\big{(}inner^{\phi}_{\stick},{\sqsubseteq}\big{)}$ 
 if $inner^{\phi}_{\stick}(\mathbb{N}) \neq \emptyset$; otherwise $minDel= null$.
\end{itemize}

Suppose $\mathbb{P}$ is a $P$-node.
One can show (using the same arguments as in the proof of Lemma \ref{lem:P-node-model-structure}) 
that $(\mathcal{J},\mathcal{J}',{\prec})$ is a good triple for the set $inner(\mathbb{P})$ if and only if there is an admissible ordering $\mathbb{N}_1,\ldots,\mathbb{N}_k$ of the children of $\mathbb{N}$ and a sequence 
$(\mathcal{K}_1,\mathcal{K}'_1,{\prec}), \ldots, (\mathcal{K}_k,\mathcal{K}'_k,{\prec})$ of good triples for the sets 
$comp(\mathbb{N}_1), \ldots,comp(\mathbb{N}_k)$ and the borders
$(B^1_L,B^1_R),\ldots,(B^k_L,B^k_R)$, respectively, such that:
\begin{itemize}
\item $\mathcal{J} = \bigcup_{i=1}^k \mathcal{K}_i$ and $\mathcal{K}_1 \prec \dots \prec \mathcal{K}_k$,
\item if $\mathcal{J}' \neq \emptyset$ then there exists a unique $d \in [k]$ such that $\mathcal{J}' = \mathcal{K}'_d$,
\item $(B^1_L,B^1_R),\ldots,(B^k_L,B^k_R)$ are as defined in Lemma~\ref{lem:P-node-model-structure}.
\end{itemize}

Let $\mathbb{N}$ be a child of $\mathbb{P}$ and let $i \in [k]$. 
Also, $(\mathcal{K},\mathcal{K}',{\prec})$ is a good triple for the set $comp(\mathbb{N})$ and the borders $(B^i_L,B^i_R)$ if and only if
there is a 
three-chain-partition $(mid_L, mid_R, mid_D)$ of the set $middle(\mathbb{N})$ and a good triple $(\mathcal{L},\mathcal{L}',{\prec})$ for the set $inner(\mathbb{N})$ such that:
\begin{itemize}
\item $\mathcal{K} = \mathcal{L} \cup mid_L \cup mid_R$, $mid_L \prec \mathcal{L} \prec mid_R$, $(mid_L,{\prec})$ equals to $(mid_L,{\sqsubset})$, and 
$(mid_R,{\prec})$ equals to the reverse of $(mid_R,{\sqsubset})$,
\item the chains $(mid_L,{\sqsubseteq})$ and $(mid_R,{\sqsubseteq})$ respect the borders $B^i_L$ and $B^i_R$,
\item $(\mathcal{L},{\prec})$ respects the borders $(C^i_L,C^i_R)$, where
$$
C^i_L = \left\{
\begin{array}{cl}
    max(mid_L,{\sqsubseteq}) & \text{if $mid_L \neq \emptyset$,} \\
    B^i_L & \text{otherwise,}\\
\end{array}
\right.
$$
and 
$$
C^i_R = \left\{
\begin{array}{cl}
    max(mid_R,{\sqsubseteq}) & \text{if $mid_R \neq \emptyset$,} \\
    B^i_R & \text{otherwise,}\\
\end{array}
\right.
$$
\item $\mathcal{K}' = mid_D \cup \mathcal{L}'$ and $mid_D \sqsubset \mathcal{L}'$.
\end{itemize}

If the above conditions hold, we say $(\mathcal{L}, \mathcal{L}', {\prec})$  \emph{fits} $(mid_L,mid_R,mid_D)$ at the position~$i$.

Now, we show how to compute the set $vbl_{inner}(\mathbb{P})$ given that the sets $vbl_{inner}(\mathbb{N})$ for every child $\mathbb{N}$ of $\mathbb{P}$ are already computed.

First, for every child $\mathbb{N}$ of $\mathbb{P}$ we compute the set $vbt_{middle}(\mathbb{N})$ of \emph{valid boundary tuples} for the set $middle(\mathbb{N})$, 
which keeps `fingerprints' of all three-chain-partitions of the set $middle(\mathbb{N})$, where
a \emph{fingerprint} of a three-chain-partition $(mid_L,mid_R,mid_D)$ is defined as a tuple 
$(Fmid_L,Lmid_L,Fmid_R,Lmid_R, Fmid_D, Lmid_D)$, where:
\begin{itemize}
 \item $Fmid_L,Lmid_L$ is the minimal and the maximal element in $(mid_L, {\sqsubseteq})$, 
 \item $Fmid_R,Lmid_R$ is the minimal and the maximal element in $(mid_R, {\sqsubseteq})$,  
 \item $Fmid_D,Lmid_D$ is the minimal and the maximal element in $(mid_D, {\sqsubseteq})$.
\end{itemize}
We assume $Fmid_L = Lmid_L = null$ if $mid_L = \emptyset$ and analogously for $Fmid_R, Lmid_R$ and $Fmid_D,Lmid_D$. 
So, the fingerprint of $(mid_L,mid_R,mid_D)$ consists of \emph{the limits} of the chains $(mid_L,{\sqsubseteq})$,
$(mid_R,{\sqsubseteq})$, and $(mid_D, {\sqsubseteq})$.

We compute the set $vbl_{middle}(\mathbb{N})$ using a poly-time algorithm solving $3$-list chain partitioning problem 
in the poset $(middle(\mathbb{N}) \cup \mathcal{D},{\sqsubseteq})$.
For this purpose, we iterate over all tuples $(Fmid_L,Lmid_L, 
Fmid_R,Lmid_R, Fmid_D, Lmid_D)$ of elements in the set $middle({\mathbb{N}})$ and we check whether there is a three-chain-partition $(mid_L,mid_R,mid_D)$ of
the set $middle({\mathbb{N}})$ with the limits given by $(Fmid_L,Lmid_L, 
Fmid_R,Lmid_R, Fmid_D, Lmid_D)$.
To this end, we first assign to every element $I$ in the poset $(middle(\mathbb{N})\cup \mathcal{D},{\sqsubseteq})$ the list of available chains $L(I)=\{L,R,D\}$
and then:
\begin{itemize}
 \item we set $L(Fmid_L)=L(Lmid_L)=\{L\}$ and we delete color $L$ from $L(I)$ for any element $I$ in $US(Lmid_L) \cup DS(Fmid_L)$,
 \item we set $L(Fmid_R)=L(Lmid_R)=\{R\}$ and we delete color $R$ from $L(I)$ for any element $I$ in $US(Lmid_R) \cup DS(Fmid_R)$,
 \item we set $L(Fmid_D)=L(Lmid_D)=\{D\}$ and we delete color $D$ from $L(I)$ for any element $I$ in $US(Lmid_D) \cup DS(Fmid_D)$,
 \item we set $L(I)=\{D\}$ for any element $I$ in $\mathcal{D}$.
\end{itemize}
One can check that there is a three-chain-partition $(mid_L,mid_R,mid_D)$ of $(middle(\mathbb{N}),{\sqsubseteq})$ with the limits given by $(Fmid_L,Lmid_L, 
Fmid_R,Lmid_R, Fmid_D, Lmid_D)$ if and only if the instance of $3$-list chain partitioning problem in the poset $(middle(\mathbb{N}) \cup \mathcal{D},{\sqsubseteq})$ 
constructed this way is satisfiable.    

Let $i \in [k]$ and let $\mathbb{N}$ be a child of $\mathbb{P}$. 
Note that we can decide whether a good triple $(\mathcal{L},\mathcal{L},{\prec})$ for the set $inner(\mathbb{N})$ and a three-chain-partition
$(mid_L,mid_R,mid_D)$ fit at the position $i$ based only on the fingerprint of $(\mathcal{L},\mathcal{L},{\prec})$ in the set $vbl_{inner}(\mathbb{N})$ 
and the fingerprint of $(mid_L,mid_R,mid_D)$ in the set $vbl_{middle}(\mathbb{N})$.
Let ${R}(\mathbb{N})$ contains all these fitting pairs of fingerprints:
In particular, $R(\mathbb{N})$ is a polynomial-size representation of all good triples for the set $comp(\mathbb{N})$ and the borders $(B^i_L,B^i_R)$.
Clearly, the set $R(\mathbb{N})$ can be computed in polynomial time.

Eventually, note that the fingerprint of a good triple $(\mathcal{J},\mathcal{J}',{\prec})$ for the set $inner(\mathbb{P})$, where 
$(\mathcal{J},\mathcal{J}',{\prec})$ is composed from  $(\mathcal{K}_1,\mathcal{K}'_1,{\prec}),\ldots,(\mathcal{K}_k,\mathcal{K}'_k,{\prec}) $ as described above, 
depends only on good triples $(\mathcal{K}_1, \mathcal{K}'_1, {\prec})$, $(\mathcal{K}_k, \mathcal{K}'_k, {\prec})$, and $(\mathcal{K}_d, \mathcal{K}'_d, {\prec})$.
So, to compute the set $vbl_{inner}(\mathbb{P})$ we iterate over all 
triples $(\mathbb{N}_F, \mathbb{N}_L, \mathbb{N}_d)$, where $\mathbb{N}_F, \mathbb{N}_L, \mathbb{N}_d$ are children of $\mathbb{P}$ such that $\mathbb{N}_F \neq \mathbb{N}_L$, 
and for any such triple $(\mathbb{N}_F, \mathbb{N}_L, \mathbb{N}_d)$ we compute:
\begin{itemize}
\item the set $R(\mathbb{N}_F)$ representing all good triples for the set $comp(\mathbb{N}_F)$ and the borders $(B^1_L,B^1_R)$.
\item the set $R(\mathbb{N}_L)$ representing all good triples for the set $comp(\mathbb{N}_L)$ and the borders $(B^k_L,B^k_R)$,
\item for every child $\mathbb{N}$ different than $\mathbb{N}_F$ and $\mathbb{N}_L$ the set $R(\mathbb{N})$ representing all good triples for the set 
$comp(\mathbb{N})$ and the borders $(B^i_L,B^i_R)$ for some $i \in [2,k-1]$ (note that the borders $(B^i_L,B^i_R)$ are the same for every $i \in [2,k-2]$).
\end{itemize}
Then, for every child $\mathbb{N}$ of $\mathbb{P}$ different than $\mathbb{N}_d$ we restrict the set $R(\mathbb{N})$ to the pairs that 
represent good triples $(\mathcal{K},\mathcal{K}',{\prec})$ for $comp(\mathbb{N})$ in which we have $\mathcal{K}' = \emptyset$.
Clearly, based on the sets $R(\mathbb{N})$ computed for every child $\mathbb{N}$ of $\mathbb{P}$, 
we can easily compute fingerprints of all good triples $(\mathcal{J},\mathcal{J}',{\prec})$ 
for the set $inner(\mathbb{P})$ such that $\mathcal{J}'$ is contained in $comp(\mathbb{N}_d)$ and 
$(\mathcal{J},{\prec})$ induces an order of the children of $\mathbb{P}$ 
with $\mathbb{N}_F$ and $N_L$ at the first and the last position, respectively.

Let $\mathbb{Q}$ be a $Q$-node in $\mathbb{T}$ and 
let $\mathbb{N}, \ldots, \mathbb{N}_k$ be an admissible order of the children of~$\mathbb{N}$.
Based on Lemma~\ref{lem:Q-node-model-structure} we can show 
that any good triple $(\mathcal{J},\mathcal{J}',{\prec})$ inducing the order $\mathbb{N}_1,\ldots,\mathbb{N}_k$ of the children $inner(\mathbb{Q})$ 
admits the same description as in the case of $P$-node, with three exceptions:
\begin{itemize}
 \item the borders $(B^1_L,B^1_R), \ldots, (B^k_L,B^k_R)$ are as defined in Lemma \ref{lem:Q-node-model-structure},
 \item a good triple $(\mathcal{K}_i,\mathcal{K}'_i,{\prec})$ for the set $comp(\mathbb{N}_i)$ and the borders $(B^i_L,B^i_R)$ 
 is composed from a three-chain-partition $(midout_L,midout_R,midout_D)$ of the set $middle(\mathbb{N}_i) \cup outer(\mathbb{N}_i)$ 
 and a good triple $(\mathcal{L}_i,\mathcal{L}'_i,{\prec})$ for the set $inner(\mathbb{N}_i)$ ,
 \item the trace $(d_1,\ldots,d_z)$ of the chain $\mathcal{J}'$ in the poset $(inner(\mathbb{Q}) \cap \mathcal{I}_a,{\sqsubseteq})$ 
 forms a zigzag consistent with the ${\sqsubseteq_{Q}}$-relation.
 \end{itemize}
Exactly the same as in the previous case, for every node $\mathbb{N}_i$ we compute the set $vbl_{midout}(\mathbb{N}_i)$ 
representing all three-chain-partitions of the set $middle(\mathbb{N}_i) \cup outer(\mathbb{N}_i)$ 
and the set $R(\mathbb{N}_i)$ representing all good triples for the set $comp(\mathbb{N}_i)$ and the borders $(B^i_L,B^i_R)$.

Now, we want to compute the fingerprints of good triples $(\mathcal{J},\mathcal{J}',{\prec})$ for the set $inner(\mathbb{Q})$ which induce the order $\mathbb{N}_1,\ldots, \mathbb{N}_k$ 
of the children of $\mathbb{Q}$.
An index $i \in [2,k-1]$ is \emph{safe} if there is a good triple 
$(\mathcal{K}_i,\mathcal{K}'_i,{\prec})$ for the set $comp(\mathbb{N}_i)$ and the borders $(B^i_{L}, B^i_R)$ such that $\mathcal{K}_i = comp(\mathbb{N}_i)$ and $\mathcal{K}'_i = \emptyset$.
We can easily decide whether index $i$ is safe based on the set $R(\mathbb{N}_i)$.
Clearly, if $i \in [2,k-1]$ is not safe, we must have $\mathcal{J}' \cap comp(\mathbb{N}_i) \neq \emptyset$ for any good triple $(\mathcal{J},\mathcal{J}',{\prec})$ for the set $inner(\mathbb{Q})$.
In particular, we may arrange all non-safe indices from $[2,k-1]$ into a ${\sqsubseteq}_Q$-consistent zigzag sequence $(e_1,\ldots,e_z)$;
otherwise the set of all good triples $(\mathcal{J},\mathcal{J}',{\prec}) $ for $inner(\mathbb{Q})$ consistent with $\mathbb{N}_1,\ldots,\mathbb{N}_k$ is empty.
Now, let 
$$
\begin{array}{lcl}
\smallskip
Zig  &=& \big{\{}(e_1,\ldots,e_z)\big{\}} \; \cup \\
& & \big{\{}(e_0,e_1,\ldots,e_z): e_0 \in [k] \text{ and } (e_0,e_1,\ldots,e_z) \text{ is a $\sqsubseteq_Q$-consitent zigzag}\big{\}}.
\end{array}
$$
Note that the fingerprints of all good triples $(\mathcal{J},\mathcal{J}',{\prec})$ for $inner(\mathbb{N})$ such that 
$(\mathcal{J},\mathcal{J}',{\prec})$ is consistent with $\mathbb{N}_1,\ldots,\mathbb{N}_k$ and the trace of $\mathcal{J'}$ is in the set $Zig$ 
represent all good triples for the set $inner(\mathbb{Q})$ consistent with $\mathbb{N}_1,\ldots,\mathbb{N}_k$.
This follows by the following exchanging property: given a good triple $(\mathcal{J},\mathcal{J}',{\prec})$ for the set $inner(\mathbb{Q})$ 
composed from $(\mathcal{K}_1,\mathcal{K}'_1,{\prec}),\ldots, (\mathcal{K}_k,\mathcal{K}'_k,{\prec})$, we can replace, 
without affecting the fingerprint of $(\mathcal{J},\mathcal{J}',{\prec})$, 
a good triple $(\mathcal{K}_i,\mathcal{K}'_i,{\prec})$ for the set $comp(\mathbb{N}_i)$ by a good triple $(comp(\mathbb{N}_i),\emptyset,{\prec})$ 
whenever $i$ a safe index which is not the first in the trace of $\mathcal{J}'$.
Performing this exchanging procedure for every safe index we obtain a good triple with the same fingerprint and the trace in the set $Zig$.  
Let $(d_1,\ldots,d_z)$ be a member of $Zig$.
Note that the representation of any good triple $(\mathcal{J},\mathcal{J'},{\prec})$, 
where $(\mathcal{J},\mathcal{J'},{\prec})$ is composed from $(\mathcal{K}_1,\mathcal{K}'_1,{\prec}),\ldots,(\mathcal{K}_k,\mathcal{K}'_k,{\prec})$ 
and the trace of $\mathcal{J}'$ equal to $(d_1,\ldots,d_z)$, 
depends only on $(\mathcal{K}_1,\mathcal{K}'_1,{\prec})$, $(\mathcal{K}_k,\mathcal{K}'_k,{\prec})$, and $(\mathcal{K}_{d_1},\mathcal{K}'_{d_1},{\prec})$.
In particular, it does not depend on $(\mathcal{K}_{d_i},\mathcal{K}'_{d_i},{\prec})$ for every $i \in [2,z]$.
Now, for every $i \in [1,z-1]$ we filter the set $R(\mathbb{N}_{d_i})$ such that it contains only the pairs representing 
good triples $(\mathcal{K}_{d_i},\mathcal{K}'_{d_i},{\prec})$ such that $\mathcal{K}'_{d_i} \subseteq outer(\mathbb{N}_{d_i})$.
Now, the pair from $R(\mathbb{N}_{d_1})$ representing $(\mathcal{K}_{d_1},\mathcal{K}'_{d_1},{\prec})$ is marked as \emph{good} 
if for every $i \in [2,z]$ there exist a pair in $R(\mathbb{N}_{d_i})$ representing $(\mathcal{K}_{d_i},\mathcal{K}'_{d_i},{\prec})$ such that
$\mathcal{K}'_{d_1} \sqsubset \mathcal{K}'_{d_2} \sqsubset \dots \sqsubset \mathcal{K}'_{d_z}$.
Now, note that the fingerprints of the good triples $(\mathcal{J},{\mathcal{J}'},{\prec})$ with the trace 
$(d_1,\ldots,d_z)$ can be computed from the sets $R(\mathbb{N}_1)$, $R(\mathbb{N}_k)$, and the set of all good entries in $R(\mathbb{N}_{d_1})$.

Finally, we can test whether there is a good triple for the set $comp(\mathbb{R})$ based on the fingerprints of all
good triples for the set $inner(\mathbb{R})$ and the fingerprints of all three-chain-partitions of the set $middle(\mathbb{R})$.

% !TeX spellcheck = en_US
 
\newcommand{\subH}{\overline{H}}

\newcommand{\eightgraph}{butterfly}
\newcommand{\Eightgraph}{Butterfly} 
\newcommand{\Hsub}{H_{\operatorname{sub}}}

\section{NP-Hardness for Butterfly-Graph Recognition}
\label{sec:butterfly-NP-hard}

This section derives the following hardness result.

\begin{theorem}[Theorem \ref{thm:recog_two_cycles_hard} restated]
\label{thm:NPC}
For every fixed graph $H$ containing two different cycles, recognition of $H$-graphs is $\NP$-complete. 
\end{theorem}

We begin by showing hardness for when $H$ is a \eightgraph.
A \eightgraph\ is the graph consisting of two \( K_3 \) joined on one node (see -- Figure~\ref{fig:minor1}).
%This section shows the following hardness result.
%\begin{theorem}
%\label{th:NPC}
%Let $H$ be a connected graph containing at least two distinct cycles (i.e. some subdivision of $H$ contains the butterfly as a minor). Then \textsc{$H$-Graph Recognition} is \NP-complete.
%\end{theorem}

%\tkcomment{Can we just write that in this section we are proving Theorem \ref{thm:recog_two_cycles_hard} -- see Section Introduction.} 
%\thcomment{Changed it accordingly}

%First, we focus on the case when $H$ is the butterfly.

\begin{theorem}
\label{thm:8:graph}
	\textsc{\Eightgraph-Graph Recognition} is \( \NP \)-complete.
%	, when $H$ consists of two circles joined at a point.
\end{theorem}

It is easy to see \NP-membership \cite{CTVZ21}.
To show \NP-hardness, we reduce from the \textsc{Bipartite 2-Track};
	that is to decide whether a given bipartite graph $G$ is 2-track.
A graph \( G \) is \emph{2-track} if
	there are $E_1,E_2$ whose union is $E(G)$
	such that $(V(G),E_1)$ and $V(G),E_2)$ are interval graphs.
Gon{\c{c}}alves \& Ochem proved its \NP-hardness \cite{Goncalves2009}.
	
\textit{Construction:}
For a given bipartite graph \( G \) we construct a graph \( G' \) that is a \eightgraph-graph if and only if \( G \) is 2-track.
Let $S$ be a star $K_{1,4}$ where every edge is subdivided once.
The vertex set $V(G')$ consists of $V(G)$, an \emph{edge-vertex} $w_{uv}$ for every edge $uv \in E(G)$ and $V(S)$.
The edge set $E(G')$ consists of $\binom{V(G)}{2}$, $E(S)$
and the edges of making $w_{uv}$ adjacent to $w \in V(G) \setminus \{u,v\}$ for every edge $uv \in E(G)$.
Finally add every edge between $V(G)$ and $V(S)$.

Given \( G \) the graph \( G' \) can be constructed in polynomial time.
It remains to show that \( G \) is 2-track if and only if \( G' \) is a \eightgraph-graph.

\begin{lemma}
Graph \( G \) is 2-track if and only if \( G' \) is a \eightgraph-graph.
\end{lemma}
\begin{proof}
	$( \Rightarrow )$
Let \( G \) be a bipartite 2-track graph.
Then there is a partition of $E(G)$ into $E_1, E_2$
	such that \( (V(G),E_1) \) and \( (V(G),E_2) \) are interval graphs.
In other words, $(V(G),E_i)$ for $i\in\{1,2\}$ is a $K_2$-graph.
Hence there is a path $P_i$ and a model $\phi_i$ mapping from $V(G)$ to a subset of $V(P_i)$ that forms a subpath,
	such that $\phi_i(u) \cap \phi_i(v) \neq \emptyset$ if and only if $uv \in E_i$.

We show that $G$ is a \eightgraph-graph by extending the models $\phi_1$ and $\phi_2$ to a model $\phi$ for a subdivision of a butterfly.
We introduce graph $X$, a star $K_{1,4}$ where each edge is subdivided once.
	Connect the endpoints of \( P_1 \) and \( P_2 \) to distinct leaves of \( X \).
	The resulting graph \( H \) is a subdivision of a \eightgraph.
For every vertex $v \in V(G)$, we define \( \phi(v) \) as the negative of the 2-track representation,
	formally $\phi(v) = V(H) \setminus ( \phi_1(v) \cup \phi_2(v) )$.
Note that \( \phi(v) \) is connected.
Further let $\phi$ represent $V(S)$ by mapping to $V(X)$.
Note that every vertex $s \in V(S)$ is adjacent to \( V(G) \).

For every edge \( uv \in E(G) \),
	there is a track \( i \in \{1,2\} \) where the old models \( \phi_i(u), \phi_i(v) \) intersect.
We set \( \phi({w_{uv}}) = \phi_i(u) \cap \phi_i(v) \),
	which then by definition is disjoint from $\phi(u)$ and $\phi(v)$
	and induces a connected subgraph in $H$.
Since $G$ is bipartite, $\phi_i(u) \cap \phi_i(v) = \phi({w_{uv}})$ does not intersect any other old model $\phi_i(x)$ for $x \notin \{u,v\}$.
Hence $\phi({w_{uv}})$ intersects $\phi(x)$ for $x \in V(G) \setminus \{u,v\}$, as required.	
To conclude, it remains to show that $\phi({w_{uv}})$ and $\phi({w_{u'v'}})$ do not intersect for distinct edges $uv , u'v' \in E(G)$.
Assuming otherwise, $\{u,v,u',v'\}$ form a clique in $G$,
	in contradiction to that $G$ is bipartite.
	
	$( \Leftarrow )$
Let \( G' \) be a \eightgraph-graph.
Let $(H^\phi, \phi)$ be a model of $G'$, where \( H^\phi \) is a subdivision of \eightgraph\.
We show that the original graph \( G \) is 2-track.
In other words, we show that
	for $i\in\{1,2\}$
	there is a path $P_i$ and a model $\phi_i$ that maps from $V(G)$ to subsets of $V(P_i)$ that form a connected subgraph,
	in such a way that
	$u,v \in V(G)$ are adjacent
		if and only if
		$\phi_1(u) \cap \phi_1(v) \neq \emptyset$ or $\phi_2(u) \cap \phi_2(v) \neq \emptyset$.
%That is, there are paths \( P_1, P_2 \) and a collection of subpaths \( M^1_v \subseteq V(P_1) \), \( M^2_v \subseteq V(P_2) \) for every vertex \( v \in V(G) \),
%	such that \( \{u,v\} \subseteq G(V) \) are adjacent if and only if \( M_{u}^1 \cap M_{v}^1 \neq \emptyset \) or  \( M_{u}^2 \cap M_{v}^2 \neq \emptyset \).
	
	Let \( x \in H^{\phi} \) be the unique node of degree \( 4 \), and let \( X := N_{H^{\phi}}[N_{H^{\phi}}[x]] \).
	Observe that the unique vertex $s \in V(S)$ of degree \( 4 \) has a representation $\phi(s)$ containing \( x \). % (compare to Chaplick et al.~\cite{DBLP:conf/wg/Chaplick0VZ17}).
Since $V(S)$ is adjacent to all $V(G)$, we may assume that \( s' \in V(S) \) is represented by \( \phi(s') \subseteq X \),
	and that \( X \subseteq \phi(v) \) for every vertex \( v \in V(G) \).
Let \( P_1, P_2 \) be the two paths in \( H^{\phi} - X \).
	
Note that the edge-vertices form an independent set, and each of them is represented by a subpath of either $P_1$ or of $P_2$.
For every vertex $v \in V(G)$,
	let $E^1(v)$ be the set of edge vertices with a representation in $P_1$ that do not intersect $\phi(v)$:
%Let $E^1$ be set the edge-vertices with a model in $P_1$ not intersecting $\phi(x)$.
Then $E^1(v)$ consists of a consecutive sequence of edge-vertices from $P_1$.
We define \( \phi_1(v) \subseteq P_1 \) as the minimum size subpath
	such that every edge-vertex $w_{uv} \in E^1(v)$ has $\phi({w_{uv}}) \subseteq \phi_1(v)$.
Analogously define the representation \( \phi_2(v) \) for every $v \in V(G)$ as a subpath in \( P_2 \).
It remains to show for distinct $u,v \in V(G)$, $u,v \in E(G)$,
	if and only if $\phi_1(u) \cap \phi_1(v) \neq \emptyset$ or $\phi_2(u) \cap \phi_2(v) \neq \emptyset$.
%We observe that adjacent $u,v \in V(G)$ have intersecting representation in $\phi$.
Then it follows that $\phi_1, \phi_2$ form a 2-track representation of $G$.

For the forward-direction let \( uv \in E(G) \).
Then there is one `track' \( i \in \{1,2\} \) such that \( \phi({w_{\{u,v\}}}) \subseteq P_i \).
Edge-vertex \( \phi(w_{\{u,v\}}) \) does not intersect \( \phi({u}) \) nor intersects \( \phi({v}) \).
By definition \( \phi(w_{\{u,v\}}) \subseteq \phi_i(u) \) and  \( \phi(w_{\{u,v\}}) \subseteq \phi_i(v) \).
Thus \(\phi_i(u) \cap \phi_i(v) \neq \emptyset \).

For the backward-direction, we show that if \(\phi_i({u}) \cap \phi_i({v}) \neq \emptyset \) for one track \( i \in \{1,2\} \),
	then \( uv \in E(G) \).
Let $i$ be such that $\phi_i({u}) \cap \phi_i({v}) \neq \emptyset$.
If the intersection \(\phi_i({u}) \cap \phi_i({v}) \) does not contain an edge-vertex, then \( \phi_i({u}) \) and \( \phi_i({v}) \) do not form a minimal size subpath.
Thus least one edge-vertex \( w_{x,y} \) is such that \( \phi({w_{x,y}}) \subseteq \phi_i({u}) \) and \( \phi({w_{x,y}}) \subseteq \phi_i({v}) \).
Representation $\phi(u)$ only contains $\phi(w_{x, y})$ for which $u=x$ or $u=y$, and, similarly, $M_v$ only contains $\phi(w_{x, y})$ for which $v=x$ or $v=y$.
Therefore we must have $\{x,y\} = \{u,v\}$.
Hence there is an edge-vertex $w_{u,v}$ implying $uv \in E(G)$.
We conclude that $\phi_1, \phi_2$ from a 2-track representation of $G$.
%
%\ \\
%
%Indeed, the model $\phi({w_{u,v}}) \subseteq P_i$ of their edge-vertex, for some $i \in \{1,2\}$,
%	implies that $\phi(u),\phi(v)$ intersect in the set of edges $E_i$. 
%\tkcomment{Here $\phi(u)$ and $\phi(v)$ always intersect. I think you want to use something different that $\phi(v)$ and $\phi(u)$}
%	\thcomment{The statement is that they intersect in $E_i$ (not in $E$). I hope it is more clear now with ``in the set of edges''}
%Vice versa, when distinct $u,v \in {V(G)}$ have intersecting representations $\phi(u),\phi(v)$, they are adjacent.
%Indeed, if $\phi(u),\phi(v)$ have some intersection in $P_i$, for $i \in \{1,2\}$,
%	at least one edge-vertex has a representation $\phi({w_{u',v'}})$ contained in $\phi(u) \cap \phi(v)$.
%Representation $\phi(u)$ only contains $\phi(w_{x, y})$ for which $u=x$ or $u=y$, and, similarly, $M_v$ only contains $\phi(w_{x, y})$ for which $v=x$ or $v=y$.
%Therefore we must have $\{u',v'\} = \{u,v\}$.
%Hence there is an edge-vertex $w_{u,v}$ implying $uv \in E(G)$.
%We conclude that $M_u^i$ for $u \in V(G)$ and $i \in \{1,2\}$ is a 2-track representation of $G$.
\end{proof}

%\tkcomment{Is this part necessary? Can not we reference here to the paper by Chaplick?}

Now we show how to generalize the above to the case when $H$ contains two distinct cycles (i.e. some subdivision of $H$ contains the butterfly as a minor).

First, observe that if $H$ contains two distinct cycles, then $H$ contains a subdivision of one of the graphs depicted on Figure~\ref{fig:minors}.

\begin{figure}[h]
\centering
	\begin{subfigure}[t]{0.32\linewidth}
		\centering
		\begin{tikzpicture}[scale=1]
			
			\node at (-1, 1) [circle,draw, fill=black, opacity=1, color=black, inner sep=0.5mm] (a) {};
			
			\node at (-1.5,1) (u) {$u$};
			
			\node at (1, 1) [circle,draw, fill=black, opacity=1, color=black, inner sep=0.5mm] (b) {};
			
			\node at (0,0) [circle,draw, fill=black, opacity=1, color=black, inner sep=0.5mm] (c) {};
			
			\node at (-0.5,0) (x) {$x$};
			
			\node at (-1,-1) [circle,draw, fill=black, opacity=1, color=black, inner sep=0.5mm] (d) {};
			
			\node at (-1.5,-1) (v) {$v$};
			
			\node at (1,-1) [circle,draw, fill=black, opacity=1, color=black, inner sep=0.5mm] (e) {};

			\draw (a) -- (c) -- (b) -- (a) node[midway, above]{};
			\draw (c) -- (e) -- (d) -- (c) node[midway, above]{};
			
		\end{tikzpicture}
		\caption{Graph $H_1$}
		\label{fig:minor1}
	\end{subfigure}
	~
	\begin{subfigure}[t]{0.32\linewidth}
		\centering
		\begin{tikzpicture}[scale=0.75]
			
			\node at (-1, 1) [circle,draw, fill=black, opacity=1, color=black, inner sep=0.5mm] (a) {};
			
			\node at (-1.5,1) (u) {$u$};
			
			\node at (1, 1) [circle,draw, fill=black, opacity=1, color=black, inner sep=0.5mm] (b) {};
			
			\node at (0,0) [circle,draw, fill=black, opacity=1, color=black, inner sep=0.5mm] (c) {};
			
			\node at (-0.5,0) (x) {$x$};
			
			\node at (0,-1) [circle,draw, fill=black, opacity=1, color=black, inner sep=0.5mm] (f) {};
			
			\node at (-0.5,-1) (y) {$y$};
			
			\node at (-1,-2) [circle,draw, fill=black, opacity=1, color=black, inner sep=0.5mm] (d) {};
			
			\node at (-1.5,-2) (v) {$v$};
			
			\node at (1,-2) [circle,draw, fill=black, opacity=1, color=black, inner sep=0.5mm] (e) {};

			\draw (a) -- (c) -- (b) -- (a) node[midway, above]{};
			\draw (f) -- (e) -- (d) -- (f) node[midway, above]{};
			\draw (f) -- (c) -- (f) node[midway, above]{};
			
		\end{tikzpicture}
		\caption{Graph $H_2$}
		\label{fig:minor2}
	\end{subfigure}
	~
		\begin{subfigure}[t]{0.32\linewidth}
		\centering
		\begin{tikzpicture}[scale=1]
			
			\node at (-1, 1) [circle,draw, fill=black, opacity=1, color=black, inner sep=0.5mm] (a) {};
			
			\node at (-1.5,1) (x) {$x$};
			
			\node at (1, 1) [circle,draw, fill=black, opacity=1, color=black, inner sep=0.5mm] (b) {};
			
			\node at (1.5,1) (y) {$y$};
			
			\node at (0,0) [circle,draw, fill=black, opacity=1, color=black, inner sep=0.5mm] (c) {};
			
			\node at (-0.5,0) (v) {$v$};
			
			\node at (0,2) [circle,draw, fill=black, opacity=1, color=black, inner sep=0.5mm] (d) {};
			
			\node at (-0.5,2) (u) {$u$};

			\draw (a) -- (c) -- (b) -- (a) node[midway, above]{};
			\draw (a) -- (b) -- (d) -- (a) node[midway, above]{};
			
		\end{tikzpicture}
		\caption{Graph $H_3$}
		\label{fig:minor3}
	\end{subfigure}
\caption{}
\label{fig:minors} 
\end{figure}

Let $C_1$ and $C_2$ be two different cycles in $H$. 
We may assume that $C_1$ and $C_2$ are disjoint or share a path (possibly a single vertex).
Let $e_1$ be an edge in $C_1$ and not in $C_2$ and let $e_2$ be an edge in $C_2$ and not in $C_1$.
The edges $e_1$ and $e_2$ are depicted in red in Figure~\ref{fig:minors_rep}.

\begin{figure}[h]
\centering
	\begin{subfigure}[t]{0.32\linewidth}
		\centering
		\begin{tikzpicture}[scale=1]
			
			\node at (-1, 1.15) [circle,draw, fill=black, opacity=1, color=black, inner sep=0.5mm] (a) {};
			
			\node at (0,1.5) [color=red] (u) {$e_1$};
			
			\node at (1, 1.15) [circle,draw, fill=black, opacity=1, color=black, inner sep=0.5mm] (b) {};
			
			\node at (0,0) [circle,draw, fill=black, opacity=1, color=black, inner sep=0.5mm] (c) {};
			
			\node at (0,-1.5) [color=red] (x) {$e_2$};
			
			\node at (-1,-1.15) [circle,draw, fill=black, opacity=1, color=black, inner sep=0.5mm] (d) {};

			\node at (1,-1.15) [circle,draw, fill=black, opacity=1, color=black, inner sep=0.5mm] (e) {};
			
			\node at (-2, 1) [circle,draw, fill=black, opacity=1, color=black, inner sep=0.5mm] (f) {};
			
			\node at (-2.5, 1.5) [circle,draw, fill=black, opacity=1, color=black, inner sep=0.5mm] (g) {};
			
			\node at (-2.5, 0.5) [circle,draw, fill=black, opacity=1, color=black, inner sep=0.5mm] (h) {};
			
			\node at (1.8, 1.5) [circle,draw, fill=black, opacity=1, color=black, inner sep=0.5mm] (i) {};
			
			\node at (1.5, 0.5) [circle,draw, fill=black, opacity=1, color=black, inner sep=0.5mm] (j) {};
			
			\node at (-2, -1) [circle,draw, fill=black, opacity=1, color=black, inner sep=0.5mm] (k) {};
			
			\node at (2, -1) [circle,draw, fill=black, opacity=1, color=black, inner sep=0.5mm] (l) {};
			
			\node at (3, -1) [circle,draw, fill=black, opacity=1, color=black, inner sep=0.5mm] (m) {};

			\draw[red, thick] (a) -- (b) -- (a) node[midway, above]{};
			\draw[red, thick] (e) -- (d) -- (e) node[midway, above]{};
			\draw[green, thick] (a) -- (c) -- (b) node[midway, above]{};
			\draw[green, thick] (d) -- (c) -- (e) node[midway, above]{};
			\draw (a) -- (f) -- (g) -- (f) -- (h) node[midway, above]{};
			\draw (i) -- (b) -- (j) node[midway, above]{};
			\draw (d) -- (k) node[midway, above]{};
			\draw (e) -- (l) -- (m) node[midway, above]{};
			
		\end{tikzpicture}
	\end{subfigure}
\caption{Graph $H$}
\label{fig:minors_rep} 
\end{figure}

Let $H'$ be the graph $H$ with every edge subdivided four times; that is, each edge of
$H$ is replaced with five edges forming a path in $H'$.
Let $C'_1$ and $C'_2$ be the subgraphs of $H'$ arisen by subdividing the edges in $C_1 \cup C_2$,
and let $e'_1$ and $e'_2$  be the middle edges arisen by subdividing $e_1$ and $e_2$, respectively.
Let $X$ be the graph $H'$ with the edges $e'_1,e'_2$ deleted.
In particular, note that $X$ is an $H$-graph.
Let $G$ be a bipartite graph.
We construct $G'$ such that $G$ is 2-track if and only if $G$ has a $H$-model the same way as earlier, with only two exceptions:
\begin{itemize}
 \item instead of using star $S$ in the construction of $G'$ we use the graph $X$,
 \item we make each vertex $v \in V(G')$ corresponding to the vertex $v \in V(G)$ adjacent to 
 all vertices in $V(C'_1) \cup V(C'_2)$ and adjacent to no other vertex of $X$. 
\end{itemize}
The rest of the construction of $G'$ is the same.
Note that for every $u,v \in V(H)$ the corresponding vertices $u',v'\in V(X)$ are non-adjacent.
Suppose $(H^{\phi},\phi)$ is an $H$-model of~$X$.
Note that:
\begin{itemize}
 \item for every vertex $u \in V(X)$ with degree $\geq 3$ in $H$ the set $\phi(u)$ covers a unique vertex in $H^{\phi}$ of degree $\geq 3$, 
 whose degree is the same as degree of $u$.  
\end{itemize}
Also, observe that any $H$-model of $G'$ can be turned into an $H$-model $(H^{\phi},\phi)$ such that:
\begin{itemize}
\item for every vertex $v \in V(G')$ corresponding to vertex $v \in V(G)$ the set $\phi(v)$ is contained in $V(C^{\phi}_1 \cup C^{\phi}_2)$ and contains $V(e^{\phi})$ for every 
$e \in (C_1 \cup C_2) \setminus \{e_1,e_2\}$,
where $C^{\phi}_i$ is a cycle of $H^{\phi}$ arisen from the subdivision of $C_i$
and $e^\phi$ is a path of $H^{\phi}$ arisen from the subdivision of the edge $e \in E(H)$
(such vertices of $G'$ are represented by sets contained in the red and green part of $H$ -- see Figure \ref{fig:minors_rep}),
\item for every edge-vertex $w$ of $G'$ the set $\phi(w)$ is contained either in $V(e^{\phi}_1)$ or in $V(e^{\phi}_2)$, where 
$e^{\phi}_i$ is a path of $H^{\phi}$ arisen from the subdivision of the edge $e_i$
(such vertices of $G'$ are represented by the sets contained in red part of $H$ -- see Figure \ref{fig:minors_rep}).
\end{itemize}
With the above observations one can easily adapt the proof of Theorem~\ref{thm:8:graph} to complete the proof of Theorem~\ref{thm:NPC}.

\bibliographystyle{plain}
\bibliography{lit_short}

\end{document}